\documentstyle[preprint,aps,rmp,epsfig,psfig]{revtex}
\def\lsim{\mathrel{\rlap{\lower4pt\hbox{\hskip1pt$\sim$}}
    \raise1pt\hbox{$<$}}}         
\def\gsim{\mathrel{\rlap{\lower4pt\hbox{\hskip1pt$\sim$}}
    \raise1pt\hbox{$>$}}}         
\def\mathrm{\rm}
\def\textbf{\bf}
\def\textit{\it}
\tightenlines
\begin{document}
\preprint{\vbox{\hbox{CERN-TH/2002-021}      \hbox{WIS/08/02-Feb-DPP}
         \hbox{hep-ph/0202058}}}
\title{Neutrino  Masses and Mixing: Evidence and Implications}
\author{M. C. Gonzalez-Garcia}
\address{ Theory Division, CERN, CH-1211, Geneva 23, Switzerland\\
and \\ 
Instituto de F\'{\i}sica Corpuscular,
Universitat de  Val\`encia -- C.S.I.C\\
Edificio Institutos de Paterna, Apt 22085, 46071 Val\`encia, Spain \\
and\\
C.N. Yang Institute for Theoretical Physics\\
State University of New York at Stony Brook\\
Stony Brook,NY 11794-3840, USA}
\author{Yosef Nir}
\address{Department of Particle Physics, Weizmann Institute of Science\\
 Rehovot 76100, Israel}

\maketitle

\begin{abstract}
Measurements of various features of the fluxes of atmospheric and solar
neutrinos have provided evidence for neutrino oscillations and therefore for 
neutrino masses and mixing. We review the phenomenology of neutrino 
oscillations in vacuum and in matter. We present the existing evidence from 
solar and atmospheric neutrinos as well as the results from laboratory 
searches, including the final status of the LSND experiment. We describe the 
theoretical inputs that are used to interpret the experimental results in terms
of neutrino oscillations. We derive the allowed ranges for the mass and
mixing parameters in three frameworks: First, each set of observations is 
analyzed separately in a two-neutrino framework; Second, the data from solar 
and atmospheric neutrinos are analyzed in a three active neutrino framework;
Third, the LSND results are added, and the status of accommodating all three 
signals in the framework of three active and one sterile light neutrinos is 
presented.
We review the theoretical implications of these results: the existence of new 
physics, the estimate of the scale of this new physics and the lessons for 
grand unified theories, for supersymmetric models with $R$-parity violation, 
for models of extra dimensions and singlet fermions in the bulk, 
and for flavor models. 
\end{abstract}
\newpage
\tableofcontents
 
\section{INTRODUCTION}
\label{sec:introduc}
In 1930 Wolfgang Pauli postulated the existence of the neutrino in order 
to reconcile data on the radioactive decay of nuclei with energy conservation. 
In radioactive decays, nuclei of atoms mutate into different nuclei when 
neutrons are transformed into slightly
lighter protons with the emission of electrons:
\begin{equation}
\rm neutron \,\rightarrow\, proton\, + \, electron \, + \,antineutrino \,.
\end{equation}
Without the neutrino, energy conservation requires that the electron and
proton share the neutron's energy. Each electron is therefore produced
with a fixed energy while experiments indicated conclusively that the
electrons were not mono-energetic but observed with a range of energies.
This energy range corresponded exactly to the many ways the three
particles in the final state of the reaction above can share energy while
satisfying its conservation law. The postulated neutrino had no
electric charge and, for all practical purposes, does not interact with
matter; it just serves as an agent to balance energy and momentum in
above reaction. In fact, Pauli pointed out that for the neutrino to do
the job, it had to weigh less than one percent of the proton mass, thus
establishing the first limit on neutrino mass.

Observing neutrinos is straightforward --- in principle. Pauli had to
wait a quarter of a century before Fred Reines and Clyde Cowan Jr.
observed neutrinos produced by a nuclear reactor. In the presence of
protons, neutrinos occasionally initiate the inverse reaction of
radioactive decay:
\begin{equation}
\rm \bar\nu\, + \,p \,\rightarrow\, n \,+\, e^+ \,.
\end{equation}
Experimentally, one exposes a material rich in protons to a neutrino beam 
and simply looks for the coincident appearance of an electron and a
neutron. In the alternative possibility where the incident neutrino
carries muon flavour, a muon appears in the final state instead. 

By the 1960's neutrino beams passed from a futuristic dream to one of the
most important tools of particle physics. The technique contributed in
important ways to the discovery of quarks, the constituents of protons
and neutrons. The technique is conceptually simple, though
technologically very challenging. A very intense beam of accelerated
protons is shot into a {\it beam-dump} which typically consists of a
kilometer-long mound of earth or a 100 meter-long block of stainless
steel. Particle physics does the rest. Protons interact with nuclei in
the dump and produce tens of pions in each collision. Charged pions 
decay into a muon and a neutrino. The material in the dump will eventually 
absorb the muons, photons and any other charged particles, so that only 
neutrinos exit at the opposite end, forming an intense and controlled beam.

Neutrinos are also produced in natural sources. Starting at the 1960's,
neutrinos produced in the sun and in the atmosphere have been observed.
As we will see in this review, these observations play an important role
in understanding the detailed features of the neutrinos. In 1987, neutrinos
from a supernova in the Large Magellanic Cloud were also detected.

The properties of the neutrino and in particular the question of its mass have 
intrigued physicists' minds ever since it was proposed (Kayser,
Gibrat-Debu and Perrier, 1989; Ramond, 1999).  In the laboratory 
neutrino masses have been searched for in two types of experiments
(Boehm and Vogel, 1987): (i) direct 
kinematic searches of neutrino mass, of which the most sensitive is the study 
of tritium beta decay, and (ii) neutrinoless double-$\beta$ decay experiments. 
Experiments achieved higher and higher precision, reaching upper limits for the
electron-neutrino mass of $10^{-9}$ the proton mass, rather than the $10^{-2}$ 
originally obtained by Pauli. This raised the question of whether neutrinos 
are truly massless like photons. 

Can one go further below the eV scale (that is $10^{-9}$ the proton mass) in
the search for neutrino masses? This is a very difficult task in
direct measurements. In 1957, however, Bruno Pontecorvo realized that the 
existence of neutrino masses implies the possibility of neutrino oscillations. 
This phenomenon is similar to what happens in the quark sector, where neutral 
kaons oscillate. Flavor oscillations of neutrinos have been searched for using
either neutrino beams from reactors or accelerators, or natural neutrinos 
generated at astrophysical sources (the Sun giving the largest flux) or in the 
atmosphere (as the byproducts of cosmic ray collisions). The longer
the distance that the neutrinos travel from their
production point to the detector, the smaller masses that can be signaled
by their oscillation. Indeed, the solar neutrinos allow us to search for
masses that are as small as $10^{-5}$ eV, that is $10^{-14}$ of the 
proton mass!

In fact, in recent years, experiments studying natural neutrino fluxes have 
provided us with the strongest evidence of neutrino masses and mixing. 
Experiments that measure the flux of atmospheric neutrinos have found results 
that suggest the disappearance of muon-neutrinos when propagating over 
distances of order hundreds (or more) kilometers. Experiments that 
measure the flux of solar neutrinos (Bahcall, 1989) found results that suggest the 
disappearance of electron-neutrinos while propagating within the Sun or between
the Sun and the Earth. The disappearance of both atmospheric $\nu_\mu$'s and 
solar $\nu_e$'s is most easily explained in terms of neutrino oscillations. As 
concerns experiments performed with laboratory beams, most have given no 
evidence of oscillations. One exception is the LSND experiment which has 
observed the appearance of electron anti-neutrinos in a muon anti-neutrino 
beam. This signal has not been confirmed so far by any other experiment. 

What can we learn from measurements of neutrino masses about our theories
of particle physics? The Standard Model (SM) of particle physics is a
mathematical description of the strong, weak and electromagnetic interactions.
Since it was conceived in the 1960's by Glashow, Salam and Weinberg, it has
successfully passed numerous experimental tests. In the absence of any direct 
evidence for their mass, neutrinos were introduced in the SM as truly massless 
fermions for which no gauge invariant renormalizable mass term can be 
constructed. Consequently, in the SM there is neither mixing nor CP violation 
in the lepton sector. Therefore, experimental evidence for neutrino masses or 
mixing or leptonic CP violation provides an unambiguous signal for 
{\it new physics} (NP).

The SM prediction of massless neutrinos is an accidental fact: unlike photons, 
no profound principle protects them from having a mass. On the contrary, modern
elementary particle theories anticipate ways in which they have small,
but definitely non-vanishing masses. In fact, there are good theoretical 
reasons to expect that neutrinos are massive but much lighter
than all the charged fermions of the SM. Specifically, it is
very likely that neutrino masses are inversely proportional to the scale 
of NP. Consequently, if neutrino masses are measured, we can
estimate the relevant {\it new scale}.

All dimensionless flavor-blind parameters of the SM, that is,
the three gauge couplings and the quartic Higgs coupling, are of order one.
In contrast, most of the flavor parameters $-$ quark and charged lepton masses 
(except for the mass of the top quark), and the three CKM mixing angles $-$ are
small and hierarchical. This situation constitutes {\it the flavor puzzle}: 
Within the SM, the hierarchical structure can be accommodated but 
is not explained. If neutrinos have Majorana (Dirac) masses, then
there are nine (seven) flavor parameters beyond the thirteen of the
SM. If some of these extra parameters are measured, we will be able to
test and refine theories that try to solve the flavor puzzle. 

In {\it grand unified theories} (GUTs), lepton masses are often related 
to quark masses. Measurements of neutrino parameters provide further 
tests of these theories.

The values of neutrino parameters that explain the anomalies observed in
atmospheric and solar neutrino fluxes can be used
to address the many theoretical questions described above: they imply NP,
they suggest an energy scale at which this NP takes place, they
provide stringent tests of flavor models, and they probe GUTs.

In this review we first present the low energy formalism for adding
neutrino masses to the SM and the induced leptonic mixing  
(Sec.~\ref{sec:standard}) and then describe the phenomenology associated 
with neutrino oscillations in vacuum and in matter (Sec.~\ref{sec:oscila}).
In Secs.~\ref{sec:solar} and~\ref{sec:atmos} we discuss the 
evidence from, respectively, solar and atmospheric neutrinos.
We review the theoretical modeling that is involved in interpreting the
experimental results in terms of neutrino oscillations. We
briefly describe the techniques used in the derivation of the
allowed ranges for the neutrino flavor parameters which, in these sections,
is performed in the framework of mixing between two neutrinos. 
Sec.~\ref{sec:lab} is devoted to the results
from searches at laboratory experiments including the final status 
of the LSND experiment. The two more robust pieces of evidence, from
solar and atmospheric neutrinos, can be
accommodated assuming masses and mixing of three standard neutrinos:
in Sec.~\ref{subsec:threemix} we derive the allowed ranges 
of parameters in this case. The present phenomenological 
status of the possibility of mixing between the three standard neutrinos
with a light sterile one, needed to accommodate also the LSND result, 
is discussed in Sec.~\ref{subsec:fourmix}.
In Secs.~\ref{sec:impli1} and~\ref{sec:impli2} we explain the various 
lessons for theory which can be drawn on the basis of these interpretations,
focusing on models that explain the atmospheric and solar neutrino data 
through mixing among three active neutrinos.  In particular, we discuss 
flavor models, GUTs, supersymmetric models with 
$R$-parity violation, and models of extra dimensions.
Our conclusions are summarized in Sec.~\ref{sec:conc}.

\section{The Standard Model and Neutrino Masses}
\label{sec:standard}
One of the most beautiful aspects of modern theories of particles physics
is the relation between forces mediated by spin-1 particles and local
(gauge) symmetries. Within the Standard Model, the strong, weak and 
electromagnetic interactions are related to, respectively, $SU(3)$,
$SU(2)$ and $U(1)$ gauge groups. Many features of the various 
interactions are then explained by the symmetry to which they are related. 
In particular, the way that the various
fermions are affected by the different types of interactions is determined
by their representations (or simply their charges in the case of Abelian gauge
symmetries) under the corresponding symmetry groups.

Neutrinos are fermions that have neither strong nor electromagnetic 
interactions. In group theory language, they are singlets of
$SU(3)_{\rm C}\times U(1)_{\rm EM}$. Active neutrinos have weak
interactions, that is, they are not singlets of $SU(2)_{\rm L}$.
Sterile neutrinos have none of the SM gauge interactions
and they are singlets of the SM gauge group.

The SM has three active neutrinos. They reside in lepton doublets,
\begin{equation}
L_\ell=\pmatrix{\nu_{L\ell}\cr \ell_L^-\cr},\ \ \ \ell=e,\mu,\tau.
\label{SMdoublets}  
\end{equation}
Here $e$, $\mu$ and $\tau$ are the charged lepton mass eigenstates. The three 
neutrino interaction eigenstates, $\nu_e$, $\nu_\mu$ and $\nu_\tau$, are 
defined as the $SU(2)_{\rm L}$-partners of these mass eigenstates. In other 
words, the charged current interaction terms for leptons read
\begin{equation}
-{\cal L}_{\rm CC}={g\over\sqrt{2}}\sum_\ell\overline{\nu_{L\ell}}\gamma^\mu
\ell^-_L W_\mu^++{\rm h.c.}.
\label{CCleptons}  
\end{equation} 
In addition, the SM neutrinos have neutral current (NC) interactions,
\begin{equation}
-{\cal L}_{\rm NC}={g\over2\cos\theta_W}\sum_\ell\overline{\nu_{L\ell}}
\gamma^\mu\nu_{L\ell} Z_\mu^0.
\label{NCneut}  
\end{equation}
Eqs.~(\ref{CCleptons}) and (\ref{NCneut}) give all the neutrino interactions
within the SM.
 
The measurement of the decay width of the $Z^0$ boson into neutrinos makes
the existence of three, and only three, light (that is, $m_\nu\lsim m_Z/2$) 
active neutrinos an experimental fact. When expressed in units of the
SM prediction for a single neutrino generation, one gets
(Groom~{\it et al,}, 2000)
\begin{eqnarray}
N_\nu&=&2.994\pm0.012\ \ ({\rm Standard\ Model\ fits\ to\ LEP\ data}),
\nonumber\\
N_\nu&=&3.00\pm0.06\ \ ({\rm Direct\ measurement\ of\ invisible\ Z\ width}).
\label{GAMinv}  
\end{eqnarray}

\subsection{The Standard Model Implies $m_\nu=0$}
\label{sec:stmo}
The SM is based on the gauge group
\begin{equation}
G_{\rm SM}=SU(3)_{\rm C}\times SU(2)_{\rm L}\times U(1)_{\rm Y},
\label{SMgagr}  
\end{equation}
with three fermion generations, where a single generation consists of five
different representations of the gauge group,
\begin{equation}
Q_L(3,2)_{+1/6}\;, \;U_R(3,1)_{+2/3}\;,\;D_R(3,1)_{-1/3}\;,
\;L_L(1,2)_{-1/2}\;,\;E_R(1,1)_{-1}.
\label{GENrep}  
\end{equation}
Our notation here means that, for example, a left-handed lepton
field $L_L$ is a singlet (1) of the $SU(3)_{\rm C}$ group, a doublet
(2) of the $SU(2)_{\rm L}$  group, and carries hypercharge $-1/2$
under the $U(1)_{\rm Y}$ group.

The vacuum expectations value (VEV) of the single Higgs doublet
$\phi(1,2)_{+1/2}$ breaks the symmetry,
\begin{equation}
\langle\phi\rangle=\pmatrix{0\cr {v\over\sqrt2}\cr}\ \ \Longrightarrow\ \ 
G_{\rm SM}\rightarrow SU(3)_{\rm C}\times U(1)_{\rm EM}.
\label{spsybr}  
\end{equation}

Since the fermions reside in chiral representations of the gauge group, there 
can be no bare mass terms. Fermions masses arise from the Yukawa interactions,
\begin{equation}
-{\cal L}_{\rm Yukawa}=Y^d_{ij}\overline{Q_{Li}}\phi D_{Rj}+
Y^u_{ij}\overline{Q_{Li}}\tilde\phi U_{Rj}+
Y^\ell_{ij}\overline{L_{Li}}\phi E_{Rj}+{\rm h.c.},
\label{yukawa}  
\end{equation}
(where $\tilde\phi=i\tau_2\phi^\star$) after spontaneous symmetry breaking. The
Yukawa interactions of Eq.~(\ref{yukawa}) lead to charged fermion masses but 
leave the neutrinos massless. One could think that neutrino masses would arise 
from loop corrections if these corrections induced effective terms
\begin{equation}
{Y^\nu_{ij}\over v}\phi\phi L_{Li}L_{Lj}.
\label{radcor}  
\end{equation}
This however cannot happen, as can be easily understood by examining the
accidental symmetries of the Standard Model. (It often happens that, as a 
consequence of the symmetries that define a model and of its particle content,
all renormalizable Lagrangian terms obey additional symmetries, that are
not a priori imposed on the model. These are called accidental symmetries.)
Within the SM, with the gauge symmetry of Eq.~(\ref{SMgagr}) and the particle
content of Eq.~(\ref{GENrep}), the following accidental global
symmetry arises (at the perturbative level): 
\begin{equation}
G_{\rm SM}^{\rm global}=U(1)_B\times U(1)_e\times U(1)_\mu\times U(1)_\tau.
\label{SMglob}  
\end{equation}
Here $U(1)_B$ is the baryon number symmetry, and $U(1)_{e,\mu,\tau}$
are the three lepton flavor symmetries, with total lepton number given by
$L=L_e+L_\mu+L_\tau$.
Terms of the form (\ref{radcor}) violate $G_{\rm SM}^{\rm global}$ and 
therefore cannot be induced by loop corrections. Furthermore, the
$U(1)_{B-L}$ subgroup of $G_{\rm SM}^{\rm global}$ is non-anomalous.
Terms of the form (\ref{radcor}) have $B-L=-2$ and therefore cannot be
induced even by nonperturbative corrections.

It follows that the SM predicts that neutrinos are precisely
massless. Consequently, there is neither mixing nor CP violation in the
leptonic sector.

\subsection{Extensions of the Standard Model Allow $m_\nu\neq0$}
There are many good reasons to think that the SM is not a
complete picture of Nature. For example, the fine-tuning problem of the
Higgs mass can be solved by Supersymmetry; gauge coupling unification
and the variety of gauge representations may find an explanation in
GUTs; baryogenesis can be initiated by decays of
heavy singlet fermions (leptogenesis); and the existence of gravity suggests 
that string theories are relevant to Nature. If any of these (or many other
proposed) extensions is indeed realized in Nature, the SM must be thought of as 
an effective low energy theory. That means that it is a valid approximation
up to the scale $\Lambda_{\rm NP}$ which characterizes the NP.

By thinking of the SM as an effective low energy theory, we
still retain the gauge group (\ref{SMgagr}), the fermionic spectrum
(\ref{GENrep}), and the pattern of spontaneous symmetry breaking (\ref{spsybr})
as valid ingredients to describe Nature at energies $E\ll\Lambda_{\rm NP}$. 
The SM predictions are, however, modified by small effects that
are proportional to powers of $E/\Lambda_{\rm NP}$. In other words, the
difference between the SM as a complete description of Nature
and as a low energy effective theory is that in the latter case we must
consider also nonrenormalizable terms.

There is no reason for generic NP to respect the accidental symmetries
of the SM (\ref{SMglob}). Indeed, there is a single set of 
dimension-five terms that is made of SM fields and is consistent 
with the gauge symmetry, and this set violates (\ref{SMglob}). It is given by
\begin{equation}
{Z^\nu_{ij}\over \Lambda_{\rm NP}}\phi\phi L_{Li}L_{Lj}.
\label{dimfiv}  
\end{equation}
(While these terms have the same form as (\ref{radcor}) we are thinking now 
not about SM radiative corrections as their source
but about some heavy fields, related to NP, which can induce
such terms by tree or loop diagrams.) In particular, (\ref{dimfiv}) violates 
$L$ (and $B-L$) by two units and leads, upon spontaneous symmetry breaking,
to neutrino masses:
\begin{equation}
(M_\nu)_{ij}={Z^\nu_{ij}\over2}{v^2\over\Lambda_{\rm NP}}.
\label{nrmass}  
\end{equation}
A few comments are in order, regarding Eq.~(\ref{nrmass}):

1) Since Eq.~(\ref{nrmass}) would arise in a generic extension of the
SM, we learn that neutrino masses are very likely to appear
if there is NP.

2) If neutrino masses arise effectively from nonrenormalizable terms, we
gain an understanding not only for the existence of neutrino masses but
also for their smallness. The scale of neutrino masses is suppressed,
compared to the scale of charged fermion masses, by $v/\Lambda_{\rm NP}$.

3) The terms (\ref{nrmass}) break not only total lepton number but also
the lepton flavor symmetry $U(1)_e\times U(1)_\mu\times U(1)_\tau$.
Therefore we should expect lepton mixing and CP violation.

The best known scenario that leads to (\ref{dimfiv}) is the {\it see-saw
mechanism} (Ramond, 1979; Gell-Mann {\it el al.}, 1979; Yanagida, 1979). Here one
assumes the existence of heavy sterile neutrinos $N_i$. Such fermions
have, in general, bare mass terms and Yukawa interactions:
\begin{equation}
-{\cal L}_N=\frac{1}{2} {M_N}_{ij}\overline{N^c_i}N_j+
Y^\nu_{ij}\overline{L_{Li}}\tilde\phi N_j +{\rm h.c.}.
\label{sinint}  
\end{equation}
The resulting mass matrix (see Sec.~\ref{diracmaj} for details)
in the basis $\left(\begin{array}{c}\nu_{Li}\\ N_j\end{array}\right)$  
has the following form:
\begin{equation}
M_\nu=\pmatrix{0&Y^\nu{v\over\sqrt2}\cr (Y^\nu)^T{v\over\sqrt2}&M_N\cr}.
\label{fumama}  
\end{equation}
If the eigenvalues of $M_N$ are all well above the electroweak breaking scale
$v$, then the diagonalization of $M_\nu$ leads to three light mass eigenstates
with a mass matrix of the form (\ref{nrmass}). In particular, the scale
$\Lambda_{\rm NP}$ is identified with the mass scale of the heavy sterile
neutrinos, that is the typical scale of the eigenvalues of $M_N$.

Two well-known examples of extensions of the SM that lead to
a see-saw mechanism for neutrino masses are SO(10) GUTs
(Ramond, 1979; Gell-Mann {\it et al.}, 1979; Yanagida, 1979)
and left-right symmetry (Mohapatra and Senjanovic, 1980).

\subsection{Dirac and Majorana Neutrino Mass Terms}
\label{diracmaj}
If the only modification that we make to the SM is to assume
that it is a low energy effective theory, that is, allowing for
non-renormalizable terms that are consistent with the gauge symmetry and
the fermionic content of the SM, then the only way that neutrinos
can gain masses is through terms of the form (\ref{dimfiv}). These are
Majorana mass terms which, in particular, violate lepton number by two units.

One can, however, open up other possibilities by adding new fields.
The most relevant extension is that where an arbitrary number $m$ of sterile 
neutrinos $\nu_{si}(1,1)_0$ is added to the three standard generations of 
Eq.~(\ref{GENrep}). Now there are, in general, two types of mass terms that
arise from {\it renormalizable} terms:
\begin{equation}
-L_{M_\nu}={M_D}_{ij}\overline{\nu_{Li}} \nu_{sj} +
\frac{1}{2} {M_N}_{ij}\overline{\nu^c_{si}}\nu_{sj}+{\rm h.c.}.
\label{massnu1}
\end{equation}
Here $\nu^c$ indicates a charge conjugated field, $\nu^c=C\overline{\nu}^T$ 
and $C$ is the charge conjugation matrix. 

The first term is a Dirac mass term. It has the following properties:

(i) Since it transforms as the doublet representation of $SU(2)_{\rm L}$, it 
is generated after spontaneous electroweak symmetry breaking from the Yukawa 
interactions $Y^\nu_{ij}\overline{L_{Li}}\tilde\phi\nu_{sj}$, similarly to the 
charged fermion masses discussed in Sec.~\ref{sec:stmo}. 

(ii) Since it has a neutrino field and an antineutrino field, it conserves
total lepton number [though it breaks the lepton flavor number symmetries
of Eq.~(\ref{SMglob})].

(iii) $M_D$ is a complex $3\times m$ matrix.

The second term in Eq.~(\ref{massnu1}) is a Majorana mass term. It is different
from the Dirac mass terms in many important aspects:

(i) It is a singlet of the SM gauge group. Therefore, it can appear as
a bare mass term. [Had we written a similar term for the active neutrinos,
it would transform as a triplet of $SU(2)_{\rm L}$. In the absence of
a Higgs
triplet, it cannot be generated by renormalizable Yukawa interactions. 
Such terms are generated for active neutrinos from the non-renormalizable
Yukawa interactions of Eq.~(\ref{dimfiv}).]

(ii) Since it involves two neutrino fields, it breaks lepton number
conservation by two units. More generally, such a term is allowed only
if the neutrinos carry no additive conserved charge. This is the reason
that such terms are not allowed for any charged fermions which, by definition,
carry $U(1)_{\rm EM}$ charges.

(iii) $M_N$ is a symmetric matrix (as follows from simple Dirac algebra)
of dimension $m\times m$.

It is convenient to define a $(3+m)$-dimensional neutrino vector $\vec\nu$,
\begin{equation}
\vec \nu=\pmatrix{\nu_{Li}\cr \nu_{sj}\cr}.
\end{equation}
That allows us to rewrite Eq.~(\ref{massnu1}) in a unified way:
\begin{equation}
-L_{M_\nu}=\frac{1}{2}\overline{\vec\nu^c} M_\nu \vec\nu +{\rm h.c.}\; ,
\end{equation}
where 
\begin{equation}
M_\nu=\pmatrix{0&M_D\cr {M_D}^T&M_N\cr}. 
\end{equation}
The matrix $M_\nu$ is complex and symmetric. It can be diagonalized by a 
unitary matrix of dimension $(3+m)$. The resulting mass eigenstates, $\nu_k$, 
obey the Majorana condition, $\nu_k^c=\nu_k$. 

There are three interesting cases, differing in the hierarchy of scales
between $M_N$ and $M_D$:

(i) One can assume that the scale of the mass eigenvalues of $M_N$ is much
higher than the scale of electroweak symmetry breaking $\langle\phi\rangle$.
This is the natural situation in various extensions of the SM
that are characterized by a high energy scale. We are back in the framework of 
the see-saw mechanism discussed in the previous section. If we simply integrate
out the sterile neutrinos, we get a low energy effective theory with three
light, active neutrinos of the Majorana-type.

(ii) One can assume that the scale of some eigenvalues of $M_N$ is not higher
than the electroweak scale. Now the SM is not even a good low energy effective 
theory: there are more than three light neutrinos, and they are mixtures of 
doublet and singlet fields. These light fields are all of the Majorana-type.

(iii) One can assume that $M_N=0$. This is equivalent to imposing lepton number
symmetry on this model. Again, the SM is not even a good low energy theory: 
both the fermionic content and the assumed symmetries are different. 
(Recall that within the SM lepton number is an accidental symmetry.)
Now only the first term in Eq.~(\ref{massnu1}) is allowed, which is a Dirac 
mass term. It is generated by the Higgs mechanism in the same way that
charged fermions masses are generated. If indeed it is the only neutrino 
mass term present and $m=3$,  we can identify the three sterile neutrinos with 
the right handed component of a four-component spinor neutrino field 
(actually with its charge conjugate). In this way, the six massive Majorana 
neutrinos combine to form three massive neutrino Dirac states, equivalently 
to the charged fermions. In this particular case the $6\times6$ diagonalizing 
matrix is block diagonal and it can be written in terms of a $3\times3$ 
unitary matrix. 

From the phenomenological point of view, it will make little difference
for our purposes whether the light neutrinos are of the Majorana- or
Dirac-type. In particular, the analysis of neutrino oscillations is the
same in both cases. Only in the discussion of neutrinoless double beta decay
will the question of Majorana versus Dirac neutrinos be crucial.

From the theoretical model building point of view, however, the two cases
are very different. In particular, the see-saw mechanism provides a natural 
explanation for the lightness of Majorana neutrinos, while for Dirac 
neutrinos there is no such generic mechanism. (The situation can be
different in models of extra dimensions; see Section \ref{subsec:extdim}.)

\subsection{Lepton Mixing}
\label{subsec:lepmix}
We briefly review here our notation for lepton mixing.
We denote the neutrino mass eigenstates by $(\nu_1,\nu_2,\nu_3,\dots,\nu_n)$ 
where $n=3+m$, and the charged lepton mass eigenstates by $(e,\mu,\tau)$. The 
corresponding interaction eigenstates are denoted by  $(e^I,\mu^I,\tau^I)$ and 
$\vec\nu=(\nu_{Le},\nu_{L\mu},\nu_{L\tau},\nu_{s1},\dots,\nu_{sm})$.
In the mass basis, leptonic charged current interactions are given by
\begin{equation}
-{\cal L}_{\rm CC}={g\over\sqrt{2}}(\overline{e_L}\ \overline{\mu_L}\ 
\overline{\tau_L})\gamma^\mu U\pmatrix{\nu_1\cr\nu_2\cr\nu_3\cr 
.\cr .\cr \nu_n\cr} W_\mu^+-{\rm h.c.}.
\label{CClepmas}  
\end{equation} 
Here $U$ is a $3\times n$ matrix (Schechter and Valle; 1980a, 1980b). 

Given the charged lepton mass matrix $M_\ell$ and the neutrino mass matrix
$M_\nu$ in some interaction basis,
\begin{equation}
-{\cal L}_{M}=(\overline{e_L^I}\ \overline{\mu_L^I}\ 
\overline{\tau_L^I})\ M_\ell \pmatrix{e_R^I\cr\mu_R^I\cr\tau_R^I\cr}
+ \frac{1}{2}\overline{\vec\nu^c} M_\nu \vec\nu +{\rm h.c.}\; ,
\end{equation}
we can find the diagonalizing matrices $V^\ell$ and $V^\nu$:
\begin{equation}
{V^\ell}^\dagger M_\ell M_\ell^\dagger V^\ell=
{\rm diag}(m_e^2,m_\mu^2,m_\tau^2),\ \ \ 
{V^\nu}^\dagger M_\nu^\dagger M_\nu V^\nu={\rm diag}(m_1^2,m_2^2,m_3^2,\dots,
m_n^2). 
\label{findia}  
\end{equation} 
Here $V^\ell$ is a unitary $3\times3$ matrix while $V^\nu$ is a unitary 
$n\times n$ matrix. The $3\times n$ mixing matrix $U$ can be found from these 
diagonalizing matrices:
\begin{equation}
U_{ij}=P_{\ell,ii}\, {V^\ell_{ik}}^\dagger \, V^\nu_{kj}\, (P_{\nu,jj}).
\label{diamat}  
\end{equation}
A few comments are in order:

(i) Note that the indices $i$ and $k$ run from 1 to 3, while $j$ runs
from 1 to $n$. In particular, only the first three lines of $V^\nu$ play
a role in Eq.~(\ref{diamat}).

(ii) $P_\ell$ is a diagonal $3\times3$ phase matrix, that is conventionally
used to reduce by three the number of phases in $U$.

(iii) $P_\nu$ is a diagonal matrix with additional arbitrary phases (chosen
to reduce the number of phases in $U$) only for Dirac states. For Majorana
neutrinos, this matrix is simply a unit matrix. The reason for that is that if 
one rotates a Majorana neutrino by a phase, this phase will appear in its
mass term which will no longer be real.

We conclude that the number of phases that can be absorbed by redefining the
mass eigenstates depends on whether the neutrinos are Dirac or Majorana 
particles. In particular, if there are only three Majorana neutrinos, 
$U$ is a $3\times 3$ matrix analogous to the CKM matrix for the
quarks (Maki, Nakagawa and Sakata, 1962; Kobayashi and Maskawa, 1973)
but due to the Majorana nature of the neutrinos it depends on 
six independent parameters: three mixing angles and three phases.
The two
Majorana phases do not affect neutrino oscillations
(Bilenky, Hosek and Petcov, 1980; Langacker, Petcov, Steigman and 
Toshev, 1987). 
[CP conservation implies that the three lepton phases are either zero
or $\pi$ (Schechter and Valle, 1981; Wolfenstein 1981).]  This is to be 
compared to the case of three Dirac neutrinos, where the number of
physical phases is one, similarly to the CKM matrix. Note, however,
that the two extra Majorana phases affect only lepton number violating
processes. Such effects are suppressed by $m_\nu/E$ and are very hard
to measure.

If no new interactions for the charged leptons are present 
we can identify their interaction eigenstates with 
the corresponding mass eigenstates after phase redefinitions.
In this case the charged current lepton mixing matrix $U$ is simply given 
by a $3\times n$ sub-matrix of the unitary matrix $V^\nu$.

\section{Neutrino Oscillations}
\label{sec:oscila}
\subsection{Neutrino Oscillations in Vacuum}
\label{subsec:osvac}
Neutrino oscillations in vacuum would arise if neutrinos were massive and 
mixed. In other words, the neutrino state that is produced by electroweak 
interactions is not a mass eigenstate. This phenomenon was first pointed
out by Pontecorvo in 1957 (Pontecorvo, 1957) while the possibility of
arbitrary mixing between two massive neutrino states was first 
introduced in Maki, Nakagawa and Sakata (1962).  

From Eq.~(\ref{CClepmas}) we see that if neutrinos have masses, the weak 
eigenstates, $\nu_\alpha$, produced in a 
weak interaction ({\it i.e.}, an inverse beta reaction or a weak decay)
are, in general, linear combinations of the mass eigenstates $\nu_i$
\begin{equation}
|\nu_\alpha\rangle =\sum_{i=1}^{n} U^*_{\alpha i} |\nu_i\rangle \;\, 
\end{equation}
where $n$ is the number of light neutrino species and $U$ is the 
the mixing matrix in Eq.(\ref{diamat}). 
(Implicit in our definition of the state $|\nu\rangle$ is its energy-momentum 
and space-time dependence). After traveling a distance $L$ (or, equivalently
for relativistic neutrinos, time $t$), a neutrino originally produced with a 
flavor $\alpha$ evolves as follows: 
\begin{equation}
{|\nu_\alpha (t)\rangle}{=\sum_{i=1}^{n}} {U^*_{\alpha i}}{|\nu_i(t)\rangle} \; .
\end{equation} 
It can be detected in the charged-current (CC) interaction $\nu_\alpha(t)
N^\prime\to\ell_\beta N$ with a probability
\begin{equation}\label{palbe}
{ P_{\alpha\beta}}=|{ \langle\nu_\beta|\nu_\alpha(t)\rangle}|^2=
|\sum_{i=1}^n \sum_{j=1}^n { U^*_{\alpha i}U_{\beta j} }
{\langle\nu_j(0)|\nu_i(t)\rangle}|^2\; ,
\end{equation}
where $E_i$ and $m_i$ are, respectively, the energy and the mass of the
neutrino mass eigenstate $\nu_i$. 

We use the standard approximation that $|\nu\rangle$ is a plane wave (for a 
pedagogical discussion of the possible quantum mechanical problems in this 
naive description of neutrino oscillations we refer the reader to 
Lipkin, 1999, 
and Kim and Pevsner, 1993), $|\nu_i(t)\rangle=e^{-i \,E_i t}|\nu_i(0)\rangle$.
In all cases of interest to us, the neutrinos are relativistic:  
\begin{equation}
{ E_i}=\sqrt{{p_i^2}+{ m_i^2}}\simeq 
{ p_i}+\frac{{m_i^2}}{2{E_i}} \; .
\end{equation}
Furthermore, we can assume that $p_i\simeq p_j\equiv p\simeq E$. Then we obtain
the following transition probability (we only include here the CP 
conserving piece):
\begin{equation}
{ P_{\alpha\beta}} ={\delta_{\alpha\beta}-4\sum_{i=1}^{n-1}\sum_{j=i+1}^n}
{ \mbox{Re}[U_{\alpha i}U^*_{\beta i} U^*_{\alpha j} U_{\beta j}]} 
\sin^2x_{ij} \; ,  
\label{pab}
\end{equation}
where
\begin{equation}
\label{defddx}
x_{ij}\equiv\Delta_{ij}L/2,\ \ \ 
\Delta_{ij}\equiv \Delta m^2_{ij}/(2E),\ \ \
\Delta m^2_{ij}\equiv m_i^2-m_j^2,
\end{equation} 
and $L=t$ is the distance between the source (that is, the production point of 
$\nu_\alpha$) and the detector (that is, the detection point of $\nu_\beta$).
In deriving Eq.~(\ref{pab}) we used
the orthogonality relation $\langle\nu_j(0)|\nu_i(0)\rangle=\delta_{ij}$.
It is convenient to use the following units:
\begin{equation}
x_{ij}=1.27 \frac{\Delta m^2_{ij}}{\rm eV^2} \frac{L/E}{m/{\rm MeV}}\; .
\label{deltaij}
\end{equation}
The transition probability  [Eq.~(\ref{pab})] has an oscillatory behavior, with
oscillation length  
\begin{equation}
L_{0,ij}^{\rm osc}=\frac{4 \pi E}{\Delta m_{ij}^2}
\label{L0}
\end{equation}
and amplitude that is proportional to elements in the mixing matrix. 
Thus, in order to have oscillations, neutrinos must have different masses 
($\Delta m^2_{ij}\neq0$) and they must mix ($U_{\alpha_i}U_{\beta i}\neq0$). 

An experiment is characterized by the typical neutrino energy $E$ and by the
source-detector distance $L$. In order to be sensitive to a given value of 
$\Delta m^2_{ij}$, the experiment has to be set up with $E/L\approx 
\Delta m^2_{ij}$ ($L\sim L_{0,ij}^{\rm osc}$). The typical values of $L/E$ for 
different types of neutrino sources and experiments are summarized in 
Table \ref{tab:lovere}.

If  $(E/L)\gg \Delta m^2_{ij}$ ($L\ll L_{0,ij}^{\rm osc}$), the oscillation 
does not have time to give an appreciable effect because $\sin^2x_{ij}\ll1$.
The case of $(E/L)\ll\Delta m^2_{ij}$ ($L\gg L_{0,ij}^{\rm osc}$) requires
more careful consideration. One must take into account that, in general, 
neutrino beams are not monochromatic. Thus, rather than measuring 
$P_{\alpha \beta}$, the experiments are sensitive to the average probability 
\begin{equation}
\begin{array}{ll}
\langle P_{\alpha \beta}\rangle &= \frac{\displaystyle \int dE_\nu  \frac{d\Phi}{d E_\nu} 
\sigma_{CC}(E_\nu) P_{\alpha \beta} (E_\nu) \epsilon(E_\nu)}
{\displaystyle \int dE_\nu  \frac{d\Phi}{d E_\nu} 
\sigma_{CC}(E_\nu)  \epsilon(E_\nu)} \\ 
& = {\displaystyle \delta_{\alpha\beta}-4\sum_{i=1}^{n-1}\sum_{j=i+1}^n
\mbox{Re}[ U_{\alpha i}U^*_{\beta i} U^*_{\alpha j} U_{\beta j}] 
\langle\sin^2x_{ij}\rangle },
\end{array}
\end{equation}
where $\Phi$ is the neutrino energy spectrum, $\sigma_{CC}$ is the cross
section for the process in which the neutrino is detected (in general, a CC
interaction), and  $\epsilon(E_\nu)$ is the detection efficiency. For 
$L\gg L_{0,ij}^{\rm osc}$, the oscillating phase goes through many cycles 
before the detection and is averaged to $\langle \sin^2x_{ij}\rangle=1/2$.

For a two-neutrino case, the mixing matrix depends on a single parameter,
\begin{equation} 
U=\left(\begin{array}{cc} \cos\theta & \sin\theta\\ 
-\sin\theta & \cos\theta \end{array} \right) \; ,
\label{mixU2}
\end{equation}
and there is a single mass-squared difference $\Delta m^2$.
Then $P_{\alpha\beta}$ of Eq.~(\ref{pab}) takes the well known form  
\begin{equation}
P_{\alpha\beta}=\delta_{\alpha\beta}- (2\delta_{\alpha\beta}-1) \sin^22\theta 
\sin^2x \;.
\label{ptwo}
\end{equation} 
The physical parameter space is covered with $\Delta m^2\geq 0$ 
and $0\leq\theta\leq\frac{\pi}{2}$ (or, alternatively,
$0\leq\theta\leq\frac{\pi}{4}$ and either sign for $\Delta m^2$).
 
Changing the sign of the mass difference, $\Delta m^2\to-\Delta m^2$, and
changing the octant of the mixing angle, $\theta\to\frac{\pi}{2}-\theta$,
amounts to redefining the mass eigenstates, $\nu_1\leftrightarrow\nu_2$:
$P_{\alpha\beta}$ must be invariant under such transformation. 
Eq.~({\ref{ptwo}) reveals, however, that $P_{\alpha\beta}$ is actually
invariant under each of these transformations separately. This situation 
implies that there is a two-fold discrete ambiguity in the interpretation
of $P_{\alpha\beta}$ in terms of two-neutrino mixing: the two different sets of
physical parameters, ($\Delta m^2, \theta$) and ($\Delta m^2, \frac{\pi}{2}
-\theta$), give the same transition probability in vacuum. One cannot tell from
a measurement of, say, $P_{e\mu}$ in vacuum whether the larger component of 
$\nu_e$ resides in the heavier or in the lighter neutrino mass eigenstate.

Neutrino oscillation experiments measure $P_{\alpha\beta}$. It is common
practice for the experiments to interpret their results in the two-neutrino
framework. In other words, the constraints on $P_{\alpha\beta}$ are translated
into allowed or excluded regions in the plane ($\Delta m^2,\; \sin^22\theta$)
by using Eq.~(\ref{ptwo}). An example is given in Fig.~\ref{fig:example}.
We now explain some of the typical features of these constraints.

When an experiment is taking data at fixed $\langle L\rangle$ and 
${\langle E \rangle}$, as is the case for most laboratory searches,
its result can always be accounted for by $\Delta m^2$ that is large enough
to be in the region of averaged oscillations, $\langle\sin^2x_{ij}\rangle=1/2$.
Consequently, no upper bound on $\Delta m^2$ can be achieved 
by such experiment.
So, for negative searches that set an upper bound on the oscillation 
probability, $\langle P_{\alpha\beta}\rangle\leq P_L$, the excluded region lies
always on the upper-right side of the $\left(\Delta m^2,\sin^22\theta\right)$
plane, limited by the following asymptotic lines:
\begin{itemize}
\item For $\Delta m^2\gg 1/\langle L/E\rangle$, a vertical line at  
$\sin^22\theta=2\,P_L$.
\item For $\Delta m^2\ll 1/\langle L/E\rangle$, the oscillating phase
can be expanded and the limiting curve takes the form 
$\Delta m^2\sin2\theta =4 \sqrt{P_L}/\langle L/E\rangle$,
which in a log-log plot gives a straight line of slope $-1/2$.
\end{itemize}

If, instead, data are taken at several values of $\langle L \rangle$ and/or 
$\langle E\rangle$, the corresponding region may be closed as it is possible to
have direct information on the characteristic oscillation wavelength. 

\subsection{Neutrinos in Matter: Effective Potentials}
When neutrinos propagate in dense matter, the interactions with the medium 
affect their properties. These effects are either coherent or incoherent. 
For purely incoherent inelastic ${\nu}$-p scattering, the characteristic 
cross section is very small: 
\begin{equation}
\sigma\sim \frac{G_F^2 s}{\pi}\sim 10^{-43} {\rm cm}^2
\left(\frac{E}{1\ {\rm MeV}}\right)^2 \; .
\label{sigmanp}
\end{equation}
The smallness of this cross section is demonstrated by the fact that if a beam 
of $10^{10}$ neutrinos with $E\sim 1$ MeV was aimed at the Earth, only one 
would be deflected by the Earth matter. It may seem then that for neutrinos 
matter is irrelevant. However, one must take into account that 
Eq.~(\ref{sigmanp}) does not contain the contribution from forward elastic 
coherent interactions. In coherent interactions, the medium remains unchanged 
and it is possible to have interference of scattered and unscattered neutrino 
waves which enhances the effect. Coherence further allows one to decouple the 
evolution equation of the neutrinos from the equations of the medium. In this 
approximation, the effect of the medium is described by an effective potential
which depends on the density and composition of the matter (Wolfenstein, 1978).

As an example we derive the effective potential for the evolution of $\nu_e$ 
in a medium with electrons, protons and neutrons. The effective low-energy 
Hamiltonian describing the relevant neutrino interactions is given by 
\begin{equation}
H_W=\frac{G_F}{\sqrt{2}} [{ J^{(+)\alpha}(x) J^{(-)}_\alpha (x)
+\frac{1}{4} J^{(N)\alpha}(x) J^{(N)}_\alpha (x)}]\; ,
\end{equation}
where the $J_\alpha$'s are the standard fermionic currents,
\begin{eqnarray}
J^{(+)}_\alpha(x)&=&\overline{\nu_e}(x)\gamma_\alpha(1-\gamma_5)e(x)\; ,\\ 
J^{(-)}_\alpha(x)&=&\overline{e}(x)\gamma_\alpha(1-\gamma_5)\nu_e(x)\; , \\
J^{(N)}_\alpha(x)&=&\overline{\nu_e}(x)\gamma_\alpha(1-\gamma_5)\nu_e(x)
-\overline{e}(x)[\gamma_\alpha(1-\gamma_5)-4\sin^2\theta_W\gamma_\alpha]e(x)
\nonumber\\
&&+\overline{p}(x)[\gamma_\alpha(1-g_A^{(p)}\gamma_5)-4\sin^2\theta_W
\gamma_\alpha]p(x)
-\overline{n}(x)\gamma_\alpha(1-g_A^{(n)}\gamma_5)n(x) \; .
\end{eqnarray}
$g^{(n,p)}_A$ are the axial couplings for neutrons and protons, respectively.  
For the sake of simplicity we concentrate on the effect of the charged current 
interactions. The effective CC Hamiltonian due to electrons in the medium is
\begin{eqnarray}
\label{HCC}
H_C^{(e)}=&\frac{G_F}{\sqrt{2}}&\int d^3p_e { f(E_e,T)} \\ &\times&
\mbox{\huge{$\langle$}}\langle e(s,p_e)|\overline{e}(x)\gamma^\alpha 
(1-\gamma_5)\nu_e(x)\overline{\nu_e}(x)\gamma_\alpha 
(1-\gamma_5)e(x)|e(s,p_e)\rangle\mbox{\huge{$\rangle$}} \nonumber \\ 
=&\frac{G_F}{\sqrt{2}}&{ \overline{\nu_e}(x)}\gamma_\alpha 
(1-\gamma_5){\nu_e(x)} \int d^3p_e { f(E_e,T)} 
\mbox{\huge{$\langle$}}{ \langle e(s,p_e)|}
{ \overline{e}(x)}\gamma_\alpha (1-\gamma_5){ e(x)}
{ |e(s,p_e)\rangle}\mbox{\huge{$\rangle$}} \; , \nonumber
\end{eqnarray}
where $s$ is the electron spin and $p_e$ its momentum. The energy distribution 
function of the electrons in the medium, $f(E_e,T)$, is assumed to be 
homogeneous and isotropic and is normalized as
\begin{equation}
\int d^3p_e { f(E_e,T)} =1\; .
\end{equation}
By $\mbox{\huge{$\langle$}}...\mbox{\huge{$\rangle$}}$ we denote the averaging 
over electron spinors and summing over all electrons in the medium. Notice that
coherence implies that $s,p_e$ are the same for initial and final electrons.

Expanding the electron fields ${e(x)}$ in plane waves we find 
\begin{eqnarray}
\langle e(s,p_e)&|&\overline{e}(x)\gamma_\alpha (1-\gamma_5)e(x)|e(s,p_e)
\rangle\nonumber\\
=\frac{1}{\sl V}\langle e(s,p_e)&|&\overline{u_{s}}(p_e) a^\dagger_s(p_e) 
\gamma_\alpha (1-\gamma_5)a_s(p_e)u_{s}(p_e)|e(s,p_e)\rangle\; ,
\end{eqnarray}
where ${\sl V}$ is a volume normalization factor. The averaging gives
\begin{equation}
\frac{1}{\sl V}\mbox{\huge{$\langle$}}{ \langle e(s,p_e)| 
a^\dagger_s(p_e)a_s(p_e) |e(s,p_e)\rangle}
\mbox{\huge{$\rangle$}} = N_e(p_e)\frac{1}{2} \sum_s\; ,
\end{equation}
where $N_e(p_e)$ is the number density of electrons with momentum $p_e$.
We assumed here that the medium has equal numbers of spin $+1/2$ and spin
$-1/2$ electrons, and we used the fact that $a^\dagger_s(p_e)a_s(p_e)=
{\cal N}^{(s)}_e(p_e)$ is the number operator. We thus obtain:
\begin{eqnarray}
\mbox{\huge{$\langle$}}\langle e(s,p_e)|\overline{e}(x)\gamma_\alpha 
(1-\gamma_5)e(x)|e(s,p_e)\rangle\mbox{\huge{$\rangle$}}&=&
{ N_e(p_e)}\frac{1}{2}{\displaystyle\sum_{s}} { \overline{u_{(s)}}(p_e)} 
\gamma_\alpha (1-\gamma_5){ u_{(s)}(p_e)}\nonumber \\
&=&\frac{\displaystyle  N_e(p_e)}{\displaystyle 2}{\rm Tr}
\Big[\frac{\displaystyle m_e + \slash\!\!p}
{\displaystyle 2E_e}\gamma_\alpha (1-\gamma_5)\Big]
={ N_e(p_e)}\frac{\displaystyle p_e^\alpha}
{\displaystyle  E_e}\label{ave} \; .
\end{eqnarray}
Isotropy implies that $\int d^3p_e\vec{p_e}f(E_e,T)=0$. Thus, only the $p^0$ 
term contributes upon integration, with $\int d^3p_ef(E_e,T)N_e(p_e)=N_e$
(the electron number density). Substituting Eq.~(\ref{ave}) in Eq.~(\ref{HCC}) 
we obtain: 
\begin{equation}
H_C^{(e)}=\frac{G_FN_e}{\sqrt{2}}\overline{\nu_e}(x)\gamma_0
(1-\gamma_5){\nu_e(x)} \; .
\end{equation}
The effective potential for ${\nu_e}$ induced by its charged current 
interactions with electrons in matter is then given by
\begin{equation}
V_C={ \langle \nu_e|}\int d^3x\ H^{(e)}_C|\nu_e\rangle 
=\frac{G_FN_e}{\sqrt{2}}\frac{2}{\sl V}\int d^3x \;u^\dagger_\nu u_\nu=
\sqrt{2} G_F N_e\; .
\label{effV}
\end{equation}
For ${\overline{\nu_e}}$ the sign of $V$ is reversed. This potential can also 
be expressed in terms of the matter density $\rho$: 
\begin{equation}
V_C=\sqrt{2} G_F { N_e}\simeq 7.6\, { Y_e}\,
\frac{ \rho}{10^{14} {\rm g/cm}^3}\ {\rm eV}\; , 
\end{equation}
where $Y_e=\frac{N_e}{N_p+N_n}$ is the relative number density. Three examples
that are relevant to observations are the following: 
\begin{itemize}
\item At the Earth core $\rho\sim 10$ g/cm$^3$ and $V_C\sim 10^{-13}$ eV; 
\item At the solar core ${\rho}\sim 100$ g/cm$^3$ and $V_C\sim 10^{-12}$ eV;
\item At a supernova ${\rho}\sim 10^{14}$ g/cm$^3$ and $V_C\sim$ eV.
\end{itemize}

Following the same procedure we can obtain the effective  potentials for 
$\nu_e$ due to interactions with different particles in the medium. The results 
are listed in Table~\ref{tab:potentials} (Kim and Pevsner, 1993).
For $\nu_\mu$ and $\nu_\tau$,  $V_C=0$ for any of these media while
$V_N$ is the same as for $\nu_e$. One can further generalize this analysis
to other types of interactions (Bergmann, Grossman and Nardi, 1999).

\subsection{Evolution Equation in Matter: Effective Mass and Mixing}
\label{subsec:oscmatter}
There are several derivations in the literature of the evolution equation of a 
neutrino system in matter (see, for instance, Halprin 1986, Mannhein 1988). We 
follow here the discussion in Baltz and Weneser, 1988. Consider a state which 
is an admixture of two neutrino species ${|\nu_e\rangle}$ and 
${|\nu_X\rangle}$ 
or, equivalently, of ${|\nu_1\rangle}$ and ${|\nu_2\rangle}$: 
\begin{equation}
\Phi(x)= \Phi_e(x){|\nu_e\rangle}+\Phi_X(x){|\nu_X\rangle}=
\Phi_1(x){|\nu_1\rangle}+\Phi_2(x){|\nu_2\rangle}
\end{equation}
The evolution of ${\Phi}$ in a medium is described by a system of 
coupled  Dirac equations:
\begin{eqnarray}
E { \Phi_1}&=&\Big[\frac{\hbar}{i} \alpha_x\frac{\partial}{\partial x} +
\beta { m_1} +{ V_{11}}\Big]{ \Phi_1}+ { V_{12}} { \Phi_2}, \nonumber \\
E { \Phi_2}&=&\Big[\frac{\hbar}{i} \alpha_x\frac{\partial}{\partial x} +
\beta { m_2} +{ V_{22}}\Big]{ \Phi_2}+ { V_{12}} { \Phi_1} \; ,
\label{diraceq}
\end{eqnarray}
where $\beta=\gamma_0$ and $\alpha_x=\gamma_0\gamma_1$. The $V_{ij}$  terms 
give the effective potential for neutrino mass eigenstates. They can be simply 
derived from the effective potential for interaction eigenstates [such as
$V_{ee}$ of Eq.~(\ref{effV})]:
\begin{equation}
V_{ij}={ \langle \nu_i|}\int d^3x { H^{\rm medium} _{int}}
{|\nu_j\rangle}=U_{i\alpha} V_{\alpha\alpha} U^*_{j\alpha} \;.
\end{equation}
We decompose the neutrino state: ${ \Phi_i(x)}={ C_i(x)}{ \phi_i(x)}$. 
Here  ${ \phi_i(x)}$  is the Dirac spinor part satisfying:
\begin{equation}
\left(\alpha_x\big\{\big[E-{ V_{ii}(x)}\big]^2-{ m_i^2}\big\}^{1/2}+
\beta { m_i}+{ V_{ii}}\right){\phi_i(x)}=E{\phi_i(x)}\;.
\end{equation}
So ${\phi_i(x)}$ has the form of free particle solutions with local energy 
${ {\cal E}_i}(x)=E-{ V_{ii}(x)}$:
\begin{equation}
{\phi_i(x)}=\left[\frac{{{\cal E}_i}+{ m_i}}{2{ {\cal E}_i}}\right]^{1/2}\times
\left[\begin{array}{c} { \chi} \\
\frac{\sqrt{{ {\cal E}_i}^2-{ m_i^2}}}{{ {\cal E}_i}+
{ m_i}}\sigma_x{\chi}\end{array}\right],
\end{equation}
where ${\chi}$ is the Pauli spinor. We make the following approximations:
\begin{itemize}
\item[$(i)$] The scale over which $V$ changes is much larger than the 
microscopic wavelength of the neutrino: $\frac{\partial V}{\partial x}/V\ll 
\hbar m/E^2$.
\item[$(ii)$]  Expanding to first order in $V$ implies that 
$V_{12}\,\alpha_x\, \phi_2\simeq\phi_1$,
$V_{12}\,\alpha_x\, \phi_1\simeq\phi_2$ and
$\{\big[E-{ V_{ii}(x)}\big]^2-{ m_i^2}\big\}^{1/2}\simeq
E-V_{ii}(x)-\frac{m_i^2}{2E}$.
\end{itemize} 
From $(i)$ we find that the Dirac equations take the form: 
\begin{eqnarray}
E C_1\phi_1&=&\frac{\hbar}{i}\alpha_x\frac{\partial C_1}{\partial x}{\phi_1}+
(\beta { m_1} +{ V_{11}}) { C_1}{\phi_1}+{ V_{12}} { C_2}{\phi_2},\nonumber \\
E C_2\phi_2&=&\frac{\hbar}{i}\alpha_x\frac{\partial C_2}{\partial x}{\phi_2}+
(\beta { m_2} +{ V_{22}}) { C_2}{\phi_2}+ { V_{12}} { C_1}{\phi_1}. 
\end{eqnarray}
Then multiplying by $\alpha_x$ and using the equation of motion of $\phi_i$ 
and $(ii)$, we can drop the dependence on the spinor $\phi$ and obtain  
\begin{eqnarray}
\frac{\hbar}{i} \frac{\partial { C_1}}{\partial x}
&=&\big[E-{ V_{11}(x)}-\frac{ m^2_1}{2E}\big]{ C_1}-{V_{12}}{C_2},\nonumber\\
\frac{\hbar}{i} \frac{\partial { C_2}}{\partial x}
&=&\big[E-{ V_{22}(x)}-\frac{ m^2_2}{2E}\big]{ C_2}-{V_{12}}{C_1}\; .
\label{schroeq1}
\end{eqnarray}
Changing notations $C_{i,\alpha}(x)\to\nu_{i,\alpha}(x)$, we rewrite 
Eq.~(\ref{schroeq1}) in a matrix form :
\begin{equation}
\frac{\hbar}{i} \frac{\partial }{\partial x}
\left (\begin{array}{c}{\nu_1}\\{\nu_2}\end{array}\right)=
\left (\hspace*{-0.2cm}\begin{array}{cc}
 E-{ V_{11}}-\frac{ m^2_1}{2E}  & -{ V_{12}}\\
-{ V_{12}} &E-{ V_{22}}-\frac{ m^2_2}{2E}  \end{array}\right)
\left (\begin{array}{c}{\nu_1}\\{\nu_2}\end{array}\right)\; .
\label{wolfmas}
\end{equation}
After removing the diagonal piece that is proportional to $E$, 
we can rotate Eq.~(\ref{wolfmas}) to the  flavor basis ($\hbar=1$) 
(Wolfenstein, 1978): 
\begin{equation}
-i \frac{\partial}{\partial x}\left(\begin{array}{c}{ \nu_e}\\{ \nu_X} 
\end{array}\right)=\left(-\frac{ M_w^2}{2E}\right)\left(\begin{array}{c}
{ \nu_e}\\ { \nu_X} \end{array}\right) \; ,
\label{evoleq}
\end{equation}
where we have defined an effective mass matrix in matter:
\begin{equation}
M_w^2=\left(\begin{array}{cc} \frac { m_1^2+m_2^2 }{2}
+2 E{ V_e}-\frac{ \Delta m^2}{2} {\cos 2\theta} &
\frac{ \Delta m^2}{2} { \sin 2\theta} \\
\frac{ \Delta m^2}{2} { \sin 2\theta} & \frac { m_1^2+m_2^2 }{2}
+2E{ V_X}+\frac{ \Delta m^2}{2} { \cos 2\theta} \end{array}\right)\; .
\label{Mw}
\end{equation}
Here ${\Delta m^2=m_2^2-m_1^2}$. 

We define the instantaneous mass eigenstates in matter, $\nu^m_i$, 
as the eigenstates of $M_w$ for a fixed value of $x$ (or $t$). They are
related to the interaction eigenstates through a unitary rotation,
\begin{equation}
\left(\begin{array}{c}{ \nu_e}\\{ \nu_X} \end{array}\right)
={ U(\theta_m)}\left(\begin{array}{c}{ \nu^m_1}\\ { \nu_2^m} \end{array}\right)
=\left(\begin{array}{ll} {\cos\theta_m} & {\sin\theta_m} \\
-{\sin\theta_m} & {\cos\theta_m} \end{array}\right)
\left(\begin{array}{c}{ \nu^m_1}\\ { \nu_2^m} \end{array}\right) \; .
\label{insmas}
\end{equation}
The eigenvalues of $M_w$, that is, the effective masses in matter are given by
(Wolfenstein, 1978; Mikheyev and Smirnov, 1985):
\begin{equation}
{ \mu_{1,2}^2}(x)=\frac { m_1^2+m_2^2 }{2}+E ({V_e+V_X})
\mp\frac{1}{2}\sqrt{\left({\Delta m^2\cos 2\theta}-A\right)^2 
+\left({\Delta m^2\sin 2\theta}\right)^2 }\; ,
\label{effmass}
\end{equation}
while the mixing angle in matter is given by
\begin{equation} 
{\tan 2\theta_{m}}= \frac{\Delta{m}^2\sin2\theta}{\Delta{m}^2\cos 2\theta-A}.
\label{effmix}
\end{equation}
The quantity $A$ is defined by
\begin{equation}
A\equiv 2 E({ V_e-V_X}).
\end{equation}
In Figs.~\ref{fig:effmass1} and \ref{fig:effmix} we plot, respectively, the 
effective masses and the mixing angle in matter as functions of the potential 
$A$, for ${ A}>0$ and ${\Delta m^2\cos 2\theta>0}$. Notice that even 
massless neutrinos acquire non-vanishing effective masses in matter.

The resonant density (or potential) $A_R$ is defined as the value of $A$ for 
which the difference between the effective masses is minimal: 
\begin{equation} 
A_R={\Delta m^2\cos 2\theta} \; .
\label{AR}
\end{equation}
Notice that once the sign of $V_e-V_X$ (which depends on the composition of 
the medium and on the state $X$) is known, this resonance condition 
can only be achieved for a given sign of $\Delta m^2\cos2\theta$, {\it i.e.} 
for mixing angles in only one of the two possible octants. We learn that 
the symmetry present in vacuum oscillations is broken by matter potentials. 
Also if the resonant condition is achieved for two neutrinos it 
cannot be achieved for antineutrinos of the same flavor and viceversa.
The mixing angle $\tan\theta_m$ changes sign at ${A_R}$. As can be seen in  
Fig.~\ref{fig:effmix}, for $A>A_R$ we have $\theta_{m}\gg{\theta}$. 

We define an oscillation length in matter: 
\begin{equation}
L^{\rm osc}=\frac{L_0^{\rm osc} \Delta m^2}{\sqrt{ ({\Delta{m}^2 \cos 2\theta} 
-{  A})^2 +({ \Delta{m}^2 \sin 2\theta} )^2 }},
\label{efflosc}
\end{equation}
where the oscillation length in vacuum, $L^{\rm osc}_0$, was defined in 
Eq.~(\ref{L0}). 
The oscillation length in matter
presents a resonant behaviour. At the resonance point the oscillation length is 
\begin{equation}
L_R^{\rm osc}=\frac{L_0^{\rm osc}}{\sin2\theta}.
\end{equation} 
The width (in distance) of the resonance, $\delta r_R$, corresponding to 
$\delta A_R=2{\Delta m^2\sin^2 2\theta}$, is given by
\begin{equation}
\delta r_R=\frac{\delta A_R}{|\frac{ d A}{ d r}|_R}=\frac{2\,\tan 2\theta}{h_R},
\label{dr}
\end{equation} 
where we have defined the resonance height: 
\begin{equation}
h_R \equiv \left|\frac{1}{ A}\frac{ d A}{ d r}\right|_R
\label{hr}
\end{equation}

For constant  $A$, {\it i.e.}, for constant matter density, the evolution of 
the neutrino system is described just in terms of the masses and mixing in 
matter. But for varying $A$, this is in general not the case.  

\subsection{Adiabatic versus Non-adiabatic Transitions}
\label{subsec:nonadiab}
Taking time derivative of Eq.~(\ref{insmas}), we find: 
\begin{equation}
\frac{\partial}{\partial t}\left(\begin{array}{c}{ \nu_e}\\{ \nu_X} \end{array}
\right)={\dot{U}(\theta_m)}\left(\begin{array}{c}{\nu^m_1}\\ {\nu_2^m}
\end{array}\right)+{{U}(\theta_m)}\left(\begin{array}{c}
{ \dot{\nu}^m_1}\\ { \dot{\nu}_2^m} \end{array}\right)\;.
\end{equation}
Using the evolution equation in the flavor basis, Eq.~(\ref{evoleq}), we get 
\begin{equation}
i\left(\begin{array}{c}{\dot{\nu}^m_1}\\ { \dot{\nu}_2^m} \end{array}\right)=
\frac{1}{2E}{ {U^\dagger(\theta_m)}} { M_w^2} { {U}(\theta_m)}
\left(\begin{array}{c}{ \nu^m_1}\\ { \nu_2^m} \end{array}\right)
-i\;{ {U^\dagger}} { \dot{U}(\theta_m)}\left(\begin{array}{c}
{ \nu^m_1}\\ { \nu_2^m} \end{array}\right)\; .
\end{equation}
For constant matter density, ${\theta_m}$ is constant and the second term 
vanishes. In general, using the definition of the effective masses ${\mu_i(t)}$
in Eq.~(\ref{effmass}), and subtracting a diagonal piece 
${(\mu_1^2+\mu_2^2)}/2E\times {I}$, we can rewrite the evolution equation as:
\begin{equation}
i\left(\begin{array}{c}{\dot{\nu}^m_1}\\ { \dot{\nu}_2^m}\end{array}\right)=
\frac{1}{4E}\left(\begin{array}{ll}-{\Delta(t)} & -4iE{\dot{\theta}_m(t)}
\\ 4iE{\dot{\theta}_m(t)}  &{ \Delta(t)}\end{array}\right)
\left(\begin{array}{c}{{\nu}^m_1}\\ { {\nu}_2^m} \end{array}\right)
\label{evoleq2}
\end{equation}
where we defined ${\Delta(t)\equiv\mu_2^2(t)-\mu_1^2(t)}$. 

The evolution equations (\ref{evoleq2}) constitute a system of coupled 
equations: the instantaneous mass eigenstates, ${\nu_i^m}$, mix in the 
evolution and are not energy eigenstates. The importance of this effect is 
controlled by the relative size of the off-diagonal piece 
$4\,E\,\dot{\theta}_m(t)$ 
with respect to the diagonal one ${\Delta(t)}$. When $\Delta(t)\gg 4 \,E\,
\dot{\theta}_m(t)$, the instantaneous mass eigenstates, $\nu^m_i$, behave 
approximately as energy eigenstates and they do not mix in the evolution.
This is the {\it adiabatic} transition approximation. 
From the definition of $\theta_m$ in Eq.~(\ref{effmix}) we find: 
\begin{equation}
\dot{\theta_m}=\frac{\Delta m^2\sin2\theta}{2{\Delta(t)^2}}{ \dot{A}}\; .
\end{equation}
The adiabaticity condition reads then
\begin{equation}
{ \Delta(t)}\gg \frac{2 E A {\Delta m^2\sin 2\theta}}{\Delta(t)^2}
\left|\frac{ \dot{A}}{{A}}\right|\; .
\label{adiacon}
\end{equation}

Since for small mixing angles the maximum of ${\dot{\theta_m}}$ occurs at 
the resonance point 
(as seen in Fig.~\ref{fig:effmix}), the strongest adiabaticity condition 
is obtained when Eq.~(\ref{adiacon}) is evaluated at the resonance
(the generalization of the condition of maximum adiabaticity violation 
to large mixings can be found in 
Friedland, 2001 and Lisi, Marrone, Montanino,
Palazzo and Petcov, 2001). 
We define the adiabaticity parameter $Q$ at the resonance as follows:
\begin{equation}
Q=\frac{\Delta m^2\sin^2 2\theta}{E {\cos2\theta}\,{ h_R}}
=\frac{4\ \pi\, { \delta r_R}} { L^{osc}_R}\; ,
\label{Q}
\end{equation}
where we used the definitions of $A_R$, $\delta r_R$, and $h_R$ in 
Eqs.~(\ref{AR}),~(\ref{dr}), and (\ref{hr}). Written in this form, we see that
the adiabaticity condition,  $Q\gg1$, implies that many oscillations take place
in the resonant region. Conversely, when $Q\lesssim 1$ the transition is 
non-adiabatic.

The survival amplitude of a ${\nu_e}$ produced in matter at {$t_0$} and exiting
the matter at $t>t_0$ can be written as follows:
\begin{equation}
A({\nu_e}\to{\nu_e};t)=\sum_{i,j}A(\nu_e(t_0) \to\nu_{i} (t_0))\;
A(\nu_{i}(t_0)\to \nu_{j}(t))\; A(\nu_{j}(t)\to \nu_{e}(t))\;\;\;
\label{aee1}
\end{equation} 
with 
\begin{eqnarray}
A(\nu_e(t_0)\to\nu_{i}(t_0))&=&\langle {\nu_i(t_0)}|{\nu_e(t_0)}
\rangle={ U^*_{ei}(\theta_{m,0})} \nonumber \\
A(\nu_{j}(t)\to \nu_{e}(t))&=&\langle{\nu_e(t)}|{\nu_j(t)}\rangle
={U_{ej}(\theta)}
\nonumber
\end{eqnarray}
where ${ U^*_{ei}(\theta_{m,0})}$ is the 
$(ei)$ element of the mixing matrix in matter at the production point and 
${U_{ej}(\theta)}$ is the $(ej)$ element
of the mixing matrix in vacuum. 

In the adiabatic approximation the mass eigenstates do not mix so
\begin{equation}
A(\nu_{i}(t_0)\to \nu_{j}(t))=\delta_{ij}\,
\langle {\nu_i(t)}|{\nu_i(t_0)}\rangle=\delta_{ij}\,
  \exp\left\{i\int_{t_0}^t     {  E_i(t')}  dt' \right\} \; .
\label{ampl2}
\end{equation}
Note that $E_i$ is a function of time because the effective mass $\mu_i$
is a function of time,
\begin{equation}
E_i(t') \simeq p+\frac{\mu_i^2(t')}{2p}\; .
\end{equation}
Thus the transition probability for the adiabatic case is given by
\begin{equation}
P({\nu_e}\rightarrow {\nu_e};t)=\left|\sum_{i}
{ U_{ei} (\theta) U^\star_{ei}(\theta_{m,0})}
 \exp\left(-\frac{i}{2E} {\int^t_{t_0} \mu^2_i(t') dt'}\right)\right|^2\; .
\label{adigen}\end{equation}
For the case of two-neutrino mixing, Eq.~(\ref{adigen}) takes the form
\begin{equation}
P({\nu_e}\rightarrow {\nu_e};t) = 
{ \cos^2\theta_m\cos^2\theta}+{ \sin^2\theta_m\sin^2\theta}
+\frac{1}{2} { \sin 2\theta_m\sin 2\theta}
\cos \left(\frac{\delta(t)}{2E}\right)\; ,
\label{aditwo}
\end{equation}
where
\begin{equation}
{ \delta(t)}=\int_{t_0}^t     {  \Delta (t')}  dt'=
\int_{t_0}^t \sqrt{({\Delta{m}^2 \cos2\theta}
-{  A(t')})^2+({ \Delta{m}^2 \sin2\theta})^2 }dt' \; ,
\end{equation}
which, in general, has to be evaluated numerically. There are some analytical
approximations for specific forms of $A(t')$: exponential, linear \dots
(see, for instance, Kuo and Pantaleone, 1989). For ${\delta(t)}\gg E$ the last 
term  in Eq.~(\ref{aditwo}) is averaged out and the survival probability takes 
the form
\begin{equation}
P({\nu_e}\rightarrow {\nu_e};t)=
\frac{1}{2}\left[1+{ \cos 2\theta_m\cos 2\theta}\right]
\label{eq:peead}
\end{equation}
In Fig.~\ref{fig:peenoad} we plot isocontours of constant survival probability 
in the parameter plane 
$(\Delta m^2,\tan^2\theta)$ for the particular case of the sun density for 
which $A>0$. Notice that, unlike $\sin^22\theta$, $\tan^2\theta$ is a single 
valued function in the full parameter range $0\leq\theta\leq \pi/2$. Therefore 
it is 
a more appropriate variable once matter effects are included and the symmetry 
of the survival probability  with respect to the change of octant for the 
mixing angle  is lost. As seen in the figure, for $\theta<\pi/4$, 
$P({\nu_e}\rightarrow {\nu_e})$ in matter can be larger or smaller than $1/2$,
in contrast to the case of vacuum oscillations where, in the averaged regime, 
${ P^{vac}_{ee}}=1-\frac{1}{2}{ \sin^22\theta}>\frac{1}{2}$. 

In Fig.~\ref{fig:peenoad} 
we also plot the limiting curve for $Q=1$. 
To the left and below of this curve the adiabatic approximation breaks down 
and the isocontours in Fig.~\ref{fig:peenoad} deviate from the 
expression in Eq.~(\ref{eq:peead}).
In this region,  the off-diagonal term ${\dot{\theta}_m}$ cannot be 
neglected and the mixing between instantaneous mass eigenstates is 
important. In this case we can write 
\begin{equation}
A(\nu_{i}(t_0)\to \nu_{j}(t))=
\langle {\nu_j(t)}|{\nu_j(t_R)}\rangle
\langle {\nu_j(t_R)}|{\nu_i(t_R)}\rangle
\langle {\nu_i(t_R)}|{\nu_i(t_0)}\rangle 
\end{equation}
where $t_R$ is the point of maximum adiabaticity violation which, 
for small mixing angles, corresponds to the resonant point.
The possibility of this {\it level crossing} can be described 
in terms of the Landau-Zener probability (Landau, 1932; Zener, 1932):
\begin{equation}
P_{LZ}=\left|\langle {\nu_j(t_R)}|{\nu_i(t_R)}\rangle\right|^2\;\;\;\;\;\ 
(i\neq j) \; .
\label{plz}
\end{equation}
Introducing this transition probability in Eq.~(\ref{aee1}) we find that 
in the non-adiabatic regime (after averaging out the oscillatory term), the
survival probability can be written as
\begin{equation}
P({\nu_e}\rightarrow {\nu_e};t)=
\frac{1}{2}\left[1+(1-2{ P_{LZ}}){ \cos 2\theta_m\cos 2\theta}\right]\; .
\label{osc:pmsw}
\end{equation}
The physical interpretation of this expression is straightforward. An electron 
neutrino produced at ${ A>A_R}$ consists of an admixture of ${\nu_1}$ with 
fraction ${\cos^2\theta_m}$ and ${\nu_2}$ with fraction ${\sin^2\theta_m}$. 
In particular, for very small mixing angles in vacuum, $\theta_m\sim \pi/2$ 
(see Fig.~\ref{fig:effmix})  so $\nu_e$ is almost a pure ${\nu_2(t_0)}$ state.
When the neutrino state reaches the resonance,  $\nu_2$ ($\nu_1$) can become 
${\nu_2}$ ($\nu_1$) with probability  $[1-P_{LZ}]$ or ${\nu_1}$ ($\nu_2$) with 
probability $P_{LZ}$. So after passing the resonance, the ${\nu_e}$ flux 
contains a fraction of ${\nu_1}$: $P_{e1}=\sin^2\theta_m P_{LZ}+ 
\cos^2\theta_m(1-{ P_{LZ}})$, and a fraction  of  ${\nu_2}$:  
$P_{e2}={ \cos^2\theta_m}{ P_{LZ}}+ {\sin^2\theta_m}(1-{ P_{LZ}})$. 
At the exit ${\nu_1}$ consists of ${\nu_e}$ with
fraction ${\cos^2\theta}$ and ${\nu_2}$  consists of ${\nu_e}$ 
with fraction ${\sin^2\theta}$ so (Parke, 1986; Haxton, 1986; Petcov, 1987) 
\begin{equation}
P_{ee}= { \cos^2\theta}P_{e1}+ { \sin^2\theta}P_{e2},
\label{osc:pmswa}
\end{equation}
which reproduces Eq.~(\ref{osc:pmsw}).

The Landau-Zener probability  can be evaluated in the 
WKB approximation (this derivation follows Kim and Pevsner, 1993):
\begin{equation}
{ P_{LZ}}=\exp\left[-2{\cal I}m\int_{t_R}^{t_0}\frac{{\mu^2_2(t)}
-{\mu^2_1(t)}}{2E}dt \right]
=\exp\left[-2 {\cal I}m\int_{A_R}^{A_0}\frac{{\mu^2_2(A)}-{\mu^2_1(A)}}
{2E|{\dot{A}}|}dA\right] \; ,
\end{equation}
where $t_0$ (or $A_0$) is a root of ${\mu^2_2}-{\mu^2_1}$=0
lying in the upper half plane of the complex variable $t$ 
(or $A$),  ${ A_0}={\Delta m^2}({\cos 2\theta}\,+\,i\,{\sin 2\theta})$.
Assuming that $|{\dot{A}}|$ is slowly varying near the resonance, we get
\begin{equation}
{\cal I}m\int_{A_R}^{A_0}\frac{{\mu^2_2(A)}-{\mu^2_1(A)}}
{2E|{\dot{A}}|}dA \sim\frac{1} {2E|{\dot{A}}|_R}
{\cal I}m\int_{A_R}^{A_0}\left[{\mu^2_2(A)}-{\mu^2_1(A)}\right]dA\; .
\end{equation} 
Shifting the integral $A\to { A}-{\Delta{m}^2}{\cos 2\theta}$, we obtain:
\begin{equation}
{\cal I}m
\int_{A_R}^{A_0}\left[{\mu^2_2(A)}-{\mu^2_1(A)}\right]dA
= {\cal I}m \int_0^{i{\Delta{m}^2}{\sin 2\theta}}dA
\sqrt{{  A}^2 +({ \Delta{m}^2 \sin 2\theta} )^2 }
=\frac{\pi}{4}({ \Delta{m}^2 \sin 2\theta} )^2\; .
\end{equation}
So we finally find  
\begin{equation}
{ P_{LZ}}=\exp\left(-\frac{\pi}{4}{{Q}}\right)
\end{equation}
where $Q$ is the adiabaticity parameter defined in Eq.~({\ref{Q}).
In this derivation we assumed that $|{\dot{A}}|$ is almost constant. This
assumption is strictly valid only for linearly growing densities 
${{A}}\propto r$. For other forms of ${{A}}$, the adiabaticity parameter 
${{Q}}$ has to be multiplied by some factor
(for a list of some of these factors see Kuo and Pantaleone, 1989). 
For instance, for an exponential density,  ${{A}}\sim \exp({-r})$,  
the factor is $(1-{\tan^2\theta})$. Moreover this approximate derivation is 
not valid for ${{Q}}\ll 1$. The general form of the Landau-Zener probability 
for an exponential density can be written as (Petcov, 1988; Krastev and
Petcov, 1988):
\begin{equation}
{ P_{LZ}}=\frac{\exp(-\gamma { \sin^2\theta})-\exp(-\gamma)}{1-\exp (-\gamma)},
\;\;\;\;\;\;\; \gamma \equiv \pi \frac{ \Delta{m}^2}{E  |{\dot{A}/A}|_R}.
\label{plzexp} 
\end{equation}

When $ {\nu_e}$ is produced at ${ A\gg A_R}$  and ${\theta}$ is small, 
${\theta_m}\sim$ 90$^\circ$. In this case the survival probability is simply 
given by
\begin{displaymath}
P({\nu_e}\rightarrow {\nu_e};t)\simeq{ P_{LZ}}\simeq 
\exp\left(-\frac{\pi}{4}{{Q}}\right) 
\end{displaymath}
Since ${{Q}}\sim \Delta{m}^2 \sin^2 2\theta/E$, the isocontours
of constant probability in this regime  correspond to diagonal lines
in the ($\Delta m^2, \tan^2\theta$) plane in a log-log plot, as 
illustrated in Fig.~\ref{fig:peenoad}. 

\subsection{Propagation in the Sun: MSW Effect} 
As an illustration of the matter effects discussed in the previous section we 
describe now the propagation of a $\nu_e-\nu_X$ neutrino system in the matter 
density of the sun. For the sake of concreteness we assume that $X$ is some 
superposition of $\mu$ and $\tau$. 

The solar density distribution decreases monotonically (see 
Fig.~\ref{fig:sundens}). For $R<0.9 R_\odot$ it can be approximated 
by an exponential $N_e(R)=N_e(0)\exp\left(-R/r_0\right)$,
with $r_0=R_\odot/10.54=6.6\times 10^7$ m $=3.3\times 10^{14}$ eV$^{-1}$. 

After traversing this density the dominant component of the exiting 
neutrino state depends on the value of the mixing angle in vacuum and 
the relative size of ${\Delta m^2 \cos 2\theta}$ versus 
${A_0}=2\,E\,G_F\,{ N_{e,0}}$ (at the neutrino production point):

(i) ${\Delta m^2\cos2\theta}\gg A_0$: matter effects are negligible
and the propagation occurs as in vacuum. The survival probability at the 
sunny surface of the Earth is
\begin{equation}
P_{ee}(\Delta m^2\cos2\theta\gg A_0)=
1-\frac{1}{2}{\sin^22\theta}>\frac{1}{2}\; .
\end{equation}

(ii) ${\Delta m^2 \cos 2\theta}\gtrsim{ A_0}$: the neutrino does 
not pass the resonance but its mixing is affected by the matter. This effect
is well described by an adiabatic propagation:
\begin{equation}
P_{ee}(\Delta m^2\cos2\theta\gsim A_0)=\frac{1}{2}\left[1+{\cos 2\theta_m
\cos 2\theta}\right]\; .
\end{equation}
Since the resonance is not crossed, $\cos 2\theta_m$ has the same sign as
$\cos 2\theta$ and the corresponding survival probability is 
also larger than 1/2. 

(iii) ${ \Delta m^2 \cos 2\theta}<{ A_0}$:  the neutrino can cross the
resonance on its way out. In this case, as discussed in the previous section, 
for small mixing angle in vacuum, ${\nu_e\sim\nu_2^m}$ at the production point
and remains ${\nu_2^m}$ till the resonance point
(for larger mixing but still in the first octant $\nu_e$ is a combination of
$\nu_1^m$ and $\nu_2^m$ with larger $\nu_2^m$ component).
It is important in this case to find whether the transition is adiabatic.
For the solar density, $Q\sim1$ corresponds to
\begin{equation}
\frac{(\Delta m^2/{\rm eV}^2)\sin^2 2\theta}{(E/{\rm MeV})
{\cos 2\theta}}\sim 3\times 10^{-9}\; .
\end{equation}
For $Q\gg1$ the transition is adiabatic and the neutrino state remains 
in the same linear combination of mass eigenstates after the resonance
determined by $\theta_m$. As seen in Fig.~\ref{fig:effmix}, 
${\theta_m}$ (that is, the $\nu_e$ component of the state) decreases
after crossing the resonance and, consequently, so does the survival
probability $P_{ee}$. In particular, for small mixing angle, 
$\nu_2$ at the exit point is almost a pure $\nu_X$ and, 
consequently, ${P_{ee}}$ can be very small. Explicitly,
\begin{equation}
P_{ee}(\Delta m^2\cos2\theta< A_0,Q\gg1)=\frac{1}{2}
\left[1+{ \cos 2\theta_{m,0}\cos 2\theta}\right]
\end{equation}
can be much smaller than $1/2$ because $\cos 2\theta_{m,0}$ and $\cos2\theta$
can have opposite signs. Note that the smaller the mixing angle in vacuum the 
larger is the deficit of electron neutrinos in the outgoing state. This is the 
MSW effect (Wolfenstein, 1978; Mikheyev and Smirnov, 1985). This behaviour is 
illustrated in Fig.~\ref{fig:peemsw} where we plot the electron survival 
probability as a function of $\Delta m^2/E$ for different values of the mixing 
angle.  

For smaller values of $\Delta m^2/E$ (right side of Fig.~\ref{fig:peemsw}) we 
approach the regime where $Q<1$  and non-adiabatic effects start playing a 
role. In this case the state can jump from $\nu_2$ into 
$\nu_1$ (or vice versa) with probability $P_{LZ}$. For small mixing angle, 
at the surface $\nu_1\sim\nu_e$ and 
the $\nu_e$ component of the exiting neutrino increases. 
This can be seen from the expression for
$P_{ee}$, 
\begin{equation}
P_{ee}(\Delta m^2\cos2\theta< A_0,Q\ll1)=\frac{1}{2}
\left[1+(1-2{ P_{LZ}}){ \cos 2\theta_m\cos2\theta}\right],
\end{equation}
and from Fig.~\ref{fig:peemsw}. For large mixing angles this expression is
still valid but  $P_{LZ}$ has to be evaluated at the point of maximum
adiabaticity violation which does not coincide with the resonant point 
as discussed in Friedland, (2001) and Lisi, Marrone, Montanino,
Palazzo and Petcov, (2001). 

\section{Solar Neutrinos}
\label{sec:solar}
Solar neutrinos are electron neutrinos produced in the thermonuclear reactions
which generate the solar energy. These reactions occur via two main chains, the
$pp$-chain and the CNO cycle, shown in Figs.~\ref{fig:pp} and~\ref{fig:cno},
respectively. There are five reactions which produce $\nu_e$ in the $pp$ 
chain and three in the CNO cycle. Both chains result in the overall 
fusion of protons into $^4$He: 
\begin{equation}
4p\ \to\  ^4{\rm He}+2 e^+ + 2\nu_e +\gamma,
\end{equation}
where the energy released in the reaction, $Q=4m_p-m_{^4{\rm He}}-2 m_e
\simeq 26$ MeV,
is mostly radiated through the photons and only a small fraction is carried
by the neutrinos, $\langle E_{2\nu_e}\rangle=0.59$ MeV.
 
In order to precisely determine the rates of the different reactions in the two
chains which would give us the final neutrino fluxes and their energy spectrum,
a detailed knowledge of the sun and its evolution is needed. Solar Models 
(Bahcall and Ulrich, 1988; Bahcall and Pinsonneault, 1992 and 1995; 
Bahcall, Basu, and Pinsonneault, 1998, 2001; Turck-Chiez {\em et al}., 1988) 
describe the properties of the sun and its evolution after entering 
the main sequence.
The models are based on a set of observational parameters: the surface 
luminosity ($3.844 \times 10^{26}$ W), the age ($4.5 \times 10^9$ years),
the radius ($6.961 \times 10^8$ m) and the mass ($1.989 \times 10^{30}$ kg),
and on several basic assumptions: spherical symmetry, hydrostatic and 
thermal equilibrium, equation of state of an ideal gas, and present surface 
abundances of elements similar to the primordial composition. Over the past 
four decades, the solar models have been 
steadily refined as the result of increased observational and experimental 
information about the input parameters (such as nuclear reaction rates and the 
surface abundances of different elements), more accurate calculations of 
constituent quantities (such as radiative opacity and equation of state), the
inclusion of new physical effects (such as element diffusion), and the
development of faster computers and more precise stellar evolution codes.

We use as Standard Solar Model (SSM) the most updated version of the model 
developed by Bahcall, Pinsonneault and Basu, (2001) (BP00).
(At present, the various solar models give essentially the same
results when interpreting the bulk of experimental data in terms of
neutrino parameters. In this sense, the implications for particle
physics discussed in this review are independent of the choice of 
solar model.)  
In Fig.~\ref{fig:bp00} we show the energy spectrum of the fluxes from the five 
$pp$ chain reactions. In what follows we refer to the neutrino fluxes by the 
corresponding source reaction, so, for instance, the neutrinos produced from 
$^8$B decay are called $^8$B neutrinos.
Most reactions produce a neutrino spectrum
characteristic of $\beta$ decay. 
For $^8$B neutrinos the energy distribution presents deviations with
respect to the maximum allowed energy because the final state, $^8$Be,  
is a wide resonance. On the other hand, the $^7$Be neutrinos
are almost monochromatic, with an energy width of about 2 keV which is
characteristic of the temperature in the core of the sun. 

As discussed in Sec.~\ref{sec:oscila}, to describe the evolution of neutrinos 
in the solar matter, one needs to know other quantities that are predicted by 
the SSM, such as the density and composition of solar matter, which give the 
solar electron number density. As discussed in Sec.~\ref{sec:oscila} and shown 
in Fig.~\ref{fig:sundens}, the solar matter density decreases monotonically and
can be approximated an exponential profile.  
Furthermore, in order to precisely determine the evolution of the neutrino
system one also needs to know the production point distribution for
the different neutrino fluxes. In Fig.~\ref{fig:ppprod} we show the BP00 
prediction for the distribution probability for the $pp$ chain.

The standard solar models have had notable successes. In particular, the 
comparison between the observed and the theoretically predicted sound speeds 
and pressures has tested the SSM and provided accurate information on the 
solar interior. The solar quantities that have been determined by 
helioseismology
include the sound velocity and matter density as a function of solar radius, 
the depth of the convective zone, the interior rotation rate, and the surface 
helium abundance. The excellent agreement between the helioseismological 
observations and the solar model calculations has made a convincing case that 
the large discrepancies between solar neutrino measurements and solar model
predictions cannot be due to errors in the solar models.

\subsection{Solar Neutrino Experiments}
\label{subsec:solexp}
\subsubsection{Chlorine experiment: Homestake}
The first result on the detection of solar neutrinos was announced
by Ray Davis Jr and his collaborators from Brookhaven in 
1968 (Davis, Harmer and Hoffman, 1968). In the gold mine of Homestake
in Lead, South Dakota, they installed a detector consisting of 
$\sim$ 615 Tons of C$_2$Cl$_4$. Solar $\nu_e$'s are captured via
 \[ ^{37}\mathrm{Cl}\ (\nu,e^-)\ ^{37}\mathrm{Ar}. \]
The energy threshold for this reaction is 0.814 MeV, so the relevant fluxes are
the $^7$Be and $^8$B neutrinos. $^7$Be neutrinos excite the Gamow-Teller (GT) 
transition to the ground state with strength known from the lifetime of the 
electronic capture of\  $^{37}$Ar.  $^8$B neutrinos can excite higher 
states of\ $^{37}$Ar, including the dominant transition to the 
isobaric state with transition energy 4.99 MeV. The strength of the GT 
transitions to these excited states are determined from the isospin related 
$\beta$ decay, $^{37}$Ca$(\beta^+)^{37}$K. The final result is that, for the 
SSM fluxes, 78\% of the expected number of events are due to $^8$B neutrinos
while 13\% arise from  $^7$Be neutrinos. The produced $^{37}$Ar is extracted 
radiochemically every three months approximately and the number of $^{37}$Ar  
decays ($t_{\frac{1}{2}}$=34.8 days) is measured in a proportional counter. 

The average event rate measured during the more than 20 years of operation is 
\begin{displaymath}
R_{\rm Cl} = 2.56 \pm 0.16 \pm 0.16 \, \mathrm{SNU} \,,
\end{displaymath}
(1 SNU = 10$^{-36}$ captures/atom/sec) which corresponds to approximately one 
third of the SSM prediction (for the latest publications see 
Cleveland {\em et al.}, 1998 and Lande {\em et al.}, 1999).

\subsubsection{Gallium experiments: SAGE and GALLEX/GNO}
In January 1990 and May 1991, two new radiochemical experiments using a 
$^{71}$Ga target started taking data, SAGE and GALLEX. The SAGE 
detector is located in Baksan, Kaberdino-Balkaria, Russia, with 30 Tons 
(increased to 57 Tons from July 1991) of liquid metallic Ga. GALLEX is located 
in Gran Sasso, Italy, and consists of 30 Tons of GaCl$_3$-HCl. In these 
experiments the solar neutrinos are captured via
 \[ ^{71}\mathrm{Ga}(\nu,e^-)^{71}\mathrm{Ge}. \]
The special properties of this target include a low threshold (0.233 MeV) and a
strong transition to the ground level of $^{71}$Ge, which gives a large cross 
section for the lower energy $pp$ neutrinos.  According to the SSM, 
approximately 54\% of the events are due to $pp$ neutrinos, 
while 26\% and 11\% arise from 
$^7$Be and $^8$B neutrinos, respectively. The extraction of $^{71}$Ge takes 
places every 3--4 weeks and the number of $^{71}$Ge decays 
($t_{\frac{1}{2}}$=11.4 days) is measured in a proportional counter.

The event rates measured by SAGE (Abdurashitov {\em et al.}, 1999; 
2002) and GALLEX (Hampel {\em et al.},  1999)
are
\begin{displaymath}
 R_{\rm SAGE} = 70.8^{+5.3}_{-5.2}~ ^{+3.7}_{-3.2} \, \mathrm{SNU} \,, 
\end{displaymath}
\begin{displaymath}
 R_{\rm GALLEX} = 77.5 \pm 6.2 ^{+4.3}_{-4.7} \, \mathrm{SNU}  \,.
\end{displaymath}
while the prediction of the SSM is 130 SNU.

The GALLEX program was completed in fall 1997 and its successor GNO started 
taking data in spring 1998. The latest combined GALLEX/GNO result is 
(Kirsten, 2002) 
\begin{displaymath}
 R_{\rm GALLEX+GNO} = 70.8 \pm 5.9 \, \mathrm{SNU} \,.
\end{displaymath}

Since the $pp$ flux is directly constrained by the solar luminosity, in all 
stationary solar models there is a theoretical minimum of the expected number 
of events of 79 SNU. Furthermore, the largest uncertainties on the capture 
cross section for $^7$Be neutrinos were considerably reduced after direct 
calibration using neutrinos from $^{51}$Cr decay as a source.

\subsubsection{Water Cerenkov experiments: Kamiokande and SuperKamiokande}
Kamiokande and its successor SuperKamiokande (SK) in Japan are water Cerenkov
detectors that are able to detect in real time the electrons scattered from the
water by elastic interaction of the solar neutrinos, 
\begin{equation}
\nu_a + e^- \to \nu_a + e^- \,,
\label{disper}
\end{equation}
The scattered electrons produce Cerenkov light which is detected by
photomultipliers. Notice that, while the detection process in
radiochemical experiments is purely a charged-current ($W$-exchange)
interaction, the detection process of Eq.~(\ref{disper}) goes through
both charged- and neutral-current  ($Z$-exchange)
interactions. Consequently, the detection process (\ref{disper}) is
sensitive to all active neutrino flavors, although $\nu_e$'s (which
are the only ones to scatter via $W$-exchange) give a contribution
that is about 6 times larger than that of $\nu_\mu$'s or $\nu_\tau$'s.

Kamiokande, with 2140 tons of water, started taking data in January 1987 and 
was terminated in February 1995. SK, with 45000 tons of water (of 
which 22500 are usable in solar neutrino measurements) started in May 1996 and 
it has analyzed so far the events corresponding to 1258 days. The detection 
threshold in Kamiokande was 7.5 MeV while SK started at a 6.5 
MeV threshold and is currently running at 5 MeV. This means that these 
experiments are able to measure only the $^8$B neutrinos (and the very small 
hep neutrino flux). Their results are presented in terms of measured $^8$B 
flux. 

The final result of Kamiokande (Fukuda {\em et al.}, 1996) and the latest
result of SK (Smy {\em et al.}, 2002) are 
\begin{eqnarray}
\Phi_{\rm Kam} &=& (2.80 \pm 0.19 \pm 0.33) \times 10^6\ 
{\rm cm}^{-2}{\rm s}^{-1} \,, \nonumber \\
\Phi_{\rm SK} &=& (2.35 \pm 0.02 \pm 0.08) \times 10^6\
{\rm cm}^{-2}{\rm s}^{-1} \,, \nonumber
\end{eqnarray}
corresponding to about 40--50\% of the SSM prediction. 

There are three features unique to the water Cerekov detectors. First, they
are real time experiments. Each event is individually recorded. Second, for 
each event the scattered electron keeps the neutrino direction within an 
angular interval which depends on the neutrino energy as 
$\sqrt{2 m_e/E_{\nu}}$. Thus, it is possible, for example, to correlate
the neutrino detection with the position of the Sun. Third, the amount of 
Cerenkov light produced by the scattered electron allows a measurement of
its energy. In summary, the experiment provides information on the time, 
direction and energy for each event. As we discuss below, signatures of 
neutrino oscillations might include distortion of the recoil electron energy 
spectrum, difference between the night-time solar neutrino flux and the
day-time flux, or a seasonal variation in the neutrino flux.
Observation of these effects would be strong evidence in support of
solar neutrino oscillations independent of absolute flux calculations.
Conversely, non-observation of these effects can constrain oscillation 
solutions to the solar neutrino problem.  

Over the years the SK collaboration has presented the information on the 
energy and time dependence of their event rates in different forms. 
In Fig.~\ref{fig:skspec} we show their 
spectrum, corresponding to 1258 days of data, relative to the predicted
spectrum (Ortiz {\it et al.}, 2000) normalized to BP00. As seen from the 
figure, no significant distortion is observed and the data is compatible with a
horizontal straight line. 

The SK Collaboration has also measured the event rates as a function of the
time of day or night or, in other 
words, of the position of the sun in the sky. The results are presented as
a nadir angle (the angle which the sun forms with the vertical) distribution 
of events. In Fig.~\ref{fig:skzen} we show their nadir angle distribution,  
corresponding to 1258 days of data.       
SK has also presented their results on the day-night
variation in the form of a day-night asymmetry,
\begin{equation}
A_{N-D}\equiv 2\frac{N-D}{D+N}=
0.021\pm 0.020 \mbox{(stat.)}\pm 0.013\mbox{(syst.)}
\end{equation}
where $D\; (N)$ is the event rate during the day (night)
period. The SK results show a small excess of events during the night but 
only at the 0.8 $\sigma$ level.

In order to simultaneously account for the energy and day-night variation,
SK present the observed energy spectrum during the day and during the night 
separately as well as in several night bins. This is the most convenient 
observable for including both time and 
energy dependence simultaneously  in the analysis. The reason for that is that
the separate distributions -- energy spectrum averaged in time and time 
dependence averaged in energy-- do not correspond to independent data samples 
and therefore have non-trivial correlations.

Finally, SK has also measured the seasonal dependence of the solar neutrino 
flux. Their results after 1258 days of data  are shown in Fig.~\ref{fig:sksea}.
The points represent the measured flux, and the curve shows the expected 
variation due to the orbital eccentricity of the Earth (assuming no new 
physics, and normalized to the measured total flux). As seen from the figure, 
the data are consistent with the expected annual variation.

\subsubsection{SNO}
The Sudbury Neutrino Observatory (SNO) was first proposed in 1987
and it started taking data in November 1999 
(Mcdonald {\em et al.}, 2000). The detector, a great
sphere surrounded by photomultipliers, contains approximately 1000 
Tons of heavy water, D$_2$O, and is located at the Creighton mine, near 
Sudbury in Canada. It is immersed in ultra pure light water to provide
shielding from radioactivity.
SNO was designed to give a model independent test of the 
possible explanations of the observed deficit in the solar neutrino flux
by having sensitivity to all flavors of active neutrinos and not just to 
$\nu_e$.

This sensitivity is achieved because energetic neutrinos can interact in the 
D$_2$O of SNO via three different reactions. Electron neutrinos may interact 
via the Charged Current (CC) reaction
\begin{equation}
\nu_e + d \to p + p + e^-\ ,
\label{CCreac}
\end{equation}
and can be detected above an energy threshold of a few MeV (presently 
$T_e>$ 5 MeV). All active neutrinos 
($\nu_a=\nu_e,\nu_\mu,\nu_\tau$) interact via the 
Neutral Current (NC) reaction
\begin{equation}
\label{eq:nunc}
\nu_a + d \to n + p + \nu_a  \ ,
\label{NCreac}
\end{equation}
with an energy threshold of 2.225 MeV. The non-sterile neutrinos can also 
interact via Elastic Scattering (ES), $\nu_a + e^- \to \nu_a + e^- $, but
with smaller cross section.

The experimental plan of SNO consists of three phases of approximately one year
duration each. In its first year of operation, SNO has concentrated on the 
measurement of the CC reaction rate while in a following phase, after the 
addition of MgCl$_2$ salt to enhance the NC signal, it will also
perform a precise measurement  of the NC rate. In the third phase, the
salt will be eliminated and a network of proportional counters filled
with $^3$He will be added with the purpose of directly measuring the
neutral current rate $^3$He(n,p)$^3$H.

SNO can also perform measurements of the energy spectrum and time variation of 
the event rates. But the uniqueness of SNO lies in its ability to directly 
test if the deficit of solar $\nu_e$ is due to changes in the flavor 
composition of the solar neutrino beam, since the ratio CC/NC compares the 
number of $\nu_e$ interactions with those from all active flavors. This 
comparison is independent of the overall flux normalization.

In June 2001, SNO published their first results (Ahmad, 2001) on the 
CC measurement and in April 2002 they published their results 
from the first phase of the experiment which include the day-night 
spectrum data in the full energy range above a threshold 
$T_e>$ 5 MeV. Their spectrum is the result of the combination of the
three possible signals.  
Assuming an undistorted energy spectrum they extract the individual rates:   
\begin{eqnarray}
\Phi^{CC}_{\rm SNO}&=&(1.76^{+0.06}_{-0.05} \pm 0.09)\times10^6\ {\rm cm}^{-2}
{\rm s}^{-1} \,, \nonumber \\
\Phi^{ES}_{\rm SNO}&=&(2.39\pm 0.24 \pm 0.12)\times 10^6\ {\rm cm}^{-2}
{\rm s}^{-1} \,, \nonumber \\
\Phi^{NC}_{\rm SNO}&=&(5.09^{+0.44}_{-0.43} \,^{+0.46}_{-0.43}) 
\times 10^6\ {\rm cm}^{-2} {\rm s}^{-1},  \nonumber
\end{eqnarray}
and a day--night asymmetry $A_{N-D}=
0.07\pm 0.049 \mbox{(stat.)}\pm 0.013\mbox{(syst.)}$.

\subsubsection{Future: Borexino and Low Energy experiments}
The Borexino experiment (Oberauer, 1999)
is designed to detect low-energy solar neutrinos in real-time through the 
observation of the ES process $\nu_a + e^- \to \nu_a + e^- $. The energy 
threshold for the recoil electrons is 250 keV. It will use 300 tons of liquid 
scintillator in an unsegmented detector with 2000 photomultiplier tubes.
The event rate predicted by the SSM for a fiducial volume of about 100 tons
is about 50 events per day, mostly generated by the 0.86 MeV monoenergetic line
of $^7$Be solar neutrinos. Since this line gives a characteristic spectral 
signature in the ES process, the flux on the earth of $^7$Be solar neutrinos 
will be determined and it will be possible to check if it is suppressed with 
respect to the one predicted by the SSM as suggested by the results of current 
experiments. The Borexino experiment is
under construction in the Laboratori Nazionali del Gran Sasso in Italy
and is scheduled to start data taking in the near future.

A new generation of experiments aiming at a high precision real time 
measurement of the low energy solar neutrino spectrum is now under study
(de Bellfon, 1999; Lanou {\em et al.}, 1993;
Arpesella, Broggini, and Cattadori, 1996; Raghavan, 1997). Some of them, such 
as HELLAZ, HERON and SUPER--MuNu, intend to detect the elastic scattering of 
the electron neutrinos with the electrons of a gas and  measure the recoil 
electron energy and its direction of emission. The proposed experiment 
LENS plans to detect the electron neutrino via its absorption in a 
heavy nuclear target with the subsequent emission of an electron and a 
delayed gamma emission. 
The expected rates at these experiments for the proposed detector sizes 
are of the order of $\sim 1-10$ $pp$ neutrinos a day. Consequently, with a 
running time of two years, they can reach a sensitivity of a few percent in the
total neutrino rate at low energy, provided that they can achieve sufficient 
background rejection.


\subsection{The Solar Neutrino Problem}
Table~\ref{tab:rates} summarizes our present knowledge of the solar neutrino 
fluxes, their contribution to the expected rates and the data from measurements
of solar neutrino experiments. The predicted event rates are linear functions 
of the seven important neutrino fluxes: $p$-$p$, $pep$, $hep$, ${\rm ^7Be}$, 
${\rm ^8B}$, ${\rm ^{13}N}$, and ${\rm ^{15}O}$. We can make the following
statements (see the last row of the table): 
\begin{itemize} 
\item Before the NC measurement at SNO all experiments observed a flux 
that was smaller than the SSM predictions, $\Phi^{\rm obs}/\Phi^{\rm
  SSM}\sim0.3-0.6$. 
\item The deficit is not the same for the various experiments, 
which may indicate that the effect is energy dependent.
\end{itemize}
These two statements constitute the solar neutrino problem 
(Bahcall, Bahcall and Shaviv, 1968; Bahcall and Davis, 1976). More 
quantitatively, a fit to the data using BP00 shows a disagreement of 6.7 
$\sigma$, which means that the probability of explaining this results as a 
consequence of a statistical fluctuation of standard particle physics 
is $P=3\times 10^{-11}$ (we follow here the latest discussion by  
Bahcall, 2002a). 

One may wonder about the solar model dependence of the solar neutrino problem.
To answer this question one can allow all the solar neutrino fluxes, with 
undistorted energy spectra, to be free parameters in fitting the measured solar
neutrino event rates, subject only to the condition that the total observed 
luminosity of the sun is produced by nuclear fusion. This luminosity constraint
can be written as a convenient linear equation in the neutrino fluxes:
\begin{equation}
1 = \sum_i \left({\alpha_i \over 10~{\rm MeV}}\right)
\left({\phi_i\over 8.532 \times 10^{10}~{\rm cm^{-2}s^{-1}}}\right),
\label{eq:lumconstraint}
\end{equation}
where $\phi_i$ are the individual neutrino fluxes ($i=pp,pep,hep,{\rm ^7Be, 
^8B, ^{13}N}$, and ${\rm ^{15}O}$) and $\alpha_i$ is the luminous energy 
released in the corresponding reaction (the most recent determination of these 
parameters can be found in Bahcall, 2002b). Since there are at present six 
experiments, one cannot use all seven of the neutrino fluxes as free parameters
and some additional constraints must be used. For instance, the ratio of the 
$pep$ to $pp$ fluxes is usually taken to be the same as in the SSM. Also the
CNO nuclear reactions are assumed to be in equilibrium and the $^{13}$N and 
$^{15}$O neutrino fluxes are taken to be exactly equal (or, in some analyses, 
both are taken to be zero). Finally the contribution from hep neutrinos is 
usually neglected.

A fit of this type (Bahcall, 2002b) shows that, previous to the SNO 
measurement,
the best fit corresponded always to an unphysical, negative $^7$Be flux. 
Forcing all fluxes to be positive, the  hypothesis of no-new-physics
was rejected at the effective $2.5\sigma$ level ($99$\% C.L.). 
Furthermore, it required that $^7$Be neutrinos be entirely missing, a result 
which many authors have argued is not physically or astrophysically reasonable 
(for instance, Bahcall and Bethe, 1990; Hata, Bludman, and Langacker, 1994;
Parke, 1995; Heeger and Robertson, 1996; Bahcall, Krastev and Smirnov, 1998).  

The results of SNO have provided further model independent evidence of the 
problem. Both SNO and SK are sensitive mainly to the $^8$B flux. 
Without NP, the measured fluxes in any reaction at these two 
experiments should be equal. 
Villante, Fiorentini and Lisi (1999) showed how one can choose the 
energy thresholds for the SK and SNO experiments in such a way 
that the response functions for the two experiments are made approximately 
equal (see also Fogli {\em et al}, 2001). 
The advantage of this method is that some of the systematic errors are 
reduced, but there is a slight loss of statistical power (Bahcall, 2002a, 
Berezinski, 2001). The first reported SNO CC result 
compared with the ES rate from SK showed that the hypothesis of
no flavour conversion was excluded at $\sim3\sigma$. 

Finally, with the NC measurement at SNO one finds that, 
\begin{equation}
\phi^{\rm SNO}_{\mu\tau}\equiv\phi_{^8B}^{\rm NC, SNO}-\phi_{^8B}^{\rm CC, SNO}=(
3.41\pm 0.45 ^{+0.48}_{-0.45}) \times 10^6\ {\rm cm}^{-2} {\rm s}^{-1}.
\label{snosk}
\end{equation}
This result provides evidence for neutrino flavor transition (from
$\nu_e$ to $\nu_{\mu,\tau}$) at the level of $5.3\sigma$. This
evidence is independent of the solar model.

\subsection{Solar Neutrino Oscillation Probabilities}
\label{subsec:solarprob}
The most generic and popular explanation of the solar neutrino anomaly is
given by the introduction of neutrino masses and mixing, leading to 
oscillations of $\nu_e$ into an active ($\nu_\mu$ and/or $\nu_\tau$) or a
sterile ($\nu_s$) neutrino. We now discuss some issues that have been
raised in the literature concerning the computation of
the corresponding neutrino survival probabilities in the full 
range of mass and mixing relevant to the solar neutrino problem.

\subsubsection{Quasivacuum oscillations and the {\sl Dark Side}}
The presence of neutrino mass and mixing implies the possibility of neutrino 
oscillations. Solar electron neutrinos can undergo oscillations either in 
vacuum or via the matter-enhanced {\it MSW mechanism}, depending on the
actual values of mass-squared differences and mixing angles. However, this 
simplified picture of solar neutrino oscillations contains a set of 
approximations which are not always valid in the context of solar neutrinos. 
In order to clarify this issue let us first review the calculation of
the solar neutrino survival probability in the two-neutrino case.

The survival amplitude for a solar $\nu_e$ of energy $E$ at a terrestrial
detector  can be written as follows:
\begin{equation}
A_{ee}=\sum_{i=1}^2 A^S_{e\,i}\,A^E_{i\,e}\,\exp[-im_i^2 (L-r)/(2E)]~. 
\end{equation}
Here $A^S_{e\,i}$ is the amplitude  of the transition $\nu_e\to\nu_i$ from the 
production point to the surface of the Sun, $A^E_{i\,e}$ is the amplitude of 
the transition  $\nu_i\to\nu_e$ from the surface of the Earth to the detector.
The propagation amplitude in vacuum from the surface of the Sun to the surface 
of the  Earth is given by the exponential term, where $L$ is the distance 
between the center of the Sun and the surface of the Earth, and $r$ is 
the radius of the Sun.  
The corresponding survival probability $P_{ee}$ is then given by:
\begin{equation} 
P_{ee}=P_1P_{1e}+P_2P_{2e}+2\sqrt{P_1P_2P_{1e}P_{2e}}\cos\xi.
\label{Pee}
\end{equation}
Here $P_i\equiv|A^S_{e\,i}|^2$ is the probability that the solar 
neutrinos reach the surface of the Sun  as $|\nu_i\rangle$, 
while $P_{ie} \equiv  |A^E_{i\,e}|^2$ is the probability of $\nu_i$ 
arriving at the surface of the Earth to be detected as a $\nu_e$. 
Unitarity implies $P_1+P_2=1$ and $P_{1e}+P_{2e}=1$. The phase $\xi$ is 
given by 
\begin{equation} 
\xi=\frac{\Delta m^2 (L-r)}{2E}+\delta\, ,
\end{equation}
where $\delta$ contains the phases due to propagation in the Sun and in the 
Earth and can be safely neglected. In the evaluation of both $P_i$ and $P_{ie}$
the effect of coherent forward interaction with the matter of the Sun and the 
Earth must be taken into account.

From  Eq.~(\ref{Pee}) one can recover more familiar expressions for 
$P_{ee}$ that hold in certain limits:  

(1) For $\Delta m^2/E\lsim 5 \times 10^{-17}$ eV, the matter effect 
suppresses flavor transitions in both the Sun and the Earth. 
Consequently, the probabilities $P_1$ and $P_{2e}$ are simply
the projections of the $\nu_e$ state onto the mass eigenstates:
$P_1 = \cos^2\theta$,    $P_{2e} = \sin^2 \theta$. 
In this case we are left with the standard vacuum oscillation formula,
\begin{equation}
P_{ee}^{\rm vac}=1-\sin^2 2\theta \sin^2[\Delta m^2 (L-r)/4E],
\label{pvac}
\end{equation}
which describes the oscillations on the way from the surface of the Sun to the 
surface of the Earth. The probability is symmetric under 
the change of octant,  
$\theta \leftrightarrow \frac{\pi}{2}-\theta$, and change of sign
$\Delta m^2\leftrightarrow -\Delta m^2$ (see Sec.~\ref{sec:oscila}). This 
symmetry implies that each point in the ($\Delta m^2$, $\sin^22\theta$) 
parameter space corresponds to two physically independent solutions, 
one in each octant.  

Averaging Eq.~(\ref{pvac}) over the Earth Orbit, $L(t)=L_0 [ 1 - 
\varepsilon\cos\frac{2\pi t}{T}]$, one gets :
\begin{equation}
\langle P^{\rm vac}_{ee}\rangle=1-\frac{1}{2}\sin^2{2\theta} 
\left[ 1 - \cos\left(\frac{{\Delta{m}^2}{L_0}}{2E}\right) 
{J_{0}}\left(\frac{{ \varepsilon \Delta{m}^2}{L_0}}{{2E}}\right)\right] ,
\label{pvacees}
\end{equation} 
where $\varepsilon=0.0167$ is the orbit eccentricity and $J_0$ is the
Bessel function. In Fig. \ref{fig:probs}.a we display the value of 
$\langle P^{\rm vac}_{ee}\rangle$ as a function of $4E/\Delta{m}^2$.
As seen in the figure, for large values of $\Delta m^2$ the probability
averages out to a constant value $1-\frac{1}{2}\sin^2(2\theta)$.

(2) For $\Delta m^2/E\gsim 10^{-14}$ eV, the last term in Eq.~(\ref{Pee})
vanishes and we recover the incoherent MSW survival probability.
In this case $P_1$ and $P_{2e}$ must be obtained by solving the
evolution equation of the neutrino states in the Sun and the Earth
respectively. As discussed in Sec.~\ref{subsec:nonadiab},
the approximate solution for the evolution in the Sun takes the well-known form 
\begin{equation}
P_1= \frac{1}{2} + \left(\frac{1}{2} - P_{LZ}\right)\cos2\theta_{m,0}.
\label{P1}
\end{equation}
Here $P_{LZ}$ denotes the standard Landau-Zener probability of Eq.~(\ref{plz})
which, for the exponential profile, takes the form shown in Eq.~(\ref{plzexp}),
and $\theta_{m,0}$ is the mixing angle in matter at the neutrino production 
point given in Eq.~(\ref{effmix}). For the approximation of exponential density
profile, $\gamma= \pi \frac{\Delta{m}^2}{E}r_0$, which is independent 
of the point in the Sun where the resonance takes place. 
Improvement over this  constant-slope  exponential density profile 
approximation can be obtained by numerically deriving the exact 
$N_e(r)$ profile at the resonant point. In this case  
$\gamma= \pi \frac{\Delta{m}^2}{E}r_0(r_{res})$. 
 
During the day there is no Earth matter effect. The survival probabilities 
$P_{ie}$ are obtained by simple projection of the $\nu_e$ state onto the mass 
eigenstates: $P_{2e,D}=1-P_{1e,D}=\sin^2\theta$. One obtains the expression in
Eq.~(\ref{osc:pmsw})
\begin{equation}
P^{\rm MSW}_{ee,D}=\frac{1}{2}+\left(\frac{1}{2}- 
P_{LZ}\right)\cos2\theta_{m,0}\cos2\theta.
\label{pmsw}
\end{equation}
In Fig.~\ref{fig:probs}.b (\ref{fig:probs}.c) we plot the this survival 
probability as a function of $4E/\Delta m^2$ for a large (small) mixing angle.
  
Let us make some remarks concerning Eq.~(\ref{pmsw}):

$(i)$ In both the limits of large and very small $E/\Delta m^2$, 
$P^{\rm MSW}_{ee}\to1-\frac{1}{2}\sin^22\theta$ (see Fig.~\ref{fig:probs}.b)
which is the same expression as for averaged vacuum oscillations.
This result, however, comes from very different reasons in the two regimes. 
For large $E/\Delta m^2$, $P_{LZ}=\cos^2\theta$ and $\cos2\theta_{m,0}=-1$. 
For small $E/\Delta m^2$, $P_{LZ}=0$ and $\cos2\theta_{m,0}=\cos2\theta$. 

$(ii)$ Due to matter effects, $P^{\rm MSW}_{ee}$ is only symmetric under 
the simultaneous transformation $(\Delta m^2,\theta)\to  
(-\Delta m^2, \frac{\pi}{2}-\theta)$. 
For ${\Delta m^2>0}$, the resonance is only possible for 
$\theta <\frac{\pi}{4}$ and MSW solutions are usually plotted in the 
$(\Delta m^2, \sin^22\theta)$ plane assuming that now each point
on this plane represents only one physical solution with $\theta$ in
the first octant. But in principle non-resonant solutions are also 
possible for $\theta >\frac{\pi}{4}$, the so called {\it dark side}
(Fogli, Lisi and Montanino, 1994 and 1996; 
de Gouvea, Friedland and Murayama, 2000; Gonzalez-Garcia and
Pe\~na-Garay, 2000; Fogli, Lisi, Montanino and Palazzo, 2000a). 

$(3)$ The analysis of the survival probability of solar neutrinos would
be simplified if there were a region where $E/\Delta m^2$ is small enough that
vacuum oscillations are averaged out and large enough to be in the extremely
non-adiabatic MSW regime. This, however, is not the case, as can be
seen form comparison of panels (a) and (b) of Fig.\ref{fig:probs}. 
In the intermediate range, $2\times 10^{14}\lsim4E/\Delta m^2\lsim10^{16}$ 
eV$^{-1}$, adiabaticity is violated and the $\cos\xi$ coherent term should be 
taken into account. The result is similar to vacuum oscillations 
but with small matter corrections. We define this case as quasi-vacuum
oscillations (QVO) 
(Petcov, 1988b; Petcov and Rich, 1989; 
Pantaleone, 1990; Pakvasa and
Pantaleone, 1990; Friedland 2000; Fogli, Lisi, Montanino, and Palazzo 2000b;
Lisi, Marrone, Montanino, Palazzo and Petcov, 2001). 
The range of $E/\Delta m^2$ for the QVO regime 
depends on the value of $E/\Delta m^2$ for which the MSW probability in 
Eq.~(\ref{pmsw}) acquires the asymptotic value 
$1-\frac{1}{2}\sin^22\theta$: 
the smaller $E/\Delta m^2$ the more separated the MSW and 
vacuum regimes are, and the narrower the QVO region is. 

It is clear from these considerations that in order to compute the survival 
probability for solar neutrinos that would be valid for any experiment and 
any value of the neutrino mass and mixing, the full expression (\ref{Pee}) has
to be evaluated. The results that we show here were obtained using
the general expression for the survival probability in Eq.~(\ref{Pee})
with $P_1$ and $P_{2e}$ found by numerically solving the evolution equation in 
the Sun and the Earth. In the Sun the evolution equation has to be 
solved from the neutrino production point to the edge of the Sun and averaged 
over the production point distribution shown in Fig.~\ref{fig:ppprod}.

\subsubsection{Evolution in the Earth}
In this section we discuss matter effects in the Earth (for some of the 
original works see J. Bouchez {\it et. al.} 1986,  
Mikheyev and Smirnov, 1987, A. J. Baltz and J. Weneser 1987 and 1988).  
Unlike the Sun, the Earth's matter density is not strongly varying. It 
consists mainly of two layers, the lower-density mantle and the higher-density 
core, each with approximately constant density. The core radius is $L=2896$ km.

When a neutrino crosses only the mantle, the Earth matter effects are 
well approximated by evolution with a constant potential. In this case, the 
resulting probability simplifies to:
\begin{equation}
P_{2e}=\sin^2\theta+\frac{4 E V_e}{\Delta m^2}\sin^2 2\theta_E
\sin^2 {\frac{\pi L}{L_m}}\,, 
\label{regen1}      
\end{equation}
where $\theta_E$ is the mixing angle in the Earth [Eq.~(\ref{effmix})], 
$L$ is the distance traveled by the neutrino within the Earth and 
$L_m$ is the oscillation length in matter [Eq~(\ref{efflosc})]. 

The right hand side of Eq.~(\ref{regen1}) consists of two terms. The first term
gives a simple projection of the mass to flavor state in vacuum (which 
corresponds to the probability during the day time). The second is the
{\it regeneration term} and will be denoted by $f_{\rm reg}$ (for details see 
Gonzalez-Garcia, Pe\~na-Garay and Smirnov, 2001). It contains the Earth matter 
effects and it is always positive. Averaging out the 
distance dependent term, we can now rewrite Eq.~(\ref{Pee}):
\begin{equation}
P^N_{ee} = P^D_{ee} -  (1 - 2 P_{LZ}) \cos2\theta_{m,0} f_{\rm reg}\,, 
\label{probtot}
\end{equation}
where $P^D_{ee}$ ($P^N_{ee}$) is the $\nu_e$ survival probability during the 
day (night). Since $f_{\rm reg}>0$ and, for the interesting 
cases of MSW transitions in the Sun, $\cos2\theta_{m,0}<0$, the relative
magnitude of $P^D_{ee}$ and $P^N_{ee}$ depends on the sign of $(1-2P_{LZ})$.
For $P_{LZ}<0.5$ (and, in particular, for adiabatic MSW transitions,
$P_{LZ} = 0$) the survival probability during the night is larger than during 
the day. The opposite holds for large adiabaticity violations, $P_{LZ} > 0.5$. 

When neutrinos cross the core, adiabaticity may be violated in their evolution.
The abrupt density change between the mantle and the core may induce a new form
of resonance, different from the MSW resonance (Petcov, 1998; Akhmedov, 1999) 
This {\it parametric resonance} (Ermilova, Tsarev and Chechin, 1986; 
Akhmedov, 1987) is relevant mainly for small mixing angles.
 

\subsection{Two-neutrino Oscillation Analysis}
\label{subsec:sol2osc}
\subsubsection{Predictions}
The expected event rate in the presence of oscillations in the experiment $i$, 
$R^{\rm th}_i$, can be written as follows:
\begin{equation}
R^{\rm th}_i=\sum_{k=1,8} \phi_k\int\! dE_\nu\, \lambda_k (E_\nu) \times 
\big[ \sigma_{e,i}(E_\nu)  \langle P_{ee} (E_\nu,t) \rangle \label{ratesth}
+\sigma_{x,i}(E_\nu) (1-\langle P_{ee} (E_\nu,t)\rangle )\big],
\end{equation}  
where $E_\nu$ is the neutrino energy, $\phi_k$ is the total neutrino flux and 
$\lambda_k$ is the neutrino energy spectrum (normalized to 1) from the solar 
nuclear reaction $k$. $\sigma_{e,i}$ ($\sigma_{x,i}$) is the $\nu_e$ ($\nu_x$, 
$x=\mu,\,\tau$) interaction cross section in the SM
with the target corresponding to 
experiment $i$, and $\langle P_{ee} (E_\nu,t)\rangle$ is the time-averaged 
$\nu_e$ survival probability. The expected signal in the absence of 
oscillations, $R^{\rm BP00}_i$, can be obtained from Eq.~(\ref{ratesth}) by 
substituting $P_{ee}=1$. In Table~\ref{tab:rates} we give the expected 
rates at the different experiments which we obtain using the fluxes of 
Bahcall, Pinsonneault, and Basu (2001).

For the Chlorine, Gallium and SNO(CC) measurements, only the electron neutrino
contributes and the $\sigma_{x,i}$-term in Eq.~(\ref{ratesth}) vanishes. For 
ES at SK or SNO  there is a possible contribution from the NC interaction 
of the other active neutrino flavors present in the beam. For the NC 
rate at SNO all active flavours contribute equally.
For absorption in chlorine and gallium, 
the cross sections $\sigma_{e,i}(E)$ can be found in  Bahcall (1997). The 
cross sections for SNO and SK(ES) can be obtained from the differential 
cross sections of (Nakamura {\em et al.}, 2001) and  
(Bahcall, Kamionkowsky, and  Sirlin, 1995), respectively, by integrating with 
the corresponding detector resolutions and for the given detection thresholds. 
In particular, for the SK and SNO energy spectrum data, to obtain 
$R^{\rm th}$ in a 
given energy bin one has to integrate within the corresponding electron recoil 
energy bin and to take into account that the finite energy resolution implies 
that the {\it measured } kinetic energy $T$ of the scattered electron is
distributed around the {\it true } kinetic energy $T^\prime$ according to a
resolution function $Res(T,\,T^\prime)$ of the form (from Bahcall, Krastev and
Lisi, 1997):
\begin{equation}
Res(T,\,T^\prime) = \frac{1}{\sqrt{2\pi}s}\exp\left[
{-\frac{(T-T^\prime)^2}{2 s^2}}\right]\ ,
\end{equation}
where $s = s_0\sqrt{T^\prime/{\rm MeV}}$, and $s_0=0.47$ MeV for 
SK (from Faid {\em et al.}, 1997). On the other hand, the 
distribution of the true kinetic energy $T^\prime$ for an interacting neutrino 
of energy $E_\nu$ is dictated by the differential cross section 
$d\sigma_\alpha(E_\nu,\,T^\prime)/dT^\prime$ and the kinematic limits are:
\begin{equation}
0\leq T^\prime\leq {\overline T}^\prime(E_\nu)=\frac{E_\nu}{1+m_e/2E_\nu}\ .
\end{equation}
For assigned values of $s_0$, $T_{\rm min}$, and $T_{\rm max}$, the
corrected cross section $\sigma_{a}(E)$ $(a = e,\,x)$ is given as
\begin{equation}
\sigma_{a}(E_\nu)=\int_{T_{\rm min}}^{T_{\rm max}}\!dT
\int_0^{{\overline T}^\prime(E_\nu)}\!dT^\prime\,Res(T,\,T^\prime)\,
\frac{d\sigma_{\alpha}(E_\nu,\,T^\prime)}{dT^\prime}\ .
\label{sigma}
\end{equation}

For data taken within a given zenith angle period $i$, the expected number 
of events in the presence of oscillations is
\begin{eqnarray} 
R^{\rm th}_i = \frac{\displaystyle 1}{\displaystyle\Delta \tau_i}
\int_{\tau(cos\Phi_{{\rm min},i})}^{\tau(cos\Phi_{{\rm max},i})}d\tau
\sum_{k=1,8} \phi_k\int\! dE_\nu\, \lambda_k (E_\nu) &\times& 
\big[ \sigma_{e,i}(E_\nu) \langle P_{ee} (E_\nu,\tau) \rangle  
\label{eq:daynight}\\ 
& &+ \sigma_{x,i}(E_\nu)(1-\langle P_{ee} (E_\nu,\tau)\rangle )\big], \nonumber
\end{eqnarray}  
where $\tau$ measures the annual averaged length of the period $i$ 
normalized to 1: $\Delta\tau_i=\tau(\cos\Phi_{{\rm max},i})-\tau
(\cos\Phi_{{\rm min},i})=$. For instance, if the period $i$ corresponds to
the entire day ($i=D$) or night ($i=N$) then
$\Delta\tau_{D(N)}=0.5$. From these predictions one can easily build the 
corresponding expected day-night asymmetry. In Fig.~\ref{fig:sk_adn} we show 
the isocontours of expected $A_{N-D}$ at SK 
in the $(\Delta m^2,\tan^2\theta)$ plane
for active neutrino oscillations (the results are very similar for
sterile neutrinos). As discussed in Sec.~\ref{subsec:solarprob}, in most of the
parameter space, the effect of traveling across the Earth is the regeneration 
of the $\nu_e$ component, resulting in a positive day-night asymmetry with the
exception of the region where the non-adiabaticity of the oscillations in the 
Sun is important ($\tan^2\theta\sim 10^{-3},\Delta m^2\sim3\times 10^{-6}$ 
eV$^2$). As discussed above, SK has observed a very small day-night asymmetry.
This observation implies that some of the oscillation parameters space 
can be excluded as we discuss below.
  
In the same fashion, integrating Eq.~(\ref{eq:daynight}) for the different  
electron recoil energy bins and for the day and night periods separately,
one can obtain the predictions for SK and SNO  day-night spectrum. 

\subsubsection{Analysis of total event rates: allowed masses and mixing} 
\label{subsubsec:solarrates}
The goal of the analysis of the solar neutrino data in terms of neutrino 
oscillation is to determine which range of mass-squared difference and mixing
angle can be responsible for the observed deficit (see for instance Hata and 
Langacker, 1997; Fogli, Lisi and Montanino, 1998; Bahcall, Krastev and Smirnov,
1998; Gonzalez-Garcia {\em et al.}, 2000). In order to answer this question in 
a statistically meaningful way one must compare the predictions in the 
different oscillation regimes with the observations, including all the sources
of uncertainties and their correlations, 
by building, for instance, the $\chi^2$ function
\begin{equation}
\chi^2_R=\sum_{i} (R^{\rm th}_i- R^{\rm exp}_i)
\sigma_{ij}^{-2} (R^{\rm th}_j- R^{\rm exp}_j).
\end{equation}
Here $R^{\rm th}_i$ is the theoretical expectation (\ref{ratesth}),
$R^{\rm exp}_i$ is the observed number of events in the experiment $i$, and 
$\sigma_{ij}$ is the error matrix which contains both the theoretical
uncertainties and the experimental statistical and systematic errors.
The main sources of uncertainty are the theoretical errors in the 
prediction of the solar neutrino fluxes for the different
reactions. These errors are due to uncertainties in the twelve basic
ingredients of the solar model, which include the nuclear reaction rates
(parametrized in terms of the astrophysical factors 
$S_{11}$, $S_{33}$, $S_{34}$, $S_{1,14}$ and $S_{17}$), 
the solar luminosity, the metalicity $Z/X$, the sun age, the opacity,
the diffusion, and the electronic capture of  $^7$Be, $C_{\rm Be}$.
These errors are strongly correlated as the same astrophysical 
factor can affect the  different neutrino production rates. 
Another source of theoretical error arises from the uncertainties in the
neutrino interaction cross section for the different detection
processes. For a detailed description of the way to include all these
uncertainties and correlations in the construction of $\sigma_{ij}$ we refer 
to the work of Fogli and Lisi (1995a).  Updated uncertainties can be found in 
Fogli {\em et al.} (2000a) and Gonzalez-Garcia and Pe\~na-Garay (2001).

The results of the analysis of the total event rates are shown in 
Fig.~\ref{fig:xirates} where we plot the allowed regions which correspond to 
90\%, 95\%, 99\% and 99.73\% (3$\sigma$) CL for 
$\nu_e$ oscillations into active neutrinos.

As seen in the figure,  for oscillations into active neutrinos
there are several oscillation regimes 
which are compatible within errors with the experimental data. These allowed 
parameter regions are denoted as {\it MSW small mixing angle} (SMA), 
{\it MSW large mixing angle} (LMA), 
{\it MSW low mass} (LOW) and {\it vacuum oscillations} (VAC). 
Before including the SNO(CC) data, the best fit corresponded to the SMA 
solution, but after SNO the best fit corresponds to the LMA solution. 
For the LMA solution, oscillations for the $^8$B neutrinos occur in the
adiabatic regime and the survival probability is higher for lower energy
neutrinos. This situation fits well the higher rate observed at gallium 
experiments. For the LOW solution, the situation is opposite but
matter effects in the Earth for pp and $^7$Be neutrinos enhance the
average annual survival probability for these lower energy
neutrinos. The combination of these effects still
allows a reasonable description of the Gallium rate. 

Oscillations into pure sterile neutrinos are strongly disfavour by the
SNO data since they  would imply very similar neutrino 
fluxes for the NC, CC and ES rates in SNO as well as the ES rate at SK. 
Schematically, in presence of oscillations,
\begin{eqnarray}
\Phi^{\rm CC}&=&\Phi_e, \nonumber \\
\Phi^{\rm ES}&=&\Phi_e\,+\, r\,\Phi_{\mu\tau}, \\
\Phi^{\rm NC}&=&\Phi_e+ \Phi_{\mu\tau}, \nonumber 
\end{eqnarray}
where 
$r\equiv \sigma_{\mu}/\sigma_{e}\simeq 0.15$ 
is the ratio of the the $\nu_e - e$ and $\nu_{\mu} - e$ elastic 
scattering cross-sections. The flux
$\Phi_{\mu\tau}$ of active no-electron neutrinos in the beam
would vanish if $\nu_e$ oscillate into a purely sterile state.
Thus, if the beam comprises of only $\nu_e$'s and $\nu_s$'s, the three observed
rates should be equal (up to effects due to spectral distortions), 
an hypothesis which is now ruled out at more than $5$-sigma by the SNO
data.
Oscillations into an admixture of active and sterile states are still allowed
provided that the $^8$B neutrino flux is allowed to be larger than the
SSM expectation (Barger, Marfatia, and Whisnant, 2002; Bahcall,
Gonzalez-Garcia and Pe\~na-Garay, 2002b). 
%
\subsubsection{Day-Night spectra: excluded masses and mixing} 
Further information on the different oscillation regimes can be
obtained from the analysis of the energy and time dependence
data from SK (for an early suggestion of the use of the day-night
spectra see Maris and Petcov, 1997),  
which is currently presented in the form
of the observed day-night or zenith spectrum  (Fukuda {\em et al}., 2001:
Smy {\em et al}., 2002).
The way to statistically treat these data has been discussed for instance 
in  Fukuda {\em et al}. (2001); Bahcall, Krastev, and Smirnov (2000);
Gonzalez-Garcia {\em et al.} (2000); Gonzalez-Garcia and Pe\~na-Garay (2001).

The observed day-night spectrum in SK is essentially undistorted 
in comparison to the SSM expectation and shows no significant
differences between the day and the night periods. Consequently,
a large region of the oscillation parameter space where these
variations are expected to be large can be excluded. 
In Fig.~\ref{fig:skspec} we show the SK spectrum corresponding to 1258 days of 
data relative to the Ortiz {\it et al.} (2000) spectrum normalized to BP00,
together with the expectations from the best fit points for the LMA,
SMA, LOW and VAC solutions.

The various solutions give different predictions for the day-night spectrum
For LMA and LOW, the expected spectrum is very little distorted. 
For SMA, a positive slope is expected, with larger slope for larger mixing 
angle within SMA. For VAC, large distortions are expected. The details are 
dependent on the precise values of the oscillation parameters.

In Fig.~\ref{fig:xispdn} we show the excluded regions at the 99\% CL
from the analysis of SK day-night spectrum data, together with the contours
corresponding to the 95\% and 99.73\% (3$\sigma$) CL. In particular, the 
central region ($2 \times 10^{-5}<\Delta m^2<3 \times 10^{-7}$,  
$\tan^2 \theta>3\times10^{-3}$) is excluded due to the small observed
day-night variation (compare to Fig.~\ref{fig:sk_adn}). The rest of the 
excluded region is due to the absence of any observed distortion of the
energy spectrum.

Superimposing Fig.~\ref{fig:xispdn} and Fig.~\ref{fig:xirates} we can
deduce the main consequences of adding the day-night spectrum information
to the analysis of the total event rates:
\begin{itemize}
\item SMA: within this region, the part with larger mixing angle 
fails to comply with the observed energy spectrum, while the part with
smaller mixing angles gives a bad fit to the total rates. 
\item VAC: the observed flat spectrum cannot
be accommodated.
\item LMA and LOW:  the small $\Delta m^2$ part of LMA and the 
large $\Delta m^2$ part of LOW are reduced because they predict a
day-night variation that is larger than observed. Both active LMA and active 
LOW solutions predict a flat spectrum in agreement with the observation.
\end{itemize}

\subsubsection{Global analysis}
\label{solar:solar} 
In order to quantitatively discuss the combined analysis of the 
full bulk of solar neutrino data, one must define a global statistical 
function. How to implement such a program is a question that 
is being discussed  intensively in the literature 
(Fogli, Lisi, Marrone, Montanino and Palazzo 2002;  
Bahcall, Gonzalez-Garcia, and Pe\~na-Garay, 
2001a, 2002a, 2002c; Garzelli and Giunti, 2001;  de Holanda and 
Smirnov, 2002; Creminelli, Segnorelli and Strumia, 2001; 
Barger, Marfatia, Whisnant and Wood, 2002;
Maltoni, Tortola, Schwetz, Valle 2002). The situation at present 
is that, although one finds slightly different globally allowed regions, 
depending on the particular prescription used in the combination, the 
main conclusions are independent of the details of the analysis.

We show  in Fig.~\ref{fig:xiglobal} and Table~\ref{tab:xiglobal} 
the results from the global analysis of Bahcall, Gonzalez-Garcia
and Pe\~na-Garay, (2002c). 
In Fig.~\ref{fig:xiglobal} we show the allowed regions which correspond to 
90\%, 95\%, 99\% and 99.73\% (3$\sigma$) CL for $\nu_e$ oscillations 
into active neutrinos. In the table we list the local minima of the allowed 
regions, the value of $\chi^2_{\rm min}$ in each local minimum. 

The results show that at present the most favoured solution is the  
LMA oscillation while the LOW solution provides a worse fit. 
There are some  small allowed {\it islands} for VAC oscillations. 
The active SMA
solution does not appear at 3$\sigma$ as a consequence of the incompatibility 
between the observed small CC rate at SNO, which would favour larger
mixing, and the flat spectrum which prefers smaller mixing. 
Oscillations of solar neutrinos into pure sterile state are also 
disfavoured at the 5.4$\sigma$ level. 
As explained below, in 
Sec.~\ref{sec:schemes}, this difference between active and 
sterile oscillations
has important implications on 4-$\nu$ mixing schemes.

\section{Atmospheric Neutrinos}
\label{sec:atmos}
Atmospheric neutrinos are produced in cascades initiated by collisions
of cosmic rays with the Earth's atmosphere. Some of
the mesons produced in these cascades, mostly pions and some kaons, decay
into electron and muon neutrinos and anti-neutrinos.
The predicted absolute fluxes of neutrinos produced by cosmic-ray
interactions in the atmosphere are uncertain at the 20\% level.  The
ratios of neutrinos of different flavor are, however, expected to be
accurate to better than 5\%.  Since $\nu_e$ is produced mainly from
the decay chain $\pi \to \mu \nu_\mu$ followed by $\mu \to e \nu_\mu\nu_e$, 
one naively expects a $2:1$ ratio of $\nu_\mu$ to $\nu_e$. (For higher energy
events the expected ratio is smaller because some of the muons arrive to Earth
before they had time to decay.) In practice, however, the theoretical 
calculation of the ratio of muon-like interactions to electron-like 
interactions in each experiment is more complicated.

Atmospheric neutrinos are observed in underground experiments using 
different techniques and leading to different type of events, which we briefly 
summarize here. They can be detected by the direct observation of their charged
current interaction inside the detector. These are the {\it contained} events. 
Contained events can be further classified into {\it fully contained} 
events, when 
the charged lepton (either electron or muon) that is produced in the neutrino 
interaction does not escape the detector, and {\it partially contained} 
muons, when 
the produced muon exits the detector. For fully contained events the flavor, 
kinetic energy and direction of the charged lepton can be best determined.
As discussed later, some experiments further divide the contained data sample 
into sub-GeV and multi-GeV events, according to whether the visible energy 
is below or above 1.2 GeV. On average, sub-GeV events arise from neutrinos of 
several hundreds of MeV while multi-GeV events are originated by neutrinos with
energies of the order of several GeV. Higher energy muon neutrinos
and antineutrinos can also be detected indirectly by observing the muons
produced in their charged current interactions in the vicinity of the
detector. These are the so called {\it upgoing muons}. Should the muon 
stop inside the detector, it is classified as a {\it stopping muon}
(which arises from neutrinos $E_\nu\sim 10$ GeV),
while if the muon track crosses the full detector the event is 
classified as a {\it through-going muon} (which is originated by neutrinos
with energies of the order of hundreds of GeV). Downgoing muons from 
$\nu_\mu$ interactions above the detector cannot be distinguished from the
background of cosmic ray muons. Higher energy $\nu_e$'s cannot be detected
this way as the produced $e$ showers immediately in the rock. In 
Fig.~\ref{fig:events} we display the characteristic neutrino energy 
distribution for these different type of events (taken from Engel, Gaisser and 
Stanev, 2000). 

Atmospheric neutrinos were first detected in the 1960's by the underground 
experiments in South Africa (Reines {\em et al.}, 1965)
and the Kolar Gold Field experiment in India (Achar {\em et al.}, 1965).
These experiments measured the flux of horizontal muons (they could
not discriminate between downgoing and upgoing directions) and 
although the observed total rate was not in full agreement with theoretical
predictions  (Zatsepin and Kuzmin, 1962; Cowsik {\em et al.}, 1965; Osborne, 
Said and Wolfendale, 1965), the effect was not statistically significant. 

A set of modern experiments were proposed and built in the 1970's and 
1980's. The original purpose was to search for nucleon
decay, for which atmospheric neutrinos constitute background, although
the possibility of using them to search for oscillation was also 
known (Ayres {\em et al}, 1984). Two different detection techniques
were employed. In water Cerenkov 
detectors the target is a large volume of water surrounded by 
photomultipliers which detect the Cerenkov-ring produced by the charged 
leptons. The event is classified as an electron-like (muon-like) event if the 
ring is diffuse (sharp). In iron calorimeters, the detector is composed of a 
set of alternating layers of iron which act as a target and some tracking 
element (such as plastic drift tubes) which allows the reconstruction of
the shower produced by the electrons or the tracks  produced by muons.
Both types of detectors allow for flavor classification of the events.

The two oldest iron calorimeter experiments, Fr\'ejus 
(Daum {\em et al.}, 1995) 
and NUSEX (Aglietta {\em et al.}, 1989), found atmospheric neutrino fluxes 
in agreement with the theoretical predictions. On the other hand, two water 
Cerenkov detectors, IMB (Becker-Szendy {\em et al.}, 1992) and  Kamiokande, 
detected a ratio of $\nu_\mu$-induced events to $\nu_e$-induced
events smaller than the expected one by a factor of about 0.6. 
Kamiokande performed separate analyses for both sub-GeV neutrinos
(Hirata {\em et al.}, 1992) and multi-GeV neutrinos 
(Fukuda {\em et al.}, 1994), which showed the
same deficit. This was the original formulation of the atmospheric 
neutrino anomaly. In Fig.~\ref{fig:R} we show the values of the
ratio $R_{\mu/e}/R^{MC}_{\mu/e}$, which denotes the double ratio of
experimental-to-expected ratio of muon-like to electron-like
events. Whether $R_{\mu/e}/R^{MC}_{\mu/e}$ is small because there is 
$\nu_\mu$ disappearance or $\nu_e$ appearance or a combination of both could 
not be determined. Furthermore, the fact that the anomaly appeared only in 
water Cerenkov and not in iron calorimeters left the window open for the 
suspicion of a possible systematic problem as the origin of the effect. 

Kamiokande also presented the zenith angular dependence of the deficit for the 
multi-GeV neutrinos (Fukuda {\em et al.}, 1994). The zenith angle,
parametrized in terms of $\cos\theta$, measures the direction
of the reconstructed charged lepton with respect to the vertical
of the detector. Vertically downgoing (upgoing) particles 
correspond to $\cos\theta=+1(-1)$. Horizontally arriving particles
come at $\cos\theta=0$. Kamiokande results seemed to indicate that the deficit
was mainly due to the neutrinos coming from below the horizon. 
Atmospheric neutrinos are produced isotropically at a distance of about 15 km
above the surface of the Earth. Therefore neutrinos coming from
the top of the detector have traveled approximately those 15 kilometers
before interacting  while those coming from the bottom of the detector 
have traversed the full diameter of the Earth, $\sim 10^4$ Km 
before reaching the detector. The Kamiokande distribution suggested
that the deficit increases with the distance between the 
neutrino production and interaction points.

In the last five years, the case for the atmospheric neutrino anomaly has 
become much stronger with the high precision and large statistics 
data from SK  and it has received important confirmation
from the iron calorimeter detectors Soudan2 and MACRO. 

SK is a 50 kiloton water Cerenkov detector constructed under Mt.
Ikenoyama located in the central part of Japan, giving it a rock over-burden of
2,700 m water-equivalent. The fiducial mass of the detector for atmospheric 
neutrino analysis is 22.5 kilotons. In June 1998, in the Neutrino98 
conference, SK presented  {\it evidence} of $\nu_\mu$ oscillations 
(Fukuda {\em et al.}, 1998) based on the angular distribution for their 
contained event data sample.  Since then SK accumulated more statistics 
and has also studied the angular dependence of the upgoing muon sample 
(Fukuda  {\em et al.}, 1999a and 1999b). 
In Fig.~\ref{fig:skatm} (from Toshito {\em et al.}, 2001) we show their data 
at summer 2001, corresponding to 79 kiloton year (1289 days) exposure for their
contained sub-GeV (2864 1-ring  $e$-like events and 2788 1-ring $\mu$-like 
events) and multi-GeV (626 1-ring $e$-like events, 558 1-ring
$\mu$-like events and 754 PC events), as well as the stopping-muons and
through-going muon samples (1416 events). 

In the figure we show the angular zenith distribution of the different samples. 
Comparing the observed and the expected (MC) distributions, we can
make the following statements:

(i) $\nu_e$ distributions are well described by the MC while $\nu_\mu$ presents
a deficit. Thus the atmospheric neutrino deficit is mainly due to disappearance
of $\nu_\mu$ and not the appearance of $\nu_e$.

(ii) The suppression of contained $\mu$-like events is stronger for larger 
$\cos\theta$, which implies that the deficit grows with the distance traveled 
by the neutrino from its production point to the detector. This effect is more 
obvious for multi-GeV events because at higher energy the direction of the 
charged lepton is more aligned with the direction of the neutrino. It
can also be described in terms of an up-down asymmetry:
\begin{equation} 
A_\mu\equiv\frac{U-D}{U+D}= -0.316\pm0.042({\rm stat}.)\pm 0.005({\rm syst}.)
\label{aud}
\end{equation} 
where $U$ $(D)$ are the contained $\mu$-like events with zenith angle
in the range $-1<\cos\theta<-0.2$  ($0.2<\cos\theta<1$). 
It deviates from the SM value, $A_\mu=0$, by 7.5 standard deviations.

(iii) The overall suppression of the flux of stopping-muons, $\Phi_{\rm ST}$,
is by a factor of about 0.6, similar to contained events. However, for 
the flux of through-going muons, 
$\Phi_{\rm TH}$, the suppression is weaker, which 
implies that the effect is smaller at larger neutrino energy. This effect is 
also parametrized in terms of the double flux ratio:
\begin{equation}
\frac{{\Phi_{\rm ST}}/{\Phi_{\rm TH}}|_{\rm obs}}
{{\Phi_{\rm ST}}/{\Phi_{\rm TH}}|_{\rm MC}}=
0.635 \pm 0.049 ({\rm stat.})\pm 0.035 ({\rm syst.})\pm 0.084 ({\rm theo.})
\label{rup}
\end{equation}
which deviates from the SM value of 1 by about 3 standard deviations.

These effects have been confirmed by the results of the iron calorimeters 
Soudan2 and MACRO which removed the suspicion that the atmospheric
neutrino anomaly is simply a systematic effect in the water detectors.
In particular Soudan2, which has mainly analyzed contained
events (Allison {\em et al.}, 1999), have measured a ratio 
$R_{\mu/e}/R^{\rm MC}_{\mu/e}=0.68\pm 0.11\pm 0.06$, in good agreement
with the results from the water Cerenkov experiments. 
The main results from MACRO concern the angular distribution for 
through-going muons (see Ambrosio {\em et al.} (2001) for their latest data)
and shows good agreement with the results from SK. The Baksan experiment 
(Boliev {\em et al.}, 1999) has also reported results on the angular 
distribution of through-going muons but their data is less precise.

To analyze the atmospheric neutrino data in terms of oscillations one needs to 
have a good understanding of the different elements entering into the 
theoretical predictions of the event rates: the atmospheric neutrino fluxes
and their interaction cross section, which we describe next.  

\subsection{Atmospheric Neutrino Fluxes}
Modern calculations of atmospheric neutrino fluxes consist
of a Monte Carlo procedure that combines the measured energy spectra and
chemical composition of the cosmic ray flux at the top of the
atmosphere with the properties of their hadronic interaction with the light
atmospheric nuclei followed by the neutrino production from secondary
$\pi$, $K$ and $\mu$ decay.

Present experiments use the neutrino flux calculations from mainly Honda 
{\em et al.} (1990,1995,1996), the Bartol group (Gaisser, Stanev and Barr, 
1988;  Barr, Gaisser and Stanev , 1989; Agraval {\em et al.}, 1996; Lipari, 
Gaisser, and Stanev, 1998) and Fiorentini, Naumov and Villante (2001).  
These calculations have in common a one-dimensional picture, in which all 
secondary particles in the showers, neutrinos included, are considered 
collinear with the primary cosmic rays (see also Bugaev and Naumov, 1989). 
The three-dimensional picture was first considered by Lee and Koh (1990) and
most recently by Batistoni {\em et al.} (2000) and Tserknyav {\em et al.} 
(1999). [For detailed comparisons between the various simulations, see Gaisser 
{\em et al.} (1996), Lipari (2000) and Battistoni (2001)].

The flux of neutrinos of flavor $i$ coming from the direction $\Omega$ can be
schematically written as
\begin{equation}
\phi_{\nu_i}(\Omega)\;=\sum_A \Phi_A\otimes R_A\otimes Y_{p,n\rightarrow\nu_i}.
\end{equation}
Here $\Phi_A$ is the primary cosmic-ray spectrum, $R_A$ is the geomagnetic 
cutoff for the protons or light nuclei incident on the atmosphere from the 
direction $\Omega$, and $Y_{p,n\rightarrow\nu_i}$ is the yield per nucleon of 
$\nu_i$. The separation to different nuclear species of the cosmic ray spectrum 
is necessary because the neutrino yield depends on the energy-per-nucleon of 
the incident cosmic rays but the geomagnetic cutoff depends on the magnetic 
rigidity  (${\cal R}=pc/Ze$) which, at the same energy per nucleon, depends 
on $A/Z$. 

We now briefly summarize the uncertainties in each of these three ingredients
of the calculation, namely the primary cosmic ray flux, the geomagnetic effects
and the hadronic interactions on light nuclei.

\subsubsection{Cosmic ray spectrum}
The cosmic radiation incident at the top of the atmosphere includes all 
stable charged particles and nuclei. Apart from particles associated 
with solar flares, the cosmic rays are assumed to come from outside 
the solar system. In Fig.~\ref{fig:crspec} (from Simpson, 1983) we show a 
compilation of the major components of the cosmic rays as a function of the 
energy per nucleon. The following features of the cosmic ray spectrum 
are relevant to the atmospheric neutrino fluxes:

(i) Composition: Most of the primaries are protons although the chemical
composition varies with energy at low energies. Above a few GeV of 
energy-per-nucleon, about 79\% of the primaries are protons while helium, the 
next most abundant component, is about 3\%. This fraction remains constant 
till beyond 100 TeV.

(ii) Energy Dependence: For energy above a few GeV, the cosmic ray flux is a 
a steeply falling function of the energy: $d\phi/dE\propto E^{-3.7}$.

(iii) Solar modulation: At energies below $\sim10$ GeV, the incoming 
charged particles are modulated by the solar wind which  
prevents the lower energy cosmic rays from the inner solar system to
reach the Earth. Because of this effect there is an anticorrelation
between the eleven-year-cycle solar activity and the intensity
of cosmic rays below 10 GeV, which makes the cosmic ray fluxes
time dependent. The flux difference at solar maximum and solar minimum is 
more than a factor of two for 1~GeV cosmic rays, and it decreases to $\sim$ 
10~\% for 10~GeV cosmic rays. The fluxes shown in Fig.~\ref{fig:crspec} 
correspond to a particular epoch of the solar cycle. 

\subsubsection{Geomagnetic effects}
Lower energy cosmic rays are affected by the geomagnetic field which they must 
penetrate to reach the top of the atmosphere. This effect depends on the point 
of the Earth where the cosmic rays arrive (being stronger near the geomagnetic
equator), and on their  direction. Near the geomagnetic poles almost all 
primary particles can reach the atmosphere moving along field lines. In 
contrast, close to the geomagnetic equator the field restricts the flux at the 
top of the atmosphere to particles with energy larger than a few GeV, the exact
value depending on the direction of the particle trajectory. The relevant 
quantity is the magnetic rigidity, ${\cal R}=pc/Ze$, which parameterizes the 
characteristic radius of curvature of the particle trajectory in the presence 
of a magnetic field: particles with ${\cal R}$ below a local cutoff are bent 
away by the Earth magnetic field and do not reach the atmosphere. 

Given a map of the magnetic field around the Earth, the value of the rigidity 
cutoff can be obtained from a computer simulation of cosmic ray trajectories 
using a backtracking technique in which one determines if a given primary  
particle three-momentum and position correspond to an allowed or a forbidden 
trajectory. To do so, one integrates back the equation of motion of the cosmic
ray particle (actually one integrates the anti-particle equation) in the 
geomagnetic field, and finds if the past  trajectory remains confined to a 
finite distance from the Earth, in which case it is a forbidden trajectory, or
it originates from large enough ($\gsim 10R_{\oplus}$) distances. The rigidity 
cutoff is calculated as the minimum ${\cal R}$ for which the trajectory is 
allowed or, in other words, the backtracked trajectory escapes from the 
magnetic field of the Earth.

The following features of the geomagnetic effects 
are relevant to the atmospheric neutrino fluxes:

(i) For a fixed direction, the cutoff rigidity grows monotonically from zero 
at the magnetic pole to a maximum  value at the magnetic equator.

(ii) The cutoff rigidities for particles traveling toward the magnetic west
are higher than those for particles traveling toward the magnetic east.

(iii) Thus, the highest cutoff corresponds to westward going, horizontal 
particles  reaching the surface of the Earth at the magnetic equator and it is 
approximately 60~GV. 

\subsubsection{The neutrino yield}
As cosmic rays propagate in the atmosphere and interact with air nuclei, they 
create $\pi$- and $K$-mesons, which in turn decay and  create atmospheric 
$\nu$'s:
\begin{eqnarray}
A_{\rm cr}+A_{\rm air}\to&\pi^\pm,K^\pm&,K^0,\cdots;\nonumber \\
&\pi^\pm, K^\pm& \to \mu^\pm \;+\; \nu_\mu(\bar\nu_\mu) \label{atmdecay}\\
&&\ \ \ \ \mu^\pm\to e^\pm\; +\; \nu_e(\bar\nu_e)\; +\; \bar\nu_\mu(\nu_\mu).
\nonumber
\end{eqnarray}
We have only displayed the dominant channels.
Since the decay distributions of the secondary mesons and muons are
extremely well known, the largest source of differences between
the various calculations for the neutrino yield is the 
use of different models for the hadronic interactions  
(Gaisser {\em et al.}, 1996). Since neutrinos produced by the decays of pions, 
kaons and muons have different energy spectra, the main features of the 
interaction model which affect the $\nu_\mu/\nu_e$ composition and the energy 
and angular dependence of the neutrino fluxes are the $K/\pi$ ratio and their 
momentum distribution. 

\subsubsection{The neutrino fluxes}
In Figs.~\ref{fig:atmflux} and \ref{fig:atmzenith} we show the atmospheric 
neutrino fluxes from the calculation of the Bartol group (Gaisser, Stanev 
and Barr, 1988;  Barr, Gaisser, and 
Stanev, 1989; Agraval {\em et al.}, 1996; Lipari, Gaisser, and Stanev, 
1998) for the location of SK and for maximum solar activity.
In the first figure we show the flux as a function of the neutrino energy 
averaged over the arrival direction. In the second figure we show 
the fluxes as a function of the zenith angle for various neutrino energies.
We emphasize the following points:

(i) Energy dependence: For $E_\nu\gtrsim 1$ GeV, the fluxes obey an approximate
power law, $d\Phi/d E\propto E^{-\gamma}$ where $\gamma\sim3(3.5)$
for muon (electron) neutrinos and antineutrinos. For $E_\nu\lesssim 1$ GeV,
the dependence on energy is weaker as a consequence of the  
{\it bending} of the 
cosmic ray spectrum by geomagnetic effects and solar modulation.

(ii) $\bar\nu/\nu$ ratio: For $E\lesssim 1$ GeV, when all pions and subsequent 
muons decay before reaching the Earth, the ratio 
$\bar\nu_e/\nu_e=\pi^-/\pi^+<1$. At these energies, the cosmic rays are mainly 
protons. The positively charged protons produce, on average, more $\pi^+$'s 
than $\pi^-$'s in their interactions. On the other hand, since $\pi^+$ ($\mu^+$)
produces a $\nu_\mu$ ($\bar\nu_\mu$) in its decay, we expect 
$\bar\nu_\mu/\nu_\mu=1$.

As energy increases, the secondary $\mu^\pm$'s do not have time to decay
before reaching the surface of the Earth and consequently
$\bar\nu_\mu/\nu_\mu$ decreases.  

(iii) $(\nu_\mu+\bar\nu_\mu)/(\nu_e+\bar\nu_e)$ ratio: At $E\lesssim 1$ GeV, 
the ratio is very close to 2 as expected from the chain decays in 
Eq.~(\ref{atmdecay}). At higher energies, the ratio decreases because, as 
mentioned above, the $\mu$'s do not have time to decay before reaching the 
Earth and less $\nu_e's$ are produced.

(iv) Up-down asymmetry: At $E\lesssim$ a few GeV, the fluxes are up-down
asymmetric due to geomagnetic effects. Geomagnetic effects are very small 
already at $E=2$ GeV (see Fig.~\ref{fig:atmzenith}) and the flux
becomes up-down symmetric above that energy.

(v) Horizontal-vertical ratio: For $E\gtrsim$ a few GeV, the fluxes are maximal
for neutrinos arriving horizontally and minimal for neutrinos arriving 
vertically (see Fig.~\ref{fig:atmzenith}). This is due to the difference
in the atmosphere density which determines whether the pions decay before 
reinteracting with the air (thereby producing neutrinos). Pions arriving 
horizontally travel longer time in less dense atmosphere and are 
more likely to decay before reinteracting. 

\subsection{Interaction Cross Sections}
In order to determine the expected event rates at the experiment, one needs to 
know the neutrino-nucleon interaction cross sections in the detector. The 
standard approach is to consider separately the contributions to this cross
section from the exclusive channels of lower multiplicity:
quasi-elastic scattering and single pion production, and include all
additional channels as part of the deep inelastic (DIS) cross section:
\begin{equation} 
\sigma_{CC}=\sigma_{QE}+\sigma_{1\pi}+\sigma_{DIS}\; .
\end{equation}
In Fig.~\ref{fig:sigma_atm} we plot the cross sections for these processes.

The cross section for the quasi-elastic interaction is given by (Smith, 1972):
\begin{equation}
\frac{d\sigma_{QE}}{d|q^2|}(\nu n \to \ell^- p)=\frac{M^2 G_F^2 \cos^2\theta_c}
{8 \pi E_\nu^2} \left[A_1(q^2)-A_2(q^2)\frac{s-u}{M^2} +A_3(q^2) 
\frac{(s-u)^2}{M^4}\right],
\end{equation}
where $s-u=4ME_\nu+q^2-m_\ell^2$, $M$ is the proton mass, $m_\ell$ is the 
charged lepton mass, and $q^2$ is the momentum transfer. For 
$\bar\nu p\to\ell^+ n$, a similar formula applies with the only change
$A_2 \to -A_2$. The functions $A_1$, $A_2$, and $A_3$ can be written in
terms of axial and vector form factors:
\begin{equation}
\begin{array}{ll}
A_1=&\frac{\displaystyle m_\ell^2-q^2}{\displaystyle 4M^2} 
\Bigl[\bigl(4-\frac{\displaystyle q^2}{\displaystyle M^2}\bigr)|F_A|^2
-\bigl(4 +\frac{\displaystyle q^2}{\displaystyle M^2}\bigr)|F_V^1|^2
-\frac{\displaystyle q^2}{\displaystyle M^2}|\xi F_V^2|^2
-\frac{\displaystyle 4 q^2}{\displaystyle M^2}
\mbox{Re}(F_V^{1\star}\xi F^2_V)\Bigr] \\
& -\frac{\displaystyle m_\ell^2}{\displaystyle M^2}\left(|F_V^1 + \xi F_V^2|^2+
|F_A|^2\right), \\
A_2= & -\frac{\displaystyle q^2}{\displaystyle M^2}
\mbox{Re} [F_A^\star (F_V^1 + \xi F_V^2)], \\
A_3=&  -\frac{\displaystyle 1}{\displaystyle 4}\left(|F_A|^2+|F_V^1|^2 -
\frac{\displaystyle q^2}{\displaystyle 4M^2}|\xi F_V^2|^2 \right),
\end{array}
\end{equation}  
where we neglected second class currents and assumed CVC. With this assumption 
all form factors are real and can be written as follows:
\begin{equation}
\begin{array}{l}
F_V^1(q^2)= \left(1-\frac{\displaystyle q^2}{\displaystyle 4 M^2}\right)^{-1}
\left(1-\frac{\displaystyle q^2}{\displaystyle M_V^2}\right)^{-2}
\left[1-\frac{\displaystyle q^2}{\displaystyle 4 M^2}(1+\mu_p-\mu_n)\right],\\
\xi F_V^2(q^2)= \left(1-\frac{\displaystyle q^2}{\displaystyle 4 M^2}
\right)^{-1}\left(1-\frac{\displaystyle q^2}{\displaystyle M_V^2}\right)^{-2}
(\mu_n-\mu_p),\\
F_A=F_A(0)\left(1-\frac{\displaystyle q^2}{\displaystyle 4 M_A^2}\right)^{-2}. 
\end{array}
\end{equation}
Here $\mu_p$ and $\mu_n$ are the proton and neutron anomalous magnetic
moments and $M_V^2=0.71$ GeV$^2$ is the vector mass which is measured with
high precision in electron scattering experiments. The largest
uncertainties in this calculation are associated with the axial form
factor. For example, the axial mass used by various
collaborations varies in the range $M_A^2=0.71-1.06$ GeV$^2$.

So far we have neglected nuclear effects. The most important of 
these effects is related
to the Pauli principle and can be included by using a simple Fermi gas model 
(Smith, 1972). In this approximation, the cross section of a bound nucleon is 
equal to the cross section of a free nucleon multiplied by a factor of
$(1-D/N)$.  For neutrons,
\begin{equation}
D=\cases{Z&$2z \leq w-v$, \cr
\frac{1}{2} A\left[1-\frac{3z}{4} (v^2+w^2)+\frac{z^3}{3}+\frac{3}{32 z}
(w^2-v^2)^2\right] & $w-v \leq 2z \leq w+v$, \cr
0&$2z \geq w+v$,\cr}
\end{equation}
with $z= \sqrt{(q^2+m_\ell^2)^2/(4M^2)-q^2}/(2 k_f)$, 
$w=(2N/A)^{1/3}$, and $v=(2Z/A)^{1/3}$. Here $A,Z,N$ are, respectively, the 
nucleon, proton and neutron numbers and $k_f$ is the Fermi momentum,
$k_f=0.225$($0.26$) GeV for oxygen (iron).  For protons, the same formula
applies with the exchange $N\leftrightarrow Z$. The effect of this
factor is to decrease the cross section. The decrease is larger for
smaller neutrino energy. For energies above 1 GeV the nuclear effects
lead to an 8\% decrease on the quasi-elastic cross section.
As seen in Fig.~\ref{fig:sigma_atm}, quasi-elastic interactions dominate
for $E_\nu\lesssim 1$ GeV and induce most of the observed contained events.

An important role in the interpretation of the zenith angle distribution of the 
contained events is played by the relative angle between the direction of 
the incoming neutrino (carrying the information on the neutrino path length) 
and the direction of the produced charged lepton (which is measured). In 
Fig.~\ref{fig:opening} we plot this angle as a function of the measured charged
lepton energy for quasi-elastic interactions. For energies below 1 GeV, 
the opening angle is rather large: for sub-GeV events the correlation between 
the measured $\ell$ direction and the distance traveled by the neutrinos
is weak. For $E\gtrsim$ a few GeV, the two directions are almost aligned (for 
example, it is in average $1^\circ$ for upgoing muons) and 
the $\ell$ direction gives a very good measurement of the neutrino path length.

Single pion production, $\nu N\to\ell^-N^\prime \pi$, is dominated by the
$\Delta(1232)$ resonance (Fogli and Nardulli, 1979; 
Nakahata {\em et. al.}, 1986). 
It is most relevant at $E_\nu\simeq 1$ GeV (see Fig.~\ref{fig:sigma_atm}).

Deep inelastic processes, $\nu N\to \ell^- X$ where $X$
represents any hadronic system, dominate atmospheric neutrino interactions
for $E_\nu\gtrsim$ a few GeV. The parton model cross section is given by
\begin{equation}
\frac{d\sigma_{DIS}(\nu)}{dx dy}=\frac{G_F^2 s x}{4 \pi}
\left[F_1-F_3 +(F_1+F_3)(1-y)^2\right], 
\end{equation}
where $y=1-E_{\ell}/E_\nu$ and $x=-q^2/(2 M E_\nu y)$. For $\bar\nu$, a similar
formula applies, with $F_3\to-F_3$. $F_1$ and $F_3$ are given in terms of the 
parton distributions. For isoscalar targets $F_1=2{\displaystyle \sum_i} 
(q_i+\bar q_i)$ and $F_3={\displaystyle \sum_i} (\bar q_i -q_i)$.  
In order to avoid double counting of the single pion production, 
the deep inelastic contribution must be integrated in the region 
of hadronic masses $W_X>W_c$  ($W_c = 1.4$ GeV), which implies 
$2 M E_\nu y (1-x) \geq W_c^2 -M^2$. 

\subsection{Two-Neutrino Oscillation Analysis}
\label{subsec:atmosc}
The simplest and most direct interpretation of the atmospheric neutrino
anomaly is that of muon neutrino oscillations (Learned, Pakvasa, and Weiler, 
1988; Barger and Whisnant, 1988; Hidaka, Honda and Midorikawa, 1988).
The estimated value of the oscillation parameters can be easily derived
in the following way: 
\begin{itemize}
\item The angular distribution of contained events shows that, for $E\sim 1$ 
GeV, the deficit comes mainly from $L\sim 10^2-10^4$ km. The corresponding 
oscillation phase must be maximal, 
$\frac{\Delta m^2 ({\rm eV}^2) L({\rm km})}{2E({\rm GeV})}\sim 1$, which
requires $\Delta m^2\sim 10^{-4}-10^{-2}$ eV$^2$.
\item Assuming that all upgoing $\nu_\mu$'s which would lead to multi-GeV 
events oscillate into a different flavor while none of the downgoing ones
do, the up-down asymmetry is given by 
$|A_\mu|=\sin^22\theta/(4-\sin^22\theta)$. The present one sigma bound reads 
$|A_\mu|>0.27$ [see Eq.~(\ref{aud})], which requires that the mixing angle is
close to maximal, $\sin^22\theta>0.85$.
\end{itemize}
In order to go beyond these rough estimates, one must compare in a 
statistically meaningful way the experimental data with the detailed 
theoretical expectations. We now describe how to obtain the allowed region. 
We consider two neutrino cases, where $\nu_\mu$ oscillates into either $\nu_e$ 
or $\nu_\tau$ or a sterile neutrino $\nu_s$.

\subsubsection{Predicted number of events}
For a given neutrino oscillation channel, the expected number of $\mu$-like and
$e$-like contained events, $N_\alpha$ ($\alpha = \mu,e$), can be computed as:
\begin{equation}
N_\mu= N_{\mu\mu} +\ N_{e\mu} \; ,  \;\;\;\;\;\ N_e= N_{ee} +  N_{\mu e} \; ,
\label{eventsnumber}
\end{equation}
where
\begin{eqnarray}
N_{\alpha\beta} &=& n_t T\int\frac{d^2\Phi_\alpha}{dE_\nu d(\cos\theta_\nu)} 
\kappa_\alpha(h,\cos\theta_\nu,E_\nu) P_{\alpha\beta} \frac{d\sigma}{dE_\beta}
\varepsilon(E_\beta)dE_\nu dE_\beta d(\cos\theta_\nu)dh\; .
\label{event0}
\end{eqnarray}
Here $P_{\alpha\beta}$ is the conversion probability of $\nu_\alpha\to
\nu_\beta$ for given values of $E_{\nu}$, $\cos\theta_\nu$ and $h$, {\em i.e.},
$P_{\alpha\beta}\equiv P(\nu_\alpha\to\nu_\beta; E_\nu,\cos\theta_\nu, h )$.  
In the SM, the only non-zero elements are the diagonal ones, 
{\em i.e.} $P_{\alpha\alpha}=1$ for all $\alpha$.  In Eq.~(\ref{event0}), $n_t$ 
denotes the number of target particles, $T$ is the experiment running time, 
$E_\nu$ is the neutrino energy, $\Phi_\alpha$ is the flux of atmospheric 
$\nu_\alpha$'s, $E_\beta$ is the final charged lepton energy,
$\varepsilon(E_\beta)$ is the detection efficiency for such a charged
lepton, $\sigma$ is the neutrino-nucleon interaction cross section,
and $\theta_\nu$ is the angle between the vertical direction and the
incoming neutrinos ($\cos\theta_\nu$=1 corresponds to the down-coming
neutrinos).  In Eq.~(\ref{event0}), $h$ is the slant distance from the
production point to the sea level for $\alpha$-type neutrinos with
energy $E_\nu$ and zenith angle $\theta_\nu$ , and $\kappa_\alpha$
is the slant distance distribution, normalized to one.  (We use in our
calculations $\kappa_\alpha$  from Gaisser and Stanev, 1998). 

To obtain the expectation for the angular distribution of contained events 
one must integrate the corresponding bins for $\cos\theta_\beta$, where 
$\theta_\beta$ is the angle of the detected lepton, taking into account the 
opening angle between the neutrino and the charged lepton directions as 
determined by the kinematics of the neutrino interaction.  

As discussed above, the neutrino fluxes, in particular in the sub-GeV range, 
depend on the solar activity.  In order to take this fact into account, one 
uses in Eq.~(\ref{event0}) a linear combination of atmospheric neutrino fluxes 
$\Phi_\alpha^{\rm max}$ and $\Phi_\alpha^{\rm min}$ which correspond to the 
most (solar maximum) and least (solar minimum) active phases of the Sun, 
respectively, with different weights which depend on the specific running 
period.

Experimental results on upgoing muons are presented in the form of  measured 
muon fluxes. To obtain the effective muon fluxes for both stopping and 
through-going muons, one must convolute the survival probabilities for 
$\nu_\mu$'s with the corresponding muon fluxes produced by the neutrino 
interactions with the Earth. One must further take into account the muon energy
loss during propagation both in the rock and in the detector, and also the 
effective detector area for both types of events, stopping and through-going.  
Schematically,
\begin{equation}
\Phi_\mu(\theta)_{S,T}=\frac{1}{A(L_{\rm min},\theta)}
\int_{E_{\mu ,{\rm min}}}^{\infty} 
\frac{d\Phi_\mu(E_\mu,\cos\theta)}{dE_\mu d\cos\theta}
A_{S,T}(E_\mu,\theta)dE_{\mu} \, \; ,
\label{upmuons1}
\end{equation}  
where
\begin{eqnarray}
\frac{d\Phi_\mu}{dE_\mu d\cos\theta}&=& N_A \int_{E_{\mu}}^\infty dE_{\mu 0}
\int_{E_{\mu 0}}^\infty dE_\nu \int_0^\infty dX  \int_0^\infty dh \;
\kappa_{\nu_\mu}(h,\cos\theta,E_\nu) \nonumber \\
& & \frac{d\Phi_{\nu_\mu}(E_\nu,\theta)}{dE_\nu d\cos\theta}P_{\mu\mu}
\frac{d\sigma(E_\nu,E_{\mu 0})}{dE_{\mu 0}}\, F_{\rm rock}(E_{\mu 0}, E_\mu, X).
\label{upmuons2}
\end{eqnarray}
Here $N_A$ is the Avogadro number, $E_{\mu 0}$ is the energy of the muon
produced in the neutrino interaction, $E_\mu$ is the muon energy
when entering the detector after traveling a distance $X$ in the rock, and
$\cos\theta$ labels both the neutrino and the muon directions which, at the 
relevant energies, are collinear to a very good approximation.  We denote by 
$F_{\rm rock}(E_{\mu 0}, E_\mu, X)$ the function which characterizes the energy
spectrum of the muons arriving the detector. The Standard practice is to use an
analytical approximation obtained by neglecting the fluctuations during muon 
propagation in the Earth. In this case the three quantities $E_{\mu 0}$, 
$E_\mu$, and $X$ are not independent:
\begin{equation}
\int_{0}^\infty F_{\rm rock}(E_{\mu 0}, E_\mu, X) dX=
\frac{1}{\langle d{\cal E}_\mu(E_\mu)/dX\rangle} \; ,
\end{equation}
where $\langle d{\cal E}_\mu(E_\mu)/dX\rangle$ is the average muon
energy loss due to ionization, bremsstrahlung, $e^+e^-$ pair
production and nuclear interactions in the rock (Lohmann, Kopp and Voss, 1985).

For SK, the pathlength traveled by the muon inside the detector 
is given by the muon range function in water:
\begin{equation}
\label{Emcut}
L(E_\mu)=\int_{0}^{E_\mu}
\frac{1}{\langle d{\cal E}_\mu(E'_\mu)/dX\rangle} \; dE'_\mu \; .
\end{equation} 
In Eq.~(\ref{upmuons1}), $A(L_{\rm min},\theta)=A_{S}(E_\mu,\theta)+A_{T}(E_\mu,
\theta)$ is the projected detector area for internal path-lengths longer than
$L_{\rm min}(= 7$ m in SK). Here $A_{S}$ and $A_{T}$ are the effective areas 
for stopping and through-going muon trajectories. These effective areas can be 
computed using a simple geometrical picture given, for instance, in Lipari and
Lusignoli (1998). For a given angle, the threshold energy cut for
SK muons is obtained by equating Eq.~(\ref{Emcut}) to
$L_{\rm min}$, {\it i.e.},  $L(E_{\mu}^{\mathrm th}) =L_{\rm min}$.

In contrast to SK, MACRO present their results as muon fluxes for 
$E_\mu>1$ GeV, after correcting for detector acceptances. Therefore in this 
case we compute the expected fluxes as in  Eqs.~(\ref{upmuons1}) and 
(\ref{upmuons2}) but without the inclusion of the effective areas.

\subsubsection{Conversion probabilities}
We consider a two-flavor scenario, $\nu_\mu\to\nu_X$ ($X=e,\tau$ or $s$). 
The oscillation probabilities are obtained by solving the  
evolution equation of the $\nu_\mu -\nu_X$ system in the matter background 
of the Earth (see section~\ref{subsec:oscmatter}):
\begin{equation}
i{\mbox{d} \over \mbox{d}t}\left(\matrix{\nu_\mu \cr\ \nu_X\cr }\right)  = 
 \left(\matrix{{H}_{\mu}& {H}_{\mu X} \cr {H}_{\mu X} & {H}_X \cr}\right)
\left(\matrix{\nu_\mu \cr\ \nu_X \cr}\right) ,
\label{evolution1}
\end{equation}
where
\begin{eqnarray}
H_\mu & \! = &  \! V_\mu-\frac{\Delta m^2}{4E_\nu}\cos2\theta \, ,\nonumber\\
H_X & \!= & V_X +  \frac{\Delta m^2}{4E_\nu} \cos2 \theta \, ,  \nonumber\\
H_{\mu X}& \!= &   \frac{\Delta m^2}{4E_\nu} \sin2 \theta \,.
\end{eqnarray}
The various neutrino potentials in matter are given by
\begin{eqnarray}
V_e& = &\frac{\sqrt{2}G_F \rho}{M} (Y_e-\frac{1}{2}Y_n)\,, \nonumber \\
V_\mu=V_\tau & = &\frac{\sqrt{2}G_F \rho}{M} (-\frac{1}{2}Y_n)\,, \nonumber\\
V_s &= & 0\, .
\label{potential}
\end{eqnarray}
Here $\rho$ is the matter density in the Earth (Dziewonski and  Anderson, 1981),
$M$ is the nucleon mass, and $Y_{e}(Y_n)$ is the electron (neutron) fraction. 
For anti-neutrinos, the signs of the potentials are reversed. 

For $X=\tau$, we have $V_\mu=V_\tau$ and consequently these potentials can be 
removed from the evolution equation. The solution of Eq.~(\ref{evolution1}) is 
then straightforward and the probability takes the well-known vacuum form 
[Eq.~(\ref{ptwo})], which is equal for neutrinos and anti-neutrinos. For $X=e$
or $s$, the effect of the matter potentials requires a numerical solution of 
the evolution equations in order to obtain $P_{\alpha\beta}$ which, 
furthermore, is different for neutrinos and anti-neutrinos. As first 
approximation, one can use a constant Earth matter density. Then (see 
section~\ref{sec:oscila}) the solutions take the same form as the vacuum 
probability [Eq.~(\ref{ptwo})], but with the mixing angle and the oscillation 
length replaced by their effective values in matter [Eqs.~(\ref{effmix}) and 
(\ref{efflosc}) with $A=2E(V_\mu-V_X)$]. 
In Fig.~\ref{fig:atm_probs} (from Lipari and Lusignoli, 1989) we show
the survival probability of $\nu_\mu$ for the different oscillation
channels for $\Delta m^2=5\times 10^{-3}$ eV$^2$, $\sin^22\theta=1$ and various
values of $E_\nu$. As seen in the figure, matter effects damp the oscillation 
amplitude. For the chosen mass difference, they are important for neutrino 
energies of few 10 GeV and therefore are relevant mainly for upgoing muons. We 
return to this point below when describing the results of the analysis.

\subsubsection{Statistical analysis}
In order to define in a statistically meaningful way the regions of neutrino 
flavor parameters that are allowed by a given set of atmospheric neutrino 
observables, one can construct, for example, a $\chi^2$ function: 
\begin{equation}
\chi^2\equiv\sum_{I,J}(N_I^{\rm data}-N_I^{\rm th}) \cdot(\sigma_{\rm data}^2 
+\sigma_{\rm th}^2)_{IJ}^{-1}\cdot(N_J^{\rm data}-N_J^{\rm th}),
\label{chi2}
\end{equation}
where $I$ and $J$ stand for any combination of experimental data sets
and event-types considered, {\em i.e}, $I = (A, \alpha)$ and $J = (B,
\beta)$. The latin indices $A, B$ stand for the different experiments
or different data samples in a given experiment. The greek indices
denote electron-type or muon-type events, {\em i.e}, $\alpha,\beta=e,\mu$.  
In Eq.~(\ref{chi2}), $N_I^{\rm th}$ stands for the predicted number of events 
(or for the predicted value of the flux, in the case of upgoing muons) 
calculated as discussed above, whereas $N_I^{\rm data}$ is the corresponding 
experimental measurement.  In Eq.~(\ref{chi2}), $\sigma_{\rm data}^2$ and 
$\sigma_{\rm th}^2$ are the error matrices containing the experimental and 
theoretical errors, respectively. They can be written as
\begin{equation}
\sigma_{IJ}^2\equiv\sigma_\alpha(A)\, \rho_{\alpha \beta}(A,B)\, 
\sigma_\beta(B),
\end{equation}
where $\rho_{\alpha \beta}(A,B)$ is the correlation matrix containing
all the correlations between the $\alpha$-like events in the $A$-type
experiment and $\beta$-like events in $B$-type experiment, whereas
$\sigma_\alpha(A)$ [$\sigma_\beta(B)$] are the errors for the number
of $\alpha$-like events in the $A$ [$\beta$-like events in the $B$] experiment.
The dimension of the error matrix varies depending on the combination of 
experiments included in the analysis. Details on the statistical analysis of 
the atmospheric neutrino data and results from the analyses with various data 
samples can be found in Fogli, Lisi and Montanino (1994,1995);
Fogli and Lisi (1995b); Gonzalez-Garcia {\em et al.} (1998,1999);
Gonzalez-Garcia, Fornengo and Valle (2000); 
Foot, Volkas and Yasuda (1998); Yasuda (1998); Akhmedov {\em et al.} (1999).

Using the definitions given above, one calculates $\chi^2$ of Eq.~(\ref{chi2}) 
as a function of the neutrino parameters. By minimizing $\chi^2$ with respect 
to $\sin^22\theta$ and $\Delta m^2$, one determines the best fit results, while
the allowed regions are determined by the conditions: $\chi^2 \equiv
\chi_{\mathrm min}^2 + 4.61 \, (6.1)\, [9.21] $ for a confidence level
(CL) of $90\,(95)\,[99]$ \%.

\subsubsection{$\nu_\mu\rightarrow\nu_e$}
At present $\nu_\mu\to\nu_e$ is excluded with high CL as the
explanation to the atmospheric neutrino anomaly for two different reasons:

(i) SK high precision data show that the $\nu_e$ contained events 
are very well described by the SM prediction both in 
normalization and in their
zenith angular dependence [see Fig.~\ref{fig:skatm}]. The $\nu_\mu$ 
distribution, however, shows an angle-dependent deficit. $\nu_\mu\to\nu_e$
oscillations can explain the angular dependence of the $\nu_\mu$ flux only at
the price of introducing angular dependence of the $\nu_e$ flux, in contrast
to the data. Furthermore, even the best fit point for $\nu_\mu\to\nu_e$
oscillations does not generate the observed up-down asymmetry in the Multi-GeV
muon sample. This is illustrated in Fig.~\ref{fig:atm_mue}, where we show the 
predicted angular distribution of contained events at SK 
for the best fit points of the different oscillation channels. For $\nu_\mu 
\to\nu_e$ oscillations, the asymmetry in the Multi-GeV muon distribution is 
much smaller than in the $\nu_\mu \to \nu_\tau$ or $\nu_\mu\to\nu_s$ channels. 

(ii) Explaining the atmospheric data with $\nu_\mu\to\nu_e$ transition has
direct implications for the $\bar\nu_e\to\bar\nu_\mu$ transition. 
In particular,
there should be a $\bar\nu_e$ deficit in the CHOOZ reactor experiment. Thus the
neutrino parameters not only give a poor fit to the atmospheric data but are
actually excluded by the negative results from the CHOOZ reactor experiment 
(see Fig.~\ref{fig:atm_mue}).

\subsubsection{$\nu_\mu\rightarrow\nu_\tau$ and $\nu_\mu\rightarrow\nu_s$}
In Fig.~\ref{fig:atm_partial} we show the values of the  oscillation 
parameters $(\Delta m^2,\sin^2\theta)$ which describe various sets of data for 
these two oscillation channels. The upper panels (labeled as FINKS) and the
central panels refer to contained events. The upper panels take into account 
only the total rates, and the central ones only the angular distribution (from 
both Kamiokande and SK). The lower panels correspond to the 
angular distribution for upgoing muons from SK. The three shaded 
regions are allowed at the 90, 95 and 99\% CL. In Fig.~\ref{fig:atm_global} we 
plot the allowed regions from the global analysis, including all the 
atmospheric neutrino data. Also shown are the expected sensitivity from 
$\nu_\mu$ disappearance at the long baseline (LBL) experiments K2K and 
MINOS discussed in ~\ref{subsec:lbl}. (These figures are an update of the 
results presented in Gonzalez-Garcia {\em et al}, 2000.) We emphasize the
following points:

(i) The allowed regions from the various data samples overlap. The oscillation 
hypothesis can then consistently explain the atmospheric neutrino data. 

(ii) The information from the total event rates alone is consistent with
arbitrarily high $\Delta m^2$ values. The reason is that no information on the 
minimum oscillation length can be inferred from the data [see section
\ref{subsec:osvac}]. 

(iii) The allowed regions for $\nu_\mu\to\nu_\tau$ transition are symmetric 
with respect to maximal mixing. This must be the case because the corresponding
probabilities take the vacuum expression and therefore depend on
$\sin^22\theta$. 

(iv) The allowed regions for $\nu_\mu\to\nu_s$ transition are asymmetric due to 
Earth matter effects. These effects are more pronounced when the condition for 
maximal matter effect, $\Delta m^2\cos2\theta\sim 2E V_s$, is fulfilled. Since 
in the chosen convention the potential difference $A=2E(V_\mu-V_s)<0$,
the matter effects enhance neutrino oscillations for $\cos 2\theta<0$ 
($\sin^2\theta>0.5$). The opposite situation holds for antineutrinos, but 
neutrino fluxes are larger and dominate in the resulting effect. As a result 
the allowed regions are {\it wider} in the $\sin^2\theta>0.5$ side of the plot.

(v) The best fit to the full data for $\nu_\mu\to\nu_\tau$ corresponds to 
$\Delta m^2=2.6\times 10^{-3}$ eV$^2$ and $\sin^22\theta=0.97$. The best fit
for  $\nu_\mu\to \nu_s$ lies at $\Delta m^2=3\times 10^{-3}$ eV$^2$ and
$\sin^2\theta=0.61$.

In order to discriminate between the $\nu_\mu\to\nu_\tau$ and 
$\nu_\mu\to\nu_s$ 
options, one can use the difference in the survival probabilities due to
the presence of matter effects for oscillations into sterile neutrinos (Lipari 
and Lusignoli, 1998). As discussed above, the effect is important mainly for 
the higher energy neutrinos which lead to through-going muons events. 
In Fig.~\ref{fig:ang_thru} we plot the expected distributions for the
best fit points for the two channels. As seen in the figure, 
the distribution is steeper for $\nu_\mu\to\nu_\tau$ 
while for  $\nu_\mu\rightarrow\nu_s$ a flattening is observed
for neutrinos coming close to the vertical due to the damping of
the oscillation amplitude [see Fig.~\ref{fig:atm_probs}].
The data favours the steeper distributions and this translates
into a better global fit for oscillations into $\nu_\tau$. For the
global analysis, an update of the results of Gonzalez-Garcia,  
{\em et al.} (2000) shows that for $\nu_\tau$ oscillations 
$\chi^2_{min}=56/63$ d.o.f while for $\nu_s$ oscillations
$\chi^2_{min}=72/63$ d.o.f. As we will see in Sec.~\ref{sec:schemes}, this 
difference has an important role in 4-$\nu$ mixing schemes.

SK has also used other methods to distinguish between 
the $\nu_\tau$ and $\nu_s$ hypotheses for explaining atmospheric 
$\nu_\mu$ disappearance. One is to examine events likely to have
been caused by NC interactions. While $\nu_\tau$'s readily undergo such
interactions, $\nu_s$'s do not, resulting in a relative suppression of
the NC signal (Vissani and Smirnov, 1998; Hall and Murayama, 1998).  
Another method attempts to observe appearance of the newly 
created $\nu_\tau$, even if only on a statistical basis, by selecting enriched 
samples. All methods strongly favour $\nu_\mu\leftrightarrow\nu_\tau$ 
oscillations over $\nu_\mu\leftrightarrow\nu_s$ (Fukuda {\em et al}, 2000; 
Toshito {\em et al.}, 2001).

\section{Laboratory Experiments}
\label{sec:lab}
Laboratory experiments to search for neutrino oscillations are performed
with neutrino beams produced at either accelerators or nuclear reactors.
In {\it disappearance} experiments, one looks for the attenuation of a neutrino
beam primarily composed of a single flavor due to the mixing with other 
flavors. In {\it appearance} experiments, one searches for interactions by 
neutrinos of a flavor not present in the original neutrino beam. 

Most of the past and present laboratory experiments did not have an oscillation
signal. In such a case, as discussed in Sec.~\ref{subsec:osvac}, the experiment
sets a limit on the corresponding oscillation probability. Appearance 
experiments set limits ${\langle P_{\alpha\beta }\rangle}<{P_L}$ for given 
flavors $\alpha\neq\beta$. Disappearance experiments set limits 
${\langle P_{\alpha\alpha }\rangle}>{1-P_L}$ for a given flavor $\alpha$ 
which, in the two neutrino case, can be translated into 
${\langle P_{\alpha\beta }\rangle}<{P_L}$ for $\beta\neq\alpha$. 
The results are usually interpreted in a two neutrino framework as exclusion 
regions in the $(\Delta m^2,\sin^22\theta)$ plane. One can take the upper bound
on the mixing angle in the asymptotic large $\Delta m^2$ range and translate it
back into the value of $P_L$: $\sin^22\theta_{\rm lim}=2 P_L$ [see discussion 
in Sec.~\ref{subsec:osvac}]. The probability $P_L$ is the relevant quantity 
when interpreting the results in the more-than-two neutrino framework.

\subsection{Short Baseline Experiments at Accelerators}
Conventional neutrino beams from accelerators are mostly produced by $\pi$ 
decays, with the pions produced by the scattering of the accelerated protons 
on a fixed target:
\begin{eqnarray}
p\,+\, \mbox{target}\to &\pi^\pm& + X  \\ \nonumber
&\pi^+& \to \mu^+ \nu_\mu  \\ \nonumber
&&\ \ \ \ \mu^+\to e^+ \nu_e \bar \nu_\mu \\ \nonumber
&\pi^-& \to \mu^- \bar\nu_\mu  \\ \nonumber
&&\ \ \ \ \mu^-\to e^- \bar\nu_e \nu_\mu. 
\end{eqnarray}
Thus the beam can contain both $\mu$- and $e$-neutrinos and antineutrinos.
The final composition and energy spectrum of the neutrino beam is determined
by selecting the sign of the decaying $\pi$ and by stopping the produced $\mu$ 
in the beam line. 

Most oscillation experiments performed so far with neutrino beams from 
accelerators have characteristic distances of the order of hundreds of meters. 
We call them {\it short baseline (SBL) experiments}. 
With the exception of the LSND
experiment, which we discuss below, all searches have been negative. In table 
\ref{tab:sbl} we show the limits on the various transition probabilities from 
the negative results of the most restricting SBL experiments. In 
Fig.~\ref{fig:mtet} (from Astier {\em et al.}, 2001) we show the excluded 
regions corresponding to (the absence of) $\nu_\mu\to\nu_\tau$ and 
$\nu_e\to\nu_\tau$ oscillations. Due to the short path length, these 
experiments are not sensitive to the low values of $\Delta m^2$ invoked
to explain either the solar or the atmospheric neutrino data but
they are relevant for 4-$\nu$ mixing schemes as we will see in
Sec.~\ref{subsec:fourmix}.

\subsection{LSND and KARMEN}
\label{subsec:lsnd}
The only positive signature of oscillations at a laboratory experiment comes 
from the Liquid Scintillator Neutrino Detector (LSND) running at Los Alamos 
Meson Physics Facility (Athanassopoulos {\em et al.}, 1995, 1996, 1998). 
The primary neutrino flux comes from $\pi^+$'s produced in a 30-cm-long water
target when hit by protons from the LAMPF linac with  800 MeV kinetic energy. 
The detector is a tank filled with 167 metric tons of dilute liquid
scintillator, located about 30 m from the neutrino source. 

Most of the produced $\pi^+$'s come to rest and decay through the sequence
$\pi^+\to\mu^+\nu_{\mu}$, followed by $\mu^+\to e^+\nu_e\bar\nu_\mu$. The
$\bar\nu_\mu$'s so produced have a maximum energy of $52.8$ MeV. 
This is called
the {\it decay at rest} (DAR) flux and is used to study $\bar\nu_\mu\to \bar\nu_e$ 
oscillations. The energy dependence of the $\bar\nu_\mu$ flux from decay at 
rest is very well known, and the absolute value is known to 7\%.
The open space around the target is short compared to the pion decay length. 
Thus only 3\%\ of the $\pi^+$'s {\it decay in flight} (DIF). 
The DIF $\nu_\mu$ flux 
is used to study $\nu_\mu\to\nu_e$ oscillations. 

In 1995, the LSND experiment published data showing candidate events that are 
consistent with $\bar\nu_\mu\to \bar\nu_e$ oscillations. Further supporting 
evidence was provided by the signal in the $\nu_\mu\to\nu_e$ channel. We 
summarize here their main results in both the DAR and DIF channels.

For DAR related measurements, $\bar\nu_e$'s are detected in the quasi 
elastic process $\bar\nu_e\,p\to e^{+}\,n$, in correlation with a 
monochromatic 
photon of $2.2$  MeV arising from the neutron capture reaction 
$np\to d\gamma$. 
The main background is due to the  $\bar\nu_e$ component in the beam that is
produced in the decay chain starting with $\pi^-$'s. This background is
suppressed by three factors. First, the $\pi^+$ production rate is about eight 
times the $\pi^-$ production rate in the beam stop. Second, 95\%\ of the 
$\pi^-$'s come to rest and are absorbed before decay in the beam stop. Third, 
88\%\ of the $\mu^-$'s from $\pi^-$'s DIF are captured from atomic orbit, a 
process which does not give a $\bar\nu_e$. Thus, the relative yield, compared 
to the positive channel, is estimated to be $\sim(1/8)\times0.05\times0.12= 
7.5 \times 10^{-4}$.

In Athanassopoulos {\em et al.} (1995) LSND report
a total of 22 events with $e^+$ energy between 36 and $60$ MeV when $4.6\pm0.6$
background events are expected. They fit the full $e^+$ event sample in the 
energy range $20 <E_e<60$ MeV and the result yields
$64.3^{+18.5}_{-16.7}$ beam-related events. Subtracting the estimated
neutrino background with a correlated gamma of $12.5\pm 2.9$ events results in 
an excess of $51.2^{+18.7}_{-16.9} \pm 8.0$ events. The interpretation of this
anomaly in terms of $\bar\nu_\mu\to\bar\nu_e$ oscillations requires 
$P_{e\mu}=(3.1\pm1.2\pm0.5)\times10^{-3}$. 

For DIF related measurements, the $\nu_e$'s are observed via the detection of 
electrons
produced in the process $\nu_e C \rightarrow e^- X$ with energy $60<E_e<200$ 
MeV. Using two independent analyses, $27.7\pm 6.4$ events are observed. The 
neutrino induced backgrounds are dominated by $\mu^+\to e^+\bar\nu_\mu\nu_e$ 
and $\pi^+\to e^+ \nu_e$ decays in flight in the beam stop area and are 
estimated to be $9.6\pm 1.9$ events. The excess above the expected background 
from conventional processes is then $18.1\pm 6.6\pm4.0$ events.
The excess events are consistent with $\nu_\mu\to\nu_e$ oscillations with
$P_{\mu e}=(2.6\pm1.0\pm0.5)\times10^{-3}$.

The LSND results have been recently updated to include the runs till 1998
(Aguilar {\em et al.}, 2001) and the total fitted excess is of 
$87.9\pm22.4\pm6$ 
events, corresponding to an oscillation probability of 
$(2.64\pm0.67\pm0.45)\times10^{-3}$. In the two-family formalism these results 
lead to the oscillation parameters shown in Fig.~\ref{fig:lsnd}. The shaded 
regions are the 90~\% and 99~\% likelihood regions from LSND. The best fit 
point corresponds to $\Delta m^2=1.2$ eV$^2$ and $\sin^22\theta=0.003$.

The region of parameter space which is favoured by the LSND observations
has been partly tested by other experiments like the old BNL E776 experiment 
(Borodovsky {\em et al.}, 1992) and more recently by the KARMEN experiment
(Gemmeke {\em et al.}, 1990). The KARMEN experiment is performed at the neutron
spallation facility ISIS of the Rutherford Appleton Laboratory. Neutrinos are 
produced by stopping the 800\,MeV protons in a massive beam stop target, 
thereby producing pions. The $\pi^-$'s are absorbed by the target nuclei 
whereas $\pi^+$'s decay at rest producing muon neutrinos via
$\pi^+\to\mu^+\nu_\mu$. The low momentum $\mu^+$'s are also stopped 
within the massive target and decay at rest, $\mu^+\to e^+\nu_e \bar\nu_\mu$.
The $\pi^+-\mu^+$ decay chain at rest gives a neutrino source with identical 
intensities for $\nu_\mu$ , $\nu_e$ and $\bar\nu_\mu$ emitted isotropically.
There is a minor fraction of $\pi^-$ decaying in flight (DIF) with the 
following $\mu^-$ DAR in the target station which 
leads to a very small contamination of $\bar\nu_e/\nu_e < 6.2\cdot 10^{-4}$.
The energy spectra of the $\nu$'s are well defined due to the DAR
of both the $\pi^+$ and $\mu^+$.

The neutrinos are detected in a rectangular tank filled with 56\,t of a liquid  
scintillator. The signature for the detection of $\bar\nu_e$  is a spatially 
correlated delayed coincidence of positrons from $\bar p(\bar\nu_e,e^+)n$ 
with energies up to $E_{e^+}=E_{\bar\nu_e}-Q=51.0$\,MeV and 
$\gamma$ emission of either of the two neutron capture processes: 
$p(n,\gamma)d$ with one $\gamma$ of $E(\gamma)=2.2$\,MeV or Gd$(n,\gamma)$ with 
3$\gamma$-quanta on average and $\sum E(\gamma)=8$\,MeV.

The raw data presented in the EPS HEP 2001 conference (Wolf {\em et al.}, 2001)
correspond to $\sim$9400\,C protons on target. Analyzing the data results in 11
sequential events which satisfy all cuts. This number is in good agreement with
the total background expectation of $12.3\pm 0.6$. Applying a Bayesian 
renormalization procedure, an upper limit of $N({\rm osc})<6.3$ at 90\%CL can 
be extracted. However using the spectral information and a maximum likelihood 
analysis, KARMEN find a best fit value $N({\rm osc})=0$ within the physically 
allowed range of parameters, which can be translated into an upper limit of 
3.8 and 3.1 oscillation events for $\Delta m^2<1$ eV$^2$ and $\Delta m^2>20$ 
eV$^2$, respectively. These numbers are based on a complete frequentist 
approach as suggested by G. Feldman and R. Cousins. The corresponding exclusion
curve in the two-neutrino parameter space is given in Fig.~\ref{fig:lsnd} 
together with the favoured region for the LSND experiment (from Wolf 
{\em et al.}, 2001). At 
high $\Delta m^2$, KARMEN  results exclude the region favored by LSND. At low 
$\Delta m^2$ , KARMEN leaves some allowed space, but the reactor experiments
at Bugey and CHOOZ add stringent limits for the larger mixing angles.
This figure represents the final status of the LSND oscillation signal.

The MiniBooNE experiment (Bazarko {\em et al.}, 2000) searches for  
$\nu_\mu \to \nu_e$ oscillations
and is specially designed to make a conclusive statement about 
the LSND's neutrino oscillation evidence.  
They use a $\nu_\mu$ beam of energy 0.5 -- 1.0 GeV initiated by a primary 
beam of 8 GeV protons from the Fermilab Booster, which contains only a small 
intrinsic $\nu_e$ component (less than 0.3\%). They search for an excess of
electron neutrino events in a detector located approximately 
500 m from the neutrino source. The MiniBooNE neutrino detector consists of 
800 tons of pure mineral oil contained in a 12.2 m diameter spherical tank.  
A structure in the tank supports phototubes, which 
detect neutrino interactions in the oil by the Cerenkov 
and scintillation light that they produce.

The $L/E$ ratio is similar to that of 
LSND, giving MiniBooNE sensitivity to the same mode of
oscillations.  However, neutrino energies are more than an order 
of magnitude higher than at LSND, so that the search at MiniBooNE 
employs different experimental techniques. In Fig.~\ref{fig:lsnd} we show 
the 90\% CL limits that MiniBooNE can achieve. Should a signal be found then
the next step would be the BooNe experiment.

\subsection{Disappearance Experiments at Reactors}
\label{subsec:reactors}
Neutrino oscillations are also searched for using neutrino beams from nuclear 
reactors. Nuclear reactors produce $\bar\nu_e$ beams with $E_\nu\sim$ MeV. Due 
to the low energy, $e$'s are the only charged leptons which can be produced in 
the neutrino CC interaction. If the $\bar\nu_e$ oscillated to another flavor, 
its CC interaction could not be observed. Therefore oscillation experiments 
performed at reactors are disappearance experiments. They have the advantage 
that smaller values of $\Delta m^2$ can be accessed due to the lower neutrino 
beam energy. In table \ref{tab:reac} we show the limits on $P_{ee}$ 
from the negative results of the reactor experiments 
Gosgen (Zacek, G., 1986), Krasnoyarsk (Vidyakin {\em et al.}, 1994), 
Bugey (Achkar {\em et al.}, 1995), and CHOOZ (Apollonio {\em et al.}, 1999). 
Gosgen, Krasnoyarks and Bugey have relatively short baselines
while CHOOZ is the first long baseline (LBL) reactor experiment.

In Fig.~\ref{fig:reactors} we show the corresponding excluded regions in the 
parameter space for two neutrino oscillations. Bugey sets the strongest 
constraint on the allowed mixing in the $\Delta m^2$ range that is interesting 
for the LSND signal. Due to its longer baseline, CHOOZ is sensitive to the 
lowest values of $\Delta m^2$,  low enough to be in the range of interest for
atmospheric neutrinos oscillations and the upper sector of the LMA
solution for solar neutrinos. 

The CHOOZ experiment searches for disappearance of $\bar{\nu}_e$'s produced in 
a power station with two pressurized-water nuclear reactors with a total 
thermal power of $8.5$ GW. At the detector, located at $L\simeq 1$ 
km from the reactors, the $\bar{\nu}_e$ reaction signature is the delayed 
coincidence between the prompt ${\rm e^+}$ signal and the signal due to the 
neutron capture in the Gd-loaded scintillator.  The ratio between the measured 
and expected fluxes averaged over the neutrino energy spectrum is given by
\begin{equation}
R = 1.01 \pm 2.8 \,\% ({\rm stat}) \pm 2.7 \,\% ({\rm syst}).
\label{rchooz}
\end{equation}
Thus no evidence was found for a deficit in the flux. The negative result 
is translated into exclusion regions in the $(\Delta m^2,\sin^22\theta)$
plane shown in Fig.~\ref{fig:reactors}. The resulting 90\% CL limits include
$\Delta m^2<7\times10^{-4}$ eV$^2$ for maximal mixing, and $\sin^22\theta<0.10$
for large $\Delta m^2$. The CHOOZ results are significant in excluding part of 
the region that corresponds to the LMA solution of the solar neutrino problem 
(see Sec.~\ref{sec:solar}). Furthermore, the CHOOZ bound rules out with high
significance the possibility that $\nu_\mu\to\nu_e$ oscillations explain the 
atmospheric neutrino deficit. The constraint on the mixing angle is also 
relevant to the interpretation of the atmospheric neutrino anomaly in the 
framework of three-neutrino mixing. We return to this issue in 
Sec.~\ref{sec:atmos}. 

Smaller values of $\Delta m^2$ can be accessed at future reactor experiments 
using longer baseline. Pursuing this idea, the KamLAND experiment 
(Piepke {\em et al.}, 2001), a 1000 ton liquid scintillation detector, is 
currently in operation in the Kamioka mine in Japan. This
underground site is conveniently located at a distance of 150-210 km 
from several Japanese nuclear power stations. The measurement of the flux and 
energy spectrum of the $\bar\nu_e$'s emitted by these reactors will 
provide a test to the LMA solution of the solar neutrino anomaly. In  
Fig.~\ref{fig:reactors} we plot the expected 90\% sensitivity for
the KamLAND experiment after 3 years of data taking (from
Piepke {\em et al.}, 2001). The experiment will, for the first time, provide a 
completely solar model independent test of this particle physics 
solution of the solar neutrino problem. After a few years of data taking, 
it should be capable of either excluding the entire LMA region or, 
not only establishing 
$\nu_{e}\leftrightarrow \nu_{\rm other}$ oscillations, but also 
measuring the oscillation parameters with unprecedented precision 
Data taking is expected to commence in 2002.

\subsection{Long Baseline Experiments at Accelerators}
\label{subsec:lbl}
Smaller values of $\Delta m^2$ can also be accessed using accelerator beams at 
long baseline (LBL) experiments. In these experiments the intense neutrino  
beam from an accelerator is aimed at a detector located underground at a 
distance of several hundred kilometers. 
The main goal of these experiments is 
to test the presently allowed solution for the atmospheric neutrino problem by 
searching for either $\nu_\mu$ disappearance or $\nu_\tau$ appearance.
At present there are three such projects approved: K2K 
(Nishikawa {\em et al.}, 1997; Ahn {\em et al.}, 2001; 
Nishikawa, {\em et al.}, 2001), which runs with a baseline of about 
235 km from KEK to SK, MINOS (Ables {\em et al.}, 1995; 
Wojcicki, 2001) under construction with a baseline of 730 km from Fermilab to 
the Soudan mine where the detector will be placed, and OPERA  
(Shibuya {\em et al.}, 1997; Cocco {\em et al.}, 2000), 
under construction with a baseline of 730 km from CERN to Gran Sasso. With  
their expected sensitivities, these experiments can cover either some fraction 
or all of the parameter region suggested by the atmospheric neutrino anomaly 
discussed in Sec.~\ref{sec:atmos}.

In the K2K experiment, a wide-band, almost pure $\nu_{\mu}$ beam from $\pi^{+}$
decays is generated in the KEK 12-GeV/c Proton Synchrotron and a neutrino 
beam-line. The detector is in SK at a distance of 250 km. Various beam 
monitors along the beam line and two different types of front detectors 
(FDs) have also been 
constructed at the KEK site. The FDs are a 1kt water Cerenkov detector, 
which is a miniature of the SK detector, and a so-called fine-grained detector
which is composed of a scintillating fiber tracker trigger counters, lead glass
counters and a muon range detector. The characteristics of the neutrino beam in
the KEK site -- direction, intensity, stability, energy spectrum and 
$\nu_e-\nu_\mu$ composition -- are examined using FDs and beam monitors.
They are then extrapolated to the SK site and used to obtain 
the expected number of events and the energy spectrum.

The K2K experiment had a successful start in early 1999, and data were recorded
during several periods in 2000/1. The accumulated beam intensity 
during the 2000 
runs was $22.9\times10^{18}$ protons on target (p.o.t.) (Ahn {\em et al.}, 
2001), increased to $\sim 38\times 10^{18}$ p.o.t. with the 2001 run till
the summer (Nishikawa, {\em et al.}, 2001) which was about 40\% 
of the goal of the experiment, $10^{20}$ p.o.t.. 
The no-oscillation prediction for this sample, based on the data from the FDs,
is $63.9^{+6.1}_{-6.6}$ events while a total of 44 events have been 
observed in 
22.5kt of the fiducial volume. The statistical probability that the observation
would be equal to or smaller than 44 is $\sim$ 8\%. In the presence of 
oscillations the expected number of events would be 41.5 for $\Delta m^2=3
\times10^{-3}$ eV$^2$ and maximal mixing. Although the central value of the
number of observed events is consistent with the data from atmospheric 
neutrinos, the discrepancy with the no-oscillation prediction is still within 
the statistical error. K2K has also studied the energy distribution 
of these events compared to the expectation based on the pion monitor 
data and a Monte-Carlo simulation and they find that the data are consistent
within statistics with an oscillation signal.

MINOS is designed to detect neutrinos delivered by the Main Injector 
accelerator at Fermilab (NuMI) with average energies of $\sim$ 5-15 GeV
depending on the beam configuration. Two detectors, functionally identical, 
will be placed in the NuMI neutrino beam: one at Fermilab and the second one in
Soudan iron mine, 732 km away. The MINOS detectors are iron/scintillator 
sampling calorimeters with a toroidal magnetic field in the iron. Observed 
interactions of $\nu_\mu$ can be divided into two classes: CC-like, with an 
identified $\mu$ track, and NC-like, muonless. The ratio of the observed 
numbers of the CC and NC-like events in the two detectors provides a sensitive 
test for oscillations. With its expected sensitivity MINOS will be able to 
precisely measure (roughly at the level of 10\%) the oscillation parameters in 
the $\nu_\mu\to \nu_\tau$ channel. The primary measurement for this
is the comparison of the rate and spectrum of the CC events in 
the Far Detector with those in the Near Detector. Comparing the
NC/CC ratios in the two detectors,  the experiment can also be sensitive to
the presence of $\nu_\mu\to \nu_s$ oscillations. MINOS
is scheduled to start data taking at the end of 2004.

Both K2K and MINOS have also the capability for detecting the appearance of
$\nu_e$'s due to $\nu_\mu\to \nu_e$ oscillations. This signal however suffers 
from large backgrounds due to both the $\nu_e$'s in the beam and the NC events 
with a topology similar to the $\nu_e$ interaction. These backgrounds can be 
partially suppressed using the information from the near detectors. 
In particular, MINOS may be able to improve the CHOOZ bounds.

OPERA is designed to search for $\nu_\mu\to\nu_\tau$ oscillations
in the Gran Sasso Laboratory. It will study the interaction of 20 GeV 
neutrinos produced at CERN. The goal is to observe the appearance
of $\nu_\tau$'s in a pure $\nu_\mu$ beam. The detector is based on a 
massive lead/nuclear emulsion target. Nuclear emulsions are exploited
for the direct observation of the decay of the $\tau$ in a
very low background environment. 

\subsection{Direct Determination of Neutrino Masses}
\label{direcdet}
Oscillation experiments have provided us with important information on
the differences between the neutrino masses-squared, $\Delta m^2_{ij}$, 
and on the leptonic mixing angles, $U_{ij}$. But they are insensitive to 
the absolute mass scale for the neutrinos, $m_i$. 

Of course, the results of an oscillation experiment do provide a lower bound
on the heavier mass in $\Delta m^2_{ij}$, $|m_i|\geq\sqrt{\Delta m^2_{ij}}$ for
$\Delta m^2_{ij}>0$. But there is no upper bound on this mass. In particular,
the corresponding neutrinos could be approximately degenerate at a mass
scale that is much higher than $\sqrt{\Delta m^2_{ij}}$. 
Moreover, there
is neither upper nor lower bound on the lighter mass $m_j$.

Information on the neutrino masses, rather than mass differences, can be 
extracted from kinematic studies of reactions in which a neutrino or an 
anti-neutrino is involved. In the absence of mixing the present limits are
(Groom {\em et al.},  2000)
\begin{eqnarray}
m_{\nu_\tau}<18.2\; {\rm MeV}\;\; (95\% \;{\rm CL}) & 
\;\;\;{\rm from}\;\;\; & \tau^-\rightarrow n\pi + \nu_\tau\\
m_{\nu_\mu}<190\; {\rm keV}\;\; (90\% \;{\rm CL})   
& \;\;\;{\rm from}\;\;\; & \pi^-\rightarrow \mu^- +\overline\nu_\mu \\
m_{\nu_e}<2.2\; {\rm eV}\;\;  (95\% \;{\rm CL})     & \;\;{\rm from}&\;\;\;  
{\rm ^3H \rightarrow\ ^3He+e^-+\overline\nu_e,} \label{nuelim}
\end{eqnarray}
where for the bound on $m_{\nu_e}$ we take 
the latest limit from the Mainz experiment (Bonn {\em et al.}, 2001).
A similar bound is obtained by Troitsk experiment  
(Lobashev {\em et al.}, 2001).
A new experimental project, KATRIN, is under consideration with an estimated 
sensitivity limit: $m_{\nu_e}\sim0.3$ eV.

In the presence of mixing these limits have to be modified and in general
they involve more than a single flavor parameter. The limit that is most 
relevant to our purposes is the most sensitive one from tritium beta decay. In 
presence of mixing, 
the electron neutrino is a combination of mass eigenstates and the 
tritium beta decay spectrum is modified as (Shrock, 1980): 
\begin{equation}
{dN \over dE}=R(E)\sum_i|U_{ei}|^2[(E_0-E)^2-{m_i}^2]^{1/2}\Theta(E_0-E-m_i),
\label{spectrum2}
\end{equation}
where $E$ is the energy of electron, $E_0$ is the total decay
energy and $R(E)$ is $m_\nu$-independent. The step function, 
$\Theta(E_0-E-m_i)$, reflects the fact that a given 
neutrino can only be produced if the available energy is larger than its mass.
According to Eq.~(\ref{spectrum2}), there are two important 
effects, sensitive to the neutrino masses and mixings, on the electron energy 
spectrum: (i) Kinks at the electron energies $E_e^{(i)}=E\sim E_0-m_i$ with
sizes that are determined by $|U_{ei}|^2$; (ii) A shift of the end point to 
$E_{\rm ep}=E_0-m_1$, where $m_1$ is the lightest neutrino mass.
The situation simplifies considerably if we are interested in constraining the 
possibility of quasi-degenerate neutrinos with mass $\sim m_\nu$. 
In this case the distortion of the spectrum can be described by a single 
parameter (Vissani, 2001a; Farzan, Peres and Smirnov, 2001),
$m_{\beta}={\sum_i m_i |U_{ei}|^2 \over \sum_i |U_{ei}|^2}\sim m_\nu$.
So the limit in Eq.~(\ref{nuelim}) applies to the unique neutrino mass
scale.

Direct information on neutrino masses can also 
be obtained from neutrinoless double 
beta decay ($2 \beta 0 \nu$) searches: 
\begin{equation}
(A,Z) \rightarrow (A,Z+2) + e^{-} + e^{-}.
\end{equation}
The rate of this process is proportional to the 
{\it effective Majorana mass of $\nu_e$},
\begin{equation}
m_{ee}=\left| \ \ \sum_i m_i U_{ei}^2 \ \ \right| 
\end{equation}
which, in addition to five parameters that affect the tritium beta decay 
spectrum, depends also on the three leptonic CP violating phases. 
Notice that in order to induce the $2\beta0\nu$ decay, $\nu_e$ must be a 
Majorana particle. 

The present strongest bound from $2\beta0\nu$-decay is obtained by
Heidelberg-Moscow group (Klapdor-Kleingrothaus {\em et al.}, 2001):
\begin{equation}
m_{ee} < 0.34 \ \  (0.26 )\  \ {\rm eV}, \ \ \
{\rm  \ \  \ \   90~\% ~~ (68 \%)  \ \ C.L.}.
\end{equation}
Taking into account systematic errors related to nuclear matrix elements,
the bound may be weaker by a factor of about 3. A sensitivity of 
$m_{ee}\sim0.1$ eV is expected to be reached by the currently running 
NEMO3 experiment (Marquet {\em et al}, 2000), while a series of new 
experiments (CUORE, EXO, GENIUS) is planned with sensitivity of up to  
$m_{ee} \sim 0.01$ eV. 

The knowledge of $m_{ee}$ will provide information on the mass
and mixing parameters that is independent of the $\Delta m^2_{ij}$'s. However,
to infer the values of neutrino masses, additional assumptions are required.
In particular, the mixing elements are complex and may lead to strong 
cancellation, $m_{ee}\ll m_1$. Yet, the combination of results
from $2\beta0\nu$ decays and Tritium beta decay can test and,
in some cases, determine the mass parameters of given
schemes of neutrino masses (Vissani, 1999; Farzan, Peres and Smirnov, 2001; 
Bilenky, Pascoli and Petcov, 2001a, 2001b;  Klapdor-Kleingrothaus, Pas and 
Smirnov, 2001; Pascoli and Petcov 2002).
\section{Three- and Four-Neutrino Mixing}
\label{sec:schemes}
In the previous sections we discussed the three pieces of evidence for neutrino
masses and mixing (solar neutrinos, atmospheric neutrinos and the LSND results)
as usually formulated in the framework of two-neutrino oscillations. The 
results are summarized in Fig.~\ref{fig:sum} where we show the ranges of masses
and mixing implied by these signals at 90 and 99\% CL for 2 d.o.f., as well as 
relevant constraints from negative searches in laboratory experiments. 

The three pieces of evidence correspond to three values of mass-squared 
differences of different orders of magnitude. Consequently, there is no
consistent explanation to all three signals based on oscillations among
the three known neutrinos. The argument for this statement is very simple.
With three neutrinos, there are only two independent mass-squared differences,
since the following relation must hold:
\begin{equation}
\Delta m^2_{21}+\Delta m^2_{32}+\Delta m^2_{13}=0.
\end{equation}
This relation cannot be satisfied by three $\Delta m^2_{ij}$ that are of
different orders of magnitude. One may wonder if this naive extrapolation 
from the two-neutrino oscillation picture holds once the full mixing 
structure of the three-neutrino oscillations in taken into account or, 
on the contrary, some special configuration of the three-neutrino parameters 
could fit the three pieces of evidence. The combined fit to 
the data performed by Fogli, Lisi, Montanino and Scioscia (1999a) shows
that this is not the case (see also Gonzalez-Garcia and Maltoni, 2002).

Whereas in the case of the solar and atmospheric neutrino indications, several 
experiments agree on the existence of the effect, the third indication is 
presently found only by the LSND experiment. Therefore, in many studies the 
LSND result is left out and the analysis of the solar and atmospheric data
is performed in the framework of mixing between the three known neutrinos.
In Sec.~\ref{subsec:threemix} we discuss the derived masses and mixing
in these scenarios. On the other hand, if all three indications in 
favour of neutrino oscillations are confirmed, one needs a minimum of four 
neutrinos. The present phenomenological status of the possibility of 
mixing between four neutrinos is discussed in Sec.~\ref{subsec:fourmix}.
Alternatively a possible
explanation of the four pieces of evidence with only three neutrinos
has been proposed assuming that CPT is violated in the neutrino sector
(see for instance, Murayama and Yanagida, 2001; 
Barenboim, Borissov, Lykken and Smirnov, 2001). 

\subsection{Three-Neutrino Mixing}
\label{subsec:threemix}

The combined description of both solar and atmospheric anomalies requires 
that all three known neutrinos take part in the oscillations. The mixing
parameters are encoded in the $3 \times 3$ lepton mixing matrix 
(Maki, Nakagawa and Sakata, 1962; Kobayashi and Maskawa, 1973). 
The two
Majorana phases do not affect neutrino oscillations
(Bilenky, Hosek and Petcov, 1980; Langacker {\em et al.}, 1987) 
. The Dirac phase (that
is the analog of the KM phase of the quark sector) does affect neutrino
oscillations in general, but for the purposes of this section we can set it
to zero.  

In this case the mixing matrix can be conveniently parametrized 
in the standard form (Groom {\em et al.},  2000):
\begin{equation}
U =\left(\begin{array}{ccc} 
c_{13} c_{12} & s_{12} c_{13} & s_{13} \\
-s_{12} c_{23} - s_{23} s_{13} c_{12} & c_{23} c_{12} - s_{23} s_{13} s_{12}
 & s_{23} c_{13} \\
s_{23} s_{12} - s_{13} c_{23} c_{12} & -s_{23} c_{12} - s_{13} s_{12} c_{23}
 & c_{23} c_{13} \end{array}\right) \;,
\label{eq:evol.2} 
\end{equation}
where $c_{ij}\equiv\cos\theta_{ij}$ and $s_{ij} \equiv \sin\theta_{ij}$.

As we have seen in the previous sections, in most of the
parameter space of solutions for solar and atmospheric oscillations, 
the required mass differences satisfy
\begin{equation}
\Delta m^2_\odot\ll \Delta m^2_{\rm atm}.
\label{deltahier}
\end{equation}
In this approximation the angles $\theta_{ij}$ can be taken without 
loss of generality to lie in the first quadrant, $\theta_{ij}\in[0,\pi/2]$.  
There are two possible mass orderings which we chose as
\begin{eqnarray}
\Delta m^2_{21}=\Delta m^2_\odot &\ll&
\Delta m^2_{32}\simeq\Delta m^2_{31}=\Delta m^2_{\rm atm}>0; \label{direct}\\
\Delta m^2_{21}=\Delta m^2_\odot &\ll&
-\Delta m^2_{31}\simeq-\Delta m^2_{32}=|\Delta m^2_{\rm atm}|>0. 
\label{inverted}
\end{eqnarray}
We refer to the first option, Eq.~(\ref{direct}), as the 
{\it direct} scheme,
and to the second one, Eq.~(\ref{inverted}), as the {\it inverted} scheme.  
The direct scheme is naturally related to hierarchical masses,
$m_1\ll m_2\ll m_3$, for which $m_2\simeq\sqrt{\Delta m^2_{21}}$ and 
$m_3\simeq\sqrt{\Delta m^2_{32}}$, or to quasi-degenerate masses, $m_1\simeq 
m_2\simeq m_3\gg \Delta m^2_{21}, \Delta m^2_{32}$. On the other hand,   
the inverted scheme implies that $m_3< m_1\simeq m_2$.  

One may wonder how good an approximation it is to set the CP violating
phases to zero. It turns out to be an excellent approximation for the analysis 
of solar, atmospheric and laboratory data if Eq.~(\ref{deltahier}) holds.
In this case, as discussed below, no simultaneous effect of the two mass 
differences is observable in any $\nu$-appearance transition. 

\subsubsection{Probabilities}
The determination of the oscillation probabilities for both solar and 
atmospheric neutrinos requires that one solves the evolution equation of the 
neutrino system in the matter background of the Sun or the Earth.
In a three-flavor framework, this equation reads:
\begin{equation}
i \frac{d\vec{\nu}}{dt} = H \, \vec{\nu},\;\;\;\;\;\;\;\; 
H = U \cdot H_0^d \cdot U^\dagger + V \;,
\label{eq:evol.1} 
\end{equation}
where $U$ is the lepton mixing matrix, $\vec{\nu}\equiv\left(\nu_e,\nu_\mu, 
\nu_\tau\right)^T$, $H_0^d$ is the vacuum hamiltonian, 
\begin{equation}
H_0^d=\frac{1}{2E_\nu}{\rm diag}
\left(-\Delta m^2_{21},0,\Delta m^2_{32}\right),
\label{eq:evol.3} 
\end{equation}
and $V$ is the effective potential that describes charged-current forward 
interactions in matter:
\begin{equation}
V = {\rm diag} \left( \pm \sqrt{2} G_F N_e, 0, 0 \right)
\equiv {\rm diag} \left( V_e, 0, 0 \right).
 \label{eq:evol.4} 
\end{equation}
In Eq.~(\ref{eq:evol.4}), the sign $+$ ($-$) refers to neutrinos 
(antineutrinos), and $N_e$ is electron number density in the Sun or the Earth.

In what follows we focus on the direct scheme of Eq.~(\ref{direct}),
for which the five relevant parameters are related to experiments in
the following way: 
\begin{eqnarray} 
   \label{oscpardef}
\Delta m^2_{\odot}&=&\Delta m^2_{21},\ \ \ 
\Delta m^2_{\rm atm}=\Delta m^2_{32},\\
\theta_{\odot}&=&\theta_{12},\ \ \ 
\theta_{\rm atm}=\theta_{23},\ \ \
\theta_{\rm reactor}=\theta_{13}\;.
\end{eqnarray}
For transitions in vacuum, the results apply also to the inverted scheme of 
Eq.~(\ref{inverted}). In the presence of matter effects, the direct and 
inverted schemes are no longer equivalent, although the difference is
hardly recognizable in the current solar and atmospheric neutrino 
phenomenology as long as  Eq.~(\ref{deltahier}) holds 
(Fogli, Lisi, Montanino and Scioscia, 1997; Gonzalez-Garcia and Maltoni, 2002)
. Under this approximation, 
the results obtained for the direct scheme can be applied to the inverted 
scheme by replacing $\Delta m^2_{32}\to -\Delta m^2_{32}$.

In general the transition probabilities present an oscillatory behaviour with 
two oscillation lengths. However, the hierarchy in the splittings, 
Eq.~({\ref{deltahier}), leads to important simplifications. 

Let us first consider the analysis of solar neutrinos. A first simplification
occurs because $L^{\rm osc}_{32}=4\pi E/\Delta m^2_{32}$ is much shorter
than the distance between the Sun and the Earth. Consequently, the oscillations
related to $L^{\rm osc}_{32}$ are averaged in the evolution from the Sun to 
the Earth. A second simplification occurs because, for the evolution
in both the Sun and the Earth, $\Delta m^2_{32}\gg2\sqrt{2}G_FN_eE_\nu
\sin^22\theta_{13}$. Consequently, matter effects on the evolution of
$\nu_3$ can be neglected. The result of these two approximations is that
the three-flavor evolution equations decouple into an effective two-flavor 
problem for the basis (Kuo and Pantaleone, 1986; Shi and Schramm, 1992)  
\begin{equation}
\nu_{e^\prime}=c_{12}\nu_1+s_{12}\nu_2\; ,\ \ \ \ 
\nu_{\mu^\prime}=-s_{12}\nu_1+c_{12} \nu_2  \; ,
\label{eigendef}
\end{equation} 
with the substitution of $N_e$ by the effective density 
\begin{equation}
N_{e}\Rightarrow N_e \cos^2 \theta_{13} \; .
\label{pote}
\end{equation}
Thus the survival probability takes the following form:
\begin{equation}
P^{3\nu}_{ee,\odot}=\sin^4\theta_{13}+\cos^4\theta_{13}
P^{2\nu}_{e^\prime e^\prime,\odot} ,
\label{p3}
\end{equation}
where $P^{2\nu}_{e^\prime e^\prime,\odot}$ is the two-flavor survival 
probability in the ($\Delta m^2_{21}, \theta_{12}$) parameter space but with 
the modified matter density of Eq.~(\ref{pote}). We conclude that the analysis 
of the solar data constrains three of the five independent oscillation 
parameters: $\Delta m^2_{21}, \theta_{12}$ and $\theta_{13}$. 

Eq.~(\ref{p3}) reveals what is the dominant effect of a non-vanishing 
$\theta_{13}$ in the solar neutrino survival probability: the energy dependent 
part of the probability, $P^{2\nu}_{e^\prime e^\prime,\odot}$, gets damped by 
the factor $\cos^4\theta_{13}$, while an energy independent term, 
$\sin^4\theta_{13}$, is added. Thus increasing $\theta_{13}$ makes the solar 
neutrino survival probability more and more energy independent. 

Let us now consider the analysis of atmospheric neutrinos. Here
$L^{\rm osc}_{21}=4\pi E/\Delta m^2_{21}$ is much larger than the relevant
distance scales. Consequently, the corresponding oscillating phase is 
negligible. In this approximation one can rotate away the corresponding angle 
$\theta_{12}$. Thus the resulting survival probabilities do not depend on
$\Delta m^2_{21}$ and $\theta_{12}$. For instance for 
constant Earth matter density,
the various $P_{\alpha\beta}$ can be written as follows:
\begin{eqnarray}
P_{ee}&=& 1-4s^2_{13,m} c^2_{13,m}  \, S_{31} ,\nonumber\\
P_{\mu\mu}&=& 1-4s^2_{13,m} c^2_{13,m} s^4_{23}\,S_{31}
          -4s^2_{13,m} s^2_{23} c^2_{23}\,S_{21}
          -4c^2_{13,m} s^2_{23} c^2_{23}\,S_{32},\nonumber\\
P_{\tau\tau} &=& 1-4s^2_{13,m} c^2_{13,m} c^4_{23}\,S_{31}
               -4s^2_{13,m} s^2_{23} c^2_{23}\,S_{21}
               -4c^2_{13,m} s^2_{23} c^2_{23}\,S_{32}\ ,\label{eq:P3atm} \\
P_{e\mu}     &=& 4s^2_{13,m} c^2_{13,m} s^2_{23}\,S_{31},\nonumber\\
P_{e\tau}    &=& 4s^2_{13,m} c^2_{13,m} c^2_{23}\,S_{31},\nonumber\\
P_{\mu\tau}  &=&-4s^2_{13,m} c^2_{13,m} s^2_{23} c^2_{23}\,S_{31}
                  +4s^2_{13,m} s^2_{23} c^2_{23}\,S_{21}
                  +4c^2_{13,m} s^2_{23} c^2_{23}\,S_{32}.\nonumber
\end{eqnarray}
Here $\theta_{13,m}$ is the effective mixing angle in matter: 
\begin{equation}\label{eq:Phi}
\sin 2\theta_{13,m}=\frac{\sin2\theta_{13}}{\sqrt{(\cos2\theta_{13}- 
2 E_\nu V_e/\Delta m^2_{32})^2+(\sin2\theta_{13})^2}}
\end{equation}
and $S_{ij}$ are the oscillating factors in matter:
\begin{equation}
S_{ij}=\sin^2\left(\frac{\Delta\mu^2_{ij}}{4E_\nu}L\right).
\label{defSij}
\end{equation}
In Eq.~(\ref{defSij}), $\Delta\mu^2_{ij}$ are the effective mass-squared 
differences in matter:
\begin{eqnarray}
\Delta\mu^2_{21}&=&\frac{\Delta m^2_{32}}{2}\left(\frac{\sin 2\theta_{13}}
{\sin 2\theta_{13,m}}-1\right)-E_\nu V_e\ ,\nonumber\\
\Delta\mu^2_{32}&=&\frac{\Delta m^2_{32}}{2}\left(\frac{\sin 2\theta_{13}}
{\sin 2\theta_{13,m}}+1\right)+E_\nu V_e,\nonumber\\
\Delta\mu^2_{31}&=&\Delta m^2_{32}\frac{\sin 2\theta_{13}}
{\sin 2\theta_{13,m}}\; .
\end{eqnarray}
and $L$ is the pathlength of the neutrino within the Earth, which depends on
its direction. We conclude that the analysis of the atmospheric data constrains
three of the five independent oscillation parameters: $\Delta m^2_{32}$, 
$\theta_{23}$ and $\theta_{13}$. 

So we find that in the approximation of Eq.~(\ref{deltahier}) 
the mixing angle $\theta_{13}$ is the only parameter common to both solar 
and atmospheric neutrino oscillations and which may potentially allow for some 
mutual influence. The main effect of the three-neutrino mixing is that now 
atmospheric neutrinos can oscillate simultaneously in both the 
$\nu_\mu\to\nu_\tau$ and $\nu_\mu\to\nu_e$ (and, similarly,  $\nu_e\to\nu_\tau$
and $\nu_e\to\nu_\mu$) channels. The oscillation amplitudes for channels 
involving $\nu_e$ are controlled by the size of $\sin^2\theta_{13}=|U_{e3}|^2$. 
We learn that in the approximation  of Eq.~(\ref{deltahier})
solar and atmospheric neutrino oscillations decouple in the limit
$\theta_{13}=0$. This angle is constrained by the CHOOZ reactor experiment.
To analyze the CHOOZ constraints we need to evaluate the survival probability 
for ${\bar \nu}_e$ of average energy $E\sim$ few MeV at a distance of $L\sim 1$
Km. For these values of energy and distance, one can safely neglect Earth 
matter effects. The survival probability takes the analytical form:
\begin{eqnarray}
P_{ee}^{\rm CHOOZ}&=& 1-\cos^4\theta_{13}\sin^22\theta_{12}
\sin^2\left(\frac{\Delta m^2_{21} L}{4 E_\nu} \right) \\ \nonumber
&  & -\sin^22\theta_{13}\left[\cos^2\theta_{12}\sin^2\left(\frac
{\Delta m^2_{31} L}{4 E_\nu}\right)+\sin^2\theta_{12}\sin^2\left(\frac
{\Delta m^2_{32} L}{4 E_\nu}\right)\right]\\\nonumber
&\simeq&1-\sin^22\theta_{13}\sin^2\left(\frac{\Delta m^2_{32}L}{4E_\nu}\right),
\label{pchooz}
\end{eqnarray}
where the second equality holds under the  approximation
$\Delta m^2_{21}\ll E_\nu/L$ which can only be safely made 
for $\Delta m^2_{21}\lesssim 3\times 10^{-4}$eV$^2$. Thus in general
the analysis of the CHOOZ reactor date involves four 
oscillation parameters: $\Delta m^2_{32}$, $\theta_{13}$, 
$\Delta m^2_{21}$, and $\theta_{12}$ (Gonzalez-Garcia, 
Maltoni, 2002;  Bilenky, Nicolo and Petcov, 2001)

\subsubsection{Allowed masses and mixing}
There are several three-neutrino oscillation analyses in the literature which 
include either solar (Fogli, Lisi and Montanino, 1996; Fogli, Lisi, Montanino 
and Palazzo, 2000a, 2000b; Gago, Nunokawa and Zukanovich, 2001)  
or atmospheric (Fogli, Lisi, Montanino and Scioscia, 1997; Fogli, Lisi, Marrone
and Scioscia, 1999b; Fogli, Lisi, Marrone, 2001b; de Rujula, Gavela
and Hernandez, 2001; Teshima and Sakai 1999) neutrino data. Combined studies 
have also been performed (Barger, Whistnant and Phillips, 1980; Fogli, Lisi and
Montanino, 1994, 1995; Barbieri {\em et al.}, 1998; 
Barger and Whisnant, 1999). 
We follow and update here the results from the analysis of 
Gonzalez-Garcia, Maltoni, Pe\~na-Garay and Valle (2001).

As discussed above, in the approximation of Eq.~(\ref{deltahier}),
solar and atmospheric neutrino oscillations decouple in the
limit $|U_{e3}|=\sin\theta_{13}=0$. In this limit the allowed values of the  
parameters can be obtained directly from the results of the analyses in terms 
of two-neutrino oscillations presented in Sec.~\ref{sec:solar} and 
Sec.~\ref{sec:atmos}. Deviations from the two-neutrino scenario are then 
determined by the size of $\theta_{13}$. Thus the first question to answer is 
how the presence of this additional angle affects the analysis of the solar
and atmospheric neutrino data. 

In Fig.~\ref{fig:solar3} we show the allowed regions for the oscillation 
parameters $\Delta m^2_{21}$ and $\tan^2\theta_{12}$ from the global analysis 
of the solar neutrino data in the framework of three-neutrino oscillations 
for different values of $\sin^2\theta_{13}$ (updated from Gonzalez-Garcia, 
Maltoni, Pe\~na-Garay and Valle, 2001). The allowed 
regions for a given CL are defined as the set of points satisfying the 
condition $\chi^2(\Delta m_{12}^2,\tan^2\theta_{12},\tan^2\theta_{13})
-\chi^2_{\rm min}\leq\Delta\chi^2 \mbox{(CL, 3~d.o.f)}$
where, for instance, $\Delta\chi^2($CL, 3~d.o.f)=6.25, 7.83, and 11.36 for 
CL=90, 95, and 99\%, respectively. The global minimum used in the construction 
of the regions lies in the LMA region and corresponds to $\tan^2\theta_{13}=0$,
that is, to the {\it decoupled} scenario. 

As seen in the figure, the modifications to the decoupled case are significant
only if $\theta_{13}$ is large. As $\sin^2\theta_{13}$ increases, all the 
allowed regions disappear, leading to an upper bound on $\sin^2\theta_{13}$ 
that is independent of the values taken by the other parameters in the 
three-neutrino mixing matrix. For instance, no region of parameter space is 
allowed (at 99\% C.L. for 3 d.o.f) for $\sin^2\theta_{13}=|U_{e3}|^2> 0.80$. 
This fact is also illustrated in Fig.~{\ref{fig:chi2t13}} where we plot the 
shift in $\chi^2$ as a function of $\sin^2\theta_{13}$ when the mass and mixing
parameters $\Delta m^2_{12}$ and $\tan^2\theta_{12}$ are chosen to minimize 
$\chi^2$. 

In Fig.~\ref{fig:atmos3} we show the allowed regions for the oscillation
parameters $\Delta m^2_{32}$ and $\tan^2\theta_{23}$ from the analysis of the 
atmospheric neutrino data in the framework of three-neutrino oscillations for 
different values of $\sin^2\theta_{13}$ 
(updated from Gonzalez-Garcia, Maltoni, Pe\~na-Garay and Valle, 2001). 
In the upper-left panel $\tan^2\theta_{13}=0$ which corresponds to pure 
$\nu_\mu \to \nu_\tau$ oscillations. Note the exact symmetry of the 
contour regions under the transformation $\theta_{23}\to\pi/4-\theta_{23}$. 
This symmetry follows from the fact that in the pure $\nu_\mu \to \nu_\tau$ 
channel matter effects cancel out and the oscillation probability depends on 
$\theta_{23}$ only through the double-valued function $\sin^22\theta_{23}$
(see Sec.~\ref{subsec:atmosc}).  For $\theta_{13}\neq0$ this symmetry breaks 
due to the three-neutrino mixing structure even if matter effects are 
neglected. The analysis of the full atmospheric neutrino data in the framework 
of three-neutrino oscillations clearly favours the $\nu_\mu \to \nu_\tau$ 
oscillation hypothesis. As a matter of fact, the best fit corresponds to a very
small value, $\theta_{13}= 6^\circ$, but it is statistically 
indistinguishable from the decoupled scenario, $\theta_{13}=0^\circ$. 
No region of parameter space is allowed (at 99\% C.L. for 3 d.o.f) for 
$\sin^2\theta_{13}=|U_{e3}|^2> 0.40$. 
The physics reason for this limit is clear from the discussion of the 
$\nu_\mu\to\nu_e$ oscillation channel in Sec.~\ref{subsec:atmosc}: large values
of $\theta_{13}$ imply a too large contribution of the $\nu_\mu \to \nu_e$ 
channel and would spoil the otherwise successful description of the angular
distribution of contained events. This situation is illustrated also in 
Fig.~{\ref{fig:chi2t13}} where we plot the shift of $\chi^2$ for the analysis 
of atmospheric data in the framework of oscillations between three neutrinos 
as a function of $\sin^2\theta_{13}$ when the mass and mixing parameters 
$\Delta m^2_{32}$ and $\tan^2\theta_{23}$ are chosen to minimize $\chi^2$. 

For any value of the mixing parameters, the mass-squared difference 
relevant for the atmospheric analysis is restricted to lie in the interval 
$1.25\times 10^{-3}<\; \Delta m^2_{32}/\mbox{\rm eV$^2$}<8\times 10^{-3}$ at 
99 \% CL. Thus it is within the range of sensitivity of the CHOOZ experiment. 
Indeed, as illustrated in Fig.~{\ref{fig:chi2t13}}, the limit on 
$\sin^2\theta_{13}$ becomes stronger when the CHOOZ data are 
combined with the atmospheric and solar neutrino results. 

One can finally perform a global analysis in the five dimensional parameter 
space combining the full set of solar, atmospheric and reactor data. Such 
analysis leads to the following allowed 3$\sigma$ ranges for individual 
parameters (that is, when the other four parameters have been chosen to 
minimize the global $\chi^2$):
\begin{eqnarray}
2.4\times 10^{-5}<& \Delta m^2_{21}/\mbox{\rm eV$^2$}&<
2.4\times 10^{-4}\;\;\;\;\;{\rm LMA}\; 
\nonumber \\
0.27<&\tan^2\theta_{12}&< 0.79 \;\;\;\;\;\;\;\;\;\;\;\;\;\;\;
{\rm LMA} \;
\nonumber \\
1.4\times 10^{-3}<& \Delta m^2_{32}/\mbox{\rm eV$^2$}&< 
6.0\times 10^{-3} 
\nonumber\\ 
0.4<&\tan^2\theta_{23}&< 3.0, 
\nonumber\\ 
\sin^2\theta_{13} &< 0.06 & 
\label{globalranges}
\end{eqnarray} 
These results can be translated into our present knowledge of the
moduli of the mixing matrix $U$:
\begin{equation}
|U| =\pmatrix{ 0.73-0.89&0.45-0.66&<0.24 \cr
0.23-0.66&0.24-0.75&0.52-0.87\cr 
0.06-0.57&0.40-0.82&0.48-0.85\cr}.   
\end{equation}
In conclusion, we learn that at present the large mixing-type solutions 
provide the best fit. As concerns $|U_{e3}|$, both solar and atmospheric data 
favour small values and this trend is strengthened by the reactor data. 

\subsection{Four-Neutrino Mixing}
\label{subsec:fourmix}
If all three indications of neutrino oscillations are confirmed, then a
minimum of four neutrinos will be required to accommodate the data. As 
discussed in Sec.~\ref{sec:standard}, the measurement of the decay width of the
$Z^0$ boson into neutrinos makes the existence of three, and only three, light 
(that is, $m_\nu\lsim m_Z/2$) active neutrinos an experimental fact 
(Eq.~{\ref{GAMinv}). Therefore, a fourth neutrino must not couple to the 
standard electroweak current, that is, it must be sterile.

One of the most important issues in the context of four-neutrino scenarios is 
the four-neutrino mass spectrum. There are six possible four-neutrino schemes,
shown in Fig.~\ref{fig:4mass}, that can accommodate the results from solar and 
atmospheric neutrino experiments as well as the LSND result. They can be 
divided in two classes: (3+1) and (2+2) (Barger {\em et al.}, 2001). 
In the (3+1) schemes, there is a group of three close-by neutrino masses that 
is separated from the fourth one by a gap of the order of 1~eV$^2$, which is 
responsible for the SBL oscillations observed in the LSND experiment. 
In (2+2) schemes, there are two pairs of close masses separated by the LSND 
gap. The main difference between these two classes is the following: if a
(2+2)-spectrum is realized in nature, the transition into the sterile neutrino
is a solution of either the solar or the atmospheric neutrino problem, 
or the sterile neutrino takes part in both, whereas with a (3+1)-spectrum the
sterile neutrino could be only slightly mixed with the active
ones and mainly provide a description of the LSND result.

As concerns the mixing parameters, we emphasize that the mixing matrix
that describes CC interactions in these schemes is a $3\times4$ matrix. 
The reason is that
there are three charged lepton mass eigenstates ($e,\mu,\tau$) and four
neutrino mass eigenstates ($\nu_1,\nu_2,\nu_3,\nu_4$). 
As discussed in Sec.~\ref{subsec:lepmix}, if we choose an
interaction basis where the charged leptons are the mass eigenstates,
then the CC mixing matrix $U$ is a sub-matrix of the $4\times4$ unitary matrix
$V^\nu$ that rotates the neutrinos from the interaction basis to the mass
basis, where the line corresponding
to $\nu_s$ is removed.

\subsubsection{Status of (3+1) schemes}
It has been argued in the literature that the (3+1)-spectra are strongly 
disfavoured by the data from SBL laboratory experiments (Okada and Yasuda, 
1997; Bilenky, Giunti, and Grimus, 1998; Barger, Pakvasa, Weiler and Whisnant, 
1998a; Bilenky, Giunti, Grimus and Schwetz, 1999). At what level they can 
actually be ruled out has been a matter of debate and the answer has also 
changed with the changes in the analysis of the LSND and KARMEN collaborations 
(Barger {\em et al.}, 2000; Giunti and Laveder, 2001; 
Peres and Smirnov, 2001). 

The arguments which disfavour the (3+1)-mass spectra are based on exclusion 
curves from SBL experiments, and on some simplified treatment of the solar and 
atmospheric neutrino data.  We summarize here these arguments 
and the present phenomenological status as given in 
Grimus and Schwetz (2001), Maltoni, Schwetz, and Valle (2001a,2001b) and
Maltoni, Tortola, Schwetz, and Valle (2002b).

The probability $P_{\mu e}$ that is relevant for LSND (and also for KARMEN and 
NOMAD) is given by 
\begin{equation}
P_{\mu e} = P_{\bar\mu\bar e}=4|U_{e4}U_{\mu4}|^2\ 
\sin^2 \frac{\Delta m^2L}{4E} \,,
\label{Pmue}
\end{equation}
where $L$ is the distance between source and detector. In this expression solar
and atmospheric splittings have been neglected as they are too small to give 
any observable effect at the $L/E$ that is relevant for LSND. In this 
approximation, for schemes (3+1)A, 
$\Delta m^2\equiv\Delta m^2_{41}=\Delta m^2_{42}=\Delta m^2_{43}$.

The LSND experiment gives then an allowed region in the 
$\Delta m^2-|U_{e4}U_{\mu4}|^2$ plane which can be directly obtained
from the two-neutrino oscillation region shown in Fig.~\ref{fig:lsnd}
with the identification $\sin^22\theta\to4|U_{e4}U_{\mu4}|^2$.
In the same way, the KARMEN experiment gives an excluded region in the
same plane which can be directly obtained from Fig.~\ref{fig:lsnd}.

Further constraints on $|U_{e4}U_{\mu4}|^2$ can be obtained by combining the 
bounds on $|U_{e4}|$ and $|U_{\mu4}|$ from reactor and accelerator experiments
in combination with the information from solar and atmospheric neutrinos. The 
strongest constraints in the relevant $\Delta m^2$ region are given by the 
Bugey (Achkar {\em et al.}, 1995) and CDHS (Dydak {\em et al.}, 1984) 
experiments. Using their limits on the survival probabilities one finds
\begin{eqnarray}
4|U_{e4}|^2(1-|U_{e4}|^2)&<&D^{\rm Bugey}_e(\Delta m^2),\nonumber\\
D^{\rm Bugey}_e&<&0.001-0.1\ ({\rm for}\ 0.1\lesssim\Delta m^2/\mbox{eV}^2
\lesssim 10),\label{bugey31}\\
4 |U_{\mu4}|^2(1-|U_{\mu4}|^2)&<&D^{\rm CDHS}_\mu(\Delta m^2),\nonumber\\
D^{\rm CDHS}_\mu&<&0.05-0.1\ ({\rm for}\ \Delta m^2\gtrsim0.5\ \mbox{eV}^2),
\label{cdhs31}
\end{eqnarray}
while for lower $\Delta m^2$ the CDHS bound weakens considerably.

In principle, the inequalities in Eq.~(\ref{bugey31}) and Eq.~(\ref{cdhs31})
can be satisfied with either small mixing parameters 
$|U_{\alpha4}|^2\lesssim  D_{\alpha}/4$, 
or close to maximal mixing, $|U_{\alpha4}|^2\gtrsim 1-D_{\alpha}/4$.
This is the generalization to these schemes of the symmetry 
$\theta\leftrightarrow \frac{\pi}{2}-\theta$ of two-neutrino 
vacuum oscillations (see Sec.~\ref{subsec:osvac}).
Solar and atmospheric data are invoked in  order to resolve this ambiguity.
Since in any of the allowed regions for solar oscillations 
$P^\odot_{ee}\lesssim 0.5$, only the small values of $|U_{e4}|$ are possible.
For atmospheric neutrinos one can use the fact that oscillations with the large
$\Delta m^2$  would wash-out the up-down asymmetry and
this wash-out grows with 
the projection of the $\nu_\mu$  over states separated by the large 
$\Delta m^2$,  which is controlled by the mixing parameter $|U_{\mu4}|$.
Consequently only the small values of  $|U_{\mu4}|$ are allowed.
Thus naively one obtains the bound 
\begin{equation}
4|U_{e4}U_{\mu4}|^2<0.25 D^{\rm Bugey}_e(\Delta m^2)
D^{\rm CDHS}_\mu(\Delta m^2),
\end{equation}
which further strengthens the exclusion curve corresponding to the 
KARMEN bound. A detailed and statistically meaningful evaluation of 
the final combined limit, including also the results from other
experiments like NOMAD and CHOOZ, leads to the results presented
in Fig.~\ref{fig:3plus1} (Grimus and Schwetz, 2001; Maltoni, Schwetz and 
Valle, 2001a). The figure shows that there is no overlap of the region allowed 
by the combined bound at 95\% CL with the region allowed by LSND at 99\% CL. 
For the combined  bound at 99\% CL there are marginal overlaps with the 99\% CL
LSND allowed region at $\Delta m^2 \sim0.9$ and 2 eV$^2$, and a very marginal
overlap region still exists around 6 eV$^2$. 

\subsubsection{(2+2) schemes: active-sterile admixtures}
The main feature of (2+2)-spectra is that either solar or atmospheric 
oscillations must involve the sterile neutrino. Such oscillations are, however,
disfavored for both the solar (see Sec.~\ref{sec:solar}) and atmospheric (see 
Sec.~\ref{sec:atmos}) neutrinos. One expects then that the (2+2) schemes are 
disfavoured. However, as first discussed by Dooling, Giunti, Kang and Kim 
(2000), within (2+2) schemes, oscillations into pure active or pure sterile 
states are only limiting cases of the most general possibility of oscillations 
into an admixture of active and sterile neutrinos. One can wonder then whether 
some admixture of active-sterile oscillations gives an acceptable 
description of both solar (Giunti, Gonzalez-Garcia and Pe\~na-Garay, 2000)
and atmospheric (Yasuda, 2000; Fogli, Lisi and Marrone, 2001a, 2001b) data.
Combined analysis have been performed by Gonzalez-Garcia, Maltoni and 
Pe\~na-Garay (2001); Maltoni, Schwetz and Valle, 2001b; 
Maltoni, Tortola, Schwetz and Valle, 2002b;

For the phenomenology of neutrino oscillations, the (2+2)A and (2+2)B schemes  
are equivalent up to the relabeling of the mass eigenstates (or, equivalently,
of the mixing angles). Thus in what follows we consider the
scheme B, where the mass spectrum presents the following hierarchy:
\begin{equation}
\Delta m^2_\odot=\Delta m^2_{21} \ll \Delta m^2_{\rm atm}=\Delta m^2_{43} \ll
\Delta m^2_{\rm LSND}=\Delta m^2_{41}\simeq \Delta m^2_{42} \simeq
\Delta m^2_{31}\simeq\Delta m^2_{32} \;.
\end{equation}

Neglecting possible CP phases, and choosing a convention that is convenient
for the study of solar and atmospheric neutrinos, the matrix  
$V^\nu_{\alpha i}$ ($\alpha=e,s,\mu,\tau$) can be written as follows:
\begin{equation}
    V^\nu = V_{24}\,V_{23}\,V_{14}\,V_{13}\,V_{34}\,V_{12} \; ,
\end{equation} 
where $V_{ij}$ represents a rotation of angle $\theta_{ij}$ in the $ij$ plane.
Since the parametrizations of the leptonic mixing matrix $U$ and of the 
corresponding lines in $V^\nu$ are the same, and in particular involve six 
mixing angles, we will concentrate  below on the allowed values 
of the $4\times4$  neutrino mixing matrix $V^\nu$.

This general form can be further simplified by taking into account the
negative results from the reactor experiments (in particular the Bugey
experiment) which in the range of $\Delta m^2_{41}$ relevant to the 
LSND experiment imply that
\begin{equation}
    |V^\nu_{e3}|^2+|V^\nu_{e4}|^2 
= c_{14}^2 s^2_{13}+s_{14}^2 \lesssim 10^{-2}.
\end{equation}
For our purposes, the two angles $\theta_{13}$ and $\theta_{14}$ can then be
safely neglected and the $U$ matrix takes the effective form:
\begin{equation}
  V^\nu=\left(\begin{array}{cccc} c_{12}& s_{12}& 0& 0\\ 
   - s_{12} c_{23} c_{24}& c_{12} c_{23} c_{24}& s_{23} c_{24} c_{34}
    - s_{24} s_{34}       & s_{23} c_{24} s_{34}+s_{24} c_{34}\\ 
      s_{12} s_{23}       & - c_{12} s_{23}& c_{23} c_{34}& c_{23} s_{34}\\ 
     s_{12} c_{23} s_{24}& - c_{12} c_{23} s_{24} & - s_{23} s_{24} c_{34}
     - c_{24} s_{34}& - s_{23} s_{24} s_{34}+ c_{24} c_{34}
        \end{array}\right)\,.
    \label{Umatrix}
\end{equation}
The full parameter space relevant to solar and atmospheric neutrino oscillation
can be covered with three of the angles $\theta_{ij}$ in the first quadrant, 
$\theta_{ij}\in[0,\pi/2]$ while one (which we choose to be $\theta_{34}$)
is allow to vary 
in the range $-\frac{\pi}{2} \leq \theta_{34} \leq \frac{\pi}{2}$.  

In this scheme solar neutrino oscillations are generated by the mass-squared 
difference between $\nu_2$ and $\nu_1$ while atmospheric neutrino oscillations 
are generated by the mass-squared difference between $\nu_3$ and $\nu_4$. 
It is clear from Eq.~(\ref{Umatrix}) that the survival of solar $\nu_e$'s 
depends mainly on $\theta_{12}$ while atmospheric $\nu_e$'s 
are not affected by the four-neutrino oscillations in the approximation 
$\theta_{13}=\theta_{14}=0$ and neglecting the effect of $\Delta m^2_{21}$ in 
the range of atmospheric neutrino energies. The survival 
probability of atmospheric $\nu_\mu$'s depends mainly on the $\theta_{34}$ . 

Thus  solar neutrino oscillations occur with a mixing angle $\theta_{12}$
between the states
\begin{equation}
    \nu_e\to \nu_\alpha \quad \mbox{\rm with} \quad
    \nu_\alpha=c_{23} c_{24}\ \nu_s + \sqrt{1-c_{23}^2 c_{24}^2}\ \nu_a \; ,
    \label{nusol}
\end{equation}
where $\nu_a$ is a linear combination of $\nu_\mu$ 
and $\nu_\tau$,
\begin{equation}
\nu_a ={1\over\sqrt{1-c_{23}^2 c_{24}^2}}(s_{23}\nu_\mu+c_{23}s_{24}\nu_\tau).
\label{defnua}
\end{equation} 
We remind the reader that $\nu_\mu$ and $\nu_\tau$ cannot be 
distinguished in solar neutrino experiments, because their matter potential 
and their interaction in the detectors are equal, due to only NC weak 
interactions. Thus solar neutrino oscillations cannot depend on the mixing 
angle $\theta_{34}$ and depend on $\theta_{23}$ and $\theta_{24}$ through the 
combination $c_{23}^2 c_{24}^2$.

Atmospheric neutrino oscillations, {\it i.e.}\ oscillations with the mass 
difference $\Delta m^2_{34}$ and mixing angle $\theta_{34}$, occur between the 
states
\begin{equation}
    \nu_\beta\rightarrow \nu_\gamma \quad \mbox{\rm with}  \quad
    \nu_\beta = s_{23} c_{24} \nu_s + c_{23}\nu_\mu -s_{23} s_{24} \nu_\tau
    \quad \mbox{\rm and} \quad
    \nu_\gamma=s_{24} \nu_s +c _{24} \nu_\tau \;.
    \label{nuatm}
\end{equation}

We learn that the mixing angles $\theta_{23}$ and $\theta_{24}$ determine two 
projections. First, the projection of the sterile neutrino onto the states 
in which the solar $\nu_e$ oscillates is given by 
\begin{equation}
  c_{23}^2c_{24}^2 = 1-|V^\nu_{a1}|^2-|V^\nu_{a2}|^2 = 
|V^\nu_{s1}|^2+|V^\nu_{s2}|^2 .
\end{equation}
Second, the projection of the $\nu_\mu$ over the solar neutrino 
oscillating states is given by
\begin{equation} 
    s^2_{23}=|V^\nu_{\mu 1}|^2+ |V^\nu_{\mu 2}|^2 
=1-|V^\nu_{\mu 3}|^2- |V^\nu_{\mu 4}|^2
    \label{muatm}
\end{equation}
One expects $s_{23}$ to be small in order to explain the atmospheric neutrino 
deficit. We will see that this is indeed the case. Furthermore, the negative 
results from the CDHS and CCFR searches for 
$\nu_\mu$-disappearance also constrain such a projection to be smaller than 
0.2 at 90\% CL for $\Delta m^2_{\rm LSND}\gtrsim 0.4$ eV$^2$.

We distinguish the following limiting cases:

(i) If $c_{23}= 1$ then $V^\nu_{\mu 1}=V^\nu_{\mu 2}=0$. The atmospheric
$\nu_\mu=\nu_\beta$ state oscillates into a state $\nu_\gamma=c_{24}\nu_\tau 
+s_{24}\nu_s$. We will denote this case as {\it restricted} (ATM$_{\rm R}$). 
In particular:
\begin{itemize}
\item If $c_{23}=c_{24}=1$, $V^\nu_{a1}=V^\nu_{a2}=0$ 
($V^\nu_{s3}=V^\nu_{s4}=0$) and we have the limit of pure two-generation
solar $\nu_e\to\nu_s$ transitions
and atmospheric $\nu_\mu\to\nu_\tau$ transitions.
\item If $c_{24}=0$ then $V^\nu_{s1}=V^\nu_{s2}=0$ and 
$V^\nu_{\tau 3}=V^\nu_{\tau 4} = 0$, corresponding to the limit of pure 
two-generation solar $\nu_e\to\nu_\tau$ transitions and
atmospheric $\nu_\mu\to\nu_s$ transitions.
\end{itemize}

(ii) If $c_{23}=0$ then $V^\nu_{s1}=V^\nu_{s2}=0$ corresponding to the limit 
of pure two-generation solar $\nu_e\to\nu_a$ with $a=\mu$ and there are no 
atmospheric neutrino oscillations as the projection of $\nu_\mu$ over the 
relevant states cancels out ($V^\nu_{\mu 3}=V^\nu_{\mu 4}=0$).

Notice that in the restricted case $\theta_{23}=0$, there is an additional 
symmetry in the relevant probabilities so that the full parameter space can be 
spanned by $0\leq\theta_{34}\leq\frac{\pi}{2}$. The reason for this 
is that we now have effectively two-neutrino oscillations for both the solar 
($\nu_e\to\nu_\alpha$) and atmospheric ($\nu_\mu\to\nu_\gamma$) cases. 

To summarize, solar neutrino oscillations depend on the new mixing angles only
through the product $c_{23}c_{24}$ and therefore the analysis of the solar 
neutrino data in four-neutrino mixing schemes is equivalent to the two-neutrino
analysis but taking into account that the parameter space is now 
three-dimensional $(\Delta m^2_{21},\tan^2\theta_{12},c^2_{23} c^2_{24})$. 
Atmospheric neutrino oscillations are affected independently by
the angles $\theta_{23}$ and $\theta_{24}$, and the analysis of 
the atmospheric neutrino data in the four-neutrino mixing schemes is equivalent
to the two-neutrino analysis, but taking into account that the parameter space
is now four-dimensional $(\Delta m^2_{43},\theta_{34}, c^2_{23},c^2_{24})$.
Furthermore the allowed ranges of active-sterile oscillations
depends on the assumed  $^8$B neutrino flux
(Barger, Marfatia, and Whisnant, 2002; Bahcall,
Gonzalez-Garcia and Pe\~na-Garay, 2002b). 
Allowing for $^8$B neutrino flux larger than the
SSM expectation result into a less stringent limit on the active-sterile
admixture.

As an illustration we show in  Fig.~{\ref{fig:chi24}} 
the shift in $\chi^2$ as a function of the active-sterile
admixture $|V^\nu_{s1}|^2+|V^\nu_{s2}|^2=c_{23}^2c_{24}^2$. 
From the figure we conclude that the solar neutrino data favour pure 
$\nu_e\to\nu_a$ oscillations but sizable active-sterile admixtures are still 
allowed. In this curve the $^8$B flux allowed to take larger values than
in the SSM which, as discussed above so the active-sterile bound is
as model independent as possible.

Similar analysis can be performed for the atmospheric neutrino data
(Gonzalez-Garcia, Maltoni and Pe\~na-Garay, 2001) to obtain the allowed
regions for the oscillation 
parameters $\Delta m^2_{43}$ and $\tan^2\theta_{34}$ from the global analysis 
for different values of $\theta_{23}$ and $\theta_{24}$
(or, equivalently, of the projections 
$|V^\nu_{\mu 1}|^2+|V^\nu_{\mu 2}|^2$ and $|V^\nu_{s1}|^2+|V^\nu_{s2}|^2$).
The global minimum corresponds to almost pure atmospheric 
$\nu_\mu-\nu_\tau$ oscillations and the 
allowed regions become considerably smaller for increasing values 
of the mixing angle $\theta_{23}$, 
which determines the size of the projection of $\nu_\mu$ over 
the neutrino states oscillating with $\Delta m^2_{43}$, 
and for increasing values of the mixing angle
$\theta_{24}$, which determines the active-sterile admixture in which the
{\it almost-$\nu_\mu$} oscillates. 
Therefore the atmospheric neutrino data give
an upper bound on both mixings which further implies a lower bound on the 
combination $c_{23}^2 c_{24}^2 =|V^\nu_{s1}|^2+|V^\nu_{s2}|^2$. 
The same combination
is limited from above by the solar neutrino data. 

Fig.~\ref{fig:chi24} shows the shift in $\chi^2$ for
the analysis of the atmospheric data as a function of the active-sterile
admixture $|V^\nu_{s1}|^2+|V^\nu_{s2}|^2=c_{23}^2c_{24}^2$ in the general
case (in which the analysis is optimized with respect to the 
parameter $s^2_{23} = |V^\nu_{\mu 1}|^2+|V^\nu_{\mu 2}|^2$)
as taken from  the latest article by Maltoni, Tortola, Schwetz, 
and Valle (2002b).

In summary, the analysis of the solar data favours the scenario in 
which the solar oscillations in the $1-2$ plane are 
$\nu_e-\nu_a$ oscillations, and gives 
an upper bound on the projection of $\nu_s$ on this plane. On the other hand, 
the analysis of the atmospheric data prefers the oscillations of
the $3-4$ states to occur between a close-to-pure $\nu_\mu$ and an active
($\nu_\tau$) neutrino and gives an upper bound on the projection of the
$\nu_s$ over the $3-4$ states or, equivalently, a lower bound on its projection
over the $1-2$ states. From Fig.~\ref{fig:chi24} we see that the exclusion 
curves from the solar and atmospheric analyses are only
marginally compatible at 3$\sigma$  CL. 

Thus the conclusion concerning the four-neutrino scenarios is that both 
3+1 and 2+2 schemes seem to be  unsatisfactory as explanations 
of the three experimental pieces of evidence 
(see also Maltoni, Tortola, Schwetz and Valle, 2002b).

\section{Implications of the Neutrino Mass Scale and Flavor Structure}
\label{sec:impli1}
\subsection{New Physics}
The simplest and most straightforward lesson of the evidence
for neutrino masses is also the most striking one: there is
NP beyond the SM. This is the first
experimental result that is inconsistent with the SM.

Most likely, the NP is related to the existence of
singlet fermions at some high energy scale that induce, at low
energies, the effective terms (\ref{dimfiv}) through the see-saw mechanism.
The existence of heavy singlet fermions is predicted by many
extensions of the SM, most noticeably GUTs 
(beyond SU(5)) and left-right symmetric theories.
 
There are of course other possibilities. One could induce neutrino
masses without introducing any new fermions beyond those of the
SM. This requires the existence of a scalar 
$\Delta_L(1,3)_{+1}$, that is a triplet of $SU(2)_L$. The smallness of
neutrino masses is then related to the smallness of the VEV
$\langle\Delta_L^0\rangle$ (required also by the success of the
$\rho=1$ relation) and does not have a generic natural explanation.

In left-right symmetric models, however, where the breaking
of $SU(2)_R\times U(1)_{B-L}\to U(1)_Y$ is induced by the VEV of an
$SU(2)_R$ triplet, $\Delta_R$, there must exist also an $SU(2)_L$ 
triplet scalar. Furthermore, the Higgs potential leads to an
order of magnitude relation between the various VEVs,
$\langle\Delta_L^0\rangle\langle\Delta_R^0\rangle\sim v^2$
(where $v$ is the electroweak breaking scale), and the smallness
of $\langle\Delta_L^0\rangle$ is related to the high scale of
$SU(2)_R$ breaking. This situation can be thought of as a see-saw
of VEVs. In this model there are, however, also singlet
fermions. The light neutrino masses arise from both the see-saw
mechanism and the triplet VEV.

Neutrino masses could also be of the Dirac type. Here, again, singlet fermions 
have to be introduced, but lepton number conservation needs to be imposed 
by hand. This possibility is disfavored by theorists since it is likely that
global symmetries are violated by gravitational effects. Furthermore, the 
lightness of neutrinos (compared to charged fermions) is again unexplained. 

Another possibility is that neutrino masses are generated by mixing
with singlet fermions but the mass scale of these fermions is not high.
Here again the lightness of neutrino masses remains a puzzle.
The best known example of such a scenario is the framework of
supersymmetry without R-parity.

Let us emphasize that the see-saw mechanism or, more generally,
the extension of the SM with non-renormalizable terms,
is the simplest explanation of neutrino masses. Models where neutrino
masses are produced by NP at low energy imply a much more
dramatic modification of the SM. Furthermore, the existence
of see-saw masses is an unavoidable prediction of various extensions of the 
SM. In contrast, many (but not all) of the low energy mechanisms 
are introduced for the specific purpose of generating neutrino masses.

In this and in the next section, where we discuss the implications of
the experimental data for theories beyond the SM, 
we choose to
focus on models that explain the atmospheric and solar neutrino data through 
mixing among three active neutrinos. In other words, we assume three-neutrino 
mixing with the oscillation parameters derived in Sec.~\ref{sec:schemes}. 
We do not review models that try to incorporate the LSND data by adding light 
sterile neutrinos and we only comment on this possibility 
in the context of theories with new extra dimensions.

\subsection{The Scale of New Physics}
Given the relation (\ref{nrmass}), $m_\nu\sim v^2/\Lambda_{\rm NP}$, it is
straightforward to use measured neutrino masses to estimate the scale
of NP that is relevant to their generation. In particular,
if there is no quasi-degeneracy in the neutrino masses, the heaviest
of the active neutrino masses can be estimated,
\begin{equation}
m_h=m_3\sim\sqrt{\Delta m^2_{\rm atm}}\approx 0.05\ {\rm eV}.
\label{estmth}
\end{equation}
(In the case of inverted  hierarchy the implied scale is 
$m_h=m_2\sim\sqrt{|\Delta m^2_{\rm atm}|}\approx 0.05\ {\rm eV}$). 
It follows that the scale in the non-renormalizable term (\ref{dimfiv})
is given by
\begin{equation}
\Lambda_{\rm NP}\sim v^2/m_h\approx10^{15}\ {\rm GeV}.
\label{estlnp}
\end{equation}
We should clarify two points regarding Eq.~(\ref{estlnp}):

1. There could be some level of degeneracy between the neutrino masses that 
are relevant to the atmospheric neutrino oscillations. In such a case 
(see Sec.~\ref{direcdet}), Eq.~(\ref{estmth}) is modified into a lower bound
and, consequently, Eq.~(\ref{estlnp}) becomes an upper bound on
the scale of NP.

2. It could be that the $Z_{ij}$ couplings of Eq.~(\ref{dimfiv}) are much
smaller than one. In such a case, again, Eq.~(\ref{estlnp}) becomes an upper 
bound on the scale of NP. On the other hand, in models of approximate flavor 
symmetries,
there are relations between the structures of the charged lepton and neutrino
mass matrices that give quite generically  $Z_{33}\gsim m_\tau^2/v^2\sim
10^{-4}$. We conclude that the likely range of $\Lambda_{\rm NP}$
that is implied by the atmospheric neutrino results is given by
\begin{equation}
10^{11}\ {\rm GeV}\lsim\Lambda_{\rm NP}\lsim10^{15}\ {\rm GeV}.
\label{ranlnp}
\end{equation}

The estimates (\ref{estlnp}) and (\ref{ranlnp}) are very exciting.
First, the upper bound on the scale of NP is well below the
Planck scale. This means that there is a new scale in Nature which
is intermediate between the two known scales, the Planck scale
$m_{\rm Pl}\sim10^{19}$ GeV and the electroweak breaking scale,
$v\sim10^2$ GeV. 

It is amusing to note in this regard that the solar neutrino problem does not 
necessarily imply such a new scale. If its solution is related to vacuum 
oscillations with $\Delta m^2_{21}\sim10^{-10}$ eV$^2$, it can be explained 
by $\Lambda_{\rm NP}\sim m_{\rm Pl}$. 
However (see Sec.~\ref{solar:solar}), the 
favoured explanation for the solar neutrino deficit is the LMA solution which 
again points towards NP scale in the range of Eq.~(\ref{ranlnp}).

Second, the scale $\Lambda_{\rm NP}\sim10^{15}$ GeV is intriguingly close
to the scale of gauge coupling unification. We will say more about this fact
when we discuss GUTs.

\subsection{Implications for Flavor Physics}
\subsubsection{The flavor parameters}
{\it Flavor physics} refers to interactions that are not universal in 
generation space. In the Standard Model interaction basis, this is the physics 
of the Yukawa couplings. 
In the mass basis, this term refers to fermion masses and mixing parameters.
There are two main reasons for the interest in flavor physics:

1. The SM has thirteen flavor parameters. These are the nine
charged fermion masses or, equivalently, Yukawa couplings, 
$Y_i={m_i\over v/\sqrt2}$:
\begin{eqnarray}
Y_t\sim1,\ \ \ &Y_c&\sim10^{-2},\ \ \ Y_u\sim 10^{-5},\nonumber\\
Y_b\sim10^{-2},\ \ \ &Y_s&\sim10^{-3},\ \ \ Y_d\sim 10^{-4},\nonumber\\
Y_\tau\sim10^{-2},\ \ \ &Y_\mu&\sim10^{-3},\ \ \ Y_e\sim 10^{-6},
\label{yukhie}
\end{eqnarray}
and the four CKM parameters:
\begin{equation}
|V_{us}|\sim0.2,\ \ \ |V_{cb}|\sim0.04,\ \ \ |V_{ub}|\sim 0.004,\ \ \ 
\sin\delta_{\rm KM}\sim1.
\label{ckmhie}
\end{equation}
One can easily see that the flavor parameters are hierarchical (that is,
they have very different magnitudes from each other) and all but two 
(the top-Yukawa and the CP violating phase) are small. The unexplained 
smallness and hierarchy pose {\it the flavor puzzle} of the SM. Its
solution may direct us to physics beyond the SM.

2. The smallness of flavor changing neutral current (FCNC) processes, such
as $\Delta m_K$ and $\mu\to e\gamma$, is explained within the SM
by the GIM mechanism. In many extensions of the SM there is,
generically, no such mechanism. In some cases, experimental bounds on flavor 
changing processes are violated. The solution of such flavor problems leads to 
refinement of the models. In other cases, the model predictions are within 
present bounds but well above the SM range. Then we can hope to 
probe this NP through future measurements of flavor changing processes.

Many mechanisms have been proposed in response to either or both of these 
flavor aspects. For example, approximate horizontal symmetries, broken by a 
small parameter, can lead to selection rules that both explain the hierarchy of
the Yukawa couplings and suppress flavor changing couplings in sectors of new 
physics.

In the extension of the SM with three active neutrinos that have
Majorana-type masses, there are nine new flavor parameters in addition
to those of the SM with massless neutrinos. These are three
neutrino masses, three lepton mixing angles and three phases in the mixing
matrix. The counting of parameters is explained in Sec.~\ref{subsec:lepmix}. 
Of these, four are determined from existing measurements of solar and 
atmospheric neutrino fluxes: two mass-squared differences and two mixing 
angles. This adds significantly to the input data on flavor physics and 
provides an opportunity to test and refine flavor models.
 
If neutrino masses arise from effective terms of the form (\ref{dimfiv}),
then the overall scale of neutrino masses is related to the scale
$\Lambda_{\rm NP}$ and, in most cases, does not teach us about flavor
physics. The more significant information for flavor models can then
be written in terms of three dimensionless parameters whose values can be read 
from the results of the global analysis in Eq.~(\ref{globalranges}); First, 
the mixing angles that are relevant to atmospheric neutrinos:
\begin{equation}
|U_{\mu3}U_{\tau3}^*|\sim 0.4-0.5 ;
\label{mnsatm}
\end{equation}
Second, the mixing angles that are relevant to solar neutrinos:
\begin{equation}
|U_{e1}U_{e2}^*|\sim 0.35-0.5 ;
\label{mnssol}
\end{equation}
Third, the ratio between the respective mass-squared differences:
\begin{equation}
{\Delta m^2_{21}\over 
|\Delta m^2_{32}|}\sim\cases{10^{-1}-10^{-3} &  
LMA,\cr 10^{-4}-10^{-5}&LOW.\cr}
\label{mssoat}
\end{equation}  
In addition, the upper bound on the third mixing angle from the
CHOOZ experiments often plays a significant role in flavor model 
building:
\begin{equation}
|U_{e3}| \lsim  0.24 .
\label{mnscho}
\end{equation}

\subsubsection{Special features of the neutrino flavor parameters}
There are several features in the numerical values of these parameters that
have drawn much attention and drove numerous investigations:

(i) Large mixing and strong hierarchy: The mixing angle that is relevant to 
the atmospheric neutrino problem, $U_{\mu3}$, is large, of order one. On the
other hand, the ratio of mass-squared differences $\Delta m^2_{21}/
|\Delta m^2_{32}|$ is small. If there is no degeneracy in the neutrino 
sector then the small ratio of mass-squared differences implies a small ratio 
between the masses themselves, $m_2/m_3\ll1$. (In the case of inverted
hierarchy, the implied hierarchy is $m_3/m_2\ll1$.) It is difficult to
explain in a natural way a situation where, in the $2-3$ generation sector,
there is large mixing but the corresponding masses are hierarchical.
Below we discuss in detail the difficulties and various possible solutions.

(ii) If the LMA solution to the solar neutrino problem holds, then the data
can be interpreted in a very different way. In this case, the two measured
mixing angles are of order one. Moreover, $\Delta m^2_{21}/|\Delta m^2_{32}|
\sim10^{-2}$ means that the ratio between the masses themselves (which, for 
fermions, are after all the fundamental parameters) is not very small,
$m_2/m_3\sim10^{-1}$. Such a value could easily be an accidentally small 
number, without any parametric suppression. If this is the correct way of
reading the data, the measured neutrino parameters may actually reflect the 
absence of any hierarchical structure in the neutrino mass matrices 
(Hall, Murayama and Weiner (2000), Haba and Murayama (2001), Berger and Siyeon 
(2001), Hirsch and King (2001)). Obviously, this interpretation 
is plausible only if the LMA solution to the solar neutrino problem holds, and 
will be excluded if the SMA (small mixing angle) or LOW (small mass ratio) 
solutions hold. Another important test of this idea will be provided by a
measurement of $|U_{e3}|$. If indeed the entries in $M_\nu$ have random values 
of the same order, all {\it three} mixing angles are expected to be of order 
one. If experiments measure $|U_{e3}|\sim10^{-1}$, that is close to the present
CHOOZ bound, it can be again argued that its smallness is accidental. The 
stronger the upper bound on this angle will become, the more difficult it will 
be to maintain this view. 

(iii) A special case of large mixing is that of {\it maximal} mixing.
In a two generation case, with a single mixing angle, maximal mixing is defined
as $\sin^22\theta=1$. In the three generation case, what we mean by maximal
mixing is that the disappearance probability is equivalent to that for maximal
two neutrino mixing at the relevant mass scale (Barger {\it et al.}, 1998b). 
Maximal atmospheric neutrino mixing corresponds to 
$4|U_{\mu3}|^2(1-|U_{\mu3}|^2)=1$, which leads to
\begin{equation}
|U_{\mu3}|^2=1/2.
\label{bimaat}
\end{equation}
Maximal solar neutrino mixing corresponds to $4|U_{e1}U_{e2}|^2=1$,
which leads to
\begin{equation}
|U_{e1}|^2=|U_{e2}|^2=1/2,\ \ \ |U_{e3}|^2=0.
\label{bimaso}
\end{equation}
As seen in Sec~\ref{sec:schemes}, present data are consistent with atmospheric 
neutrino mixing being near-maximal and solar neutrino mixing being large 
but not maximal. The possibility that both mixings are near-maximal is, 
however, not entirely excluded. 
This scenario, where both atmospheric and solar neutrino mixing are 
near-maximal, means that the structure of the leptonic mixing matrix is given 
to a good approximation by
\begin{equation}
U=\pmatrix{{1\over\sqrt2}&-{1\over\sqrt2}&0\cr
{1\over2}&{1\over2}&-{1\over\sqrt2}\cr {1\over2}&{1\over2}&{1\over\sqrt2}\cr}. 
\label{bimaxi}
\end{equation}
The case of equation (\ref{bimaxi}) is commonly called {\it bi-maximal} mixing.
We would like to make the following comments, regarding maximal or bi-maximal
mixing:

1. Theoretically, it is not difficult to construct models that explain
near-maximal solar neutrino mixing in a natural way. We will encounter
some examples below. Experimentally, it may be difficult to make a convincing
case for near-maximal (rather than just order one) solar neutrino mixing
(Gonzalez-Garcia, Pe\~na-Garay, Nir and Smirnov, 2001).

2. It is highly non-trivial to construct models that explain 
near-maximal atmospheric neutrino mixing in a natural way.  
The reason is that near-maximal mixing is often related to 
quasi-degeneracy. So when solar neutrino data are also taken into 
account, approximate degeneracy among all three 
neutrinos is required and models of three quasi-degenerate neutrinos 
are, in general, not easy to construct in a natural way.

3. The case of bi-maximal mixing is also very challenging for theory.
Many of the attempts in the literature involve fine-tuning. (Alternatively,
the term {\it bi-maximal mixing} is sometimes used to denote the case of
two large, rather than maximal, mixing angles.)

\subsubsection{Large mixing and strong hierarchy} 
In this subsection, we focus our attention on the {\it large mixing $-$ strong 
hierarchy} problem. We explain the challenge to theory from this feature and 
present various solutions that have been proposed to achieve this challenge.

A large mixing angle by itself should not be a surprise. After all, the 
naive guess would be that all dimensionless couplings [such as the $Z_{ij}$
couplings of Eq.~(\ref{dimfiv})] are naturally of order one and
consequently all mixing angles are of order one. However, the quark mixing 
angles are small, and this situation has 
led to a prejudice that so would the lepton mixing angles be. Second, and more 
important, flavor models have a built-in mechanism to naturally induce 
smallness and hierarchy in the quark parameters (and perhaps also in charged 
lepton masses). For example, a mechanism that has been intensively studied in 
the literature is that of selection rules due to an approximate symmetry. 
Within such frameworks, numbers of order one are as difficult (or as easy) to 
account for as small numbers: one can assign charges in such a way that small 
flavor parameters correspond to terms in the Lagrangian that carry charge 
different from zero, while the order one parameters correspond to terms that 
carry no charge.

The combination of a large mixing angle and strong hierarchy is however
a true puzzle. To understand the difficulty in this combination, let us
assume that both the mixing and the hierarchy can be understood from the 
relevant $2\times2$ block in the mass matrix for light (Majorana) neutrinos.
The generic form of this block is
\begin{equation}
M_\nu={v^2\over\Lambda_{\rm NP}}\pmatrix{A&B\cr B&C\cr}.
\label{twobytwo}
\end{equation}
This matrix would lead to a large mixing angle if
\begin{equation}
|B|\gsim |C-A|, 
\label{colami}
\end{equation}
and to strong mass hierarchy if (for simplicity we assume real entries):
\begin{equation}
|AC-B^2|\ll {\rm max}(A^2,B^2,C^2).
\label{comami}
\end{equation}
If we examine the two neutrino sector alone, these conditions mean fine-tuning.
Hence the challenge to find models where the hierarchy is naturally induced.

To make the problem even sharper, let us explicitly discuss models with a 
horizontal $U(1)_H$ symmetry (Froggatt and Nielsen, 1979). A horizontal
symmetry is one where different generations may carry different charges,
in contrast to the SM gauge group. We assume 
that the symmetry is broken by a small parameter $\lambda$. To be concrete,
we take $\lambda=0.2$, close to the value of the Cabibbo angle. One can
derive selection rules by attributing charge to the breaking parameter.
Our normalization is such that this charge is $H(\lambda)=-1$. While
coefficients of order one are not determined in this framework, one can derive
the parameteric suppression of all couplings and consequently have an order
of magnitude estimate of the physical parameters. Take, for example, the
case that the $H$-charges of the left-handed lepton doublets $L_i$ are 
positive. The parameteric suppression of an entry in the neutrino mass
matrix is determined by the charge that it carries, $(M_\nu)_{ij}\sim
\lambda^{H(L_i)+H(L_j)+2H(\phi)}$. The parametric suppression of the
mixing angles and mass ratios can then be estimated (see {\it e.g.}
Leurer, Nir and Seiberg, 1993)
\begin{equation}
|U_{ij}|\sim\lambda^{H(L_i)-H(L_j)},\ \ \ m(\nu_i)/m(\nu_j)
\sim\lambda^{2[H(L_i)-H(L_j)]}\ \ \ (i\leq j).
\label{phyhor}
\end{equation} 
If the various generations of left-handed fields (quarks or leptons) carry 
different $H$-charges then, in general, the (quark or lepton) mixing angles are 
suppressed. For example, $\sin\theta_C\sim\lambda$ is naturally induced if the
charges of the first two quark-doublet generations are chosen appropriately:
$H(Q_1)-H(Q_2)=1$. A mixing angle of order one can be naturally induced
if the charges of the corresponding lepton-doublet fields are equal to
each other. Equation (\ref{phyhor}) shows however that in this class of
models, independent of the charge assignments (as long as they are all
positive), we have (Grossman and Nir, 1995)
\begin{equation}
m(\nu_i)/m(\nu_j)\sim|U_{ij}|^2.
\label{posmami}
\end{equation}
Hence, for order one mixing, there is no parametric suppression of the
corresponding neutrino mass ratio, and no hierarchy induced. 

There is another possibility to induce large lepton mixing which is unique
to the case of Majorana neutrinos. Here one assigns, for example, opposite
charges, $H(L_i)=-H(L_j)$, to the relevant lepton-doublet fields. The selection
rules lead to the following structure of the mass matrix:
\begin{equation}
M_\nu\sim{v^2\over\Lambda_{\rm NP}}\pmatrix{A\lambda^{2|H(L_i)|}&B\cr B&
C\lambda^{2|H(L_i)|}\cr},\ \ \ A,B,C={\cal O}(1).
\label{opposH}
\end{equation}
(In a supersymmetric framework, the combination of supersymmetry and the 
horizontal symmetry (Leurer {\it et al.}, 1993, 1994) leads to a situation 
where either $A$ or $C$ vanish.) This mass matrix has a {\it pseudo-Dirac} 
structure and it leads to near-maximal mixing and near-degeneracy of masses
(see, for example, Nir (2000), Joshipura and Rindani (2000b)):
\begin{equation}
\sin^2\theta=1/2-{\cal O}\left(\lambda^{2|H(L_i)|}\right),\ \ \ 
{\Delta m^2\over m^2}={\cal O}\left(\lambda^{2|H(L_i)|}\right).
\label{psedir}
\end{equation}
Of course, a pseudo-Dirac structure is inconsistent with mass hierarchy.

Here are a few mechanisms that have been suggested in the literature
to induce strong hierarchy simultaneously with large mixing angle:

(i) Accidental hierarchy: the mass matrix (\ref{twobytwo}) has $A,B,C={\cal O}
(1)$ and (\ref{comami}) is accidentally fulfilled. In the context of Abelian  
horizontal symmetries, this means that the mass ratio is numerically
but not parametrically suppressed (Binetruy {\it et al.}, 1996;
Irges {\it et al.}, 1998; Elwood {\it et al.}, 1998; Vissani, 1998, 2001b; 
Ellis {\it et al.}, 1999; Sato and Yanagida, 2000).  We would like to emphasize
that this is not an unlikely resolution if the LMA solution of the solar 
neutrino problem is correct. While the oscillation experiments give the ratio
of masses-squared, the relevant quantity for theories is the ratio of masses.
For the LMA solution, $m_2/m_3\sim0.1$ which is not a particularly small
number and could easily arise from a combination of order one terms.

(ii) Several sources for neutrino masses: the leading contribution to
the neutrino mass matrix, which is responsible to the order one mixing,
is rank one. The lighter generation masses come from a different,
sub-dominant source. This is, for example, the generic situation in
supersymmetric models without R-parity (see, for example, Banks {\it
  et al.}, 1995; Borzumati {\it et al}, 1996; Davidson and Losada, 2000).
There tree level mixing with neutralinos can lead to large mixing but
gives a mass to only one neutrino generation, while the lighter masses
arise at the loop level. Another realization of this principle is the
scenario of a see-saw mechanism with a single right-handed neutrino
(Davidson and King, 1998; King, 1998, 1999, 2000;
de Gouvea and Valle, 2001). For another example, within Supersymmetric
theories, see Borzumati and Nomura, 2001.

(iii) Large mixing from the charged lepton sector: it is possible that
the neutrino mass matrix is hierarchical and nearly diagonal in the $2-3$ 
sector, and the large mixing is coming from the diagonalization of the
charged lepton mass matrix. A variety of models have been constructed
that give this structure of mass matrices, based on discrete horizontal
symmetries (Grossman {\it et al.}, 1998; Tanimoto, 1999b), holomorphic zeros 
(Grossman {\it et al.}, 1998) and $U(1)_H$ with two breaking parameters (Nir 
and Shadmi, 1999). For example, Grossman {\it et al.} (1998)  work in the 
supersymmetric framework with a horizontal $U(1)\times U(1)$
symmetry. By an appropriate choice of horizontal charges, they obtain the
following structure of lepton mass matrices (arbitrary coefficients of order 
one are omitted in the various entries):
\begin{equation}
M_\nu\sim{\langle \phi_u\rangle^2\over\Lambda_{\rm NP}}\pmatrix{
\lambda^2&\lambda&\lambda\cr\lambda&0&0\cr \lambda&0&1\cr},\ \ \  \
M_{\ell^\pm}\sim\langle \phi_d\rangle\pmatrix{\lambda^8&\lambda^6&\lambda^4\cr
\lambda^7&\lambda^5&\lambda^3\cr \lambda^7&\lambda^5&\lambda^3\cr},
\label{holzermas}
\end{equation}
where the zero entries are a consequence of holomorphy. These matrices lead to
\begin{equation}
\Delta m^2_{21}/|\Delta m^2_{32}|\sim\lambda^3,\ \ \ |U_{\mu3}|\sim1,
\label{holzerphy}
\end{equation}
where the dominant contribution to the mixing angle comes from diagonalization
of $M_{\ell^\pm}$. The solar neutrino parameters of this model correspond to
the LMA solution.

(iv) Large mixing from the see-saw mechanism: it is possible that sterile
neutrinos play a significant role in the flavor parameters of the light,
dominantly active neutrinos. In other words, an effective mass matrix of the 
form (\ref{twobytwo}) with the condition (\ref{comami}) can be a result
of selection rules applied in the full high energy theory to a mass matrix
that includes both active and sterile neutrinos and integrating out the
heavy, dominantly sterile ones (Barbieri {\it et al.}, 1998; Altarelli
and Feruglio, 1998; Eyal and Nir, 1999). It is interesting to note that in a 
large class of such models, the induced hierarchy is too strong for the LMA 
and SMA solutions, unless at least three sterile neutrinos play a role in 
determining 
the low energy parameters  (Nir and Shadmi, 1999). An interesting example
for a model of this type is presented by Altarelli and Feruglio (1998).
They consider a horizontal $U(1)$ symmetry broken by two parameters of
opposite charges. By an appropriate choice of horizontal charges for the
lepton fields, they obtain the following neutrino Dirac mass matrix $M_D$
and Majorana mass matrix for the sterile neutrinos $M_N$: 
\begin{equation}
M_D\sim\langle \phi_u\rangle\pmatrix{
\lambda^3&\lambda&\lambda^2\cr\lambda&\lambda^\prime&1\cr 
\lambda&\lambda^\prime&1\cr},\ \ \  \
M_N\sim\Lambda_{\rm NP}\pmatrix{\lambda^2&1&\lambda\cr
1&\lambda^{\prime2}&\lambda^\prime\cr \lambda&\lambda^\prime&1\cr}.
\label{seesawh}
\end{equation}
Integrating out of the three heavy neutrinos, the following Majorana
mass matrix for the light neutrinos is obtained:
\begin{eqnarray}
M_\nu\sim{\langle \phi_u\rangle^2\over\Lambda_{\rm NP}}\pmatrix{
\lambda^4&\lambda^2&\lambda^2\cr\lambda^2&A&B\cr \lambda^2&B&C\cr},\ \ \  
&& A,B,C={\cal O}(1), \label{seesawl1}\\[+.1cm]
|AC-B^2|={\cal O}(\lambda\lambda^\prime). &&\label{seesawl2}
\end{eqnarray}
If one were to apply the selection rules directly on $M_\nu$, the generic
structure of Eq.~(\ref{seesawl1}) would be reproduced, but not the relation
between the order one coefficients in Eq.~(\ref{seesawl2}).

(v) A three generation mechanism: approximate $L_e-L_\mu-L_\tau$.
One of the more interesting frameworks that produces all the observed features
of neutrino flavor parameters is intrinsically a three generation framework.
One applies an approximate $L_e-L_\mu-L_\tau$ symmetry to the mass matrices
of light, active neutrinos and of charged leptons (Barbieri {\it et al.}, 1998,
1999; Frampton and Glashow, 1999; Joshipura and Rindani, 2000a; Mohapatra 
{\it et al.}, 2000a; Cheung and Kong, 2000; Shafi and Tavartkiladze; 2000d;
Nir and Shadmi,2000; Nir, 2000). For the most general case, the symmetry
is broken by small parameters, $\epsilon_+$ and $\epsilon_-$, of charges
$+2$ and $-2$, respectively. The lepton mass matrices have the following form:
\begin{equation}
M_\nu\sim{\langle \phi_u\rangle^2\over\Lambda_{\rm NP}}\pmatrix{
\epsilon_-&1&1\cr1&\epsilon_+&\epsilon_+\cr 1&\epsilon_+&\epsilon_+\cr},\ \ \ 
M_\ell\sim\langle\phi_d\rangle\pmatrix{\lambda_e&\lambda_\mu\epsilon_-&
\lambda_\tau\epsilon_-\cr\lambda_e\epsilon_+&\lambda_\mu&\lambda_\tau\cr 
\lambda_e\epsilon_+&\lambda_\mu&\lambda_\tau\cr},  
\label{lelmltm}
\end{equation}
where the $\lambda_i$'s allow for generic approximate symmetry that acts
on the $SU(2)$-singlet charged leptons. The resulting neutrino masses
are as follows:
\begin{equation}
m_{1,2}=m\left(1\pm{\cal O}[{\rm max}(\epsilon_+,\epsilon_-)]\right),\ \ \ 
m_3=m{\cal O}(\epsilon_+).
\label{emtma}
\end{equation}
Note that the quasi-degenerate pair of mass eigenstates involved in the solar 
neutrino anomaly is heavier than the third, separated state. For the
mixing angles, one finds
\begin{equation}
|U_{\mu3}|={\cal O}(1),\ \ \ U_{e3}={\cal O}[{\rm max}(\epsilon_+,\epsilon_-)],
\ \ \ |U_{e2}|^2=1/2-{\cal O}[{\rm max}(\epsilon_+,\epsilon_-)].
\label{emtmi}
\end{equation}

The overall picture is that, somewhat surprisingly, the lepton parameters
(\ref{mnsatm}), (\ref{mnssol}) and (\ref{mssoat}) are not easy to account for
with Abelian flavor symmetries. The simplest and most predictive models
have difficulties in accommodating the large $2-3$ mixing together with the
strong $2-3$ hierarchy. One can find more complicated models that naturally
induce these parameters, but often at the cost of losing predictive power.
In particular, it may be the case that, specifically for neutrinos, one
cannot ignore the existence of heavy degrees of freedom (sterile neutrinos), 
well beyond the reach of direct experimental production, that affect the
flavor parameters of the low energy effective theory. If true, this situation
would mean that measuring the low-energy neutrino parameters cannot by itself
make a convincing case for the idea of Abelian flavor symmetries. (An exception
are models of approximate $L_e-L_\mu-L_\tau$ symmetry.)

Similar difficulties are encountered in the framework of non-Abelian 
symmetries. Again, the simplest models do not work and have to be extended in 
rather complicated ways (Barbieri {\it et al.}, 1999a,b). In some cases, the
non-Abelian symmetries can give testable (almost) exact relations between
masses and mixing angles. For example, the model of Barbieri {\it et al.} 
(1999c) predicts $\sin\theta_{12}=\sqrt{m_e/m_\mu}\cos\theta_{23}$ and
$\sin\theta_{13}=\sqrt{m_e/m_\mu}\sin\theta_{23}$. If it turns out that
all three light neutrinos are quasi-degenerate, non-Abelian symmetries
will become an unavoidable ingredient in the flavor model building, but
the task of constructing realistic models will be very challenging
(see, for example, Carone and Sher, 1998; Fukugita, Tanimoto and Yanagida, 
1998, 1999; 
Fritzsch and Xing, 1996,2000; Barbieri {\it et al.}, 1999d; Wetterich, 1999; 
Tanimoto, 1999a,2000; Tanimoto, Watari and Yanagida, 1999; Wu, 1999a,b,c; 
Perez, 2000; Ma and Rajasekaran, 2001). Radiative corrections are an important
issue when examining the naturalness of various models that account for 
quasi-degeneracy among the neutrinos (see, for example, Casas {\it et al.},
1999, 2000).

\section{Implications for Models of New Physics}
\label{sec:impli2}
In this section we review the implications of neutrino physics for various
extensions of the SM. We do not attempt to describe all relevant
models of NP, but take two representative frameworks. We choose
to focus on well motivated models that were not constructed for the special 
purpose of explaining the neutrino parameters. Thus, the neutrino parameters
either test these frameworks or provide further guidance in distinguishing 
among different options for model building within each framework. The
two frameworks are the following:

(i) GUTs: here the overall scale of the neutrino 
masses and some features of the mixing provide interesting tests.

(ii) Large extra dimensions: in the absence of a high energy scale, the
see-saw mechanism is not operative in this framework. The lightness of
the neutrinos is a challenge.

\subsection{Grand Unified Theories}
There are two significant facts about the gauge symmetries of the Standard
Model and the structure of its fermionic representations that motivate
the idea of supersymmetric grand unification. First, GUTs
provide a unification of the SM multiplets (for a recent review,
see Wilczek, 2001). Second, in the framework of the supersymmetric Standard
Model, the three gauge couplings unify 
(Dimopoulos, Raby and Wilczek, 1981; Iba\~nez and Ross, 1981).
The unification scale is given by (see, for example, Langacker and
Polonsky, 1995) 
\begin{equation}
\Lambda_{\rm GU}\sim3\times10^{16}\ {\rm GeV}.
\label{lamgu}
\end{equation}
Further support for grand unification comes from the flavor sector:
the masses of the bottom quark and the tau lepton are consistent with
equal masses at $\Lambda_{\rm GU}$, as predicted by SU(5) and its extensions.

The evidence for neutrino masses from atmospheric neutrinos, the implied
scale of $\Delta m^2_{\rm atm}$ and the required mixing of order one should
be considered as three further triumphs of the grand unification idea.

First, as mentioned above, SO(10) theories and their extensions require
that there exist singlet fermions. Neutrino masses are then a {\it prediction}
of these theories.

Second, SO(10) theories naturally give the singlet fermions heavy masses at 
the SO(10) breaking scale and, furthermore, relate the neutrino Dirac mass 
matrix $M_\nu^D$ to the up quark mass matrix $M_u$. Specifically, the naive 
SO(10) relation reads $M_\nu^D=M_u$. It follows that the mass of the heaviest 
among the three light neutrinos can be estimated:
\begin{equation}
m_3\sim m_t^2/\Lambda_{\rm SO(10)}\sim10^{-3}\ {\rm eV}.
\label{mtsot}
\end{equation}
It only takes that the lightest of the three singlet fermion masses is
a factor of fifty below the unification scale to induce neutrino masses
at the scale that is appropriate for the atmospheric neutrino anomaly.
In other words, the intriguing proximity of $\Lambda_{\rm NP}$ of Eq.
(\ref{estlnp}) to $\Lambda_{\rm GU}$ of Eq.~(\ref{lamgu}) finds a natural
explanation in the framework of SO(10). In this sense the see-saw mechanism
of Gell-Mann, Ramond and Slansky (1979), and of Yanagida (1979), predicted
neutrino masses at a scale that is relevant to atmospheric neutrinos.

The third triumph, $|U_{\mu 3}|={\cal O}(1)$, is more subtle but still quite 
impressive (Harvey, Reiss and Ramond, 1982; Babu and Barr, 1996; 
Sato and Yanagida, 1998).
Unlike the previous points that were related to SO(10) GUT,
here the consistency relates to SU(5) GUT. SU(5) theories relate the
charged lepton mass matrix $M_\ell$ to the down quark mass matrix $M_d$.
Specifically, the naive SU(5) relation reads $M_\ell=M_d^T$. It follows that
\begin{equation}
|U_{\mu 3}V_{cb}|\sim m_s/m_b.
\label{VVmsmb}
\end{equation}
Given the experimental values $|V_{cb}|\sim0.04$ and $m_s/m_b\sim0.03$
we conclude that the naive SU(5) relations predict $|U_{\mu 3}|\sim1$.
Of course, $|U_{\mu 3}|$ is also affected by the diagonalization of the
light neutrino mass matrix, but there is no reason to assume that this
contribution cancels against the charged lepton sector in such a way
that $|U_{\mu 3}|\ll1$.

Many of the other naive SO(10) relations fail, as do many of the naive
SU(5) relations. Specifically, SO(10) predicts vanishing CKM mixing
and mass ratios such as $m_c/m_t=m_s/m_b$. SU(5) predicts $m_s=m_\mu$
and $m_d=m_e$. It is possible however that these bad predictions are
corrected by subleading effects while all the successful predictions
(particularly $m_b=m_\tau$, $m_\mu/m_\tau\sim m_s/m_b$ and
$|U_{\mu 3}V_{cb}|\sim m_s/m_b$) are retained since they depend
on the leading contributions. This is demonstrated in a number of
specific GUT models (Albright, Babu and Barr, 1998; Albright and Barr, 1998, 
1999a, 1999b, 2000a, 2000b, 2001; Altarelli and Feruglio, 1998, 1999a, 1999b;
Altarelli, Feruglio and Masina, 2000a, 2000b; Berezhiani and Rossi, 1999, 2001;
Hagiwara and Okamura, 1999; Babu, Pati and Wilczek, 2000;
Shafi and Tavartkiladze, 1999, 2000a, 2000b, 2000c; Maekawa, 2001).

The flavor structure of the first two neutrino generations
depends on both the Majorana mass matrix for the singlet fermions and
the subdominant effects that correct the flavor parameters of the first
two generation quarks and charged leptons. This part of GUTs 
is therefore much more model dependent. Explicit GUT models were
constructed that accommodate the various solutions of the solar
neutrino problem. We will not describe them in any detail here.

\subsection{Extra Dimensions}
\label{subsec:extdim}
New ideas concerning the possibility of large additional dimensions and
the world on a brane can lead to sources of neutrino masses that are very
different from the ones that we discussed so far. For example, the small
mass could be related to the large volume factor of the extra dimensions
which suppresses the coupling to bulk fermions
(Arkani-Hamed {\it et al.}, 2002; Dienes {\it et al.}, 1999; 
Mohapatra {\it et al.}, 1999; Mohapatra and Perez-Lorenzana, 2000;
Lukas {\it et al.}, 2000, 2001; Dienes and Sarcevic, 2001;
Mohapatra {\it et al.}, 2000b),
to the breaking of lepton number on a distant brane
(Arkani-Hamed {\it et al.}, 2002),
or to the warp factor in the Randall-Sundrum framework
(Grossman and Neubert, 2000). In this section, we briefly describe these
three mechanisms.

The existence of large extra dimensions does not only provide new ways
of generating small neutrino masses but can also lead to interesting
phenomenological consequences. In particular, the phenomenology of matter 
oscillations in the Sun can be affected 
(Dvali and Smirnov, 1999; Barbieri {\it et al.}, 2000;
Lukas {\it et al.}, 2000; Ma {\it et al.}, 2000;
Caldwell {\it et al.}, 2001a, 2001b). Other phenomenological implications
can be used to constrain the parameters of the extra dimensions 
(Faraggi and Pospelov, 1999; Ioannisian and Pilaftsis, 2000; 
Das and Kong, 1999).

\subsubsection{Coupling to bulk fermions}
It is possible that we live on a brane with three spatial dimensions that
is embedded in a spacetime with $n$ additional large spatial dimensions
(Arkani-Hamed {\it et al}, 1998).
This idea has the potential of providing an understanding of the hierarchy
between the gravitational mass scale, $M_{\rm Pl}$, and the electroweak
scale, $m_Z$. The hope is to solve the hierarchy problem by avoiding a
fundamental high energy (that is Planck or GUT) scale altogether.
The observed Planck scale, $M_{\rm Pl}=(G_N)^{-1/2}\sim10^{19}$ GeV, is
related to the fundamental Planck scale (most likely the string scale)  
of the $4+n$ dimensional theory, $M_*$, by Gauss law:
\begin{equation}
M_{\rm Pl}^2=M_*^{n+2}V_n,
\label{gauss}
\end{equation}
where $V_n$ is the volume of the $n$-dimensional extra space. This picture
has dramatic phenomenological consequences for particle physics and cosmology.
Such a situation also poses an obvious problem for neutrino physics. 
If there is no scale of physics as high as (\ref{ranlnp}), the see-saw
mechanism for suppressing the light neutrino masses cannot be implemented.
Conversely, if there are singlet neutrinos that are confined to the
three dimensional brane where the active neutrinos live, one expects
that their mass is at the string scale, $M_*$, which in these models
is much smaller than the four dimensional $M_{\rm Pl}$, perhaps as
small as a few TeV, and the resulting light neutrino masses are well
above the scales of atmospheric or solar neutrinos.

The first implication within this framework of the evidence for neutrino masses
is that  there better be no singlet fermions confined to the brane.
Alternatively, the model must include some special ingredients to
avoid ordinary see-saw masses.

On the other hand, it is typical in this framework
that there are singlet fermions in the bulk. This would be the case,
for example, with modulinos, the fermionic partners of the moduli
fields that are generic to string theories. The crucial point is that
the Yukawa interaction between bulk singlet fermions and brane active neutrinos
is highly suppressed by a large volume factor of the $n$ extra dimensions, 
$1/\sqrt{V_n}$. This suppression factor reflects the small overlap between the 
wave functions of the sterile neutrino in the bulk and the active one
on the brane. By construction, this factor provides a suppression of the 
neutrino Dirac mass by the ratio $v/M_{\rm Pl}$. Consequently, large extra 
dimensions provide a natural source of very small Dirac mass terms for the 
active neutrinos.

The consequences for neutrino masses and mixing depend on the details
of the physics of the bulk neutrinos. The possible scenarios are clearly
described by Lukas {\it et al.} (2000). There an explicit derivation of the
effective four dimensional action from a five dimensional one is given.
The final result for the Dirac mass is as follows:
\begin{equation}
m_i^{\rm Dir}={vY_i\over\sqrt{V_n M_*^n}}=Y_iv{M_*\over M_{\rm Pl}}.
\label{smallD}
\end{equation}
Here $Y_i$ is the Yukawa coupling between the lepton doublet $L_i$ and a bulk 
singlet $M_*$ is the string scale and $V_n$
is the volume of the $n$-dimensional extra space. 
A usual Dirac mass is of order $Yv$, but in Eq.~(\ref{smallD})
we see explicitly the $(M_*/M_{\rm Pl})$ suppression factor in
the effective four-dimensional Yukawa couplings, leading to highly
suppressed Dirac mass terms.

Being a bulk state, the singlet fermion has a whole tower of Kaluza-Klein
(KK) associated states which are all coupled to the left-handed brane 
neutrinos. In the simplest scenario and for one extra dimension 
of radius $R$, the masses of all KK states are determined by the scale $1/R$: 
\begin{equation}
m^2_{n}=\frac{n^2}{R^2}\;.
\label{kkmass}
\end{equation}
In more general scenarios there can be other bulk mass 
terms and the masses of the KK states receive additional 
contributions. Then the lightest KK mass can be taken as an independent
parameter while the mass splitting between the states is still determined 
by the scale $1/R$.
 
Let us denote by $M_{\rm min}$ the lightest mass in the Kaluza-Klein 
spectrum and by $\Delta m$ the mass scale that is relevant to atmospheric 
or to solar neutrino oscillations. Then one can distinguish three cases:

1. $1/R\gg\Delta m$ and $M_{\rm min}\gg\Delta m$. The Kaluza-Klein 
states play no direct role in these oscillations. Their main effect is to give
see-saw masses to the active brane neutrinos.

2. $1/R\gg\Delta m$ and $M_{\rm min}\lsim\Delta m$. The situation is
equivalent to conventional models containing a small number of sterile
states. In particular, for $M_{\rm min}=0$ it is a possible framework  
for light Dirac neutrinos. For small but non-vanishing 
$M_{\rm min}$ it may lead to  the four-neutrino
mixing scenarios. However these scenarios are not particularly favoured.

3. $1/R\lsim\Delta m$ and $M_{\rm min}\lsim\Delta m$. A large number of
bulk modes can take direct part in the oscillation phenomena. This situation
modifies in a very interesting way the solution to the solar neutrino
problem that involves MSW resonance conversion into a sterile neutrino.
Now the $\nu_e$ can oscillate into a whole set of Kaluza-Klein states. 
However, as discussed in Sec.~\ref{sec:solar}, solar oscillations into
sterile states are now strongly disfavoured by the comparison of the
CC and NC fluxes at SNO.

In summary, large extra dimensions with sterile bulk fermions
provide a natural explanation for light neutrino masses.
In the simplest realization, in which there are no bulk mass terms
and lepton number conservation is imposed in the bulk-brane couplings,
light Dirac masses are generated. In the general case, the
light neutrinos can be either Dirac or Majorana particles,
depending on the details of the physics of the bulk neutrinos.

\subsubsection{Lepton number breaking on a distant brane}
In its simplest realization, the mechanism described above
provides a natural way of generating light Dirac masses for the neutrinos.
In the framework of extra dimensions one could also generate small
Majorana masses  by an alternative mechanism in which lepton number 
is broken on a distant brane
(Arkani-Hamed {\it et al.}, 2002). Imagine that lepton number is spontaneously
broken at the scale $M_*$ on a different brane located a distance $r$ from
our brane. Further assume that the information about this breaking is
communicated to our brane by a bulk field with a mass $m$. For the sake
of concreteness we take the case that $mr\gg1$ and there are two extra
dimensions. The resulting Majorana mass for the active neutrinos can be
estimated as follows:
\begin{equation}
m_{\nu}\sim{v^2\over M_*}e^{-mr}.
\label{smallM}
\end{equation}
We learn that the naive see-saw mass, of order $v^2/M_*$, is suppressed
by a small exponential factor. The consequences for the neutrino spectrum
depend on various model dependent features: the number of large extra
dimensions, the string scale, the distance between our brane and the
brane where lepton number is broken, and the mass of the mediating bulk field.

A variant of the above two mechanisms can be implemented if the SM
fields are confined to a thick wall (Arkani-Hamed and Schmaltz, 2000). 
Dirac masses are suppressed if left-handed and right-handed neutrinos are 
located at different points in the wall and consequently there is an
exponentially small overlap of their wave functions. Majorana masses are
suppressed if lepton number is spontaneously broken by a VEV that is localized
within the wall but at some distance from the lepton doublet fields.

\subsubsection
{The warp factor in the Randall-Sundrum [RS] scenario}
A different set-up for the extra dimensions (leading to a different
explanation of the $v/M_{\rm Pl}$ hierarchy) was proposed by Randall
and Sundrum (1999) [RS]. They considered one extra dimension parameterized by a
coordinate $y=r_c\phi$, where 
$-\pi\leq\phi\leq+\pi$.  $r_c$ is the radius of the
compact dimension, and the points $(x,\phi)$ and $(x,-\phi)$ are identified.
A {\it visible} brane is located at $\phi=\pi$ and 
a {\it hidden} one at $\phi=0$.
This set-up leads to the following non-factorizable metric:
\begin{equation}
ds^2=e^{-2kr_c|\phi|}\eta_{\mu\nu}dx^\mu dx^\nu-r_c^2d\phi^2.
\label{RSmetric}
\end{equation} 
The parameter $k$ is the bulk curvature. All dimensionful parameters in the
effective theory on the visible brane are suppressed by the warp factor,
$\epsilon\equiv e^{-kr_c\pi}$. With $kr_c\approx12$, that is $\epsilon\sim
10^{-16}$, this mechanism produces physical masses of order $v$ from 
fundamental masses of order $M_{\rm Pl}$. 

The two mechanisms described above to generate small neutrino masses do not 
work in this scenario because the extra dimensions are small and there is no 
volume suppression factor available.
Grossman and Neubert (1999) proposed a different mechanism to generate
small masses in the RS framework. (For a different mechanism, see
Huber and Shafi, 2001.) With appropriate choice of orbifold
boundary conditions, it is possible to locate the zero mode of a right-handed 
bulk neutrino on the hidden brane. If the fundamental mass scale $m$ of the 
bulk fermions is larger than half the curvature $k$ of the compact dimension,
the wave function of the right-handed zero mode on the visible brane is
power-suppressed in the ratio $v/M_{\rm Pl}$. Coupling the bulk fermions to
the Higgs and lepton doublet fields yields
\begin{equation}
m_\nu\sim v\left({v\over M_{\rm Pl}}\right)^{\frac{m}{k}-\frac{1}{2}}.
\label{RSmnu}
\end{equation}
Note that generically the relation between the neutrino mass and the weak
scale is different from the see-saw mechanism (except for the special case
$m/k=3/2$). 

To summarize, in the presence of large extra dimensions, neutrino masses
could be suppressed by the large volume factor if the left-handed neutrinos
couple to the bulk fermions (being Dirac for massless
bulk fermions), or by the distance to a brane where lepton number
is broken (which generate Majorana masses).
In the RS framework, the suppression can be induced by a power of the warp
factor. The detailed structure of neutrino mass hierarchy and mixing
is often related to the parameters that describe the extra dimensions.

\subsection{Supersymmetry without $R$-parity}
\label{subsec:norpar}
Supersymmetry is a symmetry that relates fermions and bosons. It is useful in 
solving the fine-tuning problem of the Standard Model (that is the stability 
of the small ratio between the electroweak breaking scale and the Planck scale 
against radiative corrections) and has many other virtues. Extending the 
Standard Model by imposing Supersymmetry  requires that the particle content 
is also extended. In particular, one must add scalar superpartners to the 
three generations of quarks and leptons. (The supersymmetric multiplet that
contains fermions and bosons with the same gauge quantum number is called
a superfield.) These modifications make a significant
difference as far as the accidental symmetries of the Standard Model,
Eq.~(\ref{SMglob}), are concerned. One has the option of restoring
baryon number conservation and total lepton number conservation by
imposing a discrete symmetry, $R_p=(-1)^{3B+L+2S}$. Models where this
symmetry is not imposed, and the accidental global symmetries of the
Standard Model are all violated (or, at least, either baryon number
or lepton number is not conserved) are therefore called models of
{\it Supersymmetry without R-parity}.

Supersymmetry without $R$-parity provides a very rich phenomenology.
In general, this framework has baryon number violation, lepton number
violation, and many new sources of flavor changing and CP violating
interactions. Since total lepton number and lepton flavor symmetries
are violated, it is not surprising that this framework
is relevant to the question of neutrino masses and mixing.

In some sense, the smallness of neutrino masses poses a problem for
Supersymmetry without $R_p$. If dimensionful $R_p$ violating (RPV)
couplings were at the electroweak breaking scale, and if dimensionless
$R_p$ violating couplings were of order one, then one neutrino mass would
be at the electroweak breaking scale, and the other two suppressed by
a loop factor and bottom and tau Yukawa couplings. Thus the three masses
would be well above the experimental bounds. This is part of the motivation
for considering supersymmetry {\it with} $R_p$. (Proton decay and
neutron-antineutron oscillations provide another source of motivation for
$R_p$.) In other words, even if $R_p$ is not an exact symmetry of Nature,
it better be an approximate one. Various theories have been considered that 
lead to suppression of $R_p$ violating couplings. The question of neutrino 
masses is usually considered in the framework of such special theories.
(In particular, one usually assumes that baryon number is a good symmetry
of the theory. While this assumption has no direct consequences for our
discussion below, it implies that there is no need to consider constraints
from proton decay on the lepton number violating couplings.) 

One way to think about lepton number violation in the context of supersymmetry
without $R_p$ is to notice that in such a framework, there is no quantum
number to distinguish between the down-Higgs and the 
lepton-doublet superfields.
Consequently, there are four 
{\it flavors} of the $(1,2)_{-1/2}$ representation. 
It is convenient then to denote the four doublets by $L_\alpha$, where 
$\alpha=0,1,2,3$. There are bilinear and trilinear couplings that lead to 
lepton number violation. In the superpotential, we have the bilinear 
$\mu$-terms, and the trilinear $\lambda$ and $\lambda^\prime$ terms:
\begin{equation}
W_{\not R_p}=\mu_\alpha L_\alpha\phi_u+
\lambda_{\alpha\beta k}L_\alpha L_\beta\bar\ell_k+
\lambda^\prime_{\alpha j k}L_\alpha Q_j\bar d_k.
\label{WRPV}
\end{equation}
Here $\phi_u$ is the $Y=+1/2$ Higgs superfield, $\ell_k$ ($k=1,2,3$) are the
$SU(2)_{\rm L}$ singlet charged lepton superfields, and $Q_k$ ($d_k$) are the
$SU(2)_{\rm L}$ doublet (down singlet) quark superfields.
Among the soft supersymmetry breaking terms in the scalar potential,
we have the bilinear $B$ terms and $m^2$ terms and trilinear $A$ terms:
\begin{equation}
V^{\rm soft}_{\not R_p}=B_\alpha L_\alpha\phi_u+m^2_{\alpha\beta}
L_\alpha L^\dagger_\beta+
A_{\alpha\beta k}L_\alpha L_\beta\bar\ell_k+
A^\prime_{\alpha j k}L_\alpha Q_j\bar d_k+{\rm h.c.},
\label{VRPV}
\end{equation}
where we use the same notation for superfields in Eq.~(\ref{WRPV}) 
and for their scalar components in Eq.~(\ref{VRPV}).

There are many ways to define what one means by {\it the down Higgs field},
even when $R_p$ is an approximate symmetry. These ways represent different
choices of basis in the four-flavor space of $L_\alpha$. Let us denote the 
Higgs fields by $L_0$ and the lepton fields by $L_i$ ($i=1,2,3$). Then, for 
example, one can define $L_0$ by choosing a basis where $\mu_i=0$, or by 
choosing a basis where $\langle L_i\rangle=0$ (vanishing
sneutrino VEVs), or by choosing $L_i$ to
have the three lightest charged mass eigenstates. Physical results are
of course independent of the choice of basis, but there is much confusion
in the literature resulting from the use of different conventions.
For a clarification of this subject and a very useful basis-independent
formulation, see Davidson and Losada, 2000.

The most striking feature of neutrino masses in the context of 
RPV supersymmetry is that there are many different sources for them.
Consequently, the hierarchy in neutrino masses can be the consequence of
the interplay between these different sources. Relations between mixing
and mass hierarchies that often apply when there is a single source of
masses are easily modified in this framework. Below is a list of some
of the contributions to neutrino masses that appear in a generic RPV
supersymmetry. The estimates of the sizes of the various contributions
is given in the convention that $\langle L_i\rangle=0$ and with the 
simplifying assumption that there is a single scale $m_{\rm susy}$ that 
characterizes all supersymmetry breaking and the $\mu$ term.

1. Tree level contribution:
\begin{equation}
(M_\nu)_{ij}\sim{\mu_i\mu_j\over m_{\rm susy}}.
\label{Rpmumu}
\end{equation}

2. Sneutrino-neutralino loops:
\begin{equation}
(M_\nu)_{ij}\sim {g^2\over64\pi^2\cos^2\beta}{B_i B_j\over m_{\rm susy}^3}.
\label{RpBB}
\end{equation}

3. Slepton-lepton loops:
\begin{equation}
(M_\nu)_{ij}\sim {1\over8\pi^2}\lambda_{ink}\lambda_{jkn}{m_{\ell_n}
m_{\ell_k}\over m_{\rm susy}}.
\label{Rplala}
\end{equation}

4. Squark-quark loops:
\begin{equation}
(M_\nu)_{ij}\sim {3\over8\pi^2}\lambda^\prime_{ink}\lambda^\prime_{jkn}
{m_{d_n}m_{d_k}\over m_{\rm susy}}.
\label{Rplplp}
\end{equation}

There are many additional loop diagrams that involve various two insertions of
RPV couplings. These could be two bilinear couplings, as in Eq.~(\ref{RpBB}), 
two trilinear couplings, as in Eqs.~(\ref{Rplala}) and (\ref{Rplplp}), or one 
bilinear and one trilinear coupling. For a complete list, see Davidson and 
Losada, 2000.

A few comments are in order:

(i) The relative importance of the various contributions is model dependent.

(ii) The tree contribution vanishes if the following two conditions are 
fulfilled (Hall and Suzuki, 1984; Banks {\it et al.}, 1995):
\begin{equation}
B_\alpha\propto\mu_\alpha,\ \ \ \ 
m^2_{\alpha\beta}\mu_\beta=\tilde m^2\mu_\alpha.
\label{Rpalign}
\end{equation}

(iii) Even in case that (\ref{Rpalign}) holds at a high energy scale,
misalignment between the $B$ and $\mu$ terms will be induced by RGE,
with the likely consequence that the tree contribution is dominant
at low energy (Nardi, 1997).

(iv) The tree contribution gives mass to a single neutrino state.
Thus at least two neutrino masses are induced by loop corrections.

(v) If there is a single small parameter that suppresses the leading 
RPV couplings, that is $\mu_3/\mu_0\sim B_3/B_0\sim
\lambda_{3jk}/\lambda_{0jk}\sim\lambda^\prime_{3jk}/\lambda^\prime_{0jk}$
(this is the typical case with an abelian flavor symmetry), then
the sneutrino related loop of Eq.~(\ref{RpBB}) dominates over the other
loop contributions (Grossman and Haber, 1999).  

There are many ways in which the lepton parameters deduced from
atmospheric and solar neutrino measurements can be accommodated in the
framework of RPV supersymmetry. The implications of the atmospheric
neutrino results for the tree contributions were discussed by
Bednyakov, Faessler and Kovalenko, 1998; Chun and Lee, 1999; Diaz {\it et al.},
2000; Datta, Mukhopadhyaya and Vissani, 2000; Joshipura, Vaidya and Vempati, 
2000; Suematsu, 2001. Implications, in addition, of solar neutrino results for 
the $\lambda$ and $\lambda^\prime$ dependent loops were investigated by
Chun {\it et al.}, 1999; Joshipura and Vempati, 1999a, 1999b; 
Mukhopadhyaya, Roy and 
Vissani, 1998; Choi {\it et al.}, 1999; Bhattacharyya, Klapdor-Kleingrothaus 
and Pas, 1999; Kong, 1999; Datta, Mukhopadhyaya and Roy, 2000;
Takayama and Yamaguchi, 2000; Mira {\it et al.}, 2000.
Contributions from misalignment of bilinear terms at tree and loop levels
was analyzed by Grossman and Haber, 1999, 2001; Hambye, Ma and Sarkar, 2000;
Romao {\it et al.}, 2000; Hirsch {\it et al.}, 2000.
Contributions from trilinear terms only were considered in
Rakshit, Bhattacharyya and Raychaudhuri, 1999; Drees {\it et al.}, 1998.
Combinations of bilinear and trilinear couplings in the loop contributions
were pointed out by Choi, Hwang and Chun, 1999; Kaplan and Nelson, 2000. 
Model independent descriptions of all contributions were given by
Davidson and Losada, 2000; Abada and Losada, 2000a, 2000b; 
Choi, Chun and Hwang, 2000; Kong, 2000.

Since there are many independent RPV couplings, it is easy to make an ansatz 
that would be consistent with the data. A more interesting question is whether 
one can start from some theoretical guiding principle, such as an approximate 
flavor symmetry, and then the required mass hierarchy and mixing angles would 
arise naturally (for an early attempt, see, for example, Borzumati {\it et al.},
1996). Some of the more interesting attempts in this direction are
the following:

1. When $\mu$ and $B$ are misaligned, there are contributions related to
bilinear couplings at both tree and loop level. The ratio between these
contributions is roughly ${g^2\over64\pi^2}\sim10^{-3}$. This is a little
(but certainly not far) below the hierarchy that is appropriate for 
$(\Delta m^2_{21}/|\Delta m^2_{32}|)^{1/2}$ (see, for example,
Grossman and Haber, 1999).

2. When $\mu$ and $B$ are aligned at a high scale, the ratio between the loop 
contribution from trilinear couplings and the tree contribution from 
RGE-induced misalignment is generally of the appropriate hierarchy
(see, for example, Nardi, 1997).

3. If only trilinear couplings are significant, then
the ratio between the leading contributions is likely to be of order
$m_\tau^2/3m_b^2$ or $m_s/m_b$, which is just the required hierarchy
(see, for example, Drees {\it et al.}, 1998). 

We conclude that future neutrino data will provide further guidance into
model building in the framework of Supersymmetry without $R$-parity.

\section{Conclusions}
\label{sec:conc}
Strong evidence for neutrino masses and mixing has been arising from various 
neutrino oscillation experiments in recent years. First, atmospheric neutrinos 
show deviation from the expected ratio between the $\nu_\mu$- and 
$\nu_e$-fluxes. Furthermore, the $\nu_\mu$ flux has strong zenith angle 
dependence. The simplest interpretation of these results is that there are 
$\nu_\mu-\nu_\tau$ oscillations with the following mass and mixing parameters:
\begin{equation}
\label{atmfin}
1.4\times10^{-3}\ {\rm eV}^2<\Delta m_{\rm atm}^2<6\times10^{-3}\ {\rm eV}^2,\ 
\ \ 0.4<\tan^2\theta_{\rm atm}<3.0.
\end{equation}
The ranges quoted here correspond to the results of the global three-neutrino
analysis presented in Sec.~\ref{sec:schemes} [Eq.~(\ref{globalranges})].
Second, the total rates of solar neutrino fluxes are smaller than 
the theoretical expectations. Furthermore, the suppression is different in the 
various experiments (which are sensitive to different energy ranges).
The recently announced measurements of NC and CC fluxes at SNO provide a
$5.3\sigma$ signal for neutrino flavor transition that is independent
of the solar model.  The 
simplest interpretation of these results is that there are $\nu_e-\nu_a$ 
oscillations (where $\nu_a$ is some combination of $\nu_\mu$ and $\nu_\tau$) 
with the following set of mass and mixing parameters:
\begin{eqnarray}
\label{solfin}
2.4\times10^{-5}\ {\rm eV}^2<&\Delta m^2_\odot&<2.4\times10^{-4}\ {\rm eV}^2,\ \ \ 
0.27<\tan^2\theta_\odot<0.79\ \ \ \ ({\rm LMA}),
\end{eqnarray}
where again the ranges quoted correspond to the results of the global
analysis given in Eq.~(\ref{globalranges}).
From the global analysis an upper bound on a third mixing angle arises
driven mainly by the negative results of the reactor experiments 
in combination with the deduced $\Delta m^2_{\rm atm}$ from the atmospheric
neutrino data:
\begin{equation}
\label{reafin}
\sin\theta_{\rm reactor}\leq 0.24 \;.
\end{equation}
The smallness of this angle guarantees that the results of the
three neutrino analysis combining the atmospheric and solar neutrino data
are close to the two separate two neutrino analyses.

The evidence for neutrino masses implies that the SM cannot
be a complete picture of Nature. In particular, if the SM is only
a low energy effective theory, very light neutrino masses are expected.
The scale at which the SM picture is not valid anymore is
inversely proportional to the scale of neutrino masses. Specifically 
\begin{equation}
m_\nu\gsim\sqrt{\Delta m^2_{\rm atm}}\sim0.05\ {\rm eV}\ \ \Longrightarrow\ \ 
\Lambda_{\rm NP}\lsim10^{15}\ {\rm GeV}.
\label{scalenp}\end{equation}
We learn that there is a scale of NP well below the Planck scale.

The scale of NP in Eq.~(\ref{scalenp}) is intriguingly close
to the scale of coupling unification. Indeed, since GUTs with an SO(10)
(or larger) gauge group predict neutrino masses, plausibly at the 0.1 eV
scale, the atmospheric neutrino data can be taken as another piece of
support to the GUT idea. The large mixing angle in (\ref{atmfin}) 
also finds a natural and quite generic place in GUTs.
  
The measured values of the neutrino flavor parameters are useful in
testing various ideas to explain the flavor puzzle. Quite a few of the
simplest models that explain the smallness and hierarchy in the quark
sector parameters fail to explain the neutrino parameters. The neutrino
parameters have some features that are quite unique. In particular,
the two mixing angles $\theta_{\rm atm}$ and $\theta_\odot$ are large. 
As concerns the mass hierarchy, $\Delta m^2_\odot/\Delta m^2_{\rm atm}$,
the situation is still ambiguous. If there is strong hierarchy, it is
not easy to accommodate it together with large mixing angles. If the
hierarchy is mild enough to be consistent with just accidental suppression,
and if, in addition, the third mixing angle (bounded by reactor experiments)
is not very small, it could well be that there is no hierarchy in the
neutrino parameters at all. That would call for flavor frameworks that give a
different structure for charged and neutral fermion parameters.
Other possibilities, such as quasi-degeneracy in masses or bi-maximal
mixing also call for special structure in the neutrino sector that is
very different from the quark sector.

The mass scales involved in Eqs.~(\ref{atmfin}) and (\ref{solfin}) have
implications to many other frameworks of NP. In particular,
they can help discriminate between various options in the framework of
models with extra dimensions, models of supersymmetry without $R$ parity,
left-right symmetric models, etc.

Another hint for neutrino masses comes from the LSND experiment. Here, however,
the signal is presently observed by a single experiment, and further 
experimental testing is required. The simplest interpretation of the LSND
data is that there are $\nu_e\to\nu_\mu$ oscillations with 
$\Delta m^2_{\rm LSND}={\cal O}(1\ {\rm eV}^2)$ and 
$\sin^22\theta_{\rm LSND}={\cal O}(0.003)$. The fact that 
$\Delta m^2_{\rm LSND}\gg\Delta m^2_{\rm atm},\Delta m^2_{\odot}$ means
that the three results cannot be explained with oscillations among the
three active neutrinos alone.

If the LSND result is confirmed, then more dramatic modifications of the
SM will be required. The simplest extension, that is, the addition
of a light singlet neutrino, is not excluded but it does run into difficulties
related to the fact that oscillations into purely sterile neutrinos fit
neither the atmospheric nor the solar neutrino data.

The good news is that there has been a lot of progress in neutrino 
physics in recent years. Measurements of both atmospheric and solar neutrino
fluxes make the case for neutrino masses and mixing more and more convincing.
Many theoretical ideas are being excluded, while other have at last 
experimental guidance in constructing them. The other piece of good 
news is that there
is a lot of additional experimental information concerning neutrino physics
to come in the near future. We are guaranteed to learn much more.

\section*{Note added}
The original version of this review was completed in January
2002. The year 2002 has been a very active and important year in
neutrino physics. In April, the SNO collaboration have announced the
results of their NC measurement, which established the flavour
conversion of solar neutrinos. The review was updated to include the
new effects of the NC SNO measurement. In December, the KamLAND
collaboration released their first data on the measurement of the flux
of $\bar{\nu}_e$'s from distant nuclear reactors (Eguchi {\rm et al.}
2002). Below we briefly present the main implications of this result.

\subsection*{KamLAND results}
KamLAND have measured the flux of $\bar{\nu}_e$'s from
distant nuclear reactors. In an exposure of 162 ton$\cdot$yr
(145.1 days), the ratio of the number of observed inverse
$\beta$-decay events to the number of events expected without
oscillations is 
\begin{equation}
R_{\rm KamLAND} ~=0.611 \pm 0.094\; .
\label{eq:kamlandrate}
\end{equation}
for $E_{\bar{\nu}_e}>$ 3.4 MeV. This deficit is inconsistent with the
expected rate for massless $\bar{\nu}_e$'s at the 99.95\%
confidence level.  

KamLAND have also presented the energy dependence of these events 
in the form of the prompt energy  ($E_{\rm prompt}\simeq
E_{\bar\nu_e}+m_p-m_n$) spectrum with 17 bins of width 0.425 MeV
for $E_{\rm prompt}>$ 0.9 MeV. To eliminate the background from
geo-neutrinos, only bins with $E_{\rm prompt}>$ 2.6 MeV are used. The
measured spectrum shows a clear deficit, but there is no significant
signal of energy-dependence of this effect.

The KamLAND results can be interpreted in terms of $\bar\nu_e$ 
oscillations with parameters shown in
Fig.\ref{fig:xiglobal_postkland}a (from Bahcall, Gonzalez-Garcia and 
Pe\~na-Garay, 2002d). Since KamLAND is a disappearance experiment and,
furthermore, matter effects are small, the experiment is not sensitive
to the flavour into which $\bar\nu_e$ oscillates. The allowed region
has three local minima and it is separated into `islands.' These
islands correspond to oscillations with wavelengths that are
approximately tuned to the average distance between the reactors and
the detector, $180$ km. The local minimum in the lowest mass island
($\Delta m^2 = 1.5\times 10^{-5}\,{\rm eV^2}$)
corresponds to an approximate first-maximum in the oscillation
probability (minimum in the event rate). The overall best-fit point
($\Delta m^2 = 7.1\times 10^{-5}\,{\rm eV^2}$ and $\tan^2\theta=0.52, 1.92$) 
corresponds approximately to the second maximum in the oscillation
probability. For the same $\Delta m^2$, maximal mixing is
only slightly less favored, $\Delta\chi^2=0.4$.  Thus with the present
statistical accuracy, KamLAND cannot discriminate between a
large-but-not-maximal and maximal mixing.

Because there is no significant evidence of energy distortion, the allowed
regions for the KamLAND analysis extend to high $\Delta m^2$ values 
for which oscillations would be averaged  and  the event reduction would be
energy independent. These solutions, however, are ruled out by the
non observation of a deficit at shorter baselines, and in particular
by the negative results of CHOOZ, which rule out solutions with
$\Delta m^2\gtrsim 1\ (0.8)\times 10^{-3} {\rm eV^2}$ at $3\sigma$ (99\% CL).

\subsection*{Consequences for solar neutrinos}
The most important aspect of Fig.~\ref{fig:xiglobal_postkland}a is the
demonstration by the KamLAND results that anti-neutrinos oscillate
with parameters that are consistent with the LMA solar neutrino
solution. Under the assumption that CPT is satisfied, the
anti-neutrino measurements by KamLAND apply directly to the neutrino
sector and the two sets of data can be combined to obtain the globally
allowed oscillation parameters (in fact, the KamLAND results already
show that CPT is satisfied in the neutrino sector to a good
approximation). The results of such an analysis are shown in
Fig.~\ref{fig:xiglobal_postkland}b (from  
Bahcall, Gonzalez-Garcia and Pe\~na-Garay, 2002d).

Comparing Fig~\ref{fig:xiglobal} with Fig~\ref{fig:xiglobal_postkland}b, 
we see that the impact of the KamLAND results is to narrow down the
allowed parameter space in a very significant way. In particular,
the LMA region is the only remaining allowed solution. The
LOW solution is excluded at $4.8\sigma$ and vacuum solutions
are excluded at $4.9\sigma$. The once `favored' SMA solution is now
excluded at $6.1\sigma$. Within the LMA region, the main effect of
KamLAND is the reduction of the allowed range of $\Delta m^2$ while
the impact on the determination of the mixing angle is marginal.

Furthermore, various alternative solutions to the solar neutrino problem
which predict that there should be no deficit in the KamLAND
experiment (such as neutrino decay or new flavor changing
interactions) are now excluded at 3.6$\sigma$. 

\subsection*{Effect on three-neutrino mixing scenarios}
The $\bar\nu_e$ survival probability at 
KamLAND in the three--neutrino framework is:
\begin{equation}
P(\bar{\nu}_e\leftrightarrow\bar{\nu}_e) =
\sin^4\theta_{13} + \cos^4\theta_{13} 
\left[ 1 - \sum_i f_i \sin^22\theta_{12}
\sin^2\left(\frac{1.27\Delta m^2_{21}[{\rm eV}^2]L_i[{\rm
      km}]}{E_{\nu}[{\rm GeV}]}\right)\right],
\label{eq:pkland}
\end{equation} 
where $L_i$ is the distance of reactor $i$ to KamLAND, and $f_i$ is
the fraction of the total neutrino flux which comes from reactor
$i$. Eq. (\ref{eq:pkland}) is obtained with two approximations: (a)
matter effects are neglected at KamLAND-like baselines, and
(b) for $\Delta m^2\gtrsim 10^{-4}$~eV$^2$, the KamLAND energy resolution 
is not good enough to resolve the corresponding oscillation 
wavelength so the oscillation phase associated to $\Delta m^2_{31}$ 
is averaged to $1/2$.  

Eq. (\ref{eq:pkland}) shows that a small $\theta_{13}$ does not affect
significantly the shape of the measured spectrum. On the other hand,
the overall normalization is scaled by $\cos^4\theta_{13}$ and this
factor may introduce a non-negligible effect. Within its present
accuracy, however, the KamLAND experiment cannot provide any further
significant constraint on $\tan^2\theta_{13}$ and the range in
Eq. (\ref{globalranges}) still holds (see also Fogli {\it et al.} 2002).

Thus the main effect of the KamLAND results on the global three
neutrino analysis of solar, atmospheric and reactor data
is to modify the allowed range of $\Delta m^2_{21}$
in Eq. (\ref{globalranges}) to
\begin{equation}
5.5\times 10^{-5}< \Delta m^2_{21}/\mbox{\rm eV$^2$}
< 1.9\times 10^{-4} .
\end{equation}
The allowed range of $\theta_{12}$ and the corresponding entries
of the mixing matrix $U$ are only marginally affected by the 
inclusion of the KamLAND data.

\subsection*{Effect on four-neutrino mixing scenarios}
In the 2+2 scenarios discussed in Sec.~\ref{subsec:fourmix},
the survival probability  for $\bar\nu_e$'s in KamLAND takes the same
form as for two neutrino oscillations and depends only on $\Delta
m^2_{12}$ and $\theta_{12}$ (as long as matter effects are neglected).
Therefore, KamLAND by itself does not provide any restriction on the
active-sterile admixtures in the oscillating state. Its impact on these
scenarios is indirect. By reducing the allowed parameter space for 
$\Delta m^2_{12}$ and $\theta_{12}$ in solar neutrino oscillations, 
it leaves less room for a sterile component in the solar analysis. 
The effect is, however, not very significant, as can be seen in 
Fig.~\ref{fig:chi24_postkland}. Comparing with the pre-KamLAND figure
\ref{fig:chi24}, we see that compatibility with the atmospheric
exclusion curve is achieved only at the $3.5\sigma$ ($3\sigma$) level
for the solar+KamLAND (solar alone) curve. 


\acknowledgments
We thank M. Maltoni and C. Pe\~na-Garay for providing us with updates of our 
previous published results. We thank Michael Dine and Yuval Grossman for
critical reading of the manuscript. MCG-G thanks the Weizmann institute, 
where this review was finalized, for their warm hospitality and 
the Albert Einstein Minerva Center for Theoretical Physics
for financial support. 
MCG-G is supported by the European Union fellowship
HPMF-CT-2000-00516.  This work was also supported by the Spanish
DGICYT under grants PB98-0693 and  FPA2001-3031, 
and by the ESF network 86.
YN is supported by the Israel Science Foundation founded by the
Israel Academy of Sciences and Humanities.
%

%
\begin{table}  
\caption{Characteristic values of $L$ and $E$
for various neutrino sources and experiments.} 
\label{tab:lovere}  
\begin{tabular} {|c|c|c|c|}  \hline
Experiment  &  L (m) & E (MeV) &  $\Delta m^2$ (eV$^{2}$) \\ \hline 
Solar  & $10^{10}$ & 1 & $10^{-10}$\\ \hline 
Atmospheric &  $10^4-10^7$ & $10^2$--$10^5$ & $10^{-1}-10^{-4}$ \\ \hline 
Reactor   & $10^2-10^3$ & 1 & $10^{-2}-10^{-3}$   \\ \hline 
Accelerator  &  $10^2$   & $10^3$--$10^4$ & $\gtrsim 0.1$ \\  \hline
LBL Accelerator  & $10^5-10^{6}$  & $10^4$ &  $10^{-2}-10^{-3}$ 
\end{tabular} 
\end{table} 

\begin{table}
\caption{Effective potentials for $\nu_e$ (upper sign) and $\overline{\nu_e}$ 
(lower sign) in various media.}
\label{tab:potentials}
\begin{tabular}{|c|c|c|}
\hline
{ medium} & { $V_C$} & { $V_N$}\\
\hline
{ $e^+$} and { $e^-$} 
& $\pm \sqrt{2} G_F ({ N_e}-{ N_{\bar e}})$ 
& $\mp \frac{G_F}{\sqrt{2}} 
({ N_e}-{ N_{\bar e}})(1-4\sin^2\theta_W)$\\
\hline
{ $p$} and { $\bar p$} & 0 &
$\pm \frac{G_F}{\sqrt{2}} ({ N_p}-{ N_{\bar p}})(1-4\sin^2\theta_W)$\\
\hline
{ $n$} and { $\bar n$} & 
0 & $\mp \frac{G_F}{\sqrt{2}} ({ N_n}-{ N_{\bar n}})$ \\
\hline
{ Neutral 
$(N_e=N_p)$} & 
$\pm \sqrt{2} G_F { N_e}$
& $\mp \frac{G_F}{\sqrt{2}} { N_n}$\\
\hline
\end{tabular}
\end{table}
%

%
%
\begin{table}
\caption{SSM  predictions:
solar neutrino fluxes and neutrino capture rates for the different
experiments, with $1\sigma$ uncertainties. }
\label{tab:rates}
\begin{tabular}{llccccc}
\hline
Source&\multicolumn{1}{c}{Flux}&Cl &Ga&SK&SNO(CC)& SNO(NC)
\\
&\multicolumn{1}{c}{$\left(10^{10}\ {\rm cm^{-2}s^{-1}}\right)$}&(SNU)&(SNU)
&{$\left(10^{6}\ {\rm cm^{-2}s^{-1}}\right)$}&
\multicolumn{2}{c} {$\left(10^{6}\ {\rm cm^{-2}s^{-1}}\right)$}\\
\hline
pp&$5.95 \left(1.00^{+0.01}_{-0.01}\right)$&0.0&69.7&0.0 & 
\multicolumn{2}{c}{0.0} \\
pep&$1.40 \times 10^{-2}\left(1.00^{+0.015}_{-0.015}\right)$
&0.22&2.8&0.0 & 
\multicolumn{2}{c}{0.0} \\
hep&$9.3 \times 10^{-7}$&0.04&0.1& 0.0093  & 
\multicolumn{2}{c}{0.0093}  \\
${\rm ^7Be}$&$4.77 \times
10^{-1}\left(1.00^{+0.10}_{-0.10}\right)$&1.15&34.2&0.0 &
\multicolumn{2}{c}{0.0} \\
${\rm ^8B}$&$5.05 \times
10^{-4}\left(1.00^{+0.20}_{-0.16}\right)$&6.76&12.2& 
$5.05$ & 
\multicolumn{2}{c}{$5.05$} \\
${\rm ^{13}N}$&$5.48 \times
10^{-2}\left(1.00^{+0.21}_{-0.17}\right)$&0.09&3.4&0.0 &
\multicolumn{2}{c}{0.0} \\
${\rm ^{15}O}$&$4.80 \times 
10^{-2}\left(1.00^{+0.25}_{-0.19}\right)$&0.33&5.5& 0.0 & 
\multicolumn{2}{c}{0.0} \\
${\rm ^{17}F}$&$5.63 \times
10^{-4}\left(1.00^{+0.25}_{-0.25}\right)$&0.0&0.1& 0.0 & 
\multicolumn{2}{c}{0.0} \\
\noalign{\medskip}
Total&&$7.6^{+1.3}_{-1.1}$&$128^{+9}_{-7}$&
 $5.05^{+1.01}_{-0.81}$  
&\multicolumn{2} {c}{$5.05^{+1.01}_{-0.81}$}  
 \\
Measured & & $2.56\pm0.226$
&$70.8\pm 4.3$
&$2.35 \pm 0.083 $
&$1.76\pm 0.11$ 
&$5.09\pm 0.64$ \\
$\frac{\rm Measured}{\rm SSM}$ 
& & $0.337 \pm 0.065$ & $0.55 \pm 0.048$  &$0.465 \pm 0.094$
&   $0.348 \pm 0.073$ 
&  $1.01\pm 0.23$
\\
\hline
\end{tabular}
\end{table}  
\begin{table}
\caption[b]{Best-fit global oscillation parameters with all solar neutrino 
data.  This table corresponds to the global solution illustrated in 
Fig.~\ref{fig:xiglobal}. Sterile solutions do not appear in 
Fig.~\ref{fig:xiglobal} because their $\chi^2_{\rm min}$ is too large.}
\label{tab:xiglobal}
\begin{tabular}{lcccc} 
\noalign{\bigskip}
\hline 
\noalign{\smallskip}
Solution&$\Delta m^2[{\rm eV^2}]$&$\tan^2\theta$&$f_{\rm B}$ &$\chi^2_{\rm min}$
(46 d.o.f) \\
\noalign{\smallskip}
\hline
\noalign{\smallskip}
LMA&
$5.0\times10^{-5}$  &$4.2\times10^{-1}$& 1.07
& 45.5 \\
LOW& $7.9\times10^{-8}$  &$6.1\times10^{-1}$& 0.91
& 54.3 \\
VAC& $4.6\times10^{-10}$  &$1.8\times10^{0}$& 0.77
& 52.0 \\
SMA& $5.0\times10^{-6}$  &$1.5\times10^{-3}$& 0.89
& 62.7 \\
Just So$^2$ & $5.8\times10^{-12}$  & $1.0\times10^{0} $& 0.46
& 86.3 \\
Sterile VAC & $4.6\times10^{-10}$ & $2.3\times10^{0}$ & 0.81
& 81.6 \\
Sterile SMA & $3.7\times10^{-6}$ & $4.7\times10^{-4}$ & 0.55
& 89.3 \\
\noalign{\smallskip}
\hline
\end{tabular}
\end{table}
%
%
\begin{table}
\caption[c]{90\% CL limit on the neutrino oscillation probabilities
from the negative searches at short baseline experiments.} 
\label{tab:sbl}
\begin{tabular}{|c|c|c|c|c|c|}
\hline
 Experiment& Beam & Channel & Limit (90\%) & $\Delta m^2_{\rm min}$ (eV$^2$)&
Ref. \\
\hline 
CDHSW  & CERN &$\nu_\mu\rightarrow \nu_\mu$ & $P_{\mu\mu}>0.95$ & 0.25 
& Dydak {\em et al.}, 1984 \\
E776 & BNL  & $\nu_\mu\rightarrow \nu_e$ & $P_{e\mu}<1.5\times 10^{-3}$ 
  & 0.075 & Borodovsky {\em et al.}, 1992 \\
E734 & BNL  & $\nu_\mu\rightarrow \nu_e$ & $P_{e\mu}<1.6\times 10^{-3}$ 
  & 0.4  & Ahrens {\em et al.}, 1987 \\
KARMEN2 & Rutherford  & $\bar\nu_\mu\rightarrow \bar\nu_e$ 
 & $P_{e\mu} <6.5\times 10^{-4}$ & 0.05 & Wolf et. al, 2001\\
E531& FNAL   & $\nu_\mu\rightarrow \nu_\tau$ & 
$P_{\mu\tau}<2.5\times 10^{-3}$  
& 0.9 & Ushida {\em et al.}, 1986\\ 
CCFR& FNAL   & $\nu_\mu\rightarrow \nu_\tau$ & 
   $P_{\mu \tau}<4\times 10^{-3}$  
  & 1.6 & McFarland et al, 1995\\ 
    &   &  $\nu_e\rightarrow \nu_\tau$ & 
  $P_{e\tau}<0.10$ &20 & Naples et al, 1999 \\
    &   &  $\nu_\mu\rightarrow \nu_e$ &  
  $P_{\mu e}<9.\times 10^{-4}$ &1.6 & Romosan {\em et al.}, 1997\\
Chorus& CERN &  $\nu_\mu\rightarrow \nu_\tau$ & 
  $P_{\mu\tau}<3.4\times 10^{-4}$ & 0.6 &  Eskut {\em et al.}, 2001\\
         &     &  $\nu_e\rightarrow \nu_\tau$ & 
  $P_{e\tau}<2.6\times 10^{-2}$ & 7.5 & Eskut {\em et al.}, 2001\\
Nomad& CERN &  $\nu_\mu\rightarrow \nu_\tau$ & 
  $P_{\mu\tau}<1.7\times 10^{-4}$ &0.7 &  Astier {\em et al.}, 2001\\ 
       &   &  $\nu_e\rightarrow \nu_\tau$ & 
  $P_{e\tau}<7.5\times 10^{-3}$ &5.9 & Astier   {\em et al.}, 2001 \\
       &   &  $\nu_\mu\rightarrow \nu_e$ & 
  $P_{\mu e}<6\times 10^{-4}$ &0.4 & Valuev {\em et al.}, 2001\\
\hline
\end{tabular}
\end{table}
\begin{table}[t]
\caption[d]{90\% CL lower bound on $P_{ee}$ and $\Delta m^2$ sensitivity
from searches at reactor experiments.} 
\label{tab:reac}
\begin{tabular}{|c|c|c|c|c|}
\hline
Experiment & L   & Limit (90\%) & $\Delta m^2_{\rm min}$ (eV$^2$)& Ref. \\ 
\hline
Bugey & 15,40 m  & $>0.91$ & 
$10^{-2}$ & Achkar {\em et al.}, 1995\\ 
Krasnoyarsk & 57, 230 m & $>0.93$ & 
$7\times 10^{-3}$ &Vidyakin {\em et al.}, 1994 \\ 
Gosgen &38, 48, 65 m & $>0.9$ & 
$0.02$ & Zacek, G., 1986\\ 
CHOOZ & 1 km & $>0.95$ & 
$7\times 10^{-4}$ & Apollonio {\em et al.}, 1999 \\
\hline
KamLAND & 150--210 km & $>0.85$ 
& $5 \times 10^{-5}$ & Piepke {\em et al.}, 2001
\end{tabular}
\end{table}
%
\begin{figure}
\begin{center} 
\mbox{\epsfig{file=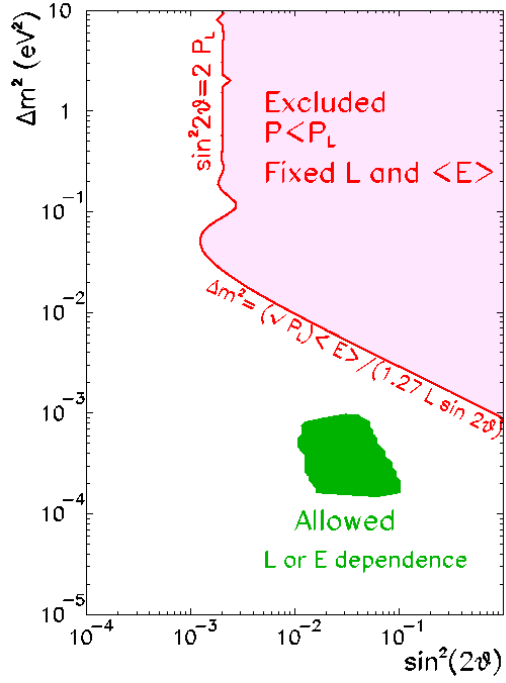,height=0.4\textheight}} 
\end{center} 
\caption{The characteristic form of an excluded region from a negative search 
with fixed $L/E$ and of an allowed region from a positive search with varying 
$L/E$ in the two-neutrino oscillation parameter plane.}
\label{fig:example} 
\end{figure}
\begin{figure}
\begin{center}
\mbox{\epsfig{file=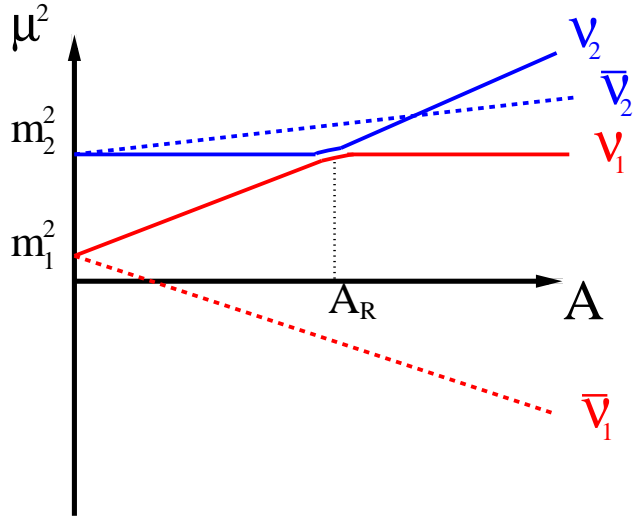,height=0.3\textheight}} 
\end{center}
\caption{Effective masses acquired in the medium by a system of two massive 
neutrinos as a function of the potential $A$ [see Eq.~(\ref{effmass})].}
\label{fig:effmass1}
\end{figure}
\begin{figure}
\begin{center}
\mbox{\epsfig{file=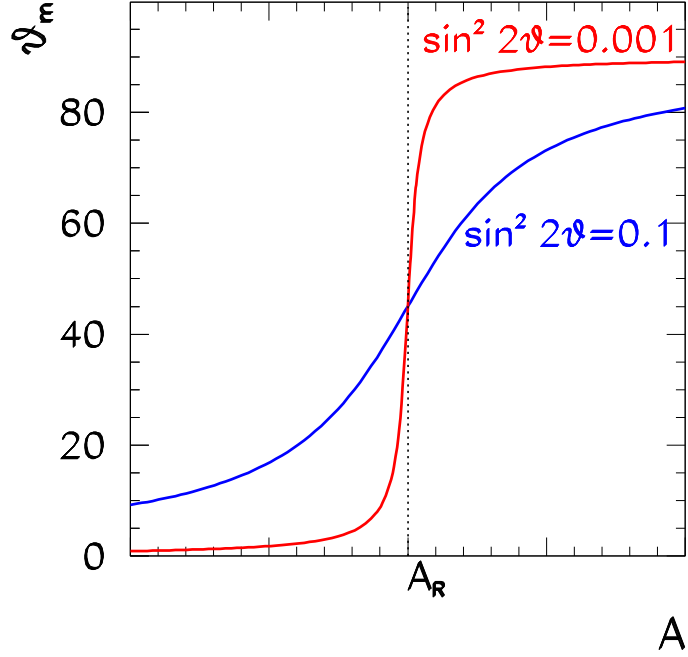,height=0.4\textheight}} 
\end{center}
\caption{The mixing angle in matter for a system of two massive neutrinos as 
a function of the potential $A$ for two different values of the mixing angle
in vacuum [see Eq.~(\ref{effmix})].}
\label{fig:effmix}
\end{figure}
\begin{figure}
\begin{center}
\mbox{\epsfig{file=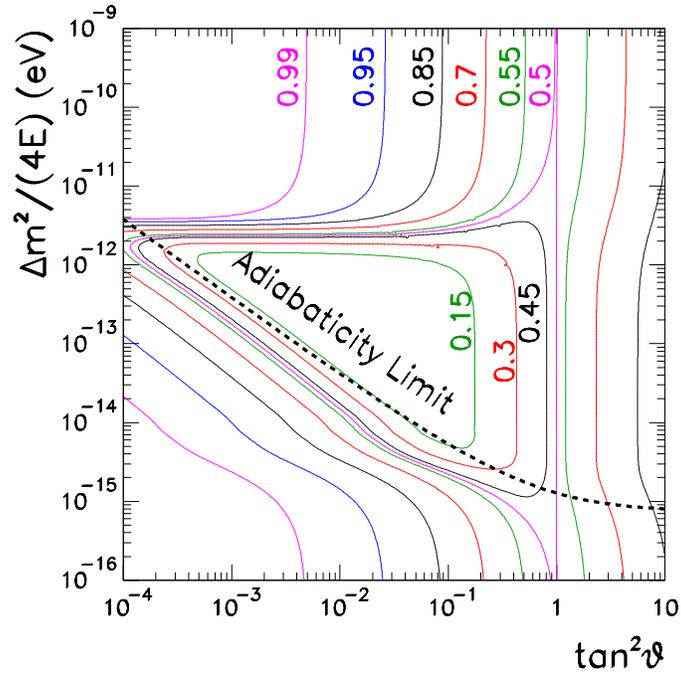,height=0.4\textheight}} 
\end{center}
\caption{Isocontours of the survival probablity $P_{ee}$ in the Sun. Also 
shown is the limit of applicability of the adiabatic approximation $Q=1$ 
(dashed line).}
\label{fig:peenoad}
\end{figure}
\begin{figure}
\begin{center}
\mbox{\epsfig{file=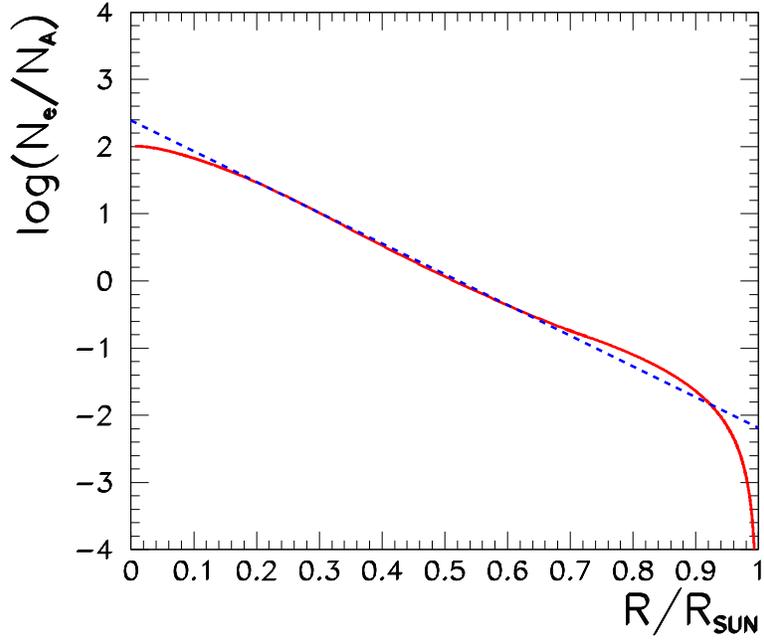,height=0.4\textheight}} 
\end{center}
\caption{The solar matter density profile for the BP00 model (full line),
and the exponential approximation (dashed line).} 
\label{fig:sundens}
\end{figure} 
\begin{figure}
\begin{center}
\mbox{\epsfig{file=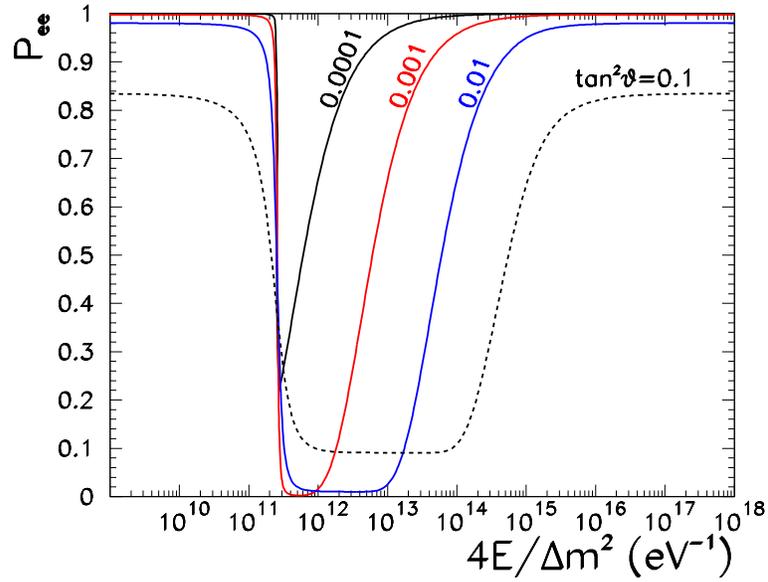,height=0.35\textheight}} 
\end{center}
\caption{The survival probablity for a $\nu_e$ state produced in the
center of the sun as a function of $E/\Delta m^2$ for various
values of the mixing angle.} 
\label{fig:peemsw}
\end{figure} 
%
%
\begin{figure}
\begin{center} 
\mbox{\epsfig{file=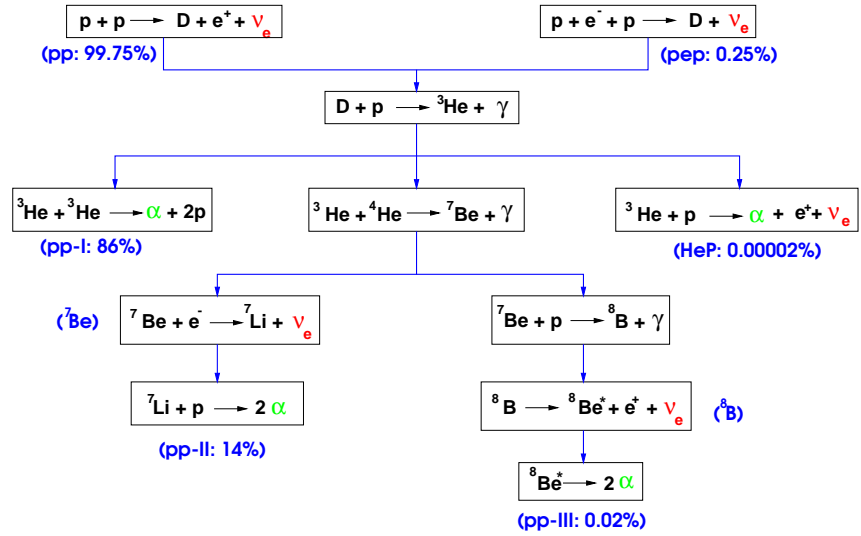,height=0.43\textheight}} 
\end{center} 
\caption{The $pp$ chain in the Sun.}
\label{fig:pp}
\end{figure}
\begin{figure}
\begin{center} 
\mbox{\epsfig{file=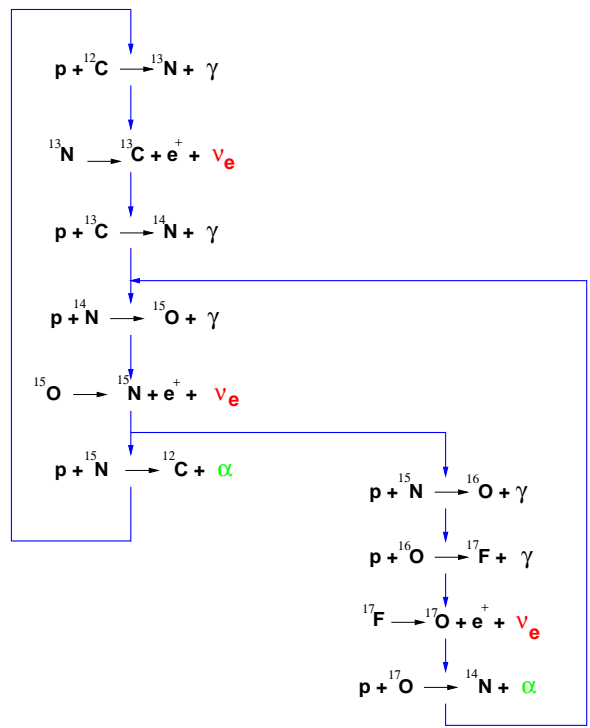,height=0.43\textheight}} 
\end{center} 
\caption{The CNO cycle in the Sun.}
\label{fig:cno}
\end{figure}
\begin{figure}[htbp]
\begin{center} 
\mbox{\epsfig{file=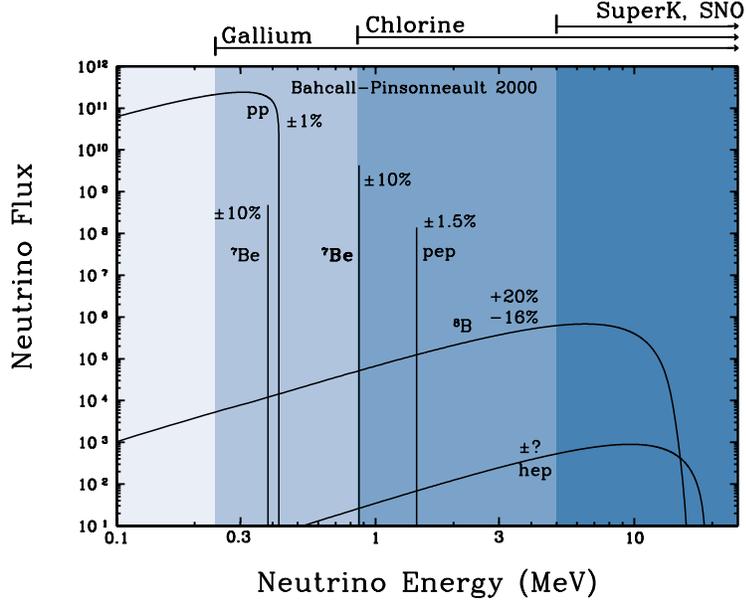,height=0.5\textheight,angle=-90}} 
\end{center} 
\caption{Neutrino fluxes from the $pp$ chain reactions as a function
of the neutrino energy.}
\label{fig:bp00}
\end{figure}
\begin{figure}
\begin{center} 
\mbox{\epsfig{file=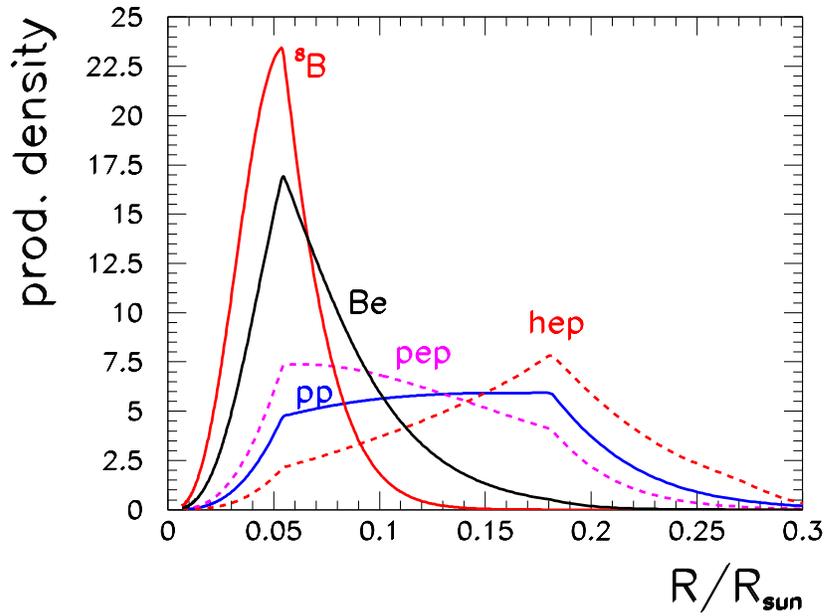,height=0.4\textheight}} 
\end{center} 
\caption{Production point distribution for the $pp$ chain neutrinos
as a function of the distance from the solar center.} 
\label{fig:ppprod}
\end{figure}
\begin{figure}[htb]
\begin{center} 
\mbox{\epsfig{file=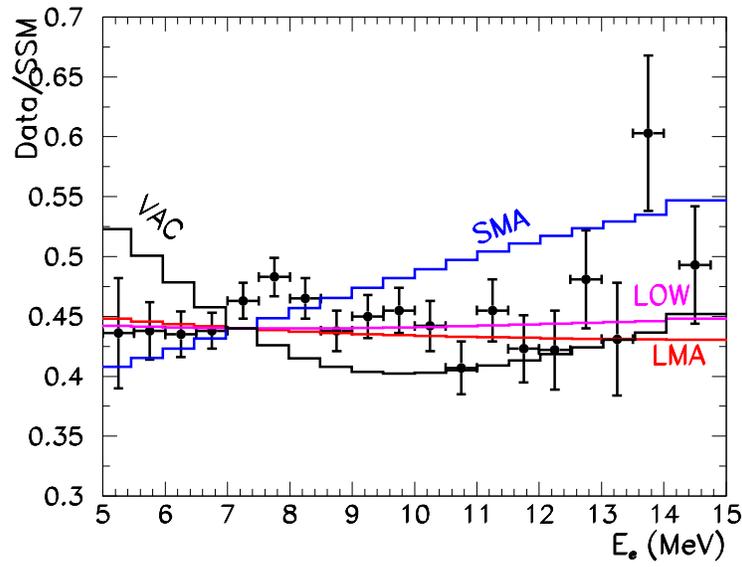,height=0.35\textheight}} 
\end{center} 
\caption[a]{The electron recoil energy spectrum measured in SK normalized 
to the SSM prediction, and the expectations for the
best fit points for the LMA, SMA, LOW and VAC solutions in Fig.~\ref{fig:xirates}.}
\label{fig:skspec}
\end{figure}
\begin{figure}
\begin{center}
\mbox{\epsfig{file=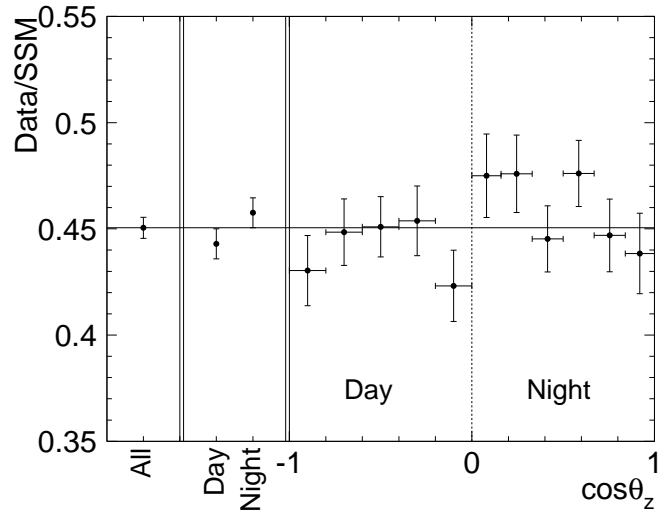,height=0.3\textheight}} 
\end{center}
\caption{The zenith angle dependence of the solar neutrino flux (statistical 
error only).  The width of the night-time bins was chosen to 
separate solar neutrinos that pass through the Earth's dense core 
($\cos\theta_z \ge 0.84$) from those that pass through the mantle 
($0<\cos\theta_z<0.84$). The horizontal line shows the average flux.}
\label{fig:skzen}
\end{figure}
\begin{figure}
\begin{center}
\mbox{\epsfig{file=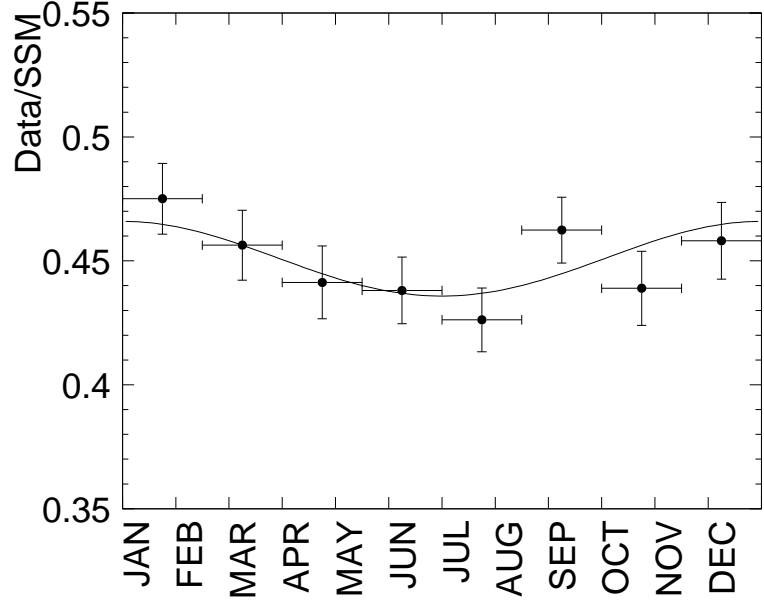,height=0.35\textheight}} 
\end{center}
\caption{Seasonal variation of the solar neutrino flux (statistical errors
only). The curve shows the expected seasonal variation of the flux induced by 
the eccentricity of the Earth's orbit.}
\label{fig:sksea}
\end{figure}
\begin{figure}[htbp]
\begin{center}
\mbox{\epsfig{file=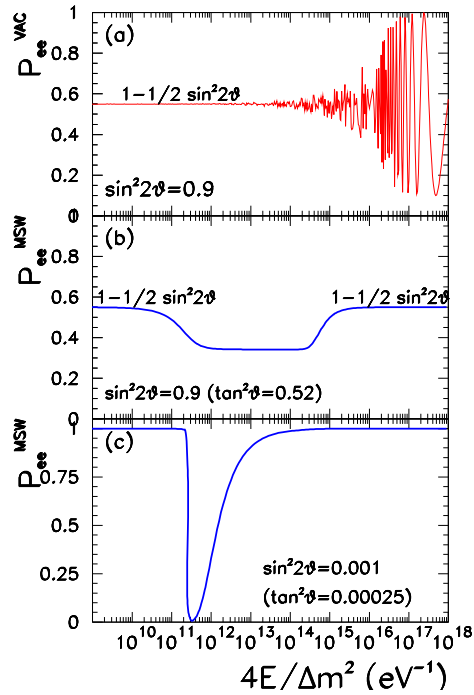,height=0.4\textheight}} 
\end{center}
\caption{$P_{ee}$ as a function of $4E/\Delta{m}^2$. In panel (a) matter 
effects are ignored [see Eq.~(\ref{pvacees})]. Panels (b) (large mixing
angle) and (c) (small mixing angle) take into account the MSW effect
[see Eq.~(\ref{pmsw})].}
\label{fig:probs}
\end{figure}
\begin{figure}[htb]
\begin{center} 
\mbox{\epsfig{file=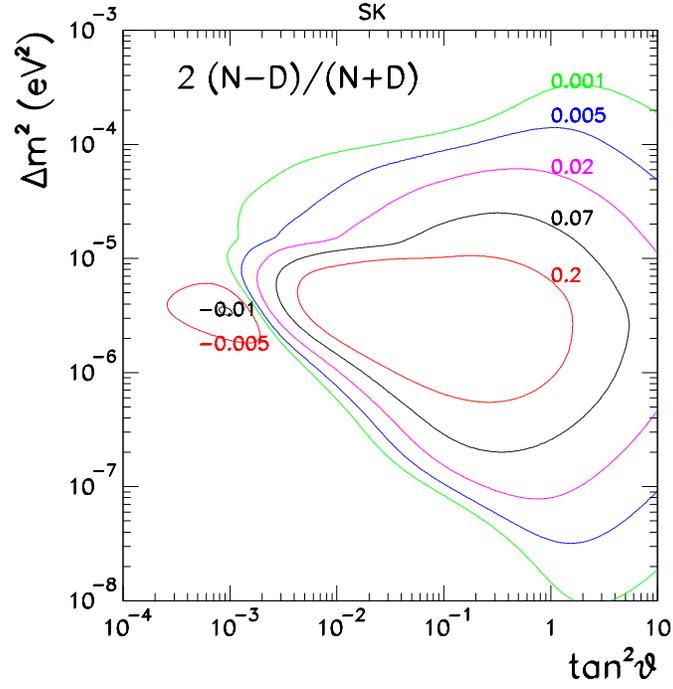,height=0.4\textheight}} 
\end{center} 
\vglue -.5cm
\caption[b]{Isocontours of the day-night asymmetry at SK.}
\label{fig:sk_adn}
\end{figure}
\begin{figure}[htb]
\begin{center} 
\mbox{\epsfig{file=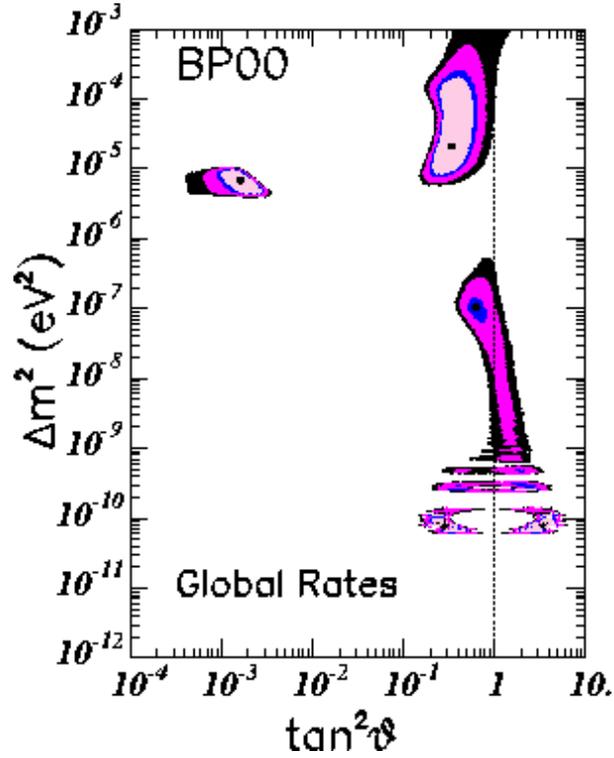,height=0.43\textheight}} 
\end{center} 
\caption[c]{Allowed oscillation parameters (at 90, 95, 99  and 99.7\% CL) 
from the analysis of the total 
event rates of the Chlorine, Gallium, SK and SNO CC experiments.
The best fit point is marked with a star.}
\label{fig:xirates}
\end{figure}
\begin{figure}[htb]
\begin{center} 
\mbox{\epsfig{file=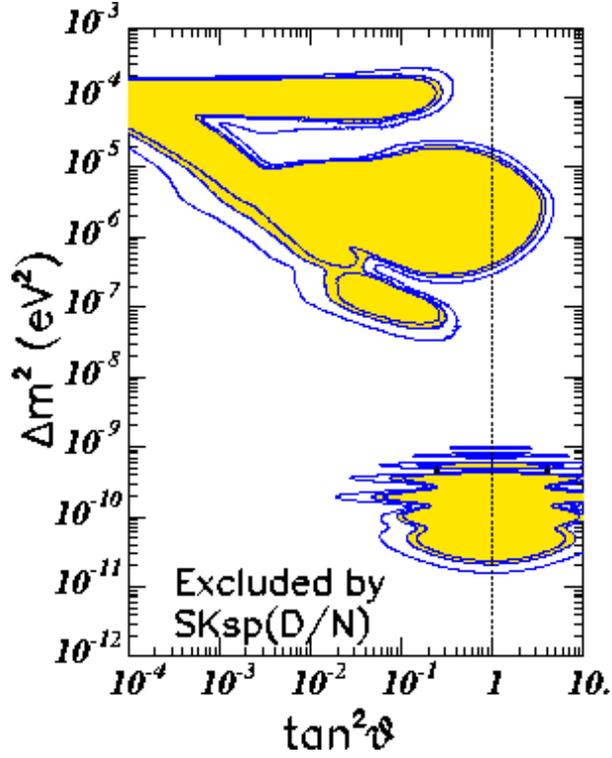,height=0.43\textheight}} 
\end{center} 
\caption[d]{Excluded oscillation parameters at 95, 99 (shadowed region) 
and 99.7\% CL from the analysis of the day-night 
spectrum data.}
\label{fig:xispdn}
\end{figure}
\begin{figure}[htb]
\begin{center} 
\mbox{\epsfig{file=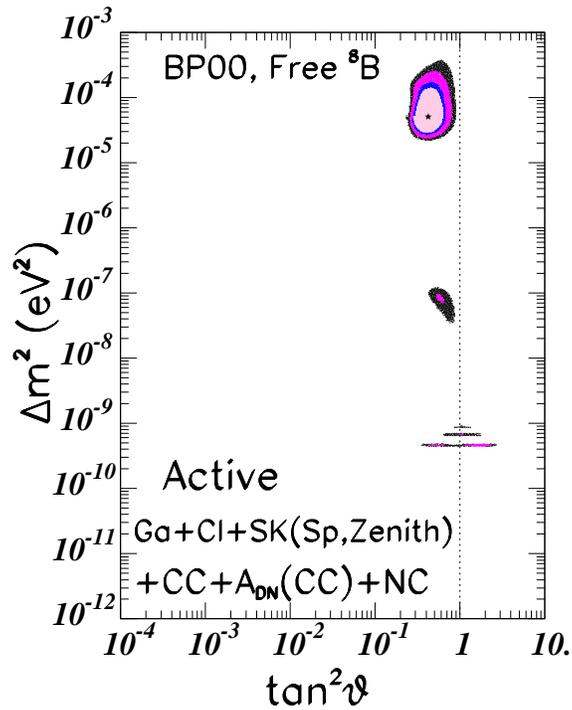,height=0.43\textheight}} 
\end{center} 
\vglue -.5cm
\caption[e]{Allowed oscillation parameters 
(at 90, 95, 99  and 99.7\% CL) from the global analysis of the 
solar neutrino data. The best fit point (LMA active) is marked with a star.}
\label{fig:xiglobal}
\end{figure}
%
%
\begin{figure}
\begin{center}
\mbox{\epsfig{file=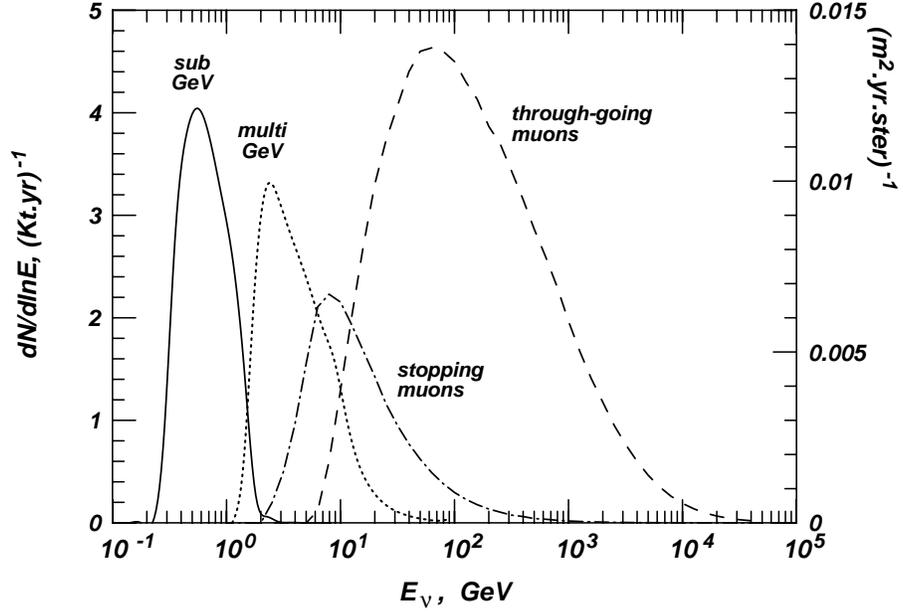,height=0.35\textheight}}
\end{center}
\caption[f]{Event rates as a function of neutrino energy for fully contained
events, stopping muons, and through-going muons at SuperKamiokande.}
\label{fig:events} 
\end{figure} 
\begin{figure}
\begin{center}
\mbox{\epsfig{file=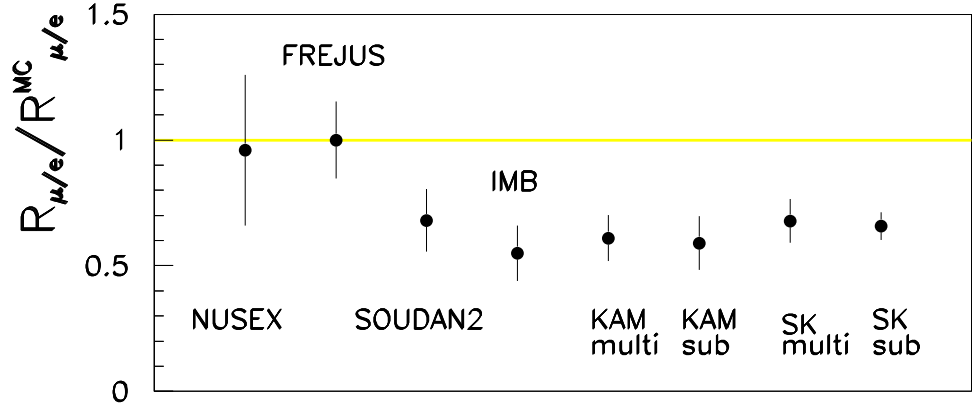,height=0.25\textheight}}
\end{center}
\caption{The double ratio of $\nu_\mu$ to $\nu_e$ events, data divided by 
theoretical predictions, for various underground atmospheric neutrino 
detectors.}
\label{fig:R} 
\end{figure} 
\begin{figure}
\begin{center}
\mbox{\epsfig{file=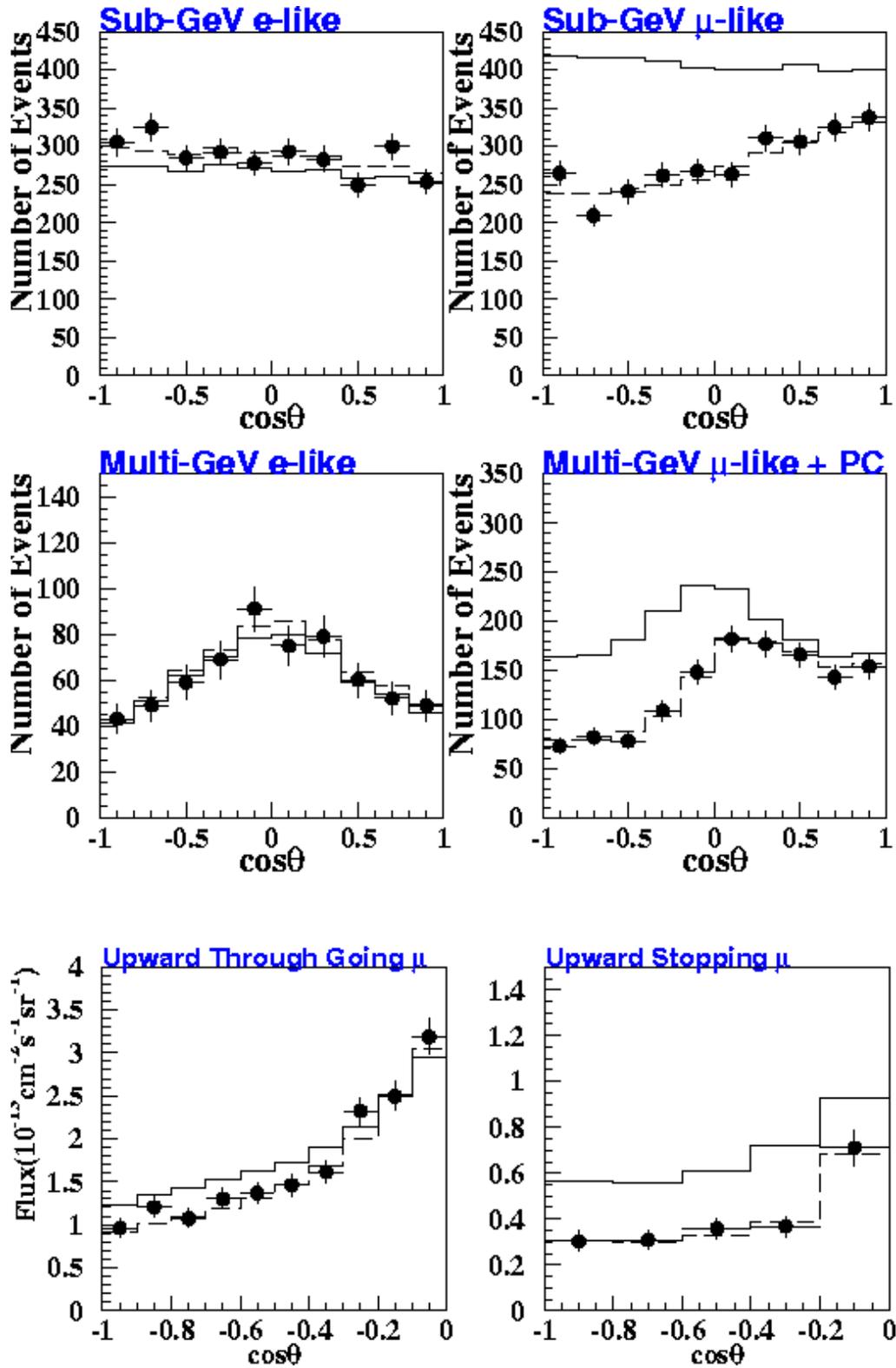,height=0.9\textheight}}
\end{center}
\caption[g]{Zenith angle distribution of SuperKamiokande 1289 days
data samples. Dots, solid line and dashed line correspond
to data, MC with no oscillation and MC with best fit oscillation parameters,
respectively.}
\label{fig:skatm} 
\end{figure} 
\begin{figure}
\begin{center}
\mbox{\epsfig{file=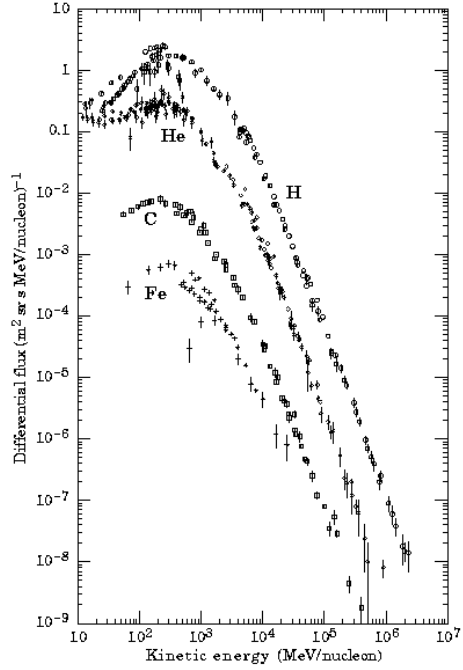,height=0.4\textheight}}
\end{center}
\caption[i]{The main components of the primary cosmic ray spectrum.}
\label{fig:crspec} 
\end{figure} 
\begin{figure}
\begin{center}
\mbox{\epsfig{file=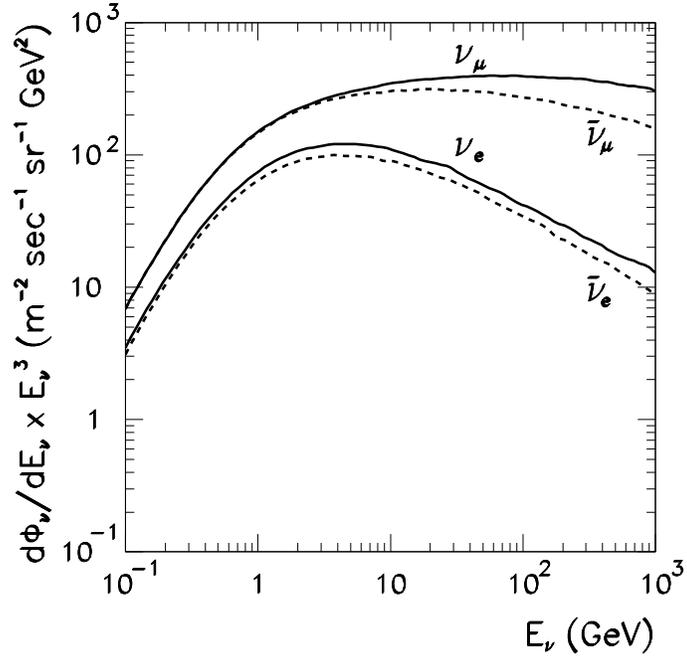,height=0.4\textheight}}
\end{center}
\caption{Atmospheric neutrino fluxes as a function of the neutrino energy.}
\label{fig:atmflux} 
\end{figure} 
\begin{figure}
\begin{center}
\mbox{\epsfig{file=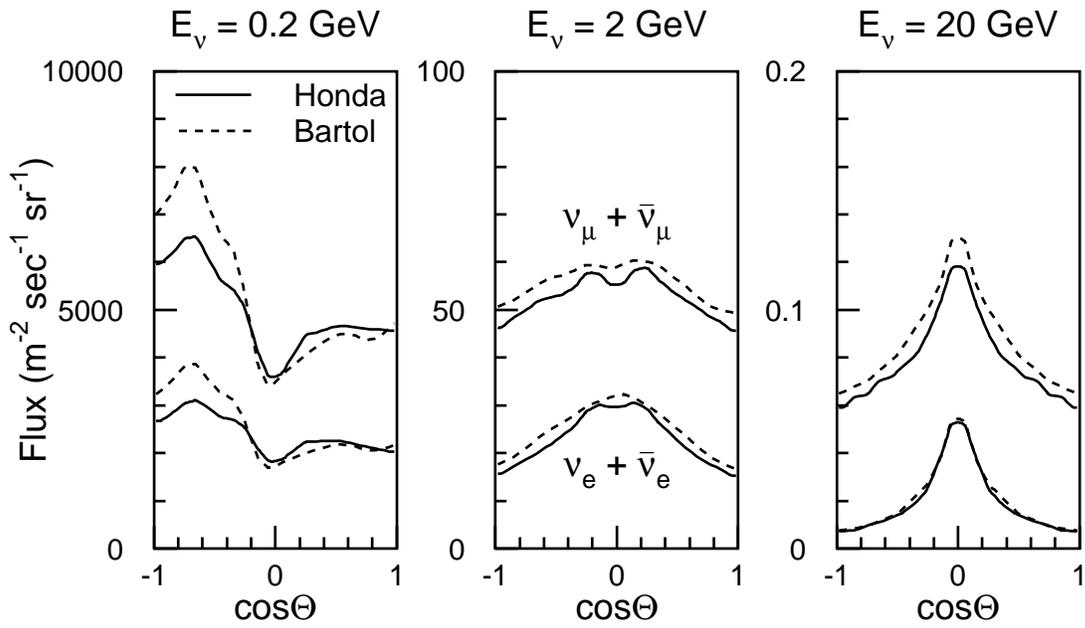,height=0.4\textheight}}
\end{center}
\caption[j]{Zenith angular dependence of the atmospheric neutrino fluxes.}
\label{fig:atmzenith} 
\end{figure} 
\begin{figure}
\begin{center}
\mbox{\epsfig{file=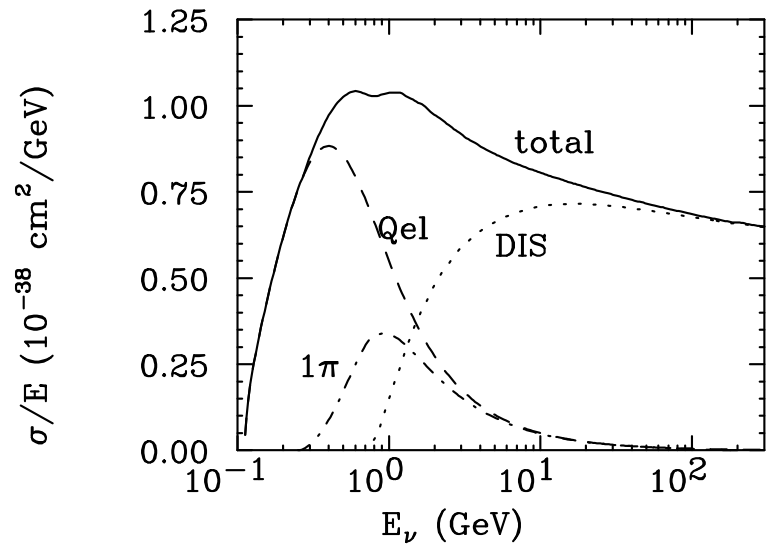,height=0.5\textheight,angle=90}}
\end{center}
\caption[k]{The various contributions to the neutrino-nucleon cross section. }
\label{fig:sigma_atm} 
\end{figure} 
\begin{figure}
\begin{center}
\mbox{\epsfig{file=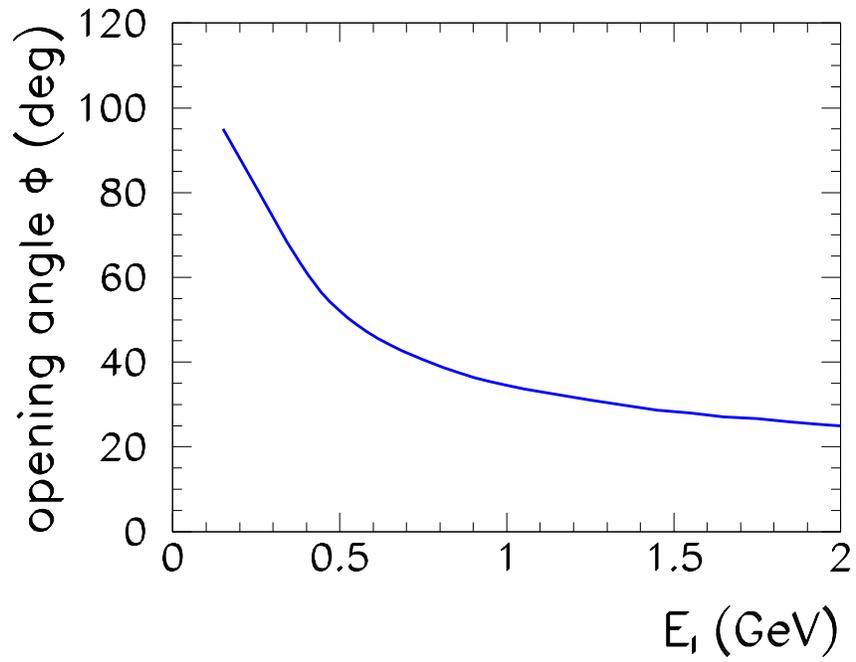,height=0.4\textheight}}
\end{center}
\caption[l]{The opening angle between the incoming neutrino and the produced
charged lepton in QE interactions as a function of the charged lepton energy.}
\label{fig:opening} 
\end{figure} 
\begin{figure}
\begin{center}
\mbox{\epsfig{file=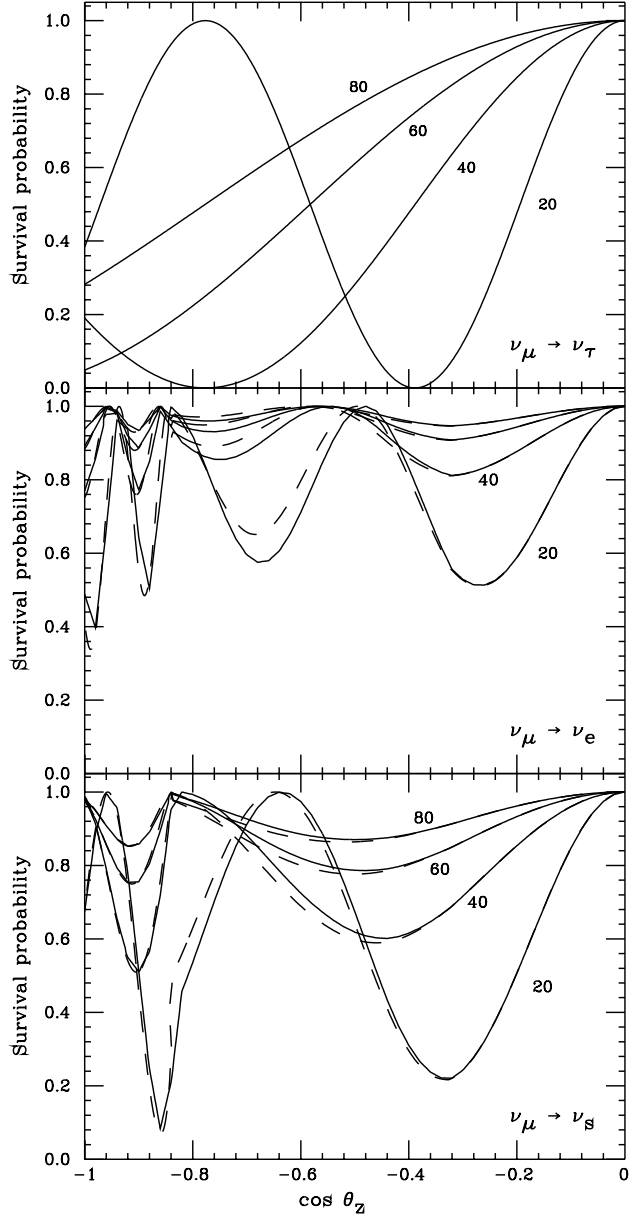,height=0.7\textheight}}
\end{center}
\caption[m]{$P_{\mu\mu}$ as a function of the zenith angle for maximal mixing 
of $\nu_\mu$ with $\nu_\tau$ (upper panel), $\nu_e$ (middle panel) and $\nu_s$ 
(lower panel). For $\Delta m^2 = 5\times 10^{-3}$~eV$^2$ the curves correspond 
to $E_\nu$ of 20, 40, 60 and 80 GeV. The dashed curves are calculated with the 
approximation of constant average densities in the mantle and in the 
core of the Earth.}
\label{fig:atm_probs} 
\end{figure} 
\begin{figure}
\begin{center}
\epsfig{file=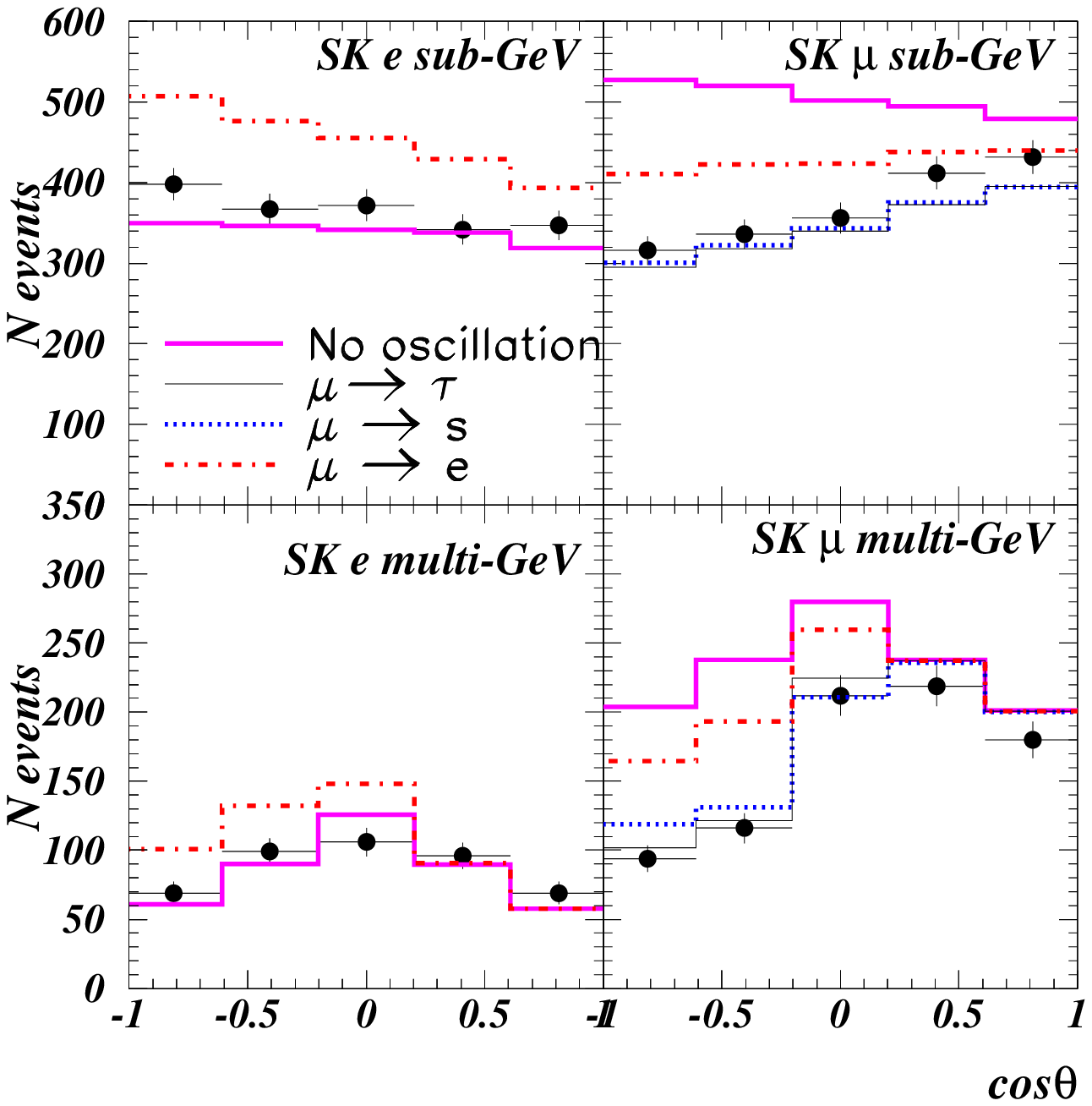,width=0.505\textwidth}
\epsfig{file=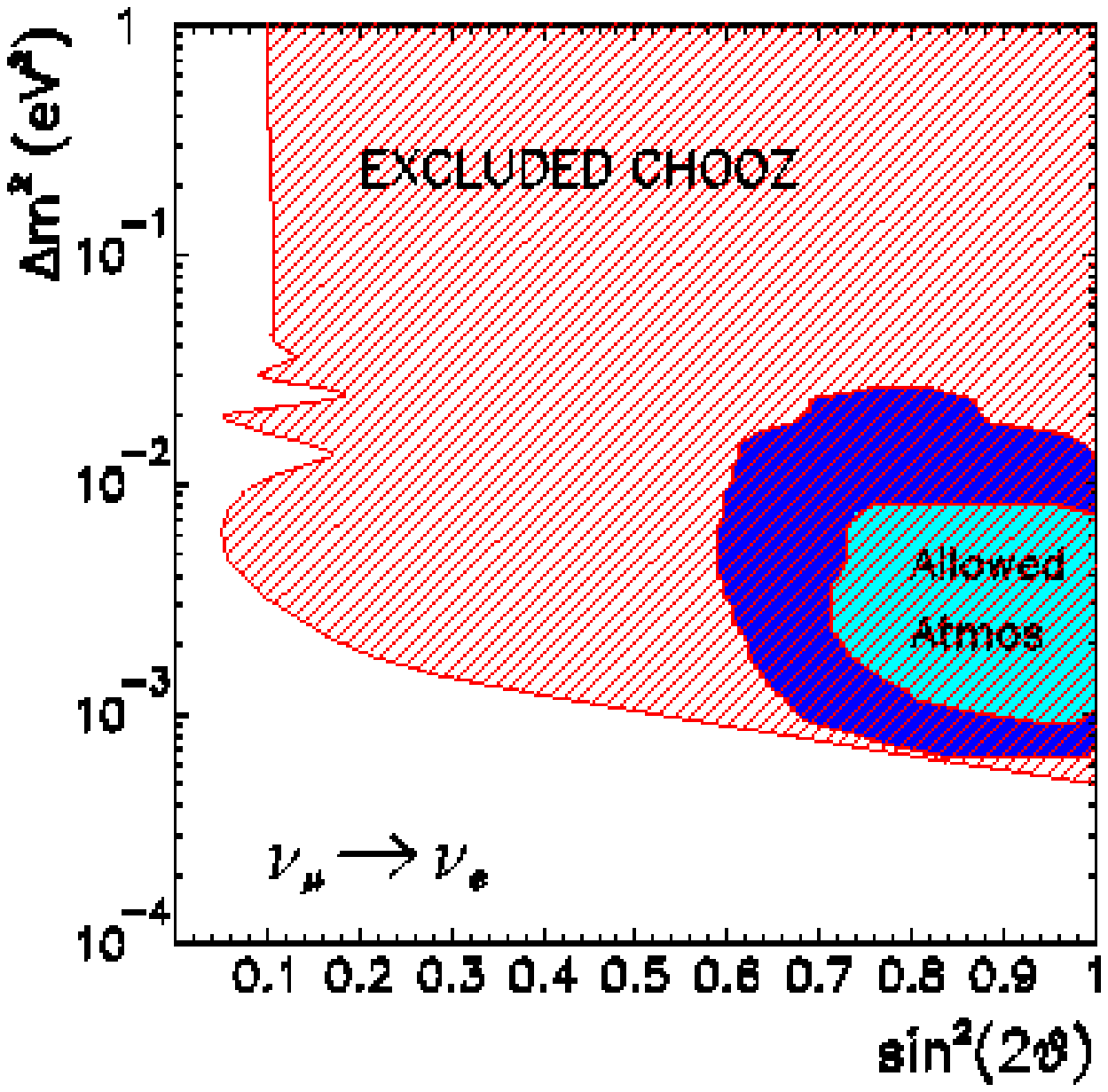,width=0.6\textwidth}
\end{center}
\caption[n]{The status of the $\nu_\mu\rightarrow\nu_e$ oscillation solution
to the atmospheric neutrino anomaly.}
\label{fig:atm_mue} 
\end{figure} 
\begin{figure}
\begin{center}
\mbox{\epsfig{file=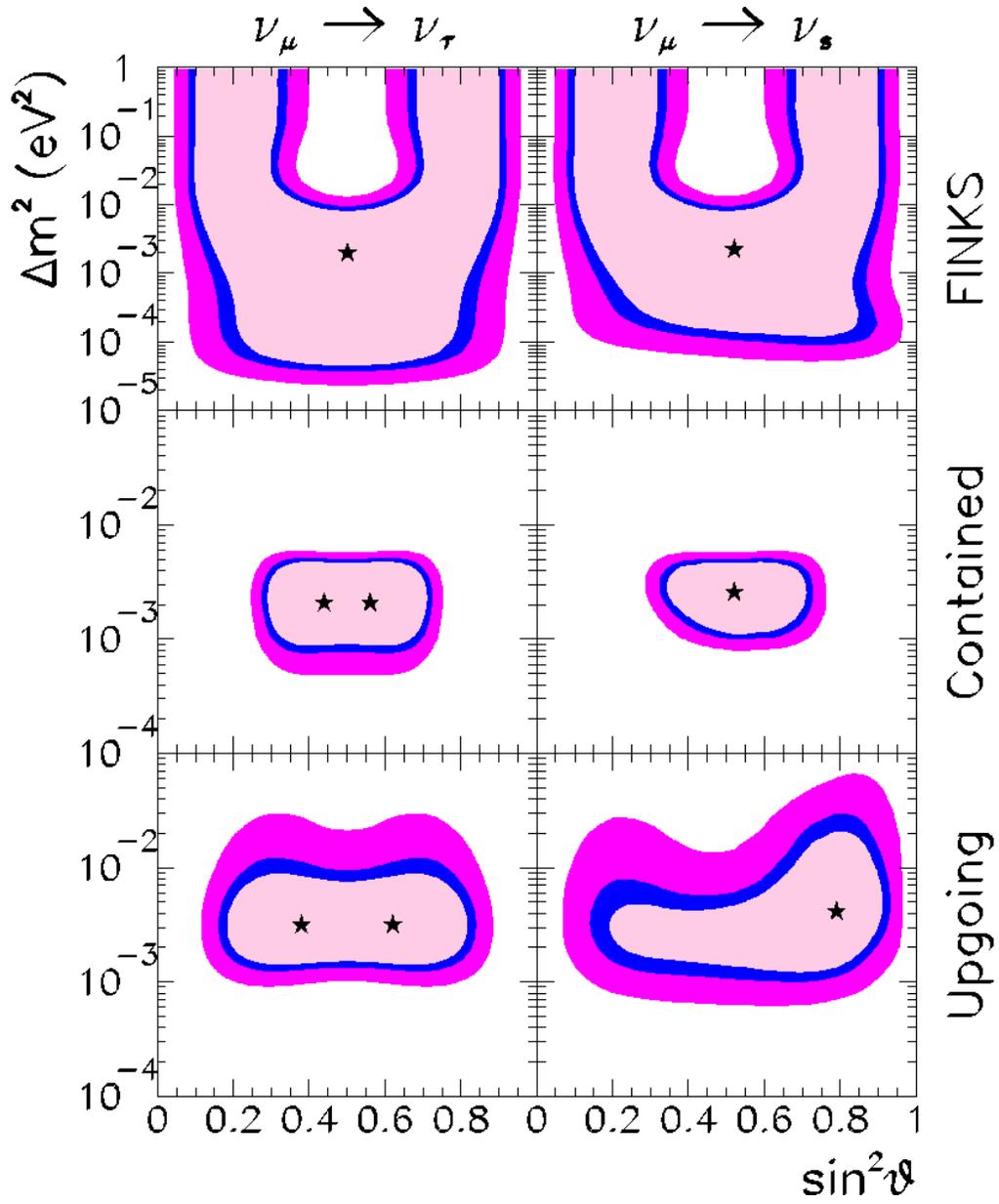,height=0.7\textheight}}
\end{center}
\caption[o]{Allowed regions (at 90, 95 and 99 \% CL) from the analysis of 
various partial data samples of atmospheric neutrinos for the oscillation 
channels $\nu_\mu\to\nu_\tau$ and $\nu_\mu\to\nu_s$. The best fit points are 
marked with a star (see text for details). }
\label{fig:atm_partial} 
\end{figure} 
\begin{figure}
\begin{center}
\mbox{\epsfig{file=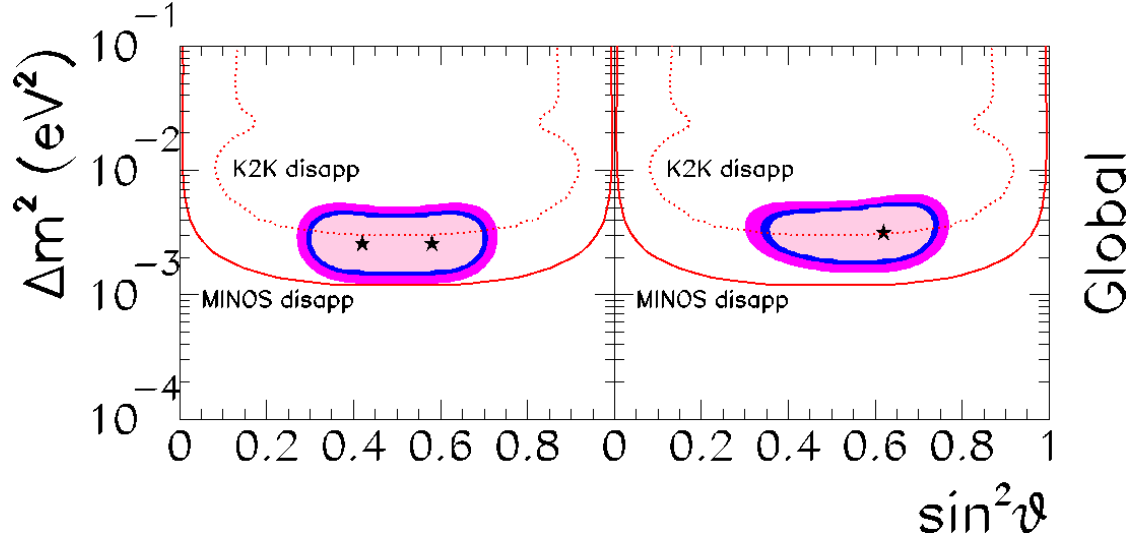,width=0.9\textwidth}}
\end{center}
\caption[p]{Allowed regions (at 90, 95 and 99 \% CL) from the analysis of 
the full data sample of atmospheric neutrinos for the oscillation channels 
$\nu_\mu\to\nu_\tau$ and $\nu_\mu\to\nu_s$. The best fit points are marked 
with a star (see text for details). Also shown are the expected sensitivities
from LBL experiments.}
\label{fig:atm_global} 
\end{figure} 
\begin{figure}
\begin{center}
\mbox{\epsfig{file=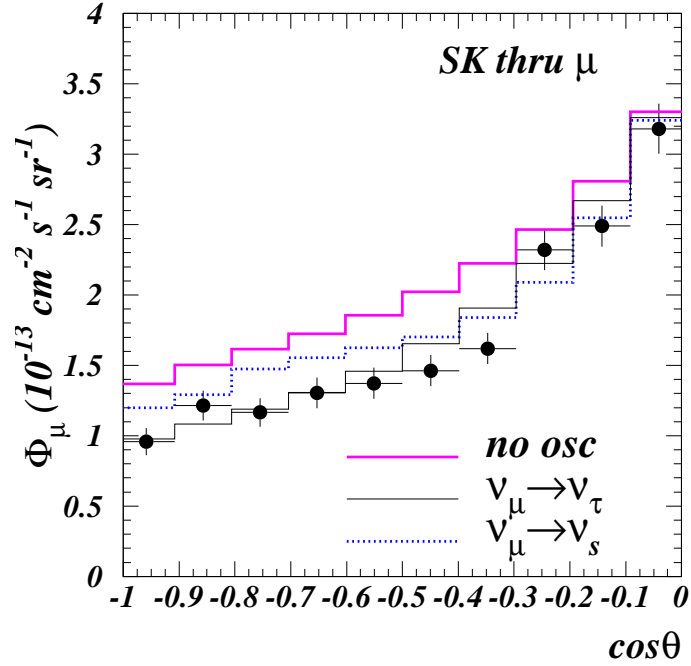,height=0.4\textheight}}
\end{center}
\caption[q]{Zenith angular dependence for through-going muon at SuperKamiokande
compared with the predictions in the case of no-oscillation and for the best 
fit points of $\nu_\mu\to\nu_\tau$ ($\Delta m^2=2.6\times 10^{-3}$ eV$^2$,
$\sin^22\theta=0.97$) and $\nu_\mu\to\nu_s$ ($\Delta m^2=3\times 10^{-3}$
eV$^2$, $\sin^2\theta=0.61$) oscillations.}
\label{fig:ang_thru} 
\end{figure} 
%
%
\begin{figure}
\begin{center}
\mbox{\epsfig{file=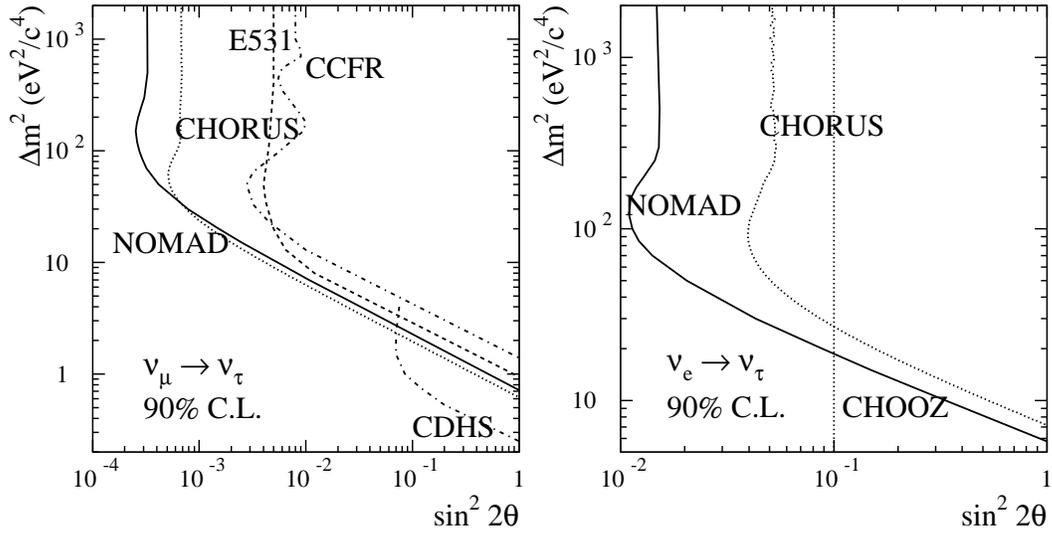,height=0.35\textheight}}
\end{center}
\caption[r]{Excluded regions at 90\% in the $\nu_\mu\to\nu_\tau$ and
$\nu_e\to\nu_\tau$ channels from SBL experiments.}
\label{fig:mtet} 
\end{figure}
\begin{figure}
\begin{center}
\mbox{\epsfig{file=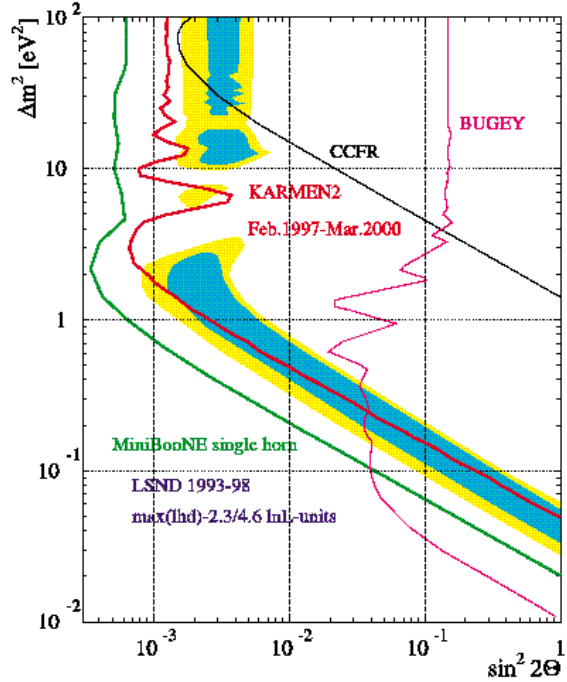,height=0.4\textheight}}
\end{center}
\caption[s]{Allowed regions (at 90 and 99 \% CL) for $\nu_e\to\nu_\mu$ 
oscillations from the LSND experiment compared with the exclusion regions 
(at 90\% CL) from KARMEN2 and other experiments. The 90 \% CL expected 
sensitivity curve for MinimBoNE is also shown.}
\label{fig:lsnd} 
\end{figure}
\begin{figure}
\begin{center}
\mbox{\epsfig{file=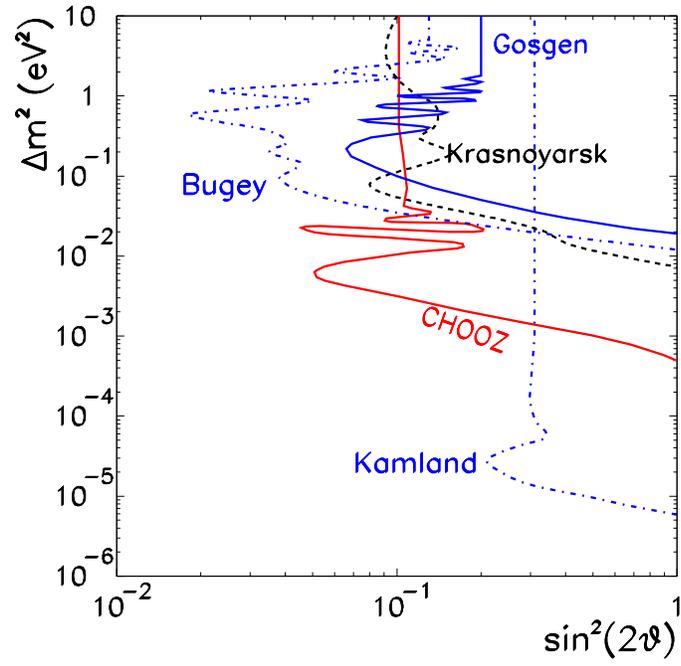,height=0.4\textheight}}
\end{center}
\caption[t]{Excluded regions at 90\% for $\nu_e$ oscillations 
from reactors experiments
and the expected sensitivity from the KamLAND experiment.} 
\label{fig:reactors} 
\end{figure}
%
%
%
\begin{figure}
\begin{center}
\mbox{\epsfig{file=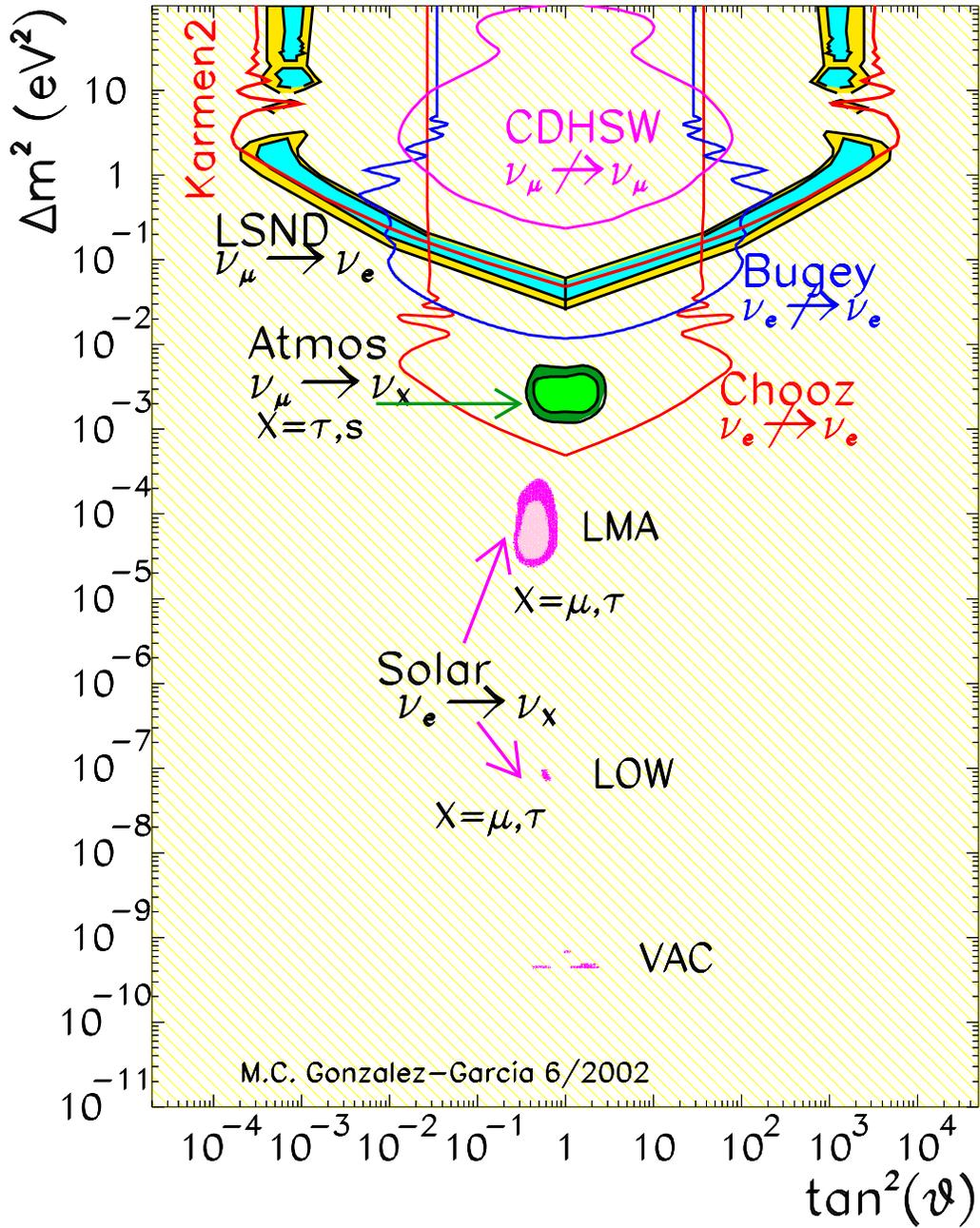,height=0.8\textheight}}
\end{center}
\caption[v]{Summary of the present pieces of evidence for neutrino masses
and mixing as obtained from 2-$\nu$ oscillation analyses. The allowed regions 
correspond to 90\% and 99\% CL for 2 d.o.f. We also show relevant bounds
from laboratory experiments.}
\label{fig:sum} 
\end{figure}
\begin{figure}
\begin{center}
\mbox{\epsfig{file=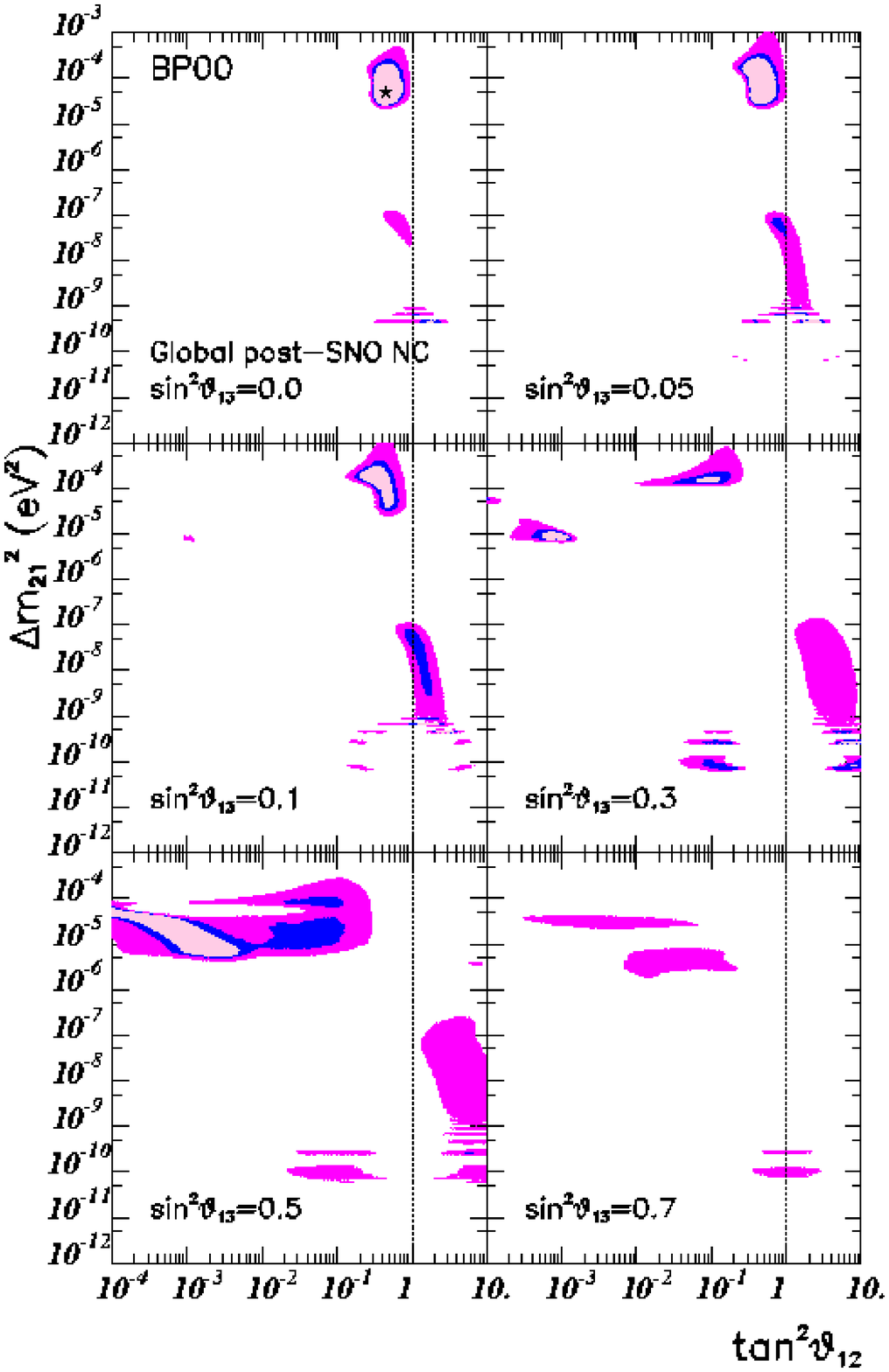,height=0.9\textheight}}
\end{center}
\caption[w]{Allowed regions (at 90, 95, and 99\% CL) in the 
($\Delta m^2_{21},\tan^2\theta_{12}$) plane from the global analysis of the 
solar neutrino data in the framework of three-neutrino oscillations
for various values of $\sin^2\theta_{13}$. The global best fit point
is marked by the star.}
\label{fig:solar3} 
\end{figure}
\begin{figure}
\begin{center}
\mbox{\epsfig{file=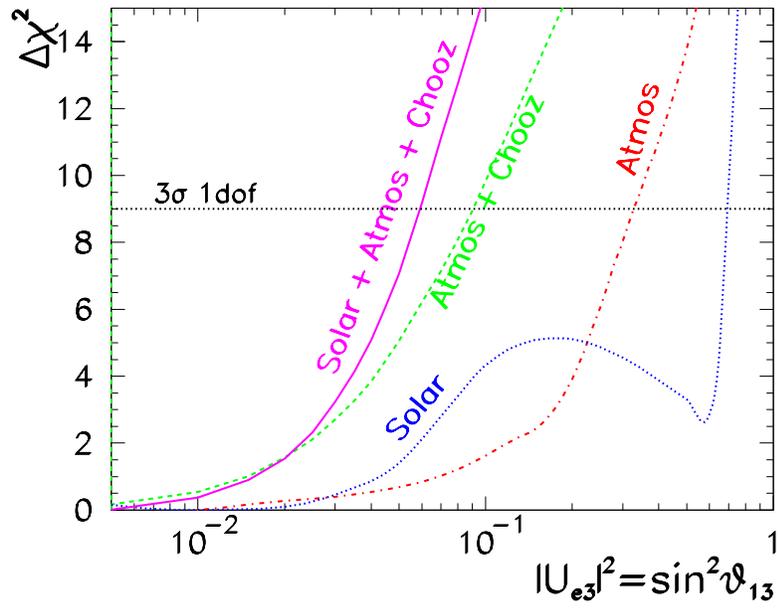,height=0.4\textheight}}
\end{center}
\caption[y]{Dependence of $\Delta\chi^2$ on $\sin^2\theta_{13}$ in the analysis
of the atmospheric, solar and CHOOZ neutrino data. The dotted horizontal line 
corresponds to the 3$\sigma$ limit for a single parameter.}
\label{fig:chi2t13}
\end{figure}
\begin{figure}
\begin{center} 
\mbox{\epsfig{file=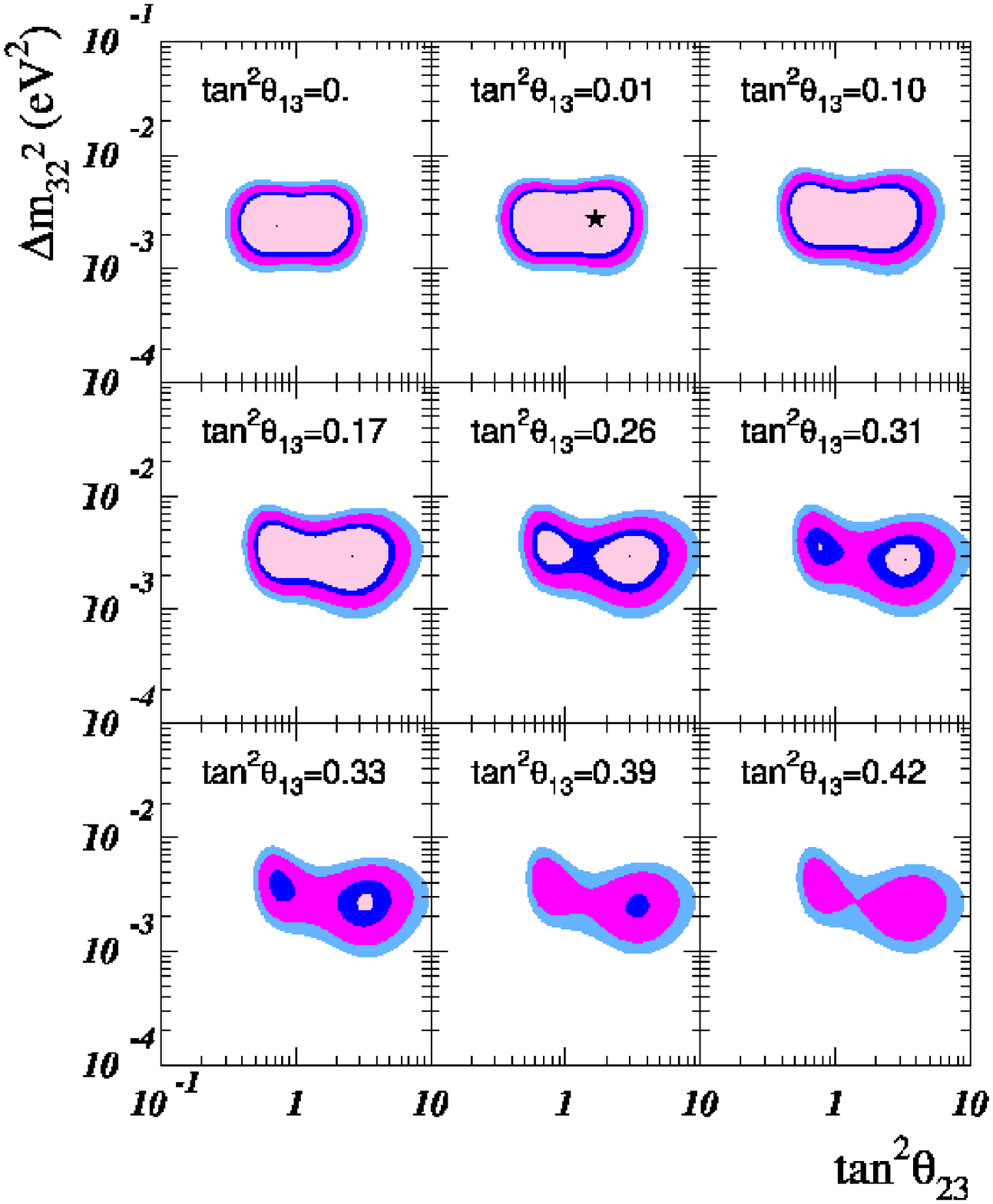,height=0.8\textheight}} 
\end{center} 
\caption[x]{Allowed regions (at 90, 95, 99, and 99.7\% CL) in the 
($\Delta m^2_{32},\tan^2\theta_{23}$) plane from the global analysis of the 
atmospheric neutrino data in the framework of three-neutrino oscillations
for various values of $\tan^2\theta_{13}$. The global best fit point
is marked by the star.}
    \label{fig:atmos3}
\end{figure}
\begin{figure}
\begin{center}
\mbox{\epsfig{file=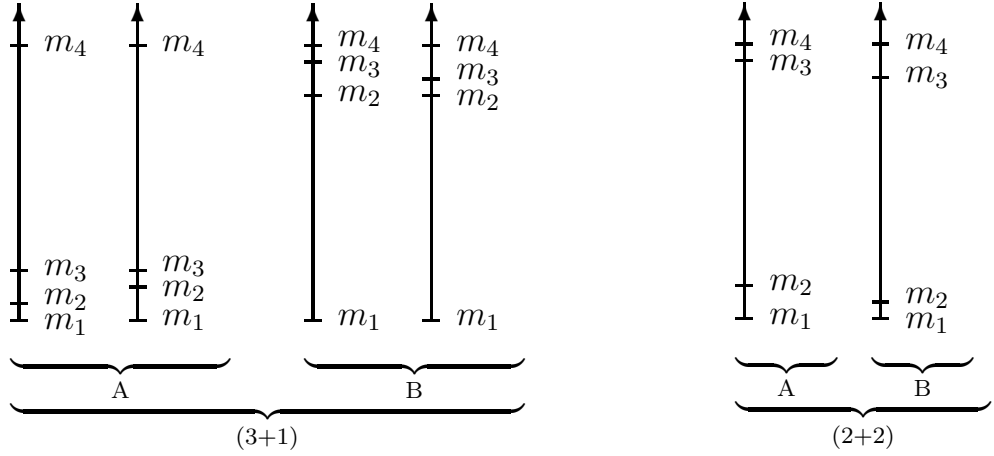,width=0.95\textwidth}}
\end{center}
 \caption[z]{The six types of 4-neutrino mass spectra. The different distances
between the masses on the vertical axes represent the different scales of
mass-squared differences required to explain solar, atmospheric and LSND
data with neutrino oscillations.}
\label{fig:4mass}
\end{figure}
\begin{figure}
\begin{center}
\mbox{\epsfig{file=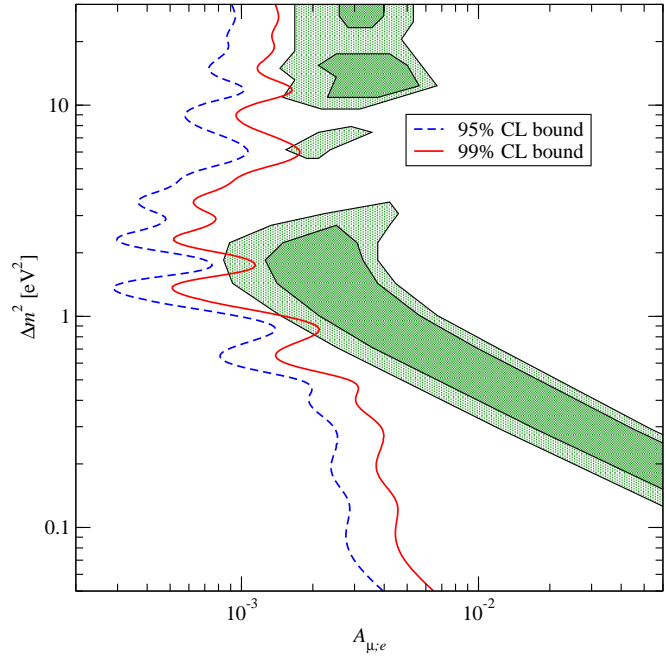,height=0.4\textheight}}
\end{center}
\caption[aa]{Upper bounds (at 95 and 99\% CL) on 
$A_{\mu;e}\equiv4|U_{e4}U_{\mu4}|^2$ in the context of (3+1)-schemes. The 
shaded regions are the regions allowed by LSND at 90 and 99\% CL.}
\label{fig:3plus1}
\end{figure}
\begin{figure}
\begin{center}
\mbox{\epsfig{file=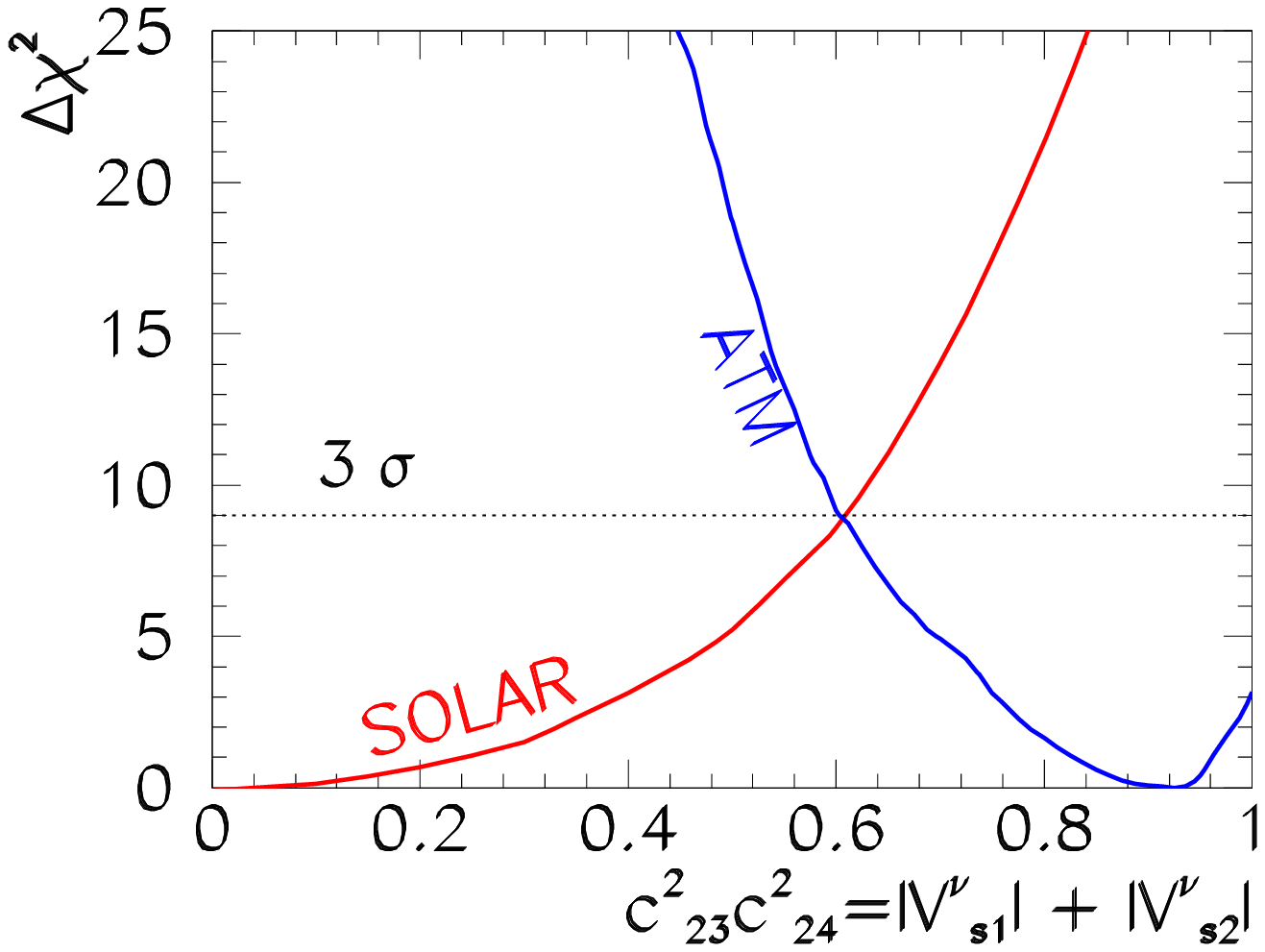,width=0.95\textwidth}}
\end{center}
\caption[ad]{
$\Delta\chi^2$ as a function of the 
active-sterile admixture  $|V^\nu_{s1}|^2+|V^\nu_{s2}|^2$ from the analysis of
solar and atmospheric data in (2+2)-schemes.}
\label{fig:chi24}
\end{figure}
\begin{figure}[htb]
\epsfig{file=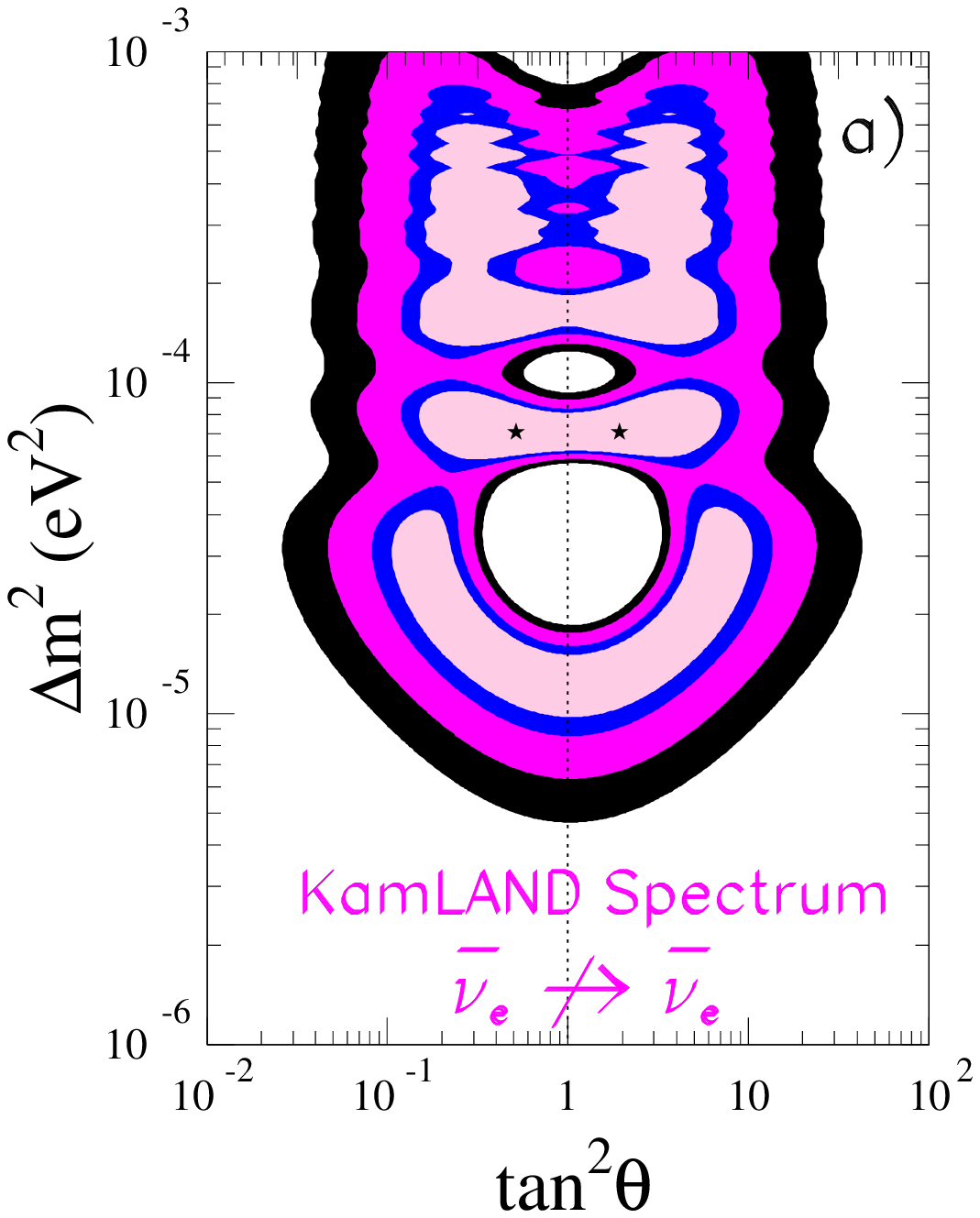,width=0.45\textwidth} 
\epsfig{file=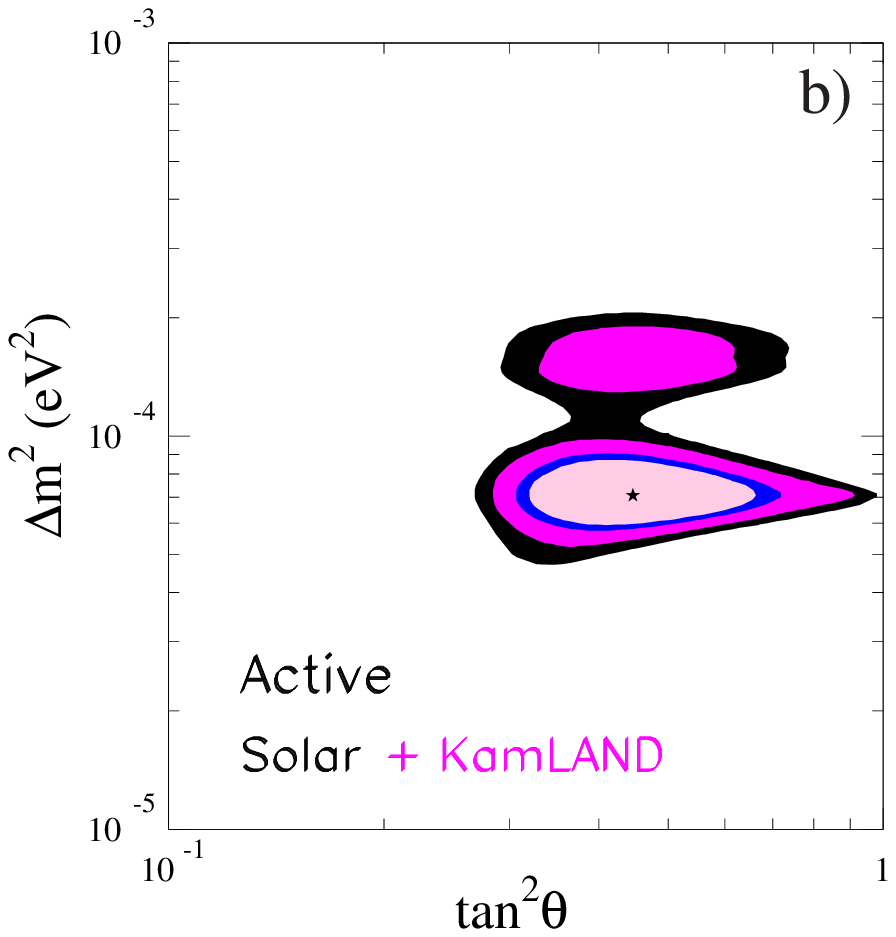,width=0.53\textwidth} 
\caption[e]{(a) 
Allowed regions ($90$\%, $95$\%, $99$\%, and $99.7$\% CL) for
anti-neutrino oscillations as implied by the KamLAND result. Results
from CHOOZ experiment are also included and rule out solutions with
$\Delta m^2\gtrsim 1(0.8)\times 10^{-3} {\rm eV^2}$ 
at $3\sigma$ (99\% CL)\,.
(b)  Allowed oscillation parameters from the combined analysis of  solar
and KamLAND data. }
\label{fig:xiglobal_postkland}
\end{figure}
\begin{figure}
\begin{center}
\mbox{\epsfig{file=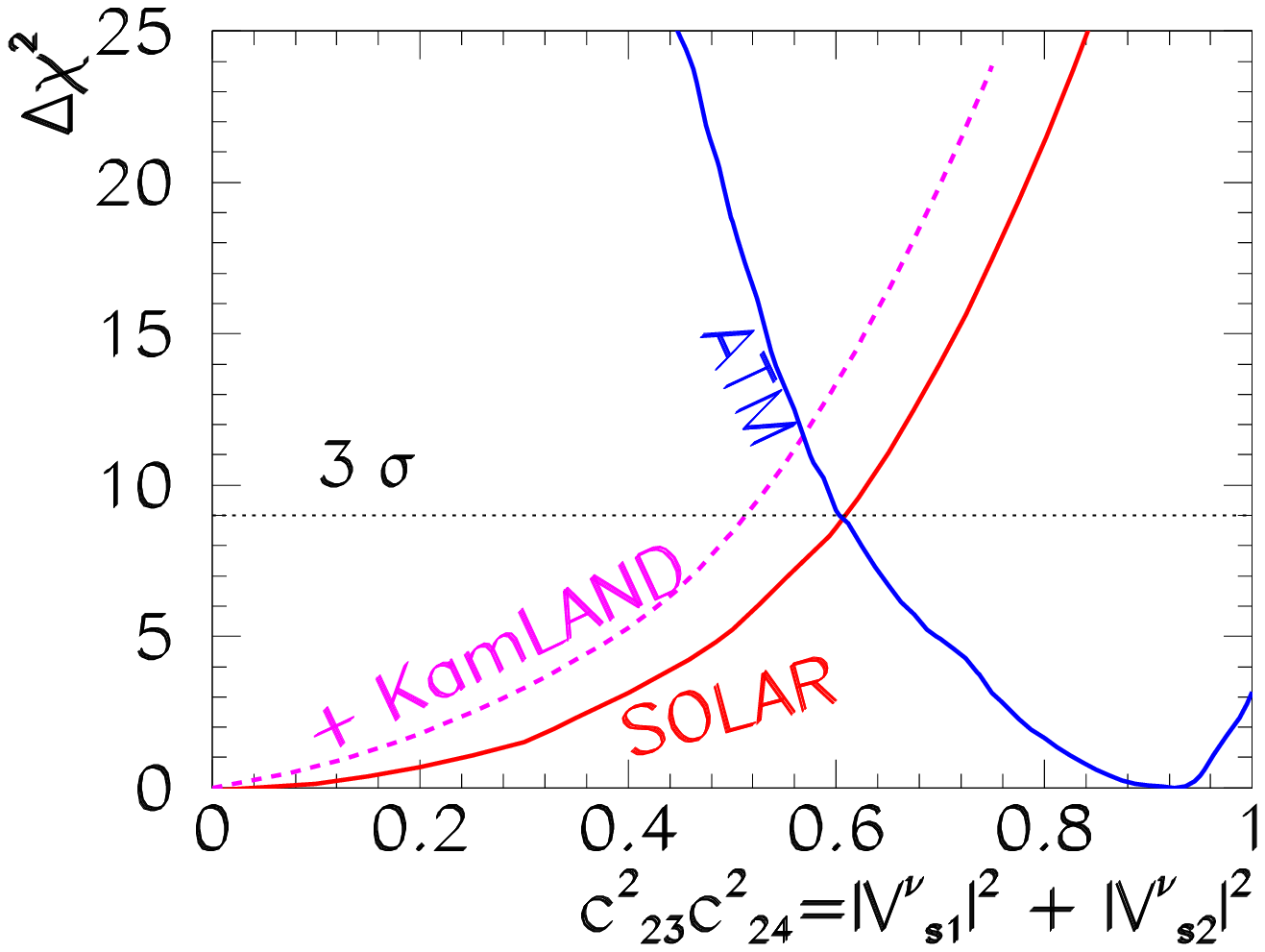,width=0.6\textwidth}}
\end{center}
\caption[ad]{
$\Delta\chi^2$ as a function of the active-sterile admixture,
$|V^\nu_{s1}|^2+|V^\nu_{s2}|^2$, from the analysis of solar,
atmospheric and solar-plus-KamLAND data in (2+2)-schemes.}
\label{fig:chi24_postkland}
\end{figure}

\end{document}